\input harvmac
\input rotate
\input epsf
\input xyv2
\input kramified.defs

\font\teneurm=eurm10 \font\seveneurm=eurm7 \font\fiveeurm=eurm5
\newfam\eurmfam
\textfont\eurmfam=\teneurm \scriptfont\eurmfam=\seveneurm
\scriptscriptfont\eurmfam=\fiveeurm
\def\eurm#1{{\fam\eurmfam\relax#1}}
 \font\teneusm=eusm10 \font\seveneusm=eusm7 \font\fiveeusm=eusm5
\newfam\eusmfam
\textfont\eusmfam=\teneusm \scriptfont\eusmfam=\seveneusm
\scriptscriptfont\eusmfam=\fiveeusm
\def\eusm#1{{\fam\eusmfam\relax#1}}
\font\tencmmib=cmmib10 \skewchar\tencmmib='177
\font\sevencmmib=cmmib7 \skewchar\sevencmmib='177
\font\fivecmmib=cmmib5 \skewchar\fivecmmib='177
\newfam\cmmibfam
\textfont\cmmibfam=\tencmmib \scriptfont\cmmibfam=\sevencmmib
\scriptscriptfont\cmmibfam=\fivecmmib
\def\cmmib#1{{\fam\cmmibfam\relax#1}}
\writedefs

\noblackbox\input rotate
\let\includefigures=\iftrue
\includefigures
\message{If you do not have epsf.tex (to include figures),}
\message{change the option at the top of the tex file.}
\def\figin{\epsfcheck\figin}\def\figins{\epsfcheck\figins}
\def\epsfcheck{\ifx\epsfbox\UnDeFiNeD
\message{(NO epsf.tex, FIGURES WILL BE IGNORED)}
\gdef\figin##1{\vskip2in}\gdef\figins##1{\hskip.5in}
\else\message{(FIGURES WILL BE INCLUDED)}%
\gdef\figin##1{##1}\gdef\figins##1{##1}\fi}
\def\DefWarn#1{}

\def\underarrow#1{\vbox{\ialign{##\crcr$\hfil\displaystyle
{#1}\hfil$\crcr\noalign{\kern1pt\nointerlineskip}\rightarrowfill\crcr}}}
\def\II{\cmmib I}

\def\EUV{\eusm V}

\def\EUW{\eusm W}

\def\EUQ{\eusm Q}

\def\EUB{\eusm B}

\def\EUG{\eusm G}

\def\EUM{\eusm M}
\def\EUN{\eusm N}

\def\EUP{\eusm P}
\def\EUQ{\eusm Q}

\def\TT{{\Bbb{T}}}


\def\figinsert{\goodbreak\midinsert}
\def\ifig#1#2#3{\DefWarn#1\xdef#1{fig.~\the\figno}
\writedef{#1\leftbracket fig.\noexpand~\the\figno}%
\figinsert\figin{\centerline{#3}}\medskip\centerline{\vbox{\baselineskip12pt
\advance\hsize by -1truein\noindent\footnotefont{\bf
Fig.~\the\figno:} #2}}
\bigskip\endinsert\global\advance\figno by1}
\else
\def\ifig#1#2#3{\xdef#1{fig.~\the\figno}
\writedef{#1\leftbracket fig.\noexpand~\the\figno}%
\global\advance\figno by1} \fi \noblackbox
\input amssym.tex
\def\hat{\widehat}
%
\overfullrule=0pt

\def\M{{\EUM}}
\def\MH{{{\EUM}_H}}

\def\tilde{\widetilde}
\def\bar{\overline}

%
\def\ad{{\rm ad}}
\def\EUBB{\cmmib B}

\def\tilde{\widetilde}
\def\bar{\overline}

\def\Tr{{\rm Tr}}

\def\neg{\negthinspace}

\def\M{{\EUM}}
\def\MH{{{\EUM}_H}}
%
\def\tilde{\widetilde}
\def\bar{\overline}
\def\Z{{\Bbb{Z}}}
\def\R{{\Bbb{R}}}
\def\C{{\Bbb{C}}}

\font\zfont = cmss10 
\font\litfont = cmr6 
\def\bigone{\hbox{1\kern -.23em {\rm l}}}
\def\ZZ{\hbox{\zfont Z\kern-.4emZ}}
\def\half{{\litfont {1 \over 2}}}


\def\ZZ{{\Bbb{Z}}}

\font\zfont = cmss10 
\font\litfont = cmr6 
\def\bigone{\hbox{1\kern -.23em {\rm l}}}
\def\ZZ{\hbox{\zfont Z\kern-.4emZ}}
\def\half{{\litfont {1 \over 2}}}

\font\zfont = cmss10 
\font\litfont = cmr6 
\def\bigone{\hbox{1\kern -.23em {\rm l}}}
\def\ZZ{\hbox{\zfont Z\kern-.4emZ}}
\def\half{{\litfont {1 \over 2}}}

\def\M{{\EUM}}
\def\MH{{{\EUM}_H}}
%
\def\tilde{\widetilde}
\def\bar{\overline}

\font\zfont = cmss10 
\font\litfont = cmr6 
\def\bigone{\hbox{1\kern -.23em {\rm l}}}
\def\ZZ{\hbox{\zfont Z\kern-.4emZ}}
\def\half{{\litfont {1 \over 2}}}


\let\includefigures=\iftrue


\def\Tr{{\rm Tr}}

 \noindent
 \Title{} {\vbox{ \centerline{
Gauge Theory, Ramification,}
\bigskip
\centerline{And The Geometric Langlands Program}}}
\smallskip
\centerline{Sergei Gukov}
\smallskip
\centerline{\it{Department of Physics, University of California}}
 \centerline{\it{Santa Barbara, CA 93106}}
\medskip
\centerline{and}
\medskip
\centerline{Edward Witten}
\smallskip
\centerline{\it{School of Natural Sciences, Institute for Advanced
Study}} \centerline{\it{Princeton, New Jersey 08540}}
\bigskip\bigskip
\noindent
In the gauge theory approach to the geometric Langlands
program, ramification can be described in terms of ``surface operators,'' which
are supported on two-dimensional surfaces somewhat as Wilson or 't Hooft operators
are supported on curves.  We describe the relevant surface operators in ${\cal N}=4$ super
Yang-Mills theory, and the parameters they depend on, and analyze how $S$-duality acts
on these parameters.  Then, after compactifying on a Riemann surface, we show that the
hypothesis of $S$-duality for surface operators leads to a natural extension of the geometric
Langlands program for the case of tame ramification.  The construction involves an action of the
affine Weyl group on the cohomology of the moduli space of Higgs bundles with ramification,
and an action of the affine braid group on $A$-branes or $B$-branes on this space.
\vskip .5cm
\noindent\Date{December, 2006} \listtoc \writetoc

\newsec{Introduction}
\seclab\intro

The Langlands program of number theory \ref\lang{R. Langlands,
``Problems In The Theory Of Automorphic Forms,'' in Lect. Notes in
Math. {\bf 170} (Springer-Verlag, 1970), pp. 18-61; ``Where Stands
Functoriality Today?'' in {\it Representation Theory And
Automorphic Forms}, Proc. Symp. Pure Math. {\bf 61} (American
Mathematical Society, 1997), pp. 457-471.}
\nref\bd{
A. Beilinson
and V. Drinfeld, ``Quantization Of Hitchin's Integrable System And
Hecke Eigensheaves,'' preprint (ca. 1995),
http://www.math.uchicago.edu/~arinkin/langlands/.}%
\nref\frthree{E. Frenkel, ``Affine Algebras, Langlands Duality,
And Bethe Ansatz,'' in the Proceedings of the XIth International
Congress of Mathematical Physics, Paris, 1994, ed. D. Iagolnitzer
(International Press, 1995) 606-642, q-alg/9506003.}%
 \nref\bdtwo{A.
Beilinson and V. Drinfeld, {\it Chiral Algebras}, American
Mathematical Society Colloquium Publications {\bf 51}
(American Mathematical Society, 2004).}%
 \nref\frenzvi{E. Frenkel
and D. Ben-Zvi, {\it Vertex Algebras And Algebraic Curves},
Mathematical Surveys And Monographs {\bf 88},
second edition (American Mathematical Society, 2004).}%
\nref\gaitfren{E. Frenkel and D. Gaitsgory, ``Local Geometric
Langlands Correspondence And Affine Kac-Moody Algebras,''
math.RT/0508382.}%
\nref\ofrenkel{E. Frenkel, ``Ramifications Of The Geometric Langlands Program,''
math.QA/0611294.}%
 relates representations of the Galois group of a
number field to automorphic forms (such as ordinary modular forms
of $SL(2,\Bbb{Z})$). It has also a geometric analog, involving
ordinary Riemann surfaces instead of number fields. This geometric
analog has turned out to be intimately related to quantum field
theory. It has been extensively studied using two-dimensional
conformal field theory \refs{\bd-\gaitfren} and more recently via
four-dimensional gauge theory and electric-magnetic duality
\ref\kapwit{
A. Kapustin and E. Witten, ``Electric-Magnetic Duality
And The Geometric Langlands Program,'' hep-th/0604151.}.
Additional explanation and references can be found in the
introduction to \kapwit\ and in a recent review article
\ref\frenkel{E. Frenkel, ``Lectures On The Langlands Program And
Conformal Field Theory,'' arXiv:hep-th/0512172.}.

The simplest version of the geometric Langlands correspondence
involves, on one side, a flat connection on a Riemann surface $C$,
and, on the other side,  a more sophisticated structure known as a
${\cal D}$-module. The problem has a generalization in which one omits
finitely many points $p_1,p_2,\dots,p_n$ from $C$, and begins with
a flat connection on $C'=C\backslash\{p_1,\dots,p_n\}$ that has a
prescribed singularity near the given points.  This situation gives a geometric
analog of what in number theory is called ramification.  Since ramification
is almost inescapable in number theory, the extension of the
geometric Langlands program to the ramified case is an important
part of  the analogy between number theory and geometry. There is
also an important local version of the problem, in which one
focuses on the behavior near a ramification point.

The goal of the present paper is to extend the gauge theory
approach to the geometric Langlands program to cover the ramified
case.  The basic idea is that allowing ramification in the
Langlands program corresponds, in gauge theory, to introducing surface operators,
somewhat analogous to Wilson or 't Hooft operators, but supported on a two-manifold
rather than a one-manifold.
The relevant surface operators are defined by specifying a
certain type of singularity on a codimension two submanifold.
Codimension two singularities in gauge theory of roughly the
relevant type have been considered in various contexts, including
the theory of cosmic strings \ref\rohm{R. M. Rohm, {\it Some Current Problems In
Physics Beyond The Standard Model}, Princeton University Ph.D. Thesis (1985), unpublished.},
Donaldson theory \ref\km{P. B. Kronheimer and T. S. Mrowka,
``Gauge Theory For Embedded Surfaces, I, II,'' Topology {\bf 32}
(1993), 773-826, {\bf 34} (1995) 37-97.}, topological field theory
(see section 5  of \ref\oldwitten{E. Witten, ``Topological Sigma
Models,'' Commun. Math. Phys. {\bf 118} (1988) 411.}),
\nref\presone{J. Preskill and
L. M. Krauss, ``Local Discrete Symmetry
And Quantum Mechanical Hair,''
Nucl. Phys. {\bf B341} (1990) 50-100.}%
\nref\prestwo{M. G. Alford, K.-M. Lee, J. March-Russell, and J. Preskill, ``Quantum Field Theory
Of Non-Abelian Strings And Vortices,'' Nucl. Phys. {\bf B384} (1992) 251-317, hep-th/9112038.}%
\nref\presthree{M. Bucher, K.-M. Lee, and J. Preskill, ``On Detecting Discrete Cheshire Charge,''
Nucl. Phys. {\bf B386} (1992) 27-42, hep-th/9112040.}%
the dynamics of gauge theory and black holes \refs{\presone-\presthree},
\nref\brav{A. Braverman, ``Instanton Counting Via Affine Lie Algebras I:
Equivariant J-Functions of (Affine) Flag Manifolds and Whittaker Vectors,'' math.AG/0401409.}%
\nref\obrav{A. Braverman and P. Etinghof, ``Instanton Counting Via Affine Lie Algebras II:
from Whittaker Vectors to the Seiberg-Witten Prepotential,'' math.AG/0409041.}%
the relation of
instantons to Seiberg-Witten theory and integrable systems \refs{\brav,\obrav},
 and string
theory, where special cases of the operators we consider can be constructed via
intersecting branes \ref\coner{N. R. Constable, J. Erdmenger, Z. Guralnik, and
I. Kirsch, ``Intersecting $D3$-Branes And Holography,''
hep-th/0211222.}, as we will describe in section \stringco.

 \nref\vafa{
M. Bershadsky, A. Johansen, V.
Sadov, and C. Vafa, ``Topological Reduction Of $4-D$ SYM  To $2-D$
Sigma Models,'' Nucl. Phys. {\bf B448} (1995) 166-186,
arXiv:hep-th/9501096.}%
\nref\hm{
J. A. Harvey, G. W. Moore, and A. Strominger, ``Reducing
$S$ Duality To $T$ Duality,'' Phys. Rev. {\bf D52} (1995) 7161-7167.}%
Now we will briefly indicate how these gauge theory singularities are related to the theory of Higgs
bundles.
The gauge theory approach to geometric Langlands is based on
${\cal N}=4$ super Yang-Mills theory twisted and  compactified on
a Riemann surface.  The theory so compactified reduces at low
energies \refs{\vafa,\hm} to a sigma model in which the target
space is a hyper-Kahler manifold that is Hitchin's moduli space of
Higgs bundles
  \ref\hitchin{
  N. Hitchin, ``The Self-Duality Equations On
A Riemann Surface,'' Proc. London Math. Soc. (3) {\bf 55} (1987)
59-126.}.
What codimension two singularities can be incorporated
in this picture? The appropriate singularities, in the basic case
that the flat connection on $C$ has only a simple pole, have been
described and analyzed by Simpson \ref\simpson{
C. Simpson,
``Harmonic Bundles On Noncompact Curves,'' J. Am. Math. Soc. {\bf
3} (1990) 713-770.}. Higgs bundles with a singularity of this type
are what we will call ramified Higgs bundles.  (They are also called Higgs bundles with
parabolic structure.)  The
associated hyper-Kahler moduli spaces have been constructed by
Konno \ref\konno{H. Konno, ``Construction Of The Moduli Space Of
Stable Parabolic Higgs Bundles On A Riemann Surface,'' J. Math.
Soc. Japan {\bf 45} (1993) 253-276.} and their topology clarified by
Nakajima \ref\nakajima{H. Nakajima, ``Hyper-Kahler Structures On
Moduli Spaces Of Parabolic Higgs Bundles On Riemann Surfaces,''
in {\it Moduli Of Vector Bundles}, Lecture Notes in Pure and Appl. Math
{\bf 179} (Marcel Dekker, New York, 1996), ed. M. Maruyama.}.  A corresponding theory for Higgs
bundles with poles of
higher order has been developed by Biquard and Boalch
\ref\biqb{O. Biquard and P. Boalch, ``Wild Non-Abelian Hodge Theory On Curves,''
Compos. Math. {\bf 140} (2004) 179-204.}. The cases of a simple pole or a
pole of higher order correspond respectively to what is called
tame ramification\foot{Sometimes, the term ``tame ramification'' is used more
narrowly to refer to the case of flat bundles with
unipotent monodromy.  We will not make this restriction
and consider connections and Higgs bundles with arbitrary simple poles, as described
in eqn. \zorgo\ and section \postpone.}
 and wild ramification in the context of the
Langlands program.  In this paper, we concentrate on tame ramification.

In sections \locsing, we consider in the context of
four-dimensional gauge theory the singularity corresponding to a
simple pole.  We make a natural proposal for how $S$-duality acts
on the parameters.  We further explore the classical geometry in
section \morecl.  The highlight of this section is the action of the affine
braid group on the cohomology of the
moduli space of ramified Higgs bundles; this action is extended
in section \tame\ to an action of the affine braid group on $A$-branes and $B$-branes.
These phenomena are close cousins of a number of structures found in representation theory,
including the Springer representations of the Weyl group \ref\springer{T. Springer,  ``A
Construction Of Representations Of Weyl Groups,''
Invent. Math. {\bf 44} (1978) 279-293.},
the Kazhdan-Lusztig theory of representations of the affine Hecke algebra
\nref\klo{D. Kazhdan and G. Lusztig, ``Proof Of The Deligne-Langlands Conjecture For Hecke
Algebras,'' Invent. Math. {\bf 87} (1987) 153-215.}%
\nref\kl{D. Kazhdan and G. Lusztig, ``Equivariant $K$-Theory And Representations Of Hecke
Algebras, II'' Invent. Math. {\bf 80} (1985) 209-231.}%
\refs{\klo,\kl}, and the recent extension of this by Bezrukavnikov
to an action of the affine braid group
on the derived category of the Springer resolution \ref\bez{R.
Bezrukavnikov, ``Noncommutative Counterparts Of The Springer Resolution,''
math.RT/0604445.}.  For an exposition of some of this material, see the book by Chriss and Ginzburg
 \ref\chrissginz{N.
Chriss and V. Ginzburg, {\it Representation Theory And Complex Geometry} (Birkhauser, Boston,
1997).}.  Interpreting such results in terms of the parameters of
a hyper-Kahler resolution (as we will
do in the case of Higgs bundle moduli space) was first suggested
by Atiyah and Bielawski \ref\atbiel{M. F. Atiyah and R. Bielawski, ``Nahm's Equations, Configuration
Spaces, and Flag Manifolds,'' Bull. Braz. Math. Soc. (NS) {\bf 33} (2002) 157-76,
math.RT/0110112.} in the context of coadjoint orbits and Slodowy slices.

Based on our duality proposal, we make in section \tame\ a
proposal for what the geometric Langlands program should say in
the case of tame ramification. (A similar proposal has been made
mathematically, based on \bez\ and other results cited in the last
paragraph.  For an exposition, see section 9.4 of the survey
\ofrenkel.  Some particular cases have been studied in detail in
forthcoming work \ref\nadben{D. Ben-Zvi and D. Nadler, to
appear.}.) The extension to wild ramification will be considered
elsewhere \ref\witwild{E. Witten, to appear.}.

\nref\kron{P. Kronheimer, ``A Hyper-Kahlerian Structure
On Coadjoint Orbits Of A Semisimple Complex Group,'' J. London Math. Soc. {\bf 42} (1990)
193-208.}%
\nref\okrontwo{P. Kronheimer, ``Instantons And The Geometry of The Nilpotent Variety,''
J. Diff. Geom. {\bf 32} (1990) 473-490.}%
\nref\biquard{O. Biquard, ``Sur les \'Equations de Nahm et la Structure de Poisson des Alg\'ebres
de Lie Semi-Simples Complexes,'' Math. Ann. {\bf 304} (1996) 253.}%
\nref\kovalev{A. G. Kovalev, ``Nahm's Equations And Complex Adjoint Orbits,'' Quart. J. Math.
Oxford Ser. (2) {\bf 47} (1996) 41-58.}%
In section \thooft, we use gauge theory to describe
the operators (generalizing the 't Hooft/Hecke operators studied in \kapwit) that can act on
branes at a ramification point.  This gives a more down-to-earth approach to some
topics treated in section \tame.
In section \local, we  give a more local description of some aspects of the behavior
at a ramification point in terms of a sigma model whose target is a complex coadjoint
orbit endowed with a hyper-Kahler metric.  Such metrics were
 first constructed for semi-simple or nilpotent orbits
 in
\refs{\kron,\okrontwo} and generalized to arbitrary orbits in \refs{\biquard,\kovalev}.
This and the related analysis in section \onahm\ are the closest we come to analyzing the
local case of geometric Langlands.
Finally, in section \local, we also briefly describe some string
theory constructions of surface operators of the type considered in this paper.

Using conformal field theory, a proposal has been made \gaitfren\
for a unified approach to the geometric Langlands program  allowing poles of any order.
This work is surveyed in \ofrenkel. Unfortunately, we make contact here neither with the use of
conformal field theory nor with this unified statement. We hope,
of course, to eventually understand more.

Some background in group theory is reviewed in Appendix A.  An index of notation appears in Appendix B.
Many facts about
Montonen-Olive duality, Hitchin's moduli space, etc., that are used in the present paper
are explained more fully in \kapwit.

We thank A. Braverman, D. Gaitsgory, E. Frenkel, and D. Kazhdan for
patient and extremely helpful explanations.
We also thank J. Andersen, P. Aspinwall, M. F. Atiyah, D.
Ben-Zvi, R. Bezrukavnikov, R. Bielawski, R. Dijkgraaf, R. Donagi,
N. Hitchin, L. Jeffrey, A. Kapustin,  P. Kronheimer, Y. Laszlo,
H. Nakajima, C. Sorger, and M. Thaddeus, among
others, for a wide variety of helpful comments and advice.
Research of SG was partly supported by
DOE grant DE-FG03-92-ER40701. Research of EW was partly supported by
NSF Grant PHY-0503584.

\newsec{Monodromy And Surface Operators}
\seclab \locsing

\subsec{Definition Of Surface Operators} \subseclab \defn

Our basic technique in this paper will be to study how ${\cal N}=4$
super Yang-Mills theory can be modified along a codimension two submanifold in
spacetime.  Thus, we consider ${\cal N}=4$
super Yang-Mills theory on a four-manifold $M$, but modified
along a two-dimensional submanifold $D$
in such a way that the four-dimensional fields will have singularities along $D$.
The construction is thus an analog for surface operators of the usual construction of
't Hooft operators via codimension three singularities.  For general background see
 \ref\kap{A. Kapustin,
``Wilson-'t Hooft Operators In Four-Dimensional Gauge Theories
And $S$-Duality,'' hep-th/0501015.} or section 6.2 of \kapwit.

Our focus will be on the GL-twisted version of ${\cal N}=4$ super
Yang-Mills theory, which is the basis of the gauge theory approach
to the geometric Langlands program. The gauge group is a compact
Lie group $G$, which we will generally assume to be simple.  The
most important bosonic fields are the gauge field $A$, which is a
connection on a $G$-bundle $E\to M$, and an ${\rm ad}(E)$-valued
one-form field $\phi$.  Our gauge theory conventions are those of
\kapwit. In particular, $A$ and $\phi$ take values in the real Lie
algebra of $G$ (and so in a unitary representation of $G$ are
represented by anti-hermitian matrices), the covariant derivative
is $D=d_A=d+A$, and the holonomy is $P\exp\left(-\int A\right)$.

The fields that will be
singular along $D$ are simply the restrictions of $A$ and $\phi$ to the normal bundle to $D$.
Locally, we can model our four-manifold as $M=D\times D'$, where $D'$ is the fiber to the normal
bundle, and the singularity will be at a point in $D'$, say the origin.

The supersymmetric equations of GL-twisted ${\cal N}=4$
super Yang-Mills theory depend on a parameter $t$, as explained in \kapwit. But upon
reduction to two dimensions this parameter disappears, and we are left with Hitchin's
equations:
\eqn\torgo{\eqalign{F-\phi\wedge \phi& = 0 \cr
                     d_A\phi & = 0 \cr
                     d_A\star \phi & = 0.\cr}} (Here $\star$ is the Hodge star operator.)
Therefore, we will define a surface operator by describing an
isolated singularity that can arise in a solution of Hitchin's
equations.

In fact, for this paper, we will only require the simplest type of
singularity. To begin with, let us take $D'=\Bbb{R}^2$, with
Euclidean coordinates $x^1,x^2$, such that $\star (dx^1)=dx^2$,
$\star(dx^2)=-dx^1$.  We also introduce polar coordinates with
$x^1+ix^2=re^{i\theta}$. We write $\TT$ for a maximal torus of $G$,
and we write $\frak g$, $\frak t$ for the Lie algebras of $G$ and
$\TT$, respectively. To describe a solution of Hitchin's equations
with an isolated singularity at the origin, we pick elements
$\alpha,\beta,\gamma\in \frak t$, and take \eqn\zorgo{\eqalign{A &
=  \alpha \,d\theta \cr
           \phi & = \beta\,{dr\over r}-\gamma\,d\theta.\cr}}
Since $\alpha$, $\beta$, and $\gamma$ commute (as $\frak t$ is
abelian), and the one-forms $d\theta $ and $dr/r$ are closed and
co-closed, Hitchin's equations are obeyed.  We define our surface
operator exactly as one usually defines 't Hooft operators (see
for example \kap\ or section 6.2 of \kapwit): we require that near
$r=0$, $A$ and $\phi$ behave as in \zorgo.  In a global situation,
for a two-dimensional submanifold $D\subset M$, we define such a
surface operator by requiring that at each point in $D$, the
fields in the normal plane look (in some gauge) like \zorgo, with
the specified values of $\alpha,\beta,$ and $\gamma$.

Clearly, we can act with the Weyl group ${\cal W}$ of $G$ on the
trio $(\alpha,\beta,\gamma)$ without changing the theory in an
essential way.  So the surface operator that we have defined
depends on  $(\alpha,\beta,\gamma)$ only modulo a Weyl
transformation.  But there is an additional freedom.  If $u\in
\frak t$ is such that $\exp(2\pi u)=1$, then a gauge
transformation by the $\TT$-valued function
\eqn\zoobmo{(r,\theta)\to \exp(\theta u)}
 shifts $\alpha$ by $u$.
Modulo this transformation, the only invariant of $\alpha$ is the
$\TT$-valued monodromy of the connection $A$ around a circle of
constant $r$; this monodromy is $\exp(-2\pi \alpha)$.  Thus, the
trio $(\alpha,\beta,\gamma)$ takes values in $\TT\times {\frak
t}\times {\frak t}$, modulo the action of ${\cal W}$.

We will often but not always use an additive notation for $\alpha$.  This corresponds to
thinking of $\TT$ as the quotient ${\frak t}/\Lambda$ for some
lattice $\Lambda$.  To identify this lattice, note that if
$\TT={\frak t}/\Lambda$, then $\Lambda=\pi_1(\TT)={\rm
Hom}(U(1),\TT)$.  We call this the cocharacter lattice of $G$,
denoted $\Lambda_{\rm cochar}$.  (See Appendix A for
more information.)  In fact, ${\rm Hom}(U(1),\TT)$
parametrizes $\TT$-valued magnetic charges, or equivalently, by the basic GNO
duality \ref\gno{P. Goddard, J. Nuyts, and D. I. Olive, ``Gauge
Theories And Magnetic Charge,'' Nucl. Phys. {\bf B125} (1977)
1-28.}, electric charge of the dual group $^L\neg G$.  $\Lambda_{\rm cochar}$
is a sublattice of the coweight lattice
$\Lambda_{\rm cowt}$.  Their quotient is the center ${\cal Z}(G)$ of $G$:
\eqn\urgu{{\cal Z}(G)=\Lambda_{\rm cowt}/\Lambda_{\rm cochar}.}

\bigskip\noindent{\it Extensions Of Bundles}

\def\char{{\rm char}}
\def\cochar{{\rm cochar}}
\def\wt{{\rm wt}}
\def\cowt{{\rm cowt}}
\def\rt{{\rm rt}}
\def\cort{{\rm cort}}
Let us try to compute the curvature at the origin of the
singular connection $A=\alpha\,d\theta$.  Since $d(d\theta)=2\pi
\delta_D$ ($\delta_D$ is a two-form delta function supported at
the origin in $D'$ and Poincar\'e dual to $D$), we seem to get
\eqn\yilzo{F=2\pi\alpha \delta_D.} This, however, cannot be a
natural formula, since we are free to shift $\alpha$ by a lattice
vector.  What has gone wrong with the definition of curvature? We
have introduced $ A$ as a connection on a $G$-bundle $E$, but
because of the singularity along $D$, this bundle is only
naturally defined on the complement of $D$ in $M$. It is possible
to pick an extension of $E$ over $D$, but there is no natural
extension. The different possible extensions of $E$ over $D$
correspond to different ways to lift $\alpha$ from $\TT={\frak
t}/\Lambda_{\rm cochar}$, where it naturally takes values, to
$\frak t$.  Once we pick an extension, it makes sense to compute
the curvature at the origin, and the result is \yilzo.

The gauge transformation $(r,\theta)\to \exp(\theta u)$ that
shifts $\alpha$ by a lattice vector, because of its singularity at
the origin, maps one extension of $E$ over $D$ to another.  Though
there is no natural way to extend $E$ over $D$ as a $G$-bundle, we
can do the following.  Near $D$, the structure group of $E$
naturally reduces to the subgroup $\TT$ that commutes with the
singular part of $A$ and $\phi$.  The singular gauge
transformation $\theta\to \exp(\theta u)$ acts trivially on $\TT$,
and hence, though there is no natural $G$-bundle over $D$, there
is a natural $\TT$-bundle over $D$.  We will assume that the
restriction of $A$ to $D$ is a connection on this $\TT$-bundle,
and that the curvature $F$, when restricted to $D$, is likewise $\frak
t$-valued.

Suppose that the gauge group $G$ is not simply-connected; for example, it may be
a group of adjoint type.  Then the gauge transformation $\exp(\theta u)$ may not
lift to a single-valued gauge transformation in the simply-connected cover $\bar G$ of
$G$; rather, under $\theta\to\theta+2\pi$, it is multiplied by an element
$y\in {\cal Z}(\bar G)$, the center of $\bar G$.
  When this is the case, this gauge transformation changes the topology of the bundle
$E$, by shifting the characteristic class $\xi\in H^2(M,\pi_1(G))$ that measures the
obstruction to lifting $E$ to a $\bar G$-bundle.  If we use an additive notation for
${\cal Z}(\bar G)$, the shift is
\eqn\nomo{\xi\to \xi+y[D],} where $[D]$ is the class Poincar\'e dual to $D$.
Thus, gauge theories with different
values of $\xi$ and suitably related values of $\alpha$ are equivalent.

A variant of this is as follows.  Suppose that the gauge group is in fact $\bar G$
or some other form in which the gauge transformation by $\exp(\theta u)$ is not single-valued.
This gauge transformation nevertheless makes sense locally
as a symmetry of ${\cal N}=4$ super Yang-Mills theory (in which
all fields are in the adjoint representation).  If the global topology is such that
a gauge transformation that looks locally like $\exp(\theta u)$ near $D$
can be extended globally over $M$,
then ${\cal N}=4$ super Yang-Mills is invariant to $\alpha\to\alpha+u$.  Here $u$
may be any  element of $\Lambda_{\rm cowt}$ for which $y[D]=0$.  More generally, if $D$
is a union of disjoint components $D_i$ near which we make gauge transformations by
$\exp(\theta_i u_i)$ with $y_i=\exp(2\pi u_i)$, then the condition is
\eqn\otimbo{\sum_iy_i[D_i]=0.}

\bigskip\noindent{\it Non-Trivial Normal Bundle}

In motivating this construction, we began with the special case of a
product $M=D\times D'$.  However, more generally, for an arbitrary
embedded two-manifold $D\subset M$, we  consider gauge fields with a
singularity like \zorgo\ in each normal plane. For simplicity, in
this paper, we only consider the case that  $M$ and $D$ are
oriented. $D$ may have a non-trivial normal bundle, and hence a
nonzero self-intersection number $D\cap D$, which can be
characterized as\foot{For this and some other assertions made
momentarily, see the description of the Thom class in \ref\bott{R.
Bott and L. Tu, {\it Differential Forms In Algebraic Topology}
(Springer, New York, 1982).}.} \eqn\yrol{D\cap D
=\int_M\delta_D\wedge \delta_D.} When the self-intersection number
is nonzero, it is not possible globally for $\alpha$ to have
arbitrary values.  We explain this first for $G=U(1)$.  A connection
$A$ which has a singularity $A=\alpha d\theta$ near $D$ will have
the property $\int_D F/2\pi = \alpha D\cap D~{\rm mod}~\Z$. Since
the integrated first Chern class $\int_D F/2\pi$ is always an
integer, it will always be that
\eqn\hobo{\alpha D\cap D\in \Z.} For any $G$, the generalization of
this is simply that the same statement holds in any $U(1)$ subgroup
of $\TT$.  So if $\alpha\to f(\alpha)$ is any real-valued linear
function on $\frak t$ that takes integer values on the lattice
$\Lambda_{\rm cochar}$, then \eqn\nobo{f(\alpha) D\cap D\in\Z.}

It is also true that if  $D\cap D\not= 0$, the twisted gauge
transformation that is defined in the normal plane in \zoobmo\
cannot always be defined globally along $D$. Only those gauge
transformations that shift $\alpha$ in a way compatible with
\nobo\ can actually be defined globally.

Singularities along surfaces with $D\cap D\not=0$ are important in
four-manifold theory \km, but will be less important in our
applications here, since  the geometric Langlands program, in its
usual form, deals with a situation in which $D\cap D=0$.

\subsec{Geometric Interpretation Of Parameters}
\subseclab\geomint

We have defined a surface operator supported on a two-manifold
$D\subset M$ by requiring that the fields behave near $D$ as in
\zorgo.  In general, quantum mechanically, Hitchin's equations or
even the second order classical field equations of the theory will
not necessarily be obeyed; the definition of the surface operator
only requires that they are obeyed near $D$. However, to understand
better the meaning of the parameters $\alpha,\beta,\gamma$ in
classical geometry, let us consider a situation in which we {\it do}
want to solve Hitchin's equations on a Riemann surface $C$ (which
corresponds to $D'$ in the above analysis), with an isolated
singularity of the above-described type near some point $p\in C$.
(We will here explain only the facts about the classical geometry
that are needed to motivate the duality conjecture of section
\duality. We will reconsider the classical geometry in much more
detail in section \morecl.)

Solutions of Hitchin's equations with the type of singularity
described in \zorgo\ have been analyzed in \refs{\simpson -
\nakajima}. Just like smooth solutions of Hitchin's equations, the
moduli space of such solutions is a hyper-Kahler manifold
$\EUM_H$, which we will call the moduli space of ramified Higgs bundles, also known
as the moduli space of Higgs bundles with parabolic structure.
 (We refer to it as $\EUM_H(G)$, $\EUM_H(C)$,
$\EUM_H(\alpha,\beta,\gamma;p)$, etc., if we want to make explicit
the gauge group, the Riemann surface, or the nature and location
of a  singularity.)

Because of the hyper-Kahler structure, solutions of Hitchin's
equations can be viewed in different ways.  {}From the standpoint of
one complex structure, usually called $I$,  a solution of
Hitchin's equations on a Riemann surface $C$ describes a Higgs
bundle, that is a pair $(E,\varphi)$, where $E$ is a holomorphic
$G$-bundle and $\varphi$ is a holomorphic section of $K_C\otimes
{\rm ad}(E)$. ($K_C$ denotes the canonical bundle of $C$.)  The
Higgs bundle is constructed as follows starting with
 a solution $(A,\phi)$ of Hitchin's equations.
One interprets the $(0,1)$ part of the gauge-covariant exterior derivative
$d_A=d+A$ as a $\bar\partial_A$ operator
that gives the
bundle $E$ a holomorphic structure.  And one defines $\varphi$ as the $(1,0)$ part of
$\phi$; that is, one decomposes $\phi$ as $\varphi+\bar\varphi$, where $\varphi $ is
of type $(1,0$) and $\bar\varphi$ is of type $(0,1)$.  Then Hitchin's equations imply
that $\varphi$ is holomorphic, and the pair $(E,\varphi)$ is a
Higgs bundle.

In the present context, setting $z=x^1+ix^2$, we find from \zorgo\
that \eqn\difo{\varphi = {1\over 2}(\beta+i\gamma){dz\over z}.}
Thus, from the point of view of complex structure $I$, the surface
operator introduces in the Higgs field a pole with
residue\foot{This statement holds if $\beta+i\gamma$ is a
``regular'' element of the Lie algebra ${\frak g}_\C$ (the subalgebra of ${\frak g}_\C$ that
commutes with it is precisely ${\frak t}_\C$).  What
happens otherwise is described in section \postpone.}
$(1/2)(\beta+i\gamma)$. The characteristic polynomial of $\varphi$
varies holomorphically in complex structure $I$, and in particular
this is so for the conjugacy class of the pole in $\varphi$.
\difo\ shows that the conjugacy class of this pole is holomorphic
in $\beta+i\gamma$, and independent of $\alpha$.  This is part of
a more general statement;
 the complex structure $I$ varies holomorphically with $\beta+i\gamma$, and is independent
 of $\alpha$.   On the other hand, the corresponding Kahler form $\omega_I$ that
is of type $(1,1)$ in complex structure $I$ has a cohomology class
that is independent of $\beta$ and $\gamma$ and is a linear
function of $\alpha$.  These statements, along with some similar
ones below, follow from the construction of the moduli space as a
hyper-Kahler quotient \refs{\konno,\nakajima}, as we will explain
in section \cpxview.  So if we look at $\MH$ from the vantage
point of complex structure $I$, then $\beta+i\gamma$ is a complex
parameter (on which $I$ depends holomorphically) and $\alpha$ is a
Kahler parameter.

In fact, though we will not require the details in this paper,
$\alpha$ has a natural meaning in pure algebraic geometry
(without mentioning Kahler metrics), but this meaning is a little
subtle. $\MH$ in complex structure $I$ parametrizes pairs
$(E,\varphi)$, where $\varphi$ has a simple pole whose conjugacy
class is determined by $\beta$ and $\gamma$ (as in \difo) and
moreover the pair $(E,\varphi)$ is ``stable.'' The appropriate
notion of stability \simpson\ depends on $\alpha$.

In complex structure $J$, the natural complex variable is the
$G_{\Bbb{C}}$-valued connection $\CA=A+i\phi$, which is flat
according to Hitchin's equations.  The monodromy of
$\CA=(\alpha-i\gamma)d\theta$ around the singularity at $p$ is
\eqn\nutella{U=\exp(-2\pi(\alpha-i\gamma)).} It depends
holomorphically on $\gamma+i\alpha$, and is independent of
$\beta$. Indeed, in complex structure $J$, $\gamma+i\alpha$ is a
complex parameter and $\beta$ is a Kahler parameter.

In complex structure $J$, $\MH$ parametrizes flat
$G_{\Bbb{C}}$-bundles over $C$ with a monodromy around the point
$p$ that is in the conjugacy class\foot{As in the previous
footnote, this statement holds if $U$ is regular; we postpone a
discussion of the more general case to section \postpone.}  containing $U$.
Like $\alpha$ in complex structure $I$, $\beta$ can be interpreted
in complex structure $J$ in purely holomorphic terms (without
mentioning Kahler metrics), but this interpretation is a little
elusive (and will play only a slight role in the present paper).
According to \simpson, $\beta$ determines in complex structure $J$
the weights of a monodromy-invariant weighted filtration
  of the flat bundle  over
$C\backslash p$ (that is, $C$ with the point $p$ omitted)
whose connection is $\CA$.

Finally, in complex structure $K=IJ$, the natural complex variable
is the $G_{\Bbb{C}}$-valued connection  $\tilde\CA=A+i\star
\phi$.  It is again flat by virtue of Hitchin's equations. Its
monodromy around the singularity at $p$ is
$\exp(-2\pi(\alpha+i\beta))$.  In complex structure $K$,
$\alpha+i\beta$ is a complex structure parameter and $\gamma$ is a
Kahler parameter.  The interpretation of $\gamma$ in complex
structure $K$ is just like the interpretation of $\beta$ in
complex structure $J$.

\bigskip
\centerline{\vbox{\offinterlineskip
\def\tablerule{\noalign{\hrule}}
\halign to 4.5truein{\tabskip=1em plus
2em#\hfil&\vrule height12pt depth5pt#&#\hfil&\vrule height12pt depth5pt#&#\hfil\tabskip=0pt\cr
\hfil Model\hfil&&\hfil Complex Modulus\hfil&&\hfil Kahler
Modulus\hfil\cr
 \tablerule
$~~~I$&&$~~~~~~~~\beta+i\gamma$ && $~~~~\alpha$
\cr  $~~~J$&&$~~~~~~~~\gamma+i\alpha$ &&
$~~~~\beta$\cr  $~~~K$ && $~~~~~~~~\alpha+i\beta$
&& $~~~~\gamma$ \cr }}}\bigskip \centerline{ \vbox{\hsize=5.1truein\baselineskip=12pt
\noindent Table 1. Complex and Kahler moduli for $\M_H$ in complex
structures $I$, $J$, and $K$.  Complex structure $I$, for
example,  depends holomorphically on
$\beta+i\gamma$, while the corresponding Kahler structure depends on $\alpha$.
}}\bigskip

These statements are summarized in  Table 1. In the table, one sees
a cyclic symmetry under permutations of $I,J,K$ together with
$\alpha,\beta,\gamma$.   (This cyclic symmetry is part of an $SO(3)$
symmetry that appears in a closely related context; see section
\onahm.) If $G=U(N)$, then $(\alpha,\beta,\gamma)$ are called
$(\alpha,2b,2c)$ in the table on p. 720 of \simpson. We have
adjusted a factor of 2 to get the cyclic symmetry.

\subsec{Theta Angles}
\subseclab\thetang

Quantum mechanically, in addition to $\alpha,\beta$, and $\gamma$,
an additional parameter is present. One may guess that this will
occur, because so far we have described in each complex structure
only a real Kahler parameter (listed in the last column of Table
1), but in supersymmetric theories, the Kahler parameters are
usually complexified.

In explaining this, let us assume for the moment that the trio
$(\alpha, \beta,\gamma)$ is regular, meaning that the subgroup of
$G$ that leaves this trio invariant is precisely $\TT$.  Requiring
the behavior \zorgo\ in each normal plane to a two-manifold
$D\subset M$ means, in particular, that along $D$ we are given a
reduction of the structure group of $E$ from $G$ to $\TT$.
Therefore, along $D$ we are doing abelian gauge theory, with gauge
group $\TT$.  (We explained at the end of section \defn\ that along $D$, there is
a natural $\TT$-bundle, though there is no natural $G$-bundle.)

In abelian gauge theory in two dimensions, an important role is
played by the ``theta-angle.'' For example, let the gauge group be
$U(1)$. A $U(1)$-bundle  ${\cal L}\to D$ is classified
topologically by its degree $d=\int_D c_1({\cal L})$, where
$c_1({\cal L})$ is the first Chern class.\foot{For simplicity, we
assume $D$ to be closed.  Otherwise, in defining the quantum field
theory, one needs a suitable boundary condition on the boundary of
$D$ (or at infinity).  With some care, one can then give a
suitable definition of $c_1({\cal L})$ and of $\theta$.} The theta
angle enters the theory by a phase $\exp(i\theta d)$ that is
included as an extra factor in the path integral. Here $\theta$
takes values in $\Bbb{R}/2\pi\Bbb{Z}\cong U(1)$.

Now let us return to our problem, involving surface operators in nonabelian gauge theory.
Suppose that $G$ has rank $r$.  Then its maximal torus
$\TT$ is isomorphic to $U(1)^r$, and a two-dimensional
gauge theory with gauge group $G$ will have $r$ theta-angles, taking values in an
$r$-dimensional torus.  Let us see exactly which torus this is.  A $\TT$-bundle over
a two-manifold $D$ can be constructed uniquely (from a topological point of view)
by starting with a $U(1)$ bundle of degree 1 and then mapping this to $\TT$ via some
homomorphism $\rho:U(1)\to \TT$.  So
$\TT$-bundles over $D$ are classified topologically by a characteristic class $\eurm m$
that takes values in
 the lattice $\Lambda_{\rm cochar}={\rm Hom}(U(1),\TT)$.  Therefore, the theta-angle of $\TT$ gauge
theory is a  homomorphism $\eta:\Lambda_{\rm cochar}\to U(1)$.

In other words, $\eta$ takes values in ${\rm Hom}(\Lambda_{\rm cochar},U(1))$.  We claim that
 \eqn\yrot{{\rm Hom}(\Lambda_{\rm cochar},U(1))={}^L\negthinspace \TT,}
 where $^L\neg \TT$ is the maximal torus
of the dual group $^L\neg G$.  In fact, by Pontryagin
duality,\foot{Pontryagin duality says that if $W$ is a locally
compact abelian group and $V={\rm Hom}(W,U(1))$, then $W={\rm
Hom}(V,U(1))$.  In our application, $W=\Lambda_{\rm cochar}={\rm Hom}(U(1),\TT)$, and
$V={}^L\neg \TT$.} since $\Lambda_{\rm cochar}={\rm Hom}(U(1),\TT)$,
\yrot\ is equivalent to \eqn\byrot{{\rm Hom}(U(1),\TT)={\rm
Hom}(^L\neg \TT,U(1)).} But this is a standard characterization of
the relation between the group and the dual group.  Indeed, the left
hand side classifies magnetic charges of $G$, and the right hand
side classifies electric charges of $^L\neg G$.  The equality of the
two is the basic GNO duality \gno.

Just as $\TT={\frak t}/\Lambda_{\rm cochar}$, we have $^L\neg
\TT={}^L{\frak t}/\Lambda_{\rm char}$, with ${}^L\frak t$ the Lie algebra of ${}^L\neg \Bbb{T}$, and
$\Lambda_{\rm char}={\rm
Hom}(T,U(1))$ the lattice of electric charges of $G$.  ${}^L\frak t$ coincides with $\frak t^\vee$,
the dual of $\frak t$.  We frequently use an
additive notation for $\eta$, thinking of it as an element of
${\frak t}^\vee/\Lambda_{\rm char}$.

\subsec{Electric-Magnetic Duality}
\subseclab\duality

${\cal N}=4$ super Yang-Mills theory has a large discrete group
$\Gamma$ of duality symmetries.  Optimistically assuming that the class of surface operators that we have
described
is mapped to itself by $\Gamma$, let us determine how $\Gamma$ acts on the parameters
$(\alpha,\beta,\gamma,\eta)$.

First we consider the fundamental electric-magnetic duality $S$.
It acts on the coupling parameter $\tau=\theta/2\pi+4\pi i/g^2$ of
the gauge theory by $S:\tau\to -1/n_{\frak g}\tau$, where
$n_{\frak g}$ is 1 for simply-laced $G$ and otherwise is 2 or 3
(the ratio of the length squared of the long and short roots of
$G$). It also maps $G$ to $^L\neg G$.  How does it transform the parameters of a surface operator?

\bigskip\noindent{\it Transformation Of $\beta $ and $\gamma$}

$S$ acts on $\beta$ and $\gamma$ in a very simple way,  because $\beta$ and $\gamma$ determine a pole
in the characteristic polynomial of the Higgs field, which transforms very simply under duality
(see \kapwit, section 5.4; however, we will adopt a different normalization from the one used there).

\nref\dorey{N.~Dorey, C.~Fraser, T.~J.~Hollowood and
M.~A.~C.~Kneipp, ``$S$-Duality in ${\EUN}=4$ Supersymmetric Gauge
Theories With Arbitrary Gauge Group,'' Phys.\ Lett.\ B {\bf 383},
422-428 (1996), arXiv:hep-th/9605069. }%
\nref\kapsei{P. Argyres, A. Kapustin, N. Seiberg, ``On $S$-Duality For Non-Simply-Laced Gauge
Groups,'' arXiv:hep-th/0603048.}%
Since $\beta$ and $\gamma$ take values in $\frak t$, while $^L\neg\beta$ and $^L\neg\gamma$ take
values in $^L\frak t$, the comparison between them depends on a choice of map from $\frak t$
to $^L\frak t$.  The vector spaces $\frak t$ and $^L\frak t$ are dual (so that we also denote $^L\frak t$
as $\frak t^\vee$), and acted on by the same Weyl
group.
Any choice of a Weyl-invariant metric on $\frak t$
gives a Weyl-invariant
identification between them.

To prepare the ground for the application to geometric Langlands in
section 4, we will describe in some detail the identification we will use.\foot{For gauge groups
$G_2$ or $F_4$, the  convention we are about to
describe differs from the most common one in the physics literature
\refs{\dorey,\kapsei}.  The relation between the two approaches is
explained at the end of Appendix A.  See also, for example, \donpan.}  First of all,
it is convenient to introduce an invariant quadratic form on $\frak g$, normalized so that a short
coroot is of length squared 2.  We write this quadratic form as $(x,y)=-\Tr\,xy$, for $x,y\in\frak g$.
(The notation is
motivated by the fact that for $G=SU(N)$, $\Tr$ is the trace in the $N$-dimensional representation.)
Similarly, we introduce a quadratic form on the Lie algebra $^L\neg\frak g$ of $^L\neg G$, normalized
so that a short root of $G$ has length squared 2.  We write this as $({}^L\neg x,{}^L\neg y)=
-{}^L\neg\Tr\,^L\neg x{}^L\neg y$.  The quadratic form on $\frak g$, when restricted to $\frak t$,
gives a Weyl-invariant map from $x\in \frak t$ to $x^*\in {}^L\frak t$ (such that $x^*(y)=-\Tr\,xy$),
and likewise, the
quadratic form on $^L\neg\frak g$, restricted to $^L\frak t$, gives a Weyl-invariant map
from $^L\neg x\in {}^L\frak t$ to $^L\neg x^*\in \frak t$.  As explained in Appendix A, the composition
of the two maps is multiplication by $n_{\frak g}$, that is,
\eqn\zok{(x^*)^*=n_{\frak g}x,} or equivalently
\eqn\trombo{^L\neg\Tr\, x^*y^*=n_{\frak g}\Tr\,xy}
for any $x,y\in \frak g$.  This relation is symmetric between $G$ and $^L\neg G$.

Now as in eqn. (2.8) of \kapwit, we normalize the scalar fields $\phi$ of ${\cal N}=4$ super Yang-Mills
theory with gauge group $G$ so that their kinetic energy in Lorentz signature is
\eqn\zrombo{ {{\rm Im}\,\tau\over 4\pi}\int d^4x\,\Tr\,D_\mu\phi D^\mu\phi, }
where ${\rm Im}\,\tau=4\pi/e^2$.  Likewise, for the $S$-dual theory with
gauge group $^L\neg G$, we normalize the scalars
$^L\neg \phi$ so that their kinetic energy is
\eqn\brombo{ {{\rm Im}\,{}^L\neg\tau\over 4\pi}\int d^4x\,\Tr\,D_\mu{}^L\neg\phi D^\mu{}^L\neg\phi, }
with
\eqn\ombo{^L\neg \tau = -{1\over n_{\frak g}\tau}.}
In general, there is no local identification between $\phi$ and $^L\neg \phi$ (but only between
gauge-invariant local operators constructed from these fields).  However, for the sake of finding
how $\beta$ and $\gamma$ transform under duality, we can abelianize the problem, as in section 5.1
of \kapwit, and go to a vacuum in which $G$ or $^L\neg G$ is spontaneously broken to its maximal
torus by expectation values of scalar fields.  In such an abelian vacuum, the light scalar
fields takes values in $\frak t$ or $^L\neg \frak t$, and duality acts on them simply by
a  linear transformation that we can choose to be Weyl-invariant (and hence a multiple of
the operation $\phi\to \phi^*$).  This transformation must preserve
the kinetic energy, so
\eqn\doobus{{{\rm Im}\,\tau\over 4\pi}\int d^4x\,\Tr\,D_\mu\phi D^\mu\phi=
 {{\rm Im}\,{}^L\neg\tau\over 4\pi}\int d^4x\,\Tr\,D_\mu{}^L\neg\phi D^\mu{}^L\neg\phi. }
Together with \ombo\ and \trombo, this implies that in an abelian vacuum
the relation between $\phi$ and $^L\neg \phi$ is
\eqn\oobus{^L\neg \phi= |\tau|\,\phi^*,}
a relation that is completely symmetric between $G$ and $^L\neg G$, as one can verify using \zok\ and
\ombo.  Since $\beta$ and $\gamma$ parametrize singularities of $\phi$, and likewise $^L\neg\beta$ and
$^L\neg\gamma$ parametrize singularities of ${}^L\neg\phi$, these parameters transform in the same
way, that is $^L\neg\beta=|\tau|\beta^*$, $^L\neg\gamma=|\tau|\gamma^*$, or more briefly
\eqn\noobus{({}^L\neg\beta,{}^L\neg\gamma)=|\tau|(\beta^*,\gamma^*).}
Since the parameters characterize the operator, not the vacuum, this formula must hold in general,
not just in an abelian vacuum.

The basic case of the geometric Langlands program is conveniently studied by setting ${\rm Re}\,\tau=0$,
in which case \noobus\ can be more conveniently written
\eqn\zoobus{({}^L\neg\beta,{}^L\neg\gamma)=({\rm Im}\,\tau)\,(\beta^*,\gamma^*).}
We can also invert this relation:
\eqn\oobus{(\beta,\gamma)=({\rm Im}\,^L\neg\tau)\,({}^L\neg\beta^*,{}^L\neg\gamma^*).}
The two formulas are compatible, since $(x^*)^*=n_{\frak g}x$ and (for imaginary $\tau$)
${\rm Im}\,^L\neg\tau=1/n_{\frak g}\,{\rm Im}\,\tau$.

$\beta$ and $\gamma$ are manifestly unaffected by a shift in the theta-angle,
which as we discuss later gives the second generator $T:\tau\to\tau+1$ of the duality
group $\Gamma$.  So their full transformation under $\Gamma$ is determined by \noobus.  The result
can be described particularly simply if $G$ is simply-laced, in which case, as explained in Appendix A,
the difference between $(\beta,\gamma)$ and $(\beta^*,\gamma^*)$ is inessential.
In that case, \noobus\ together with invariance of $(\beta,\gamma)$ under $\tau\to\tau+1$ implies that
for a general element
$\left(\matrix{a&b\cr c&d}\right)\in\Gamma\cong SL(2,\Z)$, the transformation is $(\beta,\gamma)\to
|c\tau+d|(\beta,\gamma)$.  A similar result can be written for any $G$.
If one restricts to an index 2 subgroup of $\Gamma$ that maps $G$ to itself (rather than $^L\neg G$),
\noobus\ implies that the pair $(\beta,\gamma)$ transforms by rescaling by a positive factor.  Other
elements of $\Gamma$ map $(\beta,\gamma)$ to a positive multiple of $(\beta^*,\gamma^*)$.

\bigskip\noindent{\it Transformation Of $\alpha$ and $\eta$}

The other parameters of a surface operator are $\alpha\in \TT$ and $\eta\in {}^L
\TT$. Since $S$ exchanges $G$ and $^LG$, it exchanges $\TT$ and
$^L\TT$, strongly suggesting that it exchanges $\alpha$ and
$\eta$.  This is much more interesting than the relatively trivial transformation of $\beta$ and
$\gamma$.  It will be our basic hypothesis in the present paper.

In fact, $S^2$ is a central element of the duality group $\Gamma$.
It acts trivially on $\tau$, and acts on other fields and
parameters by charge conjugation.  So $S^2$ maps $(\alpha,\eta)$
to $(-\alpha,-\eta)$. Together with the fact that $S$ exchanges
$\TT$ and $^L\neg\TT$, it follows that up to sign, $S$ must act by
\eqn\hylgo{S:(\alpha,\eta)\to (\eta,-\alpha).}

Since the duality group $\Gamma$ contains the central element
$S^2$ that reverses the sign of the pair $(\alpha,\eta)$, the
overall sign in \hylgo\ depends on precisely how we lift $S$ from
a symmetry of the upper half $\tau$-plane to an element of
$\Gamma$. We specify our lifting in eqn. \olpo\ below.

\bigskip\noindent{\it The Abelian Case}

For nonabelian $G$, we cannot prove \hylgo, but regard it as the
natural extension to surface operators of the Montonen-Olive
duality conjecture.\foot{In some special cases, \hylgo\ follows from broader
string theory duality conjectures, in view of the constructions explained in
section \stringco.}  In the case $G=U(1)$, however, we can
directly demonstrate that $S$ does act in this fashion, as we will
now explain.  We will follow the approach to abelian $S$-duality
in \ref\witugh{E. Witten, ``On $S$-Duality In Abelian Gauge
Theory,'' Selecta Math. {\bf 1} (1995) 383, hep-th/9505186.}
(which in turn was modeled on a similar approach to $T$-duality in
two dimensions  \nref\buscher{T. H. Buscher, ``Path Integral
Derivation Of Quantum Duality In Nonlinear Sigma
Models,'' Phys. Lett. {\bf 201B} (1988) 466.}%
\nref\everlinde{M. Rocek and E. Verlinde, ``Duality, Quotients,
And Currents,'' Nucl. Phys. {\bf B373} (1992) 630,
hep-th/9110053.}%
\refs{\buscher,\everlinde}).

In abelian gauge theory, the gauge field is locally a real one-form
$A$ (which we can think of as a connection on a principal
$U(1)$-bundle\foot{The Lie algebra of $U(1)$ is real, so a connection on a principal
$U(1)$-bundle is naturally represented locally by a real one-form.  (By contrast, if we view
$A$ as a connection on a unitary complex line bundle ${\cal L}$, coming from a unitary representation of
$U(1)$, we would represent it locally by an imaginary one-form.)}
${\cal R}\to M$) with curvature $F=dA$.  We take the
action to be\foot{To agree better with conventions used in much of
the physics literature as well as \kapwit\ and the present paper
(but at the cost of some tension with conventions usually used in
Donaldson theory), we have reversed the sign of $\theta$ relative to
\witugh. This has the effect of transforming $\tau\to -\bar\tau$.}
\eqn\dolgo{\eqalign{\II&={1\over 8\pi}\int_Md^4x\sqrt
h\left({4\pi\over g^2}F_{mn}F^{mn}-{i\theta\over 2\pi}{1\over
2}\epsilon_{mnpq}F^{mn}F^{pq}\right)\cr &=-{i\over
8\pi}\int_Md^4x\sqrt h\left(\tau F^+_{mn}F^{+\,mn}-\bar\tau
F_{mn}^-F^{-\,mn}\right),\cr}} where $h$ is a metric on $M$,
$\tau=\theta/2\pi+4\pi i/g^2$, $\epsilon_{mnpq}$ is the Levi-Civita
antisymmetric tensor, and $F^\pm={1\over 2}(F\pm \star F)$ are the
selfdual and anti-selfdual projections of $F$. As explained in
\witugh, $\II$ is invariant mod $2\pi i\Z$ under $\tau\to\tau+2$ for
any closed four-manifold $M$, and under $\tau\to\tau+1$ if $M$ is a
spin manifold. Quantum theory depends on the action only mod $2\pi
i\Z$ (since the action enters the path integral via a factor
$\exp(-\II)$), so the quantum theory possesses the symmetry
$\tau\to\tau+1$ or $\tau\to\tau+2$, depending on $M$.

\def\CK{\cmmib k}
\def\cv{\cmmib v}
\def\cb{ b}
Our interest here is in the more subtle symmetry $\tau\to -1/\tau$.
First we will review how to see this symmetry in the absence of
surface operators.   We add a two-form field $\CK$ (called $G$ in
\witugh) which we assume to be invariant under the usual abelian
gauge symmetry $A\to A-d\epsilon$ (for a zero-form $\epsilon$).  But
we ask for the extended gauge symmetry \eqn\xilo{\eqalign{A&\to
A+\cb\cr
                                       \CK&\to \CK+d\cb,\cr}}
where $\cb$ is any connection on a principal $U(1)$-bundle ${\cal
T}$, and $d\cb$ is its curvature.  (The ordinary Maxwell gauge
symmetry is a special case of this with $\cb=-d\epsilon$.)  If $A$
is a connection on a principal $U(1)$-bundle ${\cal R}$, then
$A+\cb$ is a connection on ${\cal R}\otimes {\cal T}$, so to get
invariance under \xilo, we will need to sum over all possible
choices of ${\cal R}$. A transformation \xilo\ can shift the periods
of $\CK$ by integer multiples of $2\pi$; thus, if $D\subset M$ is a
two-cycle, we can have \eqn\molko{\int_D\CK\to\int_D\CK+2\pi m,
~~m\in\Z.}

\def\FF{{\cal F}}
\def\CV{\cmmib V}
An obvious way to find a Lagrangian with the invariance \xilo\ is to
set $\FF=F-\CK$ and replace $F$ everywhere in the Maxwell Lagrangian
by $\FF$.  But the resulting theory is trivial, and certainly not
equivalent to Maxwell theory.  To get something interesting, we
introduce another field $\cv$ which is a connection on a principal
$U(1)$-bundle $\tilde{\cal R}$, with curvature $\CV=d\cv$. We add to
the action a term \eqn\hillox{\tilde \II={i\over 8\pi}\int_M d^4x
\sqrt h\epsilon^{mnpq}\CV_{mn}\CK_{pq}={i\over 2\pi}\int_M \CV\wedge
\CK.} To check the symmetry \xilo, note that if ${\cal T}$ is
topologically trivial, then the connection $\cb$ is globally-defined
as a one-form, and an integration by parts shows that $\tilde\II$ is
invariant under $\CK\to \CK+d\cb$. We have chosen the coefficient in
\hillox\ so that $\tilde\II$ is invariant mod $2\pi i\Z$ even if
${\cal T}$ is topologically non-trivial.

We now define an extended theory with field variables $\cv,\CK,A$
and action \eqn\zoromp{\hat\II(\cv,\CK,A)={i\over 8\pi}\int
d^4x\sqrt h\left(\epsilon^{mnpq}\CV_{mn}\CK_{pq}-\tau
\FF^+_{mn}\FF^{+\,mn}+\bar\tau \FF_{mn}^-\FF^{-\,mn}\right).} {}From
what has been said, the invariance of $\hat\II$ under \xilo\ mod
$2\pi i\Z$ should be clear. The proof of $S$-duality of abelian
gauge theory  comes by comparing two ways of studying this extended
theory.

One approach is to perform first the path integral over $\cv$.
The part of the integral that depends on $\cv$ is \eqn\linos{\sum_{\tilde{\cal R}}\int
D\cv\,\exp\left(-{i\over 2\pi}\int_M \CV\wedge \CK\right)} where
$\CV$ is the curvature of $\cv$. We must sum over $U(1)$-bundles
$\tilde{\cal R}$, and for each such bundle we must integrate over
the space of all connections on it. The integral gives a delta
function setting $d\CK=0$.  The sum over bundles gives a delta
function stating that the periods of $\CK$ take values in $2\pi\Z$.
The combined conditions precisely say that $\CK$ can be set to zero
by a transformation \xilo. After setting $\CK$ to zero, we are left
with the original abelian gauge theory \dolgo\ with $A$ as the only
field variable.

So the extended theory with action \zoromp\ is equivalent to the
original theory with coupling parameter $\tau$.  Next, let us
consider another way to study the same theory.  We use the extended
gauge invariance \xilo\ to set $A=0$.  After doing this, as the
action depends quadratically on $\CK$  without any derivatives of
$\CK$ (and the term quadratic in $\CK$ is nondegenerate), we can
``integrate out'' $\CK$ by simply solving the Euler-Lagrange
equations to determine $\CK$ in terms of $\CV$. (In \witugh, this
process is described somewhat more precisely at the quantum level.)
In this way, we get an action for $\CV$, which is
\eqn\mutox{-{i\over 8\pi}\int_Md^4x\sqrt
h\left(\left(-{1\over\tau}\right)(\CV^+)^2-\left(-{1\over\bar\tau}\right)(\CV^-)^2\right).}
This is the original abelian gauge theory, but with the connection
$\cv$ instead of $A$, and the coupling parameter $\tau$ replaced by
$-1/\tau$.  So comparing the two ways to analyze the extended theory
\zoromp\ gives us the $\tau\to -1/\tau$ symmetry of abelian gauge
theory.

Now let us introduce a surface operator, supported on a
two-manifold $D\subset M$.  To keep things simple, we will
consider the special case that this surface operator has
$\alpha=0$, $\eta\not=0$.  This means that in the path integral of
the underlying abelian gauge theory, we want to include a
factor\foot{Though $\eta$ is an angular variable, we have
normalized it to take values in $\R/\Z$, rather than $\R/2\pi\Z$,
to avoid unnatural-looking factors of $2\pi$ in the
transformations under $S$-duality.  The alternative is to modify
the definition of $\alpha$, $\beta$, and $\gamma$ by a factor of
$2\pi$.} \eqn\fony{\exp\left(i\eta\int_D{F}\right).} This is
equivalent to adding to the action a term $-i\eta\int_DF$. To
incorporate the surface operator in the extended theory, we
replace $F$ by $\FF$ and add \eqn\untu{-i\eta\int_D
{\FF}=-i\eta\int_M \delta_D\wedge {\FF}} to the action \zoromp.
Here $\delta_D$ is a two-form delta function that is Poincar\'e
dual to $D$.

The extra term does not depend on $\cv$, so if we first integrate
over $\cv$, we get back to the abelian gauge theory with the surface
operator.  But what happens if we instead use the extended gauge
symmetry to set $A=0$, and then solve for $\CK$?  The action has two
terms linear in $\CK$, which combine to \eqn\bunto{{i\over
2\pi}\int_M(\CV-2\pi\eta\delta_D)\wedge \CK.} To get this formula,
we used the last expressions given in \hillox\ and \untu.  Moreover,
this part of the action is the only part that contains $\CV$.  (The
rest of the action is quadratic in $\CK$ and independent of $\CV$.)
So the effect of having $\eta\not=0$ is precisely to replace $\CV$
by $\hat \CV=\CV-2\pi\eta\delta_D$.   Hence when we integrate out
$\CK$ again, we will get the same action as in \mutox, but with $\CV$
replaced by $\hat \CV$: \eqn\zutox{-{i\over 8\pi}\int_Md^4x\sqrt
h\left(\left(-{1\over\tau}\right)(\hat
\CV^+)^2-\left(-{1\over\bar\tau}\right)(\hat \CV^-)^2\right).}

This action is potentially divergent because of the delta function
term in $\hat \CV$.  Since the action is quadratic in $\hat \CV$,
with positive definite real part, the only way to make the action
finite is for $\CV$ to be such as to cancel the delta function
contribution in $\hat \CV$.  Thus, the connections that contribute
to the path integral must obey
\eqn\ultox{\CV=2\pi\eta\delta_D+\dots} where the ellipses refer to
terms that are regular near $D$. But this is precisely the
characterization of a surface operator with parameter $\alpha=\eta$,
as should become clear upon comparing \ultox\ to \yilzo.

So we have shown that the transformation $S:\tau\to -1/\tau$ maps a
surface operator with parameters $(\alpha,\eta)=(0,\eta)$ to one
with parameters $(\eta,0)$.  This is a special case of \hylgo.  To
get the general case, one replaces $\FF$ in \zoromp\ by
$\FF-2\pi\alpha\delta_D$, and then repeats the calculation.  After
gauging $A$ to zero and integrating out $\CK$, one gets back to an
action of the same kind, with $\tau$ replaced by $-1/\tau$ and
$\alpha$ and $\eta$ exchanged as in \hylgo.

 \subsec{Shifting The Theta Angle} \subseclab\thetangle

We now return to the case that $G$ is a simple non-abelian gauge
group.  Apart from  $S:\tau\to -1/n_{\frak g}\tau$, which we have
considered in section \duality, the other generator of the duality
group is a shift in the four-dimensional theta-angle, which enters
the four-dimensional action via a term
\eqn\migo{\II_\theta=-i\theta {\eurm N}} where \eqn\migox{{\eurm
N}=-{1\over 8\pi^2}\int_M \Tr\,F\wedge F} is the instanton number.
$\Tr$ is a negative-definite quadratic form on $\frak g$ such that
$\Tr\, x^2=-2$ for $x$ a short coroot. (The notation is motivated
by the fact that for $G=SU(N)$, $\Tr$ is the trace in the
$N$-dimensional representation.) The normalization ensures that
 if $M$ is a closed four-manifold without surface
operators, and $G$ is simply-connected, then ${\eurm N}$ is
integer-valued.  For example, if $G=SU(N)$, then ${\eurm N}
=-\int_Mc_2(E)$, where $c_2$ is the second Chern class. When ${\eurm N}$ is
integer-valued, there is a symmetry $T:\theta\to \theta+2\pi$, or
$T:\tau\to\tau+1$. This expresses the familiar fact that quantum
field theory depends on the action $\II$ only modulo $2\pi i\Z$.

If $G$ is not simply-connected, then ${\eurm N}$ takes values in
${{1\over k}}\Z$ for some integer $k$, and instead of $T$ we
consider the symmetry $T^k:\tau\to\tau+k$.

Assuming for notational simplicity that $G$ is simply-connected,
let us study the symmetry $T$ in the presence of a surface
operator supported on a two-manifold $D$. First, in the presence
of the singularity associated with the surface operator, we have
to define precisely what we mean by the integral defining ${\eurm
N}$.  The integral has a bulk contribution, coming from the
integration over the complement of $D$: \eqn\zigox{{\eurm
N}_0=-{1\over 8\pi^2}\int'_M\Tr\,F\wedge F.} The symbol $\int'_M$
means that we integrate over the complement of $D$, ignoring
possible delta function contributions at $D$. According to \yilzo,
once we pick an extension of the bundle $E$ over $D$, there is
also a delta function contribution to the integral.  This
contribution is \eqn\bigox{{\eurm N}'=-{1\over 2\pi}\int_D \Tr
\,\alpha F-{1\over 2}D\cap D\,\Tr\,\alpha^2 .}  To obtain \bigox\
(which corresponds to Proposition 5.7 in \km), we use the fact
that the delta function contribution to $F$ near $D$ is
$2\pi\alpha\delta_D$. We have also used \yrol.

The sum \eqn\hombo{{\eurm N}={\eurm N}_0+{\eurm N}' =-{1\over
8\pi^2}\int'_M\Tr\,F\wedge F-{1\over 2\pi}\int_D \Tr \,\alpha
F-{1\over 2}D\cap D\,\Tr\,\alpha^2 }
 is
integer-valued. However, ${\eurm N}$ is not natural in the sense
that it depends on a choice of lifting of $\alpha$ from ${\frak t}/
\Lambda_{\rm cochar}$, where it naturally takes values, to $\frak
t$. There is a simple reason for this; the integer-valued invariant
$\int_M c_2(E)$ (or its analog for groups other than $SU(N)$)
is not determined by the restriction of $E$ to
$M\backslash D$, but depends on the choice of an extension of $E$
over $D$. Integrality of $\eurm N$, however, implies that upon reduction mod $\Bbb{Z}$, we get
\eqn\zigox{{\eurm N}_0=-{\eurm N}'={1\over 2\pi}\int_D \Tr\,\alpha
F+{1\over 2}D\cap D\,\Tr\,\alpha^2~{\rm mod}~\Bbb{Z}.} Since ${\eurm
N}_0$ does not depend on a lifting of $\alpha$, the same must be
true mod $\Z$
 of the right hand side of \zigox.  We can verify this as follows.
The restriction of $F$ to $D$ is ${\frak t}$-valued, and its
integral ${\eurm m}=\int_D F/2\pi$ is a ``magnetic charge,'' an
element of the lattice $\Lambda_{\rm cochar}$.  This lattice
actually coincides with the root lattice of $^LG$ (since we are
assuming $G$ to be simply-connected) or in other words the coroot
lattice of $G$. Because of the way the trace was normalized, the
bilinear function ${\eurm m},\,{\eurm m}'\to \Tr\,{\eurm m}{\eurm
m}'$ takes integer values for ${\eurm m},\,{\eurm m}'$ in this
lattice, and takes even integer values if ${\eurm m}={\eurm m}'$.
  Once an extension of $E$ is picked,
$\alpha$ takes values in $\frak t$, which is the same as
$\Lambda_{\rm cochar} \otimes_{\Bbb Z}{\Bbb R}$. So we can write
\eqn\filgox{{\eurm N}'=-\Tr\,\alpha{\eurm m} -{1\over 2}D\cap
D\,\Tr\,\alpha^2} and hence \eqn\dilgox{{\eurm
N}_0=\Tr\,\alpha{\eurm m}+{1\over 2}D\cap D\,\Tr\,\alpha^2~{\rm
mod}~\Z.} This statement is invariant under shifts of $\alpha$ by a lattice
vector, given  the integrality properties of the trace, together
with \nobo.

We want to define $\II_\theta$ to be independent of any choice of
extension of the bundle.  The only obvious way to do this is to
omit the delta function contribution from $D$, and set
\eqn\xigox{\II_\theta=-i\theta {\eurm N}_0.} This does not mean
that we will ignore the delta function contribution. Rather, we
will have such a contribution from the two-dimensional theta-like
parameter $\eta$ that was introduced in section \thetang.  In the
same notation, we take the contribution of $\eta$ to the action to
be \eqn\relfo{\II_\eta=-2\pi i \Tr\,\eta{\eurm m}-\pi i D\cap
D\,\Tr\,\alpha\eta .}  The term $\Tr\,\eta{\eurm m}$ is the
expected term for the theta-like angles $\eta$.  To this, we have
added a $c$-number term that depends only on $\alpha$ and $\eta$
and not on any of the field variables of the theory.

The sum of the two contributions to the action is therefore
\eqn\elfo{\hat \II = \II_\theta+\II_\eta= -i\theta {\eurm
N}_0-2\pi i\Tr\,\eta{\eurm m}-\pi iD\cap D\, \Tr\,\alpha\eta .} At
this stage, we can more fully justify our definition of
$\II_\theta$. Adding to $\II_\theta$  a multiple of
$\Tr\,\alpha{\eurm m}$
 would have no essential effect,
since one can compensate for this by shifting $\eta$.  So we may
as well define $\II_\theta$ as we have.

We will now see that the choice we have made leads to a simple
result for how $\alpha$ and $\eta$ must transform  under
$\theta\to\theta+2\pi$. The change in $ \II_\theta$ is
\eqn\refrog{\Delta \II_\theta=-2\pi i {\eurm N}_0=-2\pi
i\Tr\,\alpha{\eurm m}-\pi i D\cap D\,\Tr\,\alpha^2 ~{\rm mod}~2\pi
i \Z ,} where \dilgox\ has been used.

To get a symmetry of the theory, $\hat \II$ must be invariant mod
$2\pi i\Bbb{Z}$.  For this, we let $T:\theta\to\theta+2\pi$ act on
$(\alpha,\eta)$ by \eqn\imax{\eqalign{\eta&\to\eta-\alpha\cr
\alpha&\to \alpha.\cr}}  The variation of $\II_\eta$ then
precisely cancels the variation of $\II_\theta$ mod $2\pi i\Z$,
and $\hat \II$ is invariant.

The constant term $-\pi i  D\cap D\,\Tr\,\alpha\eta$ that we
included in the action is not invariant mod $2\pi i$ under lattice
shifts of $\alpha$ or $\eta$.  However, the non-invariance is,
like this term itself, independent of the quantum fields.
Geometrically, this means that when $D\cap D$ is non-zero, the
partition function is not a complex-valued function of $\alpha$
and $\eta$ but a section of a complex line bundle over $\TT\times
{}^L\neg\TT$.
  Of course, it would be
possible to omit the $c$-number term from the action and instead
claim that the symmetry $T:\tau\to\tau+1$ holds up to a
$c$-number.

This discussion generalizes straightforwardly to the case that $G$
is not simply-connected and the instanton number takes values in
${1\over k}\Z$.  One considers the symmetry $T^k:\tau\to\tau+k$,
and the same derivation shows that it acts as
\eqn\zimax{\eqalign{\eta&\to\eta-k\alpha\cr \alpha&\to
\alpha.\cr}}

Let us combine the result \imax\ with our previous result \hylgo\
for the action of electric-magnetic duality.  First we consider
the case that $G$ is selfdual and simply-laced, so that in
particular the duality group is simply $\Gamma=SL(2,\Z)$. The only
simple Lie group that actually has these properties is $E_8$. In
this situation,
 $\Gamma$ is generated by the elements \eqn\olpo{S=\left(\matrix{0&
1\cr -1& 0\cr}\right), ~~~T=\left(\matrix{1&1\cr 0&1\cr}\right).}
The formulas \hylgo\ and \imax\ tell us that for ${\eurm M}$ equal
to $S$ or $T$, $\alpha$ and $\eta$ transform under ${\eurm M}$ by
\eqn\zolpo{(\alpha,\eta)\to (\alpha,\eta){\eurm M}^{-1}.} This is
therefore true for any ${\eurm M}\in SL(2,\Bbb{Z})$.  In
particular, the pair $(\alpha,\eta)$ transform naturally under
$SL(2,\Z)$, and our results for the action of $S$ and $T$, though
motivated independently, are compatible with each other.

The generalization to any simple $G$ is as follows.  First recall
that electric charge takes values in the character lattice\foot{In
much of the physics literature, root and coroot lattices are taken here,
because the theory is considered only on $\R^{3,1}$ or $\R^4$, where
the physical electric and magnetic charges take values in the root and coroot
lattices. However, the fact that the theory could be probed with
Wilson and 't Hooft operators as external charges shows that it must
be possible to refine the usual discussion of duality to the case
that the charges take values in the character and cocharacter lattices.  This
refinement is the one relevant here, roughly because the same
topological issues arise either by allowing Wilson and 't Hooft
operators, working on a general four-manifold, or admitting surface
operators such as those considered here. } $\Lambda_{\rm char}$ of
$G$, and magnetic charge takes values in the cocharacter lattice
$\Lambda_{\rm cochar}$. So the full set of charges takes values in
the lattice $\hat\Lambda=\Lambda_{\rm cochar}\oplus \Lambda_{\rm
char}$, which has a natural  non-degenerate skew pairing since
$\Lambda_{\rm char}$ and $\Lambda_{\rm cochar}$ are dual lattices.
The duality group $\Gamma$ acts linearly on this lattice, preserving
the skew pairing, as well as acting naturally on $\tau=\theta/2\pi
+4\pi i/g^2$. For example, for $E_8$ we have
$\hat\Lambda=\Lambda_{\rm char}\otimes \Z^2$, where $\Z^2$ is a rank
two lattice, and $\Gamma=SL(2,\Z)$ acts on $\hat\Lambda$ via its
natural action on  $\Z^2$.  At any rate, $\Gamma$ always acts on the
full set of electric and magnetic charges, and thus on
$\hat\Lambda$. The details are a little complex, however, especially
\refs{\dorey,\kapsei} if  $G$ is not simply-laced.

At any rate, the linear action of $\Gamma$ on $\hat\Lambda$
induces an action on $\hat\Lambda\otimes_\Z\R$, and hence on
$(\hat\Lambda\otimes_\Z\R)/\hat\Lambda$.  But
$(\hat\Lambda\otimes_\Z\R)/\hat\Lambda$ is the same as $\TT\times
{}^L\neg\TT$.  So the action of $\Gamma$ on the charges determines
an action on $\TT\times {}^L\neg\TT$, where the pair
$(\alpha,\eta)$ take values. Thus it determines a natural action
on $(\alpha,\eta)$.  For any $G$, the meaning of \hylgo\ and
\imax\ is that the action of $\Gamma$ on $(\alpha,\eta)$ is
precisely the natural action determined by its action on the
electric and magnetic charges. The hypothesis \hylgo\ asserts that
this is true for $S$, and it is true for $T:\tau\to\tau+1$ since
the derivation of \imax\ was actually a close cousin of the
computation \ref\elwitten{E. Witten, ``Dyons Of Electric Charge
$e\theta/2\pi$,'' Phys. Lett. {\bf B86} (1979) 283-7.} of the action of
$T:\theta\to\theta+2\pi$ on the charges.

\bigskip\noindent{\it The Dirac String}

\ifig\welco{\bigskip A surface operator whose support $D$ has a boundary $L$, which turns
out to be the world-line of a magnetic monopole or dyon. }
{\epsfxsize=2in\epsfbox{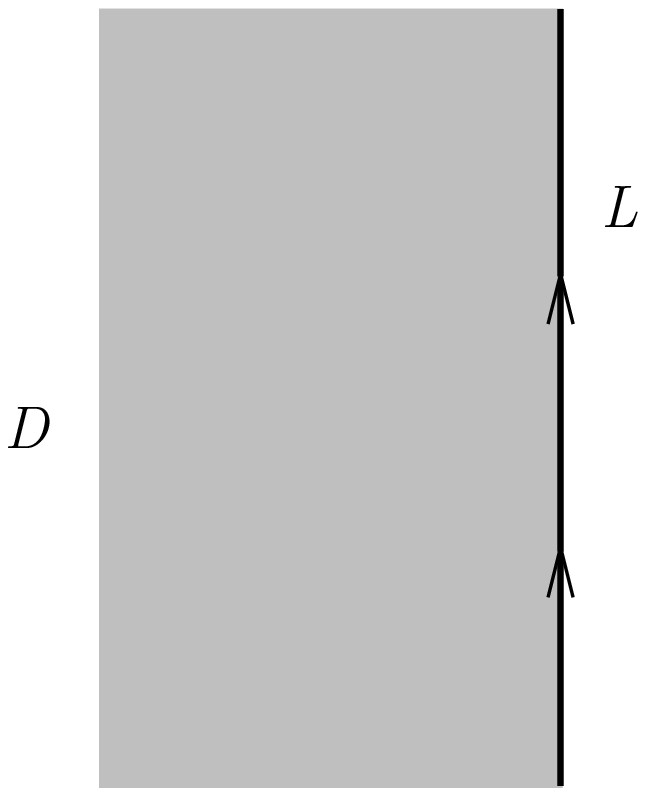}}
Now, leaving some details to the reader, and making use of some ideas in \prestwo,
 we are going to explain an informal interpretation of  this result.

Let us ask whether the support $D$ of a surface operator can have a boundary $L$ as in
\welco.
A little thought will show that for this to occur,
$L$ must be the world-line of a magnetic monopole
with magnetic charge $\alpha$.  The gauge field holonomy around $D$ must unwind at the boundary
$L$, and this unwinding of the holonomy characterizes magnetic charge.

However, we have not imposed Dirac quantization on the magnetic charge $\alpha$.  As we consider
it to be defined modulo a lattice vector, we are really interested in the case that Dirac
quantization is {\it not} obeyed.  According to Dirac, in this case, for gauge-invariance,
$L$ must be the boundary of the world-volume of what is commonly called a ``Dirac string.''
In our context, the surface $D$ can be regarded as the world-volume of the string.  Thus,
our surface operator can be regarded as representing the Dirac string associated
with improperly quantized magnetic charge.  Dirac defined the Dirac string in terms of the
monodromy of the gauge field around it, so our surface operator indeed has the right property
to be the world-volume of a  Dirac string.

More generally, if  $\eta\not=0$,
the monopole at the end of the string also carries electric charge.
It is thus a dyon.

Our claim that $(\alpha,\eta)$ transform under $S$-duality just like magnetic
and electric charge  can be interpreted as a statement that Dirac strings associated with
improperly quantized charges transform under duality just like properly quantized
charges.

We conclude with a related comment, also anticipated in \prestwo.
Physically, a surface operator might arise if ${\cal N}=4$ super Yang-Mills theory
(or any gauge theory of interest, such as the Standard Model) is embedded in some more
complete theory that reduces to it at low energies.  The embedding in a more complete theory
might give rise to what in other contexts are called cosmic strings.  Suppose that such a string
is heavy enough that we can consider it to be frozen in position, with a known orbit in spacetime.
It is then appropriate to consider
 the ``low energy'' ${\cal N}=4$ dynamics in the presence of the string.
This will involve studying the ${\cal N}=4$ theory coupled to a surface operator of some kind.
The details depend on the particular type of cosmic string considered.
Strings that produce an Aharonov-Bohm effect, as first explored in \rohm, will lead to surface
operators of the sort considered in the present paper.
In section \stringco, we will consider some surface operators defined
 by embedding ${\cal N}=4$ super Yang-Mills theory in
a more complete theory
(with the more complete theory being Type IIB superstring theory).
In the theory of cosmic strings, it is familiar that some kinds of string can break,
terminating on magnetic monopoles \ref\hooft{G. 't Hooft, ``Magnetic Monopoles In Unified
Gauge Theories,'' Nucl. Phys. {\bf B79} (1974) 276-284.}.

\bigskip\noindent{\it An Illustration}

We will now explain in more detail how $S$-duality acts on the charges in the case of
$G=SU(2)$, $^LG=SO(3)$.  This will enable us to spell out a few points that have been hidden
in our analysis above.
 Following \ref\vafawitten{C. Vafa and E. Witten, ``A Strong
Coupling Test Of $S$-Duality,'' Nucl. Phys. {\bf B431} (1994) 3-77, hep-th/9408074.},
we explain two different
ways to view the problem.  This discussion is not used in the rest
of the paper, and we will be rather brief on some points.

$SU(2)$ is simply-connected, so for $SU(2)$ gauge theory, the
instanton number is integer-valued and there is a symmetry
$T:\tau\to\tau+1$, acting as in \imax.  On the other hand, for
$SO(3)$, the instanton number takes values in ${1\over 4}\Bbb{Z}$,
or in ${1\over 2}\Bbb{Z}$ if $M$ is a spin manifold.  So the basic
shift symmetry of the theta-angle is $T^4:\tau\to\tau+4$, or
$T^2:\tau\to\tau+2$ if $M$ is spin. These symmetries act as in
\zimax. In addition, there is the symmetry $S:\tau\to -1/\tau$,
which exchanges $SU(2)$ and $SO(3)$. The group  of duality
symmetries of $SU(2)$ gauge theory is generated if $M$ is not spin
by \eqn\undi{T=\left(\matrix{1&1\cr 0&1\cr}\right),
~~~ST^4S^{-1}=\left(\matrix{1&0\cr-4&1\cr}\right).} ($ST^4S^{-1}$
is a duality symmetry for $SU(2)$, since $S^{-1}$ maps $SU(2)$ to
$SO(3)$, $T^4$ is a duality transformation of $SO(3)$, and $S$
maps back to $SU(2)$.) This duality group, which is known as
$\Gamma_0(4)$, is a congruence subgroup of $SL(2,\Z)$. It is a
group of duality symmetries of $SU(2)$ gauge theory on a general
$M$, and acts on $\alpha$ and $\eta$ according to \zolpo. The
group of duality symmetries of $SO(3)$ gauge theory is likewise a
copy of $\Gamma_0(4)$, generated by $STS^{-1}$ and $T^4$.  And
finally, of course, we have the symmetry $S$ that exchanges the
two groups. If $M$ is spin, one can replace 4 by 2 everywhere, and
the duality groups are isomorphic to $\Gamma_0(2)$.

$G=SU(2)$ and $^LG=SO(3)$ have the same Lie algebra, so we can think of
$\TT=\frak t/\Lambda_{\rm rt}$ and $^L\neg\TT=\frak t/\Lambda_{\rm wt}$ as
quotients of the same space by different lattices, which are respectively the
root and weight lattices $\Lambda_{\rm rt}$ and $\Lambda_{\rm wt}$.
$\Lambda_{\rm rt}$ is of index two in $\Lambda_{\rm wt}$.
 The identity map on $\frak t$
therefore projects to a natural, two-to-one map from $\TT$, where $\alpha$ takes values, to
$^L\neg\TT$, where $\eta$ takes values.
This map is implicit in the transformation $T:(\alpha,\eta)\to
(\alpha,\eta+\alpha)$.  The identity map on $\frak t$ does not
project to a natural map of $^L\neg\TT$ to $\TT$, but the map of
multiplication by 2 (or any even integer) does so project. So
there is a natural map $ST^2S^{-1}:(\alpha,\eta)\to
(\alpha-2\eta,\eta)$, which is realized as a duality
transformation if $M$ is spin. Similarly the operation
$ST^4S^{-1}:(\alpha,\eta)\to (\alpha-4\eta,\eta)$, which appears
as a duality transformation for any $M$, is naturally defined.

This way of describing things is in some tension with the physics
literature, where the duality group is generally considered to be
$SL(2,\Z)$ for a simply-laced group such as $SU(2)$.  However, the
usual analysis is made only for $M=\R^4$, and without detailed
consideration of Wilson and 't Hooft operators.  Under these
circumstances, the $SO(3)$ and $SU(2)$ gauge theories coincide.
One can distinguish the two theories on $\R^4$ by adding a Wilson
operator in the two-dimensional representation to specialize to
$SU(2)$, or an 't Hooft operator with minimal charge to specialize
to $SO(3)$. (One cannot add operators of both types
simultaneously, as they are not mutually local.)  If the $SU(2)$
and $SO(3)$ theories are elaborated in this way, the appropriate
duality groups on $\R^4$ are precisely those that we have
just described on a general spin manifold.

Actually, there is another formulation of the problem, of which we
will only give an outline, that really gives more information, and
in which  the full group $SL(2,\Z)$ plays a role. In this
approach, we always take the gauge group to be the adjoint group
$SO(3)$ (not its cover $SU(2)$), but we specify the second
Stieffel-Whitney class $\xi=w_2(E)$ of the bundle $E$. Summing
over all $SO(3)$-bundles with fixed $\xi$, we define a partition
function\foot{Along with a full set of operators, correlation functions,
quantum states, branes, etc., making up a  quantum field theory.  In a fuller description,
the whole quantum field theory, not just the set of partition functions, is transformed
by $SL(2,\Z)$.} $Z_\xi(\tau)$ for each $\xi$.  One then shows
\vafawitten\ that the $Z_\xi(\tau)$ transform in a unitary
representation of $SL(2,\Z)$.

To  incorporate in this approach a surface operator with
parameters $\alpha,\eta$, supported on a surface $D$, one defines
$\alpha$ and $\eta$ to take values in $\TT$, the maximal torus of
the simply-connected group $SU(2)$. Then we have for each $\xi$ a
partition function
$Z_\xi(\tau;\alpha,\eta)$, and the claim is
that this family of partition functions transforms as a
representation of $SL(2,\Z)$ (which acts on $\tau$ in the usual
fashion, on $\xi$ as described in \vafawitten, and on $\alpha$ and
$\eta$ by the natural action described in \zolpo).

The description by parameters $\xi,\alpha,\eta$ is slightly
redundant. To describe the redundancy, we use a multiplicative
notation for the maximal torus $\TT$ (where $\alpha$ and $\eta$
take values) and recall that this torus contains the element $-1$
of the center of $SU(2)$. Then shifting $\alpha$ to $-\alpha$ is
equivalent to replacing $\xi$ with $\xi+[D]$ according to \nomo, and
shifting $\eta$ to $-\eta$ multiplies the partition function
$Z_\xi(\tau;\alpha,\eta)$ by $(-1)^{(\xi,D)}$ (this is essentially explained
at the end of section \ramcase).  This redundancy is
compatible with the action of $SL(2,\Z)$, but it seems hard to
eliminate it without making the action of $SL(2,\Z)$ look less
natural.

\subsec{Levi Subgroups And More General Surface Operators}
\subseclab\levi

Now we will describe a simple but important generalization of our definition of surface
operators in which
the maximal torus $\TT$ is replaced by a more general subgroup $\Bbb{L}$ of $G$ that contains $\TT$.
We assume also that $\Bbb{L}$ can be characterized as the subgroup of $G$ that
commutes with some $\alpha\in \frak t$.  Such an $\Bbb{L}$ is called a subgroup of $G$ of Levi type.
We consider two such groups to be equivalent if they are conjugate in $G$.
The usual case is that $\alpha\not= 0$ and $\Bbb{L}$ is a proper subgroup of $G$, but we also will
consider the case $\alpha=0$ and $\Bbb{L}=G$.  Any Levi subgroup contains $\TT$, so $\TT$ is a minimal
Levi subgroup. A typical example of a non-minimal Levi subgroup is the subgroup
of $SU(3)$ of the form
\eqn\xuntyx{\left(\matrix{* & * & 0 \cr
                        * & * & 0 \cr
                        0 & 0 & * \cr}\right).}
This is  isomorphic to $U(2)$, and is actually a next-to-minimal Levi subgroup, since
a smaller Levi subgroup would have to be  $\TT$ itself.  In general, for any $G$ of rank $r$,
a next-to-minimal Levi subgroup is isomorphic to $SU(2)\times U(1)^{r-1}$
or a quotient of this by $\Z_2$.

We now will define what we will call a surface operator of type $\Bbb{L}$.  In this language,
the surface operator that we defined originally is an operator of type $\TT$; we will
also call it the generic surface operator.
To define a surface operator of type $\Bbb{L}$,  we consider  a two-manifold $D$ with a gauge
theory singularity labeled classically
by the usual parameters $(\alpha,\beta,\gamma)$.  But now we require the parameters
to be $\Bbb{L}$-invariant.  Moreover, when we perform the quantum
path integral, we divide by gauge transformations that are $\Bbb{L}$-valued when restricted to $D$,
not just those that are $\TT$-valued.
It is because of this last step that the surface operator of type $\Bbb{L}$ is not
just a special case -- for $\Bbb{L}$-invariant parameters -- of the surface operator that we defined
originally, the operator of type $\TT$  in which $(\alpha,\beta,
\gamma)$ are simply $\frak t$-valued.  The surface operator of type $\Bbb{L}$
 is something new, associated with a different
path integral.  Actually, we will learn in section \topology\  that if $\Bbb{L}$ contains $\TT$
as a proper subgroup, then the surface operator of type
$\TT$ becomes singular when the parameters become $\Bbb{L}$-invariant.
So for such special parameters, the operator that makes sense is
the new surface operator of type $\Bbb{L}$.

At the quantum level, there is an additional parameter $\eta$, the two-dimensional theta-angle
$\eta$ of section $\thetang$.  In the present case, with the group $\TT$ replaced by $\Bbb{L}$,
$\eta$ takes values not in $^L\neg \TT$ but in a subgroup.
In fact, we want to define $\eta$ as a theta-angle of the abelian part of $\Bbb{L}$.
To give a convenient description of where $\eta$ takes values,
it is useful to observe that Levi subgroups $\Bbb{L}\subset G$ are in natural correspondence
with Levi subgroups $^L\neg \Bbb{L}\subset{} ^L\neg G$.  One way to describe the correspondence
is to make use of the fact that the Weyl group ${\cal W}$ of $G$ naturally coincides with the
Weyl group $^L\neg{\cal W}$ of $^L\neg G$. (See Appendix A.)
 Moreover, the Weyl groups of $\Bbb{L}$ and $^L\neg\Bbb{L}$
 are subgroups of ${\cal W}$ and $^L\neg {\cal W}$. The correspondence
between $\Bbb{L}$ and $^L\neg\Bbb{L}$ is simply that they have the same Weyl group, which we
will denote as ${\cal W}_\Bbb{L}$.
Another way to state the correspondence between $\Bbb{L}$ and $^L\neg \Bbb{L}$ is to say that the coroots
of $^L\neg \Bbb{L}$ (which are a subset of the coroots of $^L\neg G$) are multiples of the coroots
of $\Bbb{L}$ -- with different multiples for long or short coroots.  With all this understood,
a surface operator of type $\Bbb{L}$ depends on parameters $(\alpha,\beta,\gamma,\eta)$ which
take values in the ${\cal W}_\Bbb{L}$-invariant part of $\TT\times \frak t\times \frak t\times {}^L\neg
\TT$.  We are using here the fact that $\Bbb{L}$-invariance of $(\alpha,\beta,\gamma)$ is equivalent
to ${\cal W}_\Bbb{L}$-invariance.  The formulation in terms of ${\cal W}_\Bbb{L}$-invariance has an important
advantage: it enables us to treat $\eta$ on the same footing
as the other variables.  This is more satisfactory than saying that $(\alpha,\beta,\gamma)$
are $\Bbb{L}$-invariant and $\eta$ is $^L\neg \Bbb{L}$-invariant.  The condition
that $\eta$ must be ${\cal W}_\Bbb{L}$-invariant is the right one, because it means that $\eta$ is
a character of the abelian magnetic fluxes of the gauge bundle $E$ restricted to $D$;
the structure group of this bundle is $\Bbb{L}$, and its characteristic classes are
${\cal W}_\Bbb{L}$-invariant.

We extend the duality conjecture of section \duality\ to say that a surface operator
of type $\Bbb{L}$ maps under duality to a surface operator of type $^L\neg \Bbb{L}$, with the parameters
transforming in the familiar fashion $(\alpha,\beta,\gamma,\eta)\to(\eta,\beta,\gamma,-\alpha)$.

\bigskip\noindent{\it Some More Group Theory}

Now we describe some more group theory that will be useful in the rest of the paper.

Levi subgroups are closely related to what are called parabolic subgroups of $G_\C$.
Let us pick a particular $\alpha\in \frak t$ which commutes precisely with $\Bbb{L}$.  We say
that such an $\alpha$ is $\Bbb{L}$-regular (and if $\Bbb{L}=\TT$, we simply say that $\alpha$ is regular).
We let $\EUP$ be the subgroup of $G_\C$ whose Lie algebra $\frak p$
is spanned by elements $\psi\in \frak g$ that obey \eqn\lorp{[\alpha,\psi]=i\lambda\psi,
~\lambda\geq 0.}  A group of this form is called a parabolic subgroup of $G_\C$.
If in \lorp\ we replace the condition $\lambda\geq 0$ by $\lambda>0$, we get the Lie algebra
$\frak n$ of a subgroup $\EUN\subset \EUP$ that is known as the unipotent radical of $\EUP$.

For example, for $G_\C=SL(3,\C)$ and
$\Bbb{L}=\TT$, we can take $\alpha=i\,{\rm diag}(y_1,y_2,y_3)$ with the $y_i$ real
and $y_1>y_2>y_3$.  In this case, the parabolic group we get is the group of upper triangular
matrices
\eqn\buntyx{\left(\matrix{*&*&*\cr 0&*&*\cr 0&0&*\cr}\right),}
and is called a Borel subgroup $\EUB$.  In this example, the unipotent radical $\EUN$
consists of matrices of this form:
\eqn\buntyx{\left(\matrix{1&*&*\cr 0&1&*\cr 0&0&1\cr}\right),}

A different choice of $\TT$-regular (or simply regular)
$\alpha$ would be obtained from the one we used
by permuting the eigenvalues by a Weyl transformation.  This leads to a Weyl-conjugate Borel
subgroup $\EUB'$.
More generally, for any $G$,
the parabolic subgroup associated with a pair $(\TT,\alpha)$
is called a Borel subgroup $\EUB$, and is unique up to a Weyl transformation.
  $\EUB$ is  a minimal parabolic subgroup, just as $\TT$ is a minimal Levi subgroup.

A non-minimal Levi subgroup $\Bbb{L}$ may be associated with several inequivalent parabolic subgroups.
For instance,
in the example of \xuntyx, we can take
\eqn\buntyx{\alpha=iy\left(\matrix{1&0&0\cr 0&1&0\cr 0&0&-2\cr}\right),}
with real nonzero $y$.  If $y$ is positive, we get the parabolic subgroup $\EUP$
of matrices of the form
\eqn\untyxo{\left(\matrix{*& * & * \cr
                        * & * & * \cr
                        0 & 0 & * \cr}\right),} but
                        if $y$ is negative, we get an inequivalent parabolic subgroup $\EUP'$
                        of matrices of the form
                        \eqn\punty{\left(\matrix{* & * & 0 \cr
                        * & * & 0\cr
                        * & * & * \cr}\right).}
The unipotent radicals $\EUN$ and $\EUN'$ consist of matrices of the form
\eqn\untyxt{\left(\matrix{1 & 0 & * \cr
                        0 & 1 & * \cr
                        0 & 0 & 1 \cr}\right),}
                        or
                        \eqn\untyx{\left(\matrix{1 & 0 & 0 \cr
                        0 & 1 & 0 \cr
                        * & * & 1 \cr}\right).}

\bigskip\noindent{\it Maximal Levi Subgroup}

If $\alpha=0$, we get a special case of the above construction in which $\Bbb{L}=G$, $\EUP=G_\C$, and
$\EUN=1$. It is most convenient to allow this special case and to regard $G$ itself as a maximal
Levi subgroup.

In the above, we have been a little imprecise about whether $\alpha$ and $\eta$ are Lie
algebra-valued or torus-valued.  The torus-valued case leads to a more general construction, since
the space of ${\cal W}_\Bbb{L}$-invariant $\alpha$ or $\eta$ may not be connected. We will spell
this out for the case $\Bbb{L}=G$.
The condition for $\alpha$ to be $G$-invariant
says that it takes values in the center of $G$, which we denote ${\cal Z}(G)$.  Thus, what
it means to have a surface operator of type $G$ supported on a surface $D$ is that the gauge
field is defined on $M\backslash D$
and has monodromy around $D$ labeled by a prescribed element
$\alpha\in {\cal Z}(G)$.  The dual of the choice of $\alpha$ is that
a $G$-bundle on $D$ is classified topologically by a characteristic class
$\xi\in H^2(D,\pi_1(G))\cong
\pi_1(G)$, and the path integral can be weighted
by the choice of a discrete $\eta$-angle valued in ${\rm Hom}(\pi_1(G),U(1))$.
The finite groups ${\cal Z}(G)$ and ${\rm Hom}(\pi_1(G),U(1))$ are exchanged by duality,
and the natural duality conjecture states that the discrete versions of $\alpha$ and $\eta$
are exchanged in the usual way.

Though our duality conjecture for surface operators of type $\Bbb{L}$
is meant to include the disconnected components, we will not consider them in
any detail in this paper.

\newsec{More On The Classical Geometry}
\seclab\morecl

In this section, we will reconsider the construction of section \geomint\ and describe
aspects that require
understanding the classical geometry of $\MH$
in more depth.
We  aim for a relatively simple -- but certainly not mathematically
complete -- introduction
to aspects that are or  may become useful for the geometric
Langlands program.  Most of the topics we consider have been treated much more fully in the
mathematical literature, but a few points may be new, notably
 the action of the affine Weyl group on the cohomology
of $\MH$ (section \actweyl), the linearity of the cohomology classes of the symplectic
forms (section \cpxview), and the nature of the local singularities of $\MH$
at non-regular points (section \topology).

Many of the most significant phenomena that we will describe
have local models involving hyper-Kahler
metrics on complex coadjoint orbits, constructed in \refs{\kron-\kovalev}.
We defer this to section \onahm.

\subsec{Hyper-Kahler Quotient} \subseclab\hyperquo

One of the most basic properties of the moduli space $\MH(G,C)$ of
Higgs bundles on a Riemann surface $C$ is that it can be
constructed as a hyper-Kahler quotient \hitchin. We recall first
the construction in the absence of singularities. Picking a smooth
$G$-bundle $E\to C$, we denote as $\EUW$ the space of pairs
$(A,\phi)$, with $A$ a connection on $E$ and $\phi\in
\Omega^1(C,{\rm ad}(E))$. $\EUW$ is a flat hyper-Kahler manifold;
for a detailed account see section 4.1 of \kapwit.  Let $\EUG$ be
the group of gauge transformations of $E$. Then $\EUG$ acts on
$\EUW$, preserving the hyper-Kahler structure, with hyper-Kahler
moment map $\vec\mu=(\mu_I,\mu_J,\mu_K)$:
\eqn\joget{\eqalign{\mu_I & =-{1\over
2\pi}\int_C\,\Tr\,\epsilon(F-\phi\wedge \phi),\cr \mu_J &
=-{1\over 2\pi}\int_C|d^2z|\,\Tr\,\epsilon\left(D_z\phi_{\bar
z}+D_{\bar z}\phi_z\right), \cr \mu_K& = -{i\over
2\pi}\int_C|d^2z|\,\Tr\,\epsilon\left(D_z\phi_{\bar z}-D_{\bar
z}\phi_z\right).\cr}} Here $\epsilon\in \Omega^0(C,{\rm ad}(E))$
is an element of the Lie algebra of $\EUG$; $\vec\mu$ is linear in
$\epsilon$. Given a hyper-Kahler manifold $\EUW$ and a group
$\EUG$ that acts on $\EUW$, preserving the hyper-Kahler structure,
with moment map $\vec\mu$, the hyper-Kahler quotient
$\EUW/\neg/\neg/\EUG$  is defined in general \ref\rochitchin{N. J.
Hitchin, A. Karlhede, U. Lindstrom, and M. Rocek, ``Hyperkahler
Metrics and Supersymmetry,'' Commun. Math. Phys. {\bf 108} (1987)
535-589.} as $\vec\mu^{-1}(0)/\EUG$. This hyper-Kahler quotient is
a hyper-Kahler manifold. In the present case,  the equations
$\vec\mu=0$ (for all $\epsilon$) are precisely Hitchin's equations
\torgo. The hyper-Kahler quotient $\EUW/\neg/\neg/\EUG$ is hence
precisely the moduli space $\MH$ of solutions of Hitchin's
equations.

This story can be repeated for Higgs bundles with an isolated singularity of the familiar sort
\refs{\konno,\nakajima}. We pick a point $p\in C$, and a reduction
of the structure group of the bundle $E$ at $p $ to a torus $\TT$,
and we denote as $\EUG_p$ the subgroup of $\EUG$ consisting of
gauge transformations whose restriction to $p$ lies in $\TT$. We
also pick a trio\foot{As explained in section \defn, $\MH$ really depends on $\alpha$
only via its image in $\frak t/\Lambda_{\cochar}=\TT$.}
$(\alpha,\beta,\gamma)\in \frak t$, the Lie
algebra of $\TT$. Picking coordinates $r,\theta$ near $p$, we denote
as $\EUW(\alpha,\beta,\gamma;p)$ the space of pairs $(A,\phi)$
with the familiar sort of singularity near $p$:
\eqn\ozorgo{\eqalign{A & =\alpha\,d\theta+\dots\cr
 \phi & = \beta\,{dr\over r}-\gamma\,d\theta+\dots.\cr}}
 The ellipses refer to terms less singular than $1/r$.
 And we denote as $\EUG_p$ the group of gauge transformations of
 the bundle $E$ which at $p$ take values in $\TT$. These are the gauge transformations
 that preserve the condition \ozorgo.
 Then the moduli
 space $\M_H(\alpha,\beta,\gamma;p)$ is the hyper-Kahler quotient
 $\EUW(\alpha,\beta,\gamma;p)/\neg/\neg/\EUG_p$.  We call this the moduli space of Higgs
 bundles with ramification at the point $p$.  One can similarly construct as
 a hyper-Kahler quotient the analogous moduli space of Higgs
 bundles with ramification at several points, that is, the
 moduli space of solutions of Hitchin's equations
 with singularities at several points $p_1,\dots,p_s\in C$,
 labeled by parameters $(\alpha_i,\beta_i,\gamma_i)\in \frak t$,
 $i=1,\dots,s$.  To keep the notation simple, we formulate the
 present section mainly for the case of ramification at one
 point, but all statements have direct analogs for the more
 general case.

A hyper-Kahler manifold, with complex structures $I,J,$ and $K$,
has corresponding Kahler forms $\omega_I$, $\omega_J$, and
$\omega_K$.  In the present case, these can be written, just as in the
absence of singularities,\foot{As in \kapwit, we write $\delta$ for the exterior derivative on
$\EUW$, and $d$ for the exterior derivative on $C$.} \eqn\tomog{\eqalign{\omega_I & =
-{i\over 2\pi}\int_C|d^2z|\,\Tr\left(\delta A_{\bar z}\wedge
\delta A_z-\delta \phi_{\bar z}\wedge \delta\phi_z\right)\cr
&=-{1\over 4\pi}\int_C\,\Tr\left(\delta A\wedge \delta A
-\delta\phi\wedge\delta \phi\right) \cr
                      \omega_J & ={1\over 2\pi}\int_C |d^2z|\,\Tr\left(
                      \delta\phi_{\bar z}\wedge \delta A_z+\delta\phi_z\wedge
                      \delta A_{\bar z} \right)\cr
                      \omega_K & =
                      {i\over 2\pi}\int_C|d^2z|\,\Tr\left(\delta\phi_{\bar z}
                      \wedge\delta A_z-\delta\phi_z\wedge\delta A_{\bar z}\right)\cr
                      & = {1\over 2\pi}\int_C \,\Tr\,\delta\phi
                      \wedge\delta A.\cr}}

Just as in the absence of singularities (see section 4.1 of
\kapwit), $\omega_I$ is cohomologous to \eqn\zixo{\omega_I'  =
-{i\over 2\pi}\int_C|d^2z|\,\Tr\left(\delta A_{\bar z}\wedge
\delta A_z\right)=-{1\over 4\pi}\int_C\,\Tr\left(\delta A\wedge
\delta A \right),} which depends only on $A$ and so is a pullback from the moduli
space $\M$ of $G$-bundles.  Indeed, let $\lambda_I={1\over 4\pi}\int_C\Tr
\,\phi\wedge\delta \phi$. $\lambda_I$ is gauge-invariant, and
vanishes (after integration by parts and use of Hitchin's
equations) if contracted with a generator of
gauge-transformations, $\delta A=-d_A\epsilon,
\,\delta\phi=[\epsilon,\phi]$.  So it is the pullback of a
one-form on $\MH$, and the formula
$\omega_I-\omega_I'=\delta\lambda_I$ shows that $\omega_I$ and
$\omega_I'$ are cohomologous.

In the absence of singularities, the cohomology classes of
$\omega_J$ and $\omega_K$ vanish.  For example, one proves this for
$\omega_K$ by writing $\omega_K=\delta\lambda_K$, where
$\lambda_K={1\over 2\pi}\int_C\Tr \,\phi\wedge\delta A$.  This still
works in the presence of singularities as long as $\gamma=0$, but
for $\gamma\not=0$, $\lambda_K$ does not vanish if contracted with
the generator of a gauge transformation so is not the pullback of a
one-form on $\MH$. In fact, contraction with the generator of a
gauge transformation maps $\omega_K$ to ${1\over
2\pi}\int_C\Tr\,\phi\wedge (-d_A\epsilon)= \Tr\,\gamma \epsilon(p)$,
where we have integrated by parts and used the fact that
$d_A\phi=-2\pi \gamma \delta_p$.  So the cohomology class of
$\omega_K$ is nonzero for $\gamma\not=0$. Similarly, the cohomology
class of $\omega_J$ is nonzero for $\beta\not=0$.  We will describe
all of these cohomology classes in section \topology.

\bigskip\noindent{\it Generalization Of Type $\Bbb{L}$}

We can readily extend this to incorporate a general Levi subgroup $\Bbb{L}$.

We restrict the parameters $(\alpha,\beta,\gamma)$ to be $\Bbb{L}$-invariant.
We define $\EUG_p$ to be the group
of gauge transformations whose restriction to $p$ lies in $\Bbb{L}$ (rather than in $\TT$, as above).
The hyper-Kahler quotient $\EUW(\alpha,\beta,\gamma;p)/\neg/\neg/\EUG_p$,
carried out exactly as above,
now gives us what we will call $\M_{H,\Bbb{L}}$, the moduli space of Higgs bundles with ramified
structure of type $\Bbb{L}$.
See \konno\ for a rigorous
construction. When we do not indicate the ramified type explicitly, this means
that we are taking $\Bbb{L}=\TT$.

One has to be careful in defining $\M_{H,\Bbb{L}}$ and require
that the solution deviate from the asymptotic form determined by $(\alpha,\beta,\gamma)$
 by terms that
are small compared to $1/r\ln r$, not just small compared to $1/r$.  The purpose of
this is to avoid the type of asymptotic behavior shown in eqn. \yeflo\ below.

The reason that we have introduced a special notation for
$\M_{H,\Bbb{L}}(\alpha,\beta,\gamma;p)$ is that, as will
hopefully be clear in section \topology, it is not the limit of
$\M_H(\alpha,\beta,\gamma;p)$ (defined for a regular triple
$(\alpha,\beta,\gamma)$) as $(\alpha,\beta,\gamma)$ approach $\Bbb{L}$-invariant values.
The two spaces do not even have the
same dimension.  Rather, their relation turns out to be that, when the trio
$(\alpha,\beta,\gamma)$
becomes $\Bbb{L}$-invariant,
$\M_{H,\Bbb{L}}$ is a locus of
singularities of $\MH$.

A special case of this definition is the case $\Bbb{L}=G$. Ramified
structure of type $G$ can be no ramification at all, since the
group of gauge transformations that at $p$ take values in $G$ is
simply the group of all gauge transformations.  However, as discussed at the end of
section \levi, if $G$ has a non-trivial center
${\cal Z}$,
our definitions lead to a slight generalization.  Indeed, the triple $(\alpha,\beta,\gamma)
\in \TT\times \frak t\times \frak t$
is $G$-invariant precisely if $\beta=\gamma=0$ and $\alpha$ is an element of
${\cal Z}$.  Thus $\M_{H,G}$ is a union of  components labeled by
${\cal Z}$.  One of these components (corresponding to the identity element of
$\cal Z$) is the moduli space $\MH$ of ordinary Higgs bundles with
no singularity at $p$.

The opposite extreme
is  $\Bbb{L}=\TT$. In this case,
$\M_{H,\Bbb{L}}(\alpha,\beta,\gamma;p)$ is what we usually denote simply as
$\M_H(\alpha,\beta,\gamma;p)$.  This is the generic case that we
have in mind when we speak of ramified Higgs bundles without specifying
$\Bbb{L}$.

Our point of view is that a surface operator supported on a surface $D$
is defined by the choice of the group $\Bbb{L}$ of
Levi type.  A particular surface operator leads to a particular quantum field theory problem
and a particular moduli space $\M_{H,\Bbb{L}}$ of ramified Higgs bundles.  In the theory of
$\M_{H,\Bbb{L}}$,
an emphasis is sometimes placed on the parabolic structure, a notion that we will explain in
section \parabolic.
As we will see (and as one may anticipate from the definition of parabolic subgroups
in section \levi), the interpretation in terms of parabolic structure depends on
$\alpha$ having an $\Bbb{L}$-regular value.  One can \nakajima\ interpolate from
one parabolic type to another,  keeping $\Bbb{L}$
fixed, without meeting a singularity, by varying $\beta$ and $\gamma$ as well as $\alpha$.
So the same surface operator can lead to different kinds of parabolic structure,
and it is better to label the surface operators by the Levi type rather than the parabolic type.
That is why we prefer to speak of the moduli space of ramified Higgs bundles, rather than the
moduli space of parabolic Higgs bundles.  We will, however, speak of parabolic Higgs bundles
when emphasizing the role of complex structure $I$, where this terminology is standard.

\subsec{Complex Viewpoint} \subseclab\cpxview

One can often  \rochitchin\ learn more about a hyper-Kahler quotient by focusing
on one of the complex structures, for example $I$. The function
 $\nu_I=\mu_J+i\mu_K$ is holomorphic in complex structure $I$.  The
hyper-Kahler quotient $\EUW/\neg/\neg/\EUG_p$ is the same as the
symplectic quotient of $\nu_I^{-1}(0)$ by $\EUG_{p}$. (This symplectic
quotient is defined by setting to zero the ordinary moment map
$\mu_I$, restricted to $\nu_I^{-1}(0)$, and then dividing by $\EUG_p$. At this point, one
has set to zero both $\nu_I$ and $\mu_I$, giving the hyper-Kahler quotient.)
Via the usual relation of symplectic and complex analytic quotients,
the symplectic quotient of $\nu_I^{-1}(0)$ by $\EUG_{p}$ can also be obtained as
a geometric invariant theory quotient $\nu_I^{-1}(0)/\EUG_{p,\C}$,
where $\EUG_{p,\C}$ is the complexification of $\EUG_p$. (Because
$\EUW$ is an affine space with linear action of $\EUG_p$, the
complexification $\EUG_{p,\C}$ of $\EUG_p$ has a natural action on $\EUW$ once
one selects one of the complex structures of $\EUW$.)  All of
these statements have analogs if complex structure $I$ is replaced
by any one of the complex structures that form part of the
hyper-Kahler structure of $\EUW$.

In the present case, $\nu_I^{-1}(0)/\EUG_{p,\C}$ can be given a
holomorphic description, as a moduli space of stable parabolic Higgs bundles $(E,\varphi)$.
This is explained  in
\refs{\simpson,\konno}.  For our purposes, we do not really need the details,
but we do need the fact that such a description exists.
Roughly, a parabolic Higgs bundle, with parabolic structure
 at a specified point $p\in C$, is a Higgs bundle
$(E,\varphi)$, where $E$ is a holomorphic $G_\C$-bundle over $C$, and $\varphi$
is a  section of $K_C\otimes {\rm ad}(E)$ that is holomorphic away from $p$, and has
a simple pole at $p$ that obeys a certain condition.
If $\sigma=\half(\beta+i\gamma)$ is regular, meaning that it commutes precisely with $\TT$
(or  more generally with $\Bbb{L}$),
the condition can be anticipated from eqn. \difo\ and
is that the polar part of $\varphi$ is conjugate to $\sigma\,dz/z$.  (Here $z$ is a local
holomorphic parameter near $p$.)
We postpone to section \postpone\ an explanation of what happens if $\sigma$ is not regular.

The condition for a parabolic Higgs bundle to be stable depends on
$\alpha$ in general, but is independent of $\alpha$ if $\sigma$ is
generic \nakajima.  In Kahler geometry, one would expect wall-crossing phenomena at special
values of $\alpha$, but in hyper-Kahler geometry this is avoided for generic $\sigma$,
as we explain in section \topology.

The moduli space of stable parabolic Higgs bundles, which we
temporarily denote $\tilde\M_H$,  is a complex symplectic manifold,
with holomorphic symplectic form $\Omega_I=\omega_J+i\omega_K$.
Moreover, the complex structure of $\tilde\M_H$ and the cohomology
class of $\Omega_I$ are manifestly holomorphic in $\beta+i\gamma$.
They are also manifestly independent of $\alpha$ (because $\alpha$
only affects the stability condition) if $\sigma$ is generic.

Since
$\MH(\alpha,\beta,\gamma;p)$, viewed as a complex manifold in
complex structure $I$, is the same as $\tilde\M_H$, all of these statements
have immediate implications for $\MH(\alpha,\beta,\gamma;p)$.
They explain certain claims made
in the table in section \geomint. In particular,
complex structure $I$ on $\MH$ varies holomorphically with
$\beta+i\gamma$, and is independent of $\alpha$.  Moreover, if we
write $[\,\omega]$ for the cohomology class of a closed differential
form $\omega$, then the cohomology class $[\Omega_I]$ varies
holomorphically with $\beta+i\gamma$ and is independent of
$\alpha$.  In particular, the classes $[\,\omega_J]$ and
$[\,\omega_K]$ are independent of $\alpha$.

\bigskip\noindent{\it Analog In Other Complex Structures}

Of course, there is a similar story in complex structure
$J$. The quotient $\nu_J^{-1}(0)/\EUG_{p,\C}$ can be given a
holomorphic interpretation as the moduli space of stable parabolic
(or filtered) local systems.  A parabolic local system is a
 $G_{\C}$-valued flat connection on $C\backslash p$ ($C$ with the point $p$
omitted), with a constraint on the monodromy around $p$.  We call
this monodromy $V$. If the element $U=\exp(-2\pi(\alpha-i\gamma))$
of $G_\C$ is regular (which means that the subgroup of $G_\C$ that
commutes with it is precisely the torus $\TT$),  the constraint on
the monodromy is that $V$ must be conjugate to $U$, as we would
expect from eqn. \nutella. Otherwise, one needs a more careful
description, which we postpone to section \postpone. The condition
for a parabolic local system to be stable depends on $\beta$ in
general, but is independent of $\beta$ if $\gamma+i\alpha$ is
generic \nakajima.

The moduli space $\hat \M_H$ of stable parabolic local systems is a complex
symplectic manifold, with holomorphic symplectic form
$\Omega_J=\omega_K+i\omega_I$.  We now make the same argument as in complex structure $I$.
The fact that  $\M_H$, viewed as a complex
manifold in complex structure $J$, is the same as $\hat \M_H$,
makes obvious certain claims made
in the table in section \geomint.  In particular, the
complex structure $J$ on $\MH$ varies holomorphically with
$\gamma+i\alpha$ and is independent of $\beta$.  Similarly, the
cohomology class $[\Omega_J]$ is holomorphic in $\gamma+i\alpha$,
and independent of $\beta$.  Therefore $[\,\omega_K]$ and
$[\,\omega_I]$ are independent of $\beta$.

We can apply the same reasoning, of course, in complex structure
$K$, to show that the complex structure $K$ and symplectic
structure $\Omega_K$ of $\MH$ are holomorphic in $\alpha+i\beta$
and independent of $\gamma$.  In particular, therefore, the
cohomology classes $[\,\omega_I]$ and $[\,\omega_J]$ are independent
of $\gamma$.  This completes the justification of the claims made
in the table in section \geomint.

\bigskip\noindent{\it Linearity of The Cohomology Classes}

We can get considerably farther using the fact that the same reasoning
applies for any of the complex structures that make up the hyper-Kahler
structure of $\MH$.
These are complex structures of the form $\hat I=pI+qJ+rK$,
with real parameters $p,q,$ and $r$ obeying
$p^2+q^2+r^2=1$.
All statements in the table of section \geomint\ have analogs for any $\hat I$.
Indeed, these statements remain valid if we make an
$SO(3)$ rotation of the space spanned by the complex structures
$I,J,K$, along with the same rotation of the spaces spanned by the
three symplectic structures $\omega_I$, $\omega_J$, and $\omega_K$
and by the three variables $\alpha,$ $\beta$, and $\gamma$.  This
$SO(3)$ is not a symmetry of $\MH$, but it is a symmetry of the
reasoning we have used in deducing the statements in the table.

This makes possible some simple inferences about the
cohomology classes $[\,\omega_I]$, $[\,\omega_J]$, and $[\,\omega_K]$.  Let us write $x_i$,
$i=1,2,3$, for those cohomology classes.
Thus the $x_i$ take values in the vector space
$H^2(\MH,\R)$. They are functions of $\alpha,\beta$, and $\gamma$,
which we  will denote as $y_i$, $i=1,2,3$. Then what we have
established so far is that \eqn\jus{{\partial x_i\over\partial
y_j}=0,~~i\not= j.} However, by making the same argument in a generic complex
structure $\hat I$, we learn that
\jus\ also holds after making an $SO(3)$ rotation on $\vec
x=(x_1,x_2,x_3)$ along with $\vec y=(y_1,y_2,y_3)$.  This implies
that \eqn\bus{{\partial x_i\over\partial y_j}=\delta_{ij}v} for
some function $v$ (which takes values in $H^2(\MH,\R)$).\foot{
This is equivalent to the statement that for any complex structure $\hat I=pI+qJ+rK$, the
cohomology class of the
 corresponding holomorphic $(2,0)$-form $\Omega_{\hat I}$ (which is obtained from
$\Omega_I$ by a suitable
$SO(3)$ rotation) is independent of the Kahler form (which is a multiple of
$p\alpha+q\beta+r\gamma$).}  By
differentiating again, we learn from \bus\ that \eqn\nus{\delta_{ij}{\partial
v\over \partial x_k}=\delta_{ik}{\partial v\over\partial x_j}.}
But this implies (by considering the case $i=j$,
$i\not= k$) that $\partial v/\partial x_k=0$ for all $k$, so that
$v$ is constant.

Since $v$ is a constant, \bus\ implies that the $x_i$ are linear
functions of the $y_j$. In particular, $[\,\omega_I]$, which we
already know to depend only on $\alpha$, is actually a {\it linear}
function of $\alpha$. Thus \eqn\xoco{[\,\omega_I]=a+\Tr\,\alpha h}
with some constants $a$ and $h$.  ($a$ takes values in
$H^2(\MH,\R)$, and $h$ in $H^2(\MH,\R)\otimes {\frak t}$.)
Similarly, we already know that $[\,\omega_J]$ only depends on $\beta$
and vanishes at $\beta=0$; we can now deduce from \bus\ that
\eqn\oco{[\,\omega_J]=\Tr\,\beta h} with the same $h$ as in \xoco. And
by the same token, \eqn\noco{[\,\omega_K]=\Tr\,\gamma h.} We will
describe $a$ and $h$ in sections \topres\ and \topology.

\bigskip\noindent{\it $\MH$ As A Symplectic Variety}

  As we have just seen, the cohomology
class $[\,\omega_K]$ is independent of $\alpha$ and $\beta$.  If $\MH$ were compact, it would
follow that $\MH$ as a real symplectic variety with symplectic structure $\omega_K$ is independent
of $\alpha$ and $\beta$.  Since $\MH$ is noncompact, this conclusion does not follow just
from constancy of the cohomology class of the symplectic form, but it does follow
by considering more carefully the facts we used to prove this constancy.

To change $\alpha$ without changing the real symplectic variety $(\MH,\omega_K)$, we simply
view $\MH$ in complex structure $I$.  Because of the interpretation via parabolic Higgs
bundles, this complex structure and the corresponding  holomorphic
two-form $\Omega_I=\omega_J+i\omega_K$ are independent of $\alpha$, as long as certain
singularities are avoided.  So we can vary $\alpha$ keeping fixed the real symplectic variety
$(\MH,\omega_K)$.

Similarly, to vary $\beta$ without changing the real symplectic variety $(\MH,\omega_K)$,
we view $\MH$ in complex structure $J$.  Because of the interpretation via filtered local
systems, this complex structure and the corresponding   holomorphic
two-form $\Omega_J=\omega_K+i\omega_I$ are independent of $\beta$, as long as certain
singularities are avoided.  So we can vary $\beta$ keeping fixed the real symplectic variety
$(\MH,\omega_K)$.

The singularities that have to be avoided in this process are described in section \topology\
and are of real codimension at least two.  So we conclude that the real symplectic
variety $(\MH,\omega_K)$ is independent of $\alpha$ and $\beta$, an important result for
applications to the geometric Langlands program.

An important point is that the isomorphism given by this argument is not canonical but depends
on the path by which $\alpha$ and $\beta$ are varied.  The reason is simply that we keep
one structure fixed in varying $\alpha$ and a different structure fixed in varying $\beta$.
If we vary both $\alpha$ and $\beta$, no structure except $\omega_K$ is held fixed, and
$\MH$ varies by a symplectomorphism.
In varying $\alpha$ and $\beta$ around a closed loop that avoids the singularities,
one will in general get a  symplectomorphism of $\MH$ that is ``topologically'' non-trivial,
that is, it cannot be deformed to the identity by a family of symplectomorphisms.
We will develop this idea
systematically in sections \actweyl\ and \opmon\
to get an action of the affine braid group on branes on $\MH$.

\bigskip\noindent{\it $\C^*$ Action}

The $SO(3)$ group that rotates the space of complex or symplectic
structures of $\MH$ is not a symmetry of $\MH$.  However, a
subgroup of it is a symmetry, just as for ordinary Higgs bundles
without parabolic structure, where this is described in \hitchin,
pp. 107-8.

Let us first consider the case that $\beta=\gamma=0$.  Let the group
${\cal U}_1\cong U(1)$ act on $(A,\phi)$ by leaving $A$ invariant
and transforming $\varphi\to\lambda\varphi$, with $|\lambda|=1$.
Since $\phi=\varphi+\bar\varphi$ (where $\varphi$ and $\bar\varphi$
are of type $(1,0)$ and $(0,1)$), this determines the transformation
of $\phi$: \eqn\zonk{\phi\to\lambda \varphi+\bar\lambda
\bar\varphi.}
   The action of ${\cal U}_1$
   leaves invariant the characterization \zorgox\ of the singularity
(as long as $\beta=\gamma=0$).  It leaves invariant the moment map $\mu_I$, while
rotating the $\mu_J-\mu_K$ plane.  So it gives a manifest symmetry of the hyper-Kahler
moment map construction, and a group of symmetries of the hyper-Kahler metric on $\MH$.
In fact, ${\cal U}_1$ is an $SO(2)$ subgroup of the $SO(3)$ that rotates the three
complex structures.  It acts on the family of complex structures $\hat I=pI+qJ+rK$ by
leaving $p$ fixed and rotating the $q-r$ plane.  Alternatively, if we parametrize
the family of complex structures by a complex variable $w$,
setting
\eqn\wyre{I_w={1-\bar w w\over 1+\bar w
w}I+{i(w-\bar w)\over 1+\bar w w}J+{w+\bar w\over 1+\bar w w}K,}
then $\C^*$ acts on the parameter $w$ by $w\to\lambda^{-1}w$.

If we relax the condition $|\lambda|=1$, we no longer get a symmetry of the hyper-Kahler
metric of $\MH$.  However, exactly as in \hitchin, we get a group ${\cal U}\cong \C^*$
that acts on $\MH$ preserving the complex structure $I$ and transforming the family
$I_w$ by $w\to\lambda^{-1}w$, just as for $|\lambda|=1$.  The fixed
points are $w=0$, $I_w=I$, and $w=\infty$, $I_w=-I$.  All other complex structures
$I_w$ are equivalent
under the action of ${\cal U}$.

All this carries over to the case $\beta,\gamma\not=0$, except that the
parameters $\beta$ and $\gamma$ must be transformed by $(\beta+i\gamma)\to\lambda(\beta
+i\gamma)$.  This follows for $|\lambda|=1$
by observing that the transformation \zonk\ leaves fixed the singularity
\zorgox\ if $\beta$ and $\gamma$ are transformed as claimed.

\subsec{The Non-Regular Case} \subseclab\postpone

Now we will describe what happens to some of the above statements when the pair
$(\alpha,\gamma)$ or the pair $(\beta,\gamma)$ is non-regular.
This may help the reader understand the
constructions of \simpson, and is useful background in some of the
applications to the geometric Langlands program.

\bigskip\noindent{\it Complex Structure $J$}

We begin with complex structure $J$, in which a
solution of Hitchin's equations corresponds to a parabolic
local system.  Also, for simplicity, we consider first the basic case that the Levi
subgroup used to define our surface operator is $\Bbb{L}=\TT$.

 We consider a solution of Hitchin's equations with an isolated singularity
at a point $p\in C$.   We pick coordinates near $p$ as in section
\defn\ (so $p$ is defined by $z=0$ where $z=x_1+ix_2=re^{i\theta}$),
and we assume a solution that behaves near $r=0$ as
\eqn\zorgox{\eqalign{A & = \alpha \,d\theta +\dots\cr
           \phi & = \beta\,{dr\over r}-\gamma\,d\theta+\dots,\cr}}
where the ellipses refer to terms that are less singular than $1/r$
as $r\to 0$.  Hitchin's equations ensure that the $G_\C$-valued
connection $\CA=A+i\phi$ is flat.

For $b>0$ (but small enough so that the coordinates $r,\theta$ are
defined for $r\leq b$), let $C_b$ be the circle $r=b$, and let $V_b$
be the monodromy around $C_b$ of the flat connection $\CA$. The
$V_b$ have two basic properties: (1) The conjugacy
class of $V_b$ is independent of $b$.  This is so because the
connection is flat. Since we only care about the monodromy up to
conjugacy, we pick any one of the $V_b$ and call it $V$, the
monodromy around the singularity. (2) If we set
$U=\exp\left(-2\pi(\alpha-i\gamma)\right)$, then
\eqn\doofus{\lim_{b\to 0}V_b=U.} This is so because in the limit of
$b\to 0$, we can compute the holonomy just from the most singular
terms in $A$ and $\phi$, and if we do so then the result is $U$.

If $U$ is regular, the two properties imply that $V$ is conjugate to
$U$. For $SL(2,\C)$, we can prove this as follows.  The two properties imply that
$\Tr \,U=\Tr\, V$ (where the trace is taken in the two-dimensional representation); if $U$ is
regular, this implies that $U$ and $V$ are conjugate.  Similar reasoning holds for any $G$,
with the trace replaced by a full set of invariant functions.
 If $U$ is not regular, $U$ and $V$ need not be conjugate.  For example, for
$G=SL(2,\C)$, let us consider the non-regular element $U=1$, which
corresponds to $\alpha=\gamma=0$.  For $U=1$, we can satisfy the two
conditions with \eqn\etry{V_b=\left(\matrix{1& b\cr 0 &
1\cr}\right).}  The $V_b$ for $b>0$ are all conjugate, and
$\lim_{b\to 0}V_b=1$.  An element $V$ of this form is called
unipotent (this means simply that $V-1$ is nilpotent).

So it is
possible for a solution of Hitchin's equations with
$\alpha=\gamma=0$, $\beta\not=0$ to have monodromy that is unipotent
but not equal to 1.  Not only is this possible, but it is the
generic behavior, simply because the condition for $V$ to be
unipotent is one complex condition (which one can formulate as
$\Tr\,V=2$), while for $V$ to equal 1 is three complex conditions.
Comparing these dimensions, one might think  that if
$\alpha=\gamma=0$, then those Higgs bundles for which $V$ is
actually 1 would be a family of complex codimension 2.  A more
careful analysis shows that this is correct if $\beta=0$ (in which
case the Higgs bundles with $V=1$ are a locus of $A_1$ orbifold
singularities), but that for $\beta\not=0$ the locus with $V=1$ is
``blown up,'' and is of complex codimension 1. The statements about the generic
behavior, the singularity,
and the blowup should become clearer below, especially in section \topology.  The blowup is
described by specifying what Simpson \simpson\ calls a
``filtration'' of the local system.

Let $\frak C$ be the conjugacy class in $G_\C$ containing $U$.
There is, for any
$G$ and any choice of ${\frak C}$, a finite set of conjugacy classes
${\frak C}_\lambda$, $\lambda=1,\dots, s$, with the property that a family of
elements $V_b\in {\frak C}_\lambda$ can have a limit in ${\frak C}$ for $b\to
0$. Differently put, $\frak C$ is in the closure of ${\frak C}_\lambda$.
We call the ${\frak C}_\lambda$ the conjugacy classes that are affiliated to
${\frak C}$.  For example, if $U$ is regular, the only affiliated
conjugacy class is ${\frak C}$ itself.  At the other extreme, if $U=1$,
the affiliated conjugacy classes are precisely the ones that
parametrize unipotent elements of $G$. For $G$ of large rank, there
are many such classes (given for $SL(N,\C)$ by block triangular
matrices with 1's on the diagonal and blocks of different sizes).

In any event, there is always a unique affiliated
conjugacy class ${\frak C}^*$ of maximal dimension, in fact of dimension
${\rm dim}\,G-r$, where $r$ is the rank of $G$. It is called a
regular conjugacy class, because it parametrizes elements that are
regular, that is they commute with only an $r$-dimensional subgroup
of $G_\C$.

The class $\frak C$ may not be regular, but it has another
distinguishing property. Among the affiliated conjugacy classes,
$\frak C$ is the unique one that is ``semi-simple,'' that is, it
parametrizes group elements that are semi-simple (they can be
conjugated to the maximal torus).  $\frak C$ is certainly
semi-simple, since it contains $U=\exp(-2\pi(\alpha-i \gamma))$,
which is an element of $\TT$.

But if $U$  is not regular, then ${\frak C}\not={\frak C}^*$,
and elements of the regular
conjugacy class affiliated to $U$ are not semi-simple.   For example, if
$G=SU(N)$ and $U=1$, then the regular conjugacy class affiliated to
$U$ contains the ``principal unipotent element''
\eqn\yfo{V=\left(\matrix{1& 1& 0&\dots&0\cr
                         0&1&1&\dots&0\cr
                           & & &\vdots&\cr
                           0&0&0&\dots&1\cr}\right),}
with $1$'s on and just above the main diagonal and zeroes elsewhere.
For any $\alpha$ and $\gamma$, the generic Higgs bundle gives a
local system whose monodromy is in the regular conjugacy class
associated to $U$.

\bigskip\noindent{\it Complex Structure $I$}

Now we consider complex structure $I$, in which a solution of Hitchin's equations is
a parabolic Higgs bundle.

Here it will be helpful to begin by considering in detail the example of $SL(2,\C)$.
We take $\beta=\gamma=0$ (this being the only non-regular choice of $\beta$ and $\gamma$
for $SL(2,\C)$), and as we want to assume that $\alpha$ is regular, we take
\eqn\neuro{\alpha=iy\left(\matrix{1 & 0\cr 0&-1\cr}\right),}
with $0<y<1/2$.  (The reason for the factor of $i$ is that as $\alpha$ takes values in the
real Lie algebra of $SU(2)$, it is anti-hermitian in a unitary representation.)
The limiting form of the solution of Hitchin's equations for $r\to 0$ is therefore
\eqn\buro{\eqalign{A&=iy\,d\theta\left(\matrix{1 & 0\cr 0&-1\cr}\right)\cr
                   \phi&=0.}}
We want to see what sort of pole $\varphi$, defined as the $(1,0)$ part of $\phi$,
can acquire when we replace \buro\ with a more general solution of Hitchin's equations
that has the same asymptotic behavior as $r\to 0$.

At first sight, it might appear that no pole at all is possible.  By definition, we want
to perturb the limiting solution \buro\ by terms that are less singular than $1/r$.  So
it may seem that $\varphi$ will not be sufficiently singular to have a pole. However,
to decide whether $\varphi$ has a pole, we need to trivialize the holomorphic structure of
the bundle $E$ near the singular point $p$.  It turns out that once this is done, $\varphi$
can have a pole.

When we expand Hitchin's equations around the solution \buro, the linearized
equations have a solution
\eqn\truro{\phi=\epsilon {dz\over z}(\bar z z)^{y}\left(\matrix{0&1\cr 0&0\cr}\right),}
with $\epsilon$ a small parameter.  This solution is less singular than $1/r$, so including
this perturbation is compatible with the asymptotic behavior
\buro.  (Of course, to get a real solution for $\phi$,
one must subtract the hermitian conjugate solution.
The $(1,0)$ part of $\phi$ must be upper triangular, since otherwise an analogous solution
is more singular than $1/r$ at $r=0$.)

To trivialize
the holomorphic bundle $E$ near $r=0$, we write down the appropriate
 $\bar\partial_A$ operator that defines the holomorphic structure of $E$:
\eqn\nuro{\bar\partial_A=d\bar z\left({\partial\over\partial\bar
z}+A_{\bar z}\right)= d\bar z\left({\partial\over\partial\bar
z}-{y\over 2\bar z}\left(\matrix{1&0\cr 0&-1\cr}\right) \right).} We
can write this as \eqn\pillo{\bar\partial_A=f\bar\partial f^{-1},}
where $\bar\partial =d\bar z\partial/\partial\bar z$ is the standard
$\bar\partial$ operator, and \eqn\zillo{f=\left(\matrix{(\bar z
z)^{y/2}& 0\cr 0& (\bar z z)^{-y/2}\cr}\right).} So if
$\bar\partial_A\varphi=0$ (which is part of Hitchin's equations),
then $\bar\partial(f^{-1}\varphi f)=0.$ But \eqn\illo{f^{-1}\varphi
f= \epsilon {dz\over z}\left(\matrix{0&1\cr 0&0\cr}\right).} So the
conclusion is that $\varphi$ can have a pole, relative to the
trivialization of the bundle $E$, but the residue of this pole is
strictly upper triangular (in a basis in which $-i\alpha$ is
diagonal with decreasing eigenvalues along the diagonal).

In this discussion, we started with $\beta=\gamma=0$, so $\sigma={1\over 2}(\beta+i\gamma)$
also vanishes.  The residue of the pole in $\varphi$ turned out to be
\eqn\cillo{\tau=\epsilon\left(\matrix{0 & 1\cr 0 & 0 \cr}\right).}
The conjugacy class of $\tau$ is independent of $\epsilon$, and the limit of $\tau$ for
$\epsilon\to 0$ is $\sigma$.  These properties imply that all invariant polynomials take the
same value for $\sigma$ and $\tau$; for $SL(2,\C)$, this simply means that $\Tr\,\sigma^2=
\Tr\,\tau^2=0$.

In general, for any $G$, let $\frak c$ be the orbit or conjugacy class in the
Lie algebra $\frak g_\C$ that contains $\sigma={1\over
2}(\beta+i\gamma)$.  This conjugacy class parametrizes semi-simple
elements of $\frak g_\C$ (that is, elements  that can be conjugated
to a maximal torus) since $\sigma$ itself is semi-simple. We say
that a conjugacy class ${\frak c}_\lambda$ is affiliated to $\frak
c$ if a sequence of elements of ${\frak c}_\lambda$ can converge to
an element of $\frak c$, or in other words if $\frak c$
is in the closure of ${\frak c}_\lambda$.
In general, there are finitely many
conjugacy classes affiliated to $\frak c$.
The residue $\tau$ of the pole of the
Higgs field always takes values in an affiliated conjugacy class.
 This is true by reasoning
similar to what we have explained in the above example.

If $\sigma$ is regular as well as semi-simple, then $\frak c$ itself
is the only affiliated conjugacy class, and in particular $\tau$ is
conjugate to $\sigma$.  At the opposite extreme, if $\sigma=0$, then
the conjugacy classes affiliated to $\frak c$ are precisely the
classes of nilpotent elements of $\frak g_\C$.

For every $\frak c$, there is a unique affiliated conjugacy class
${\frak c}^*$ of maximal dimension, in fact dimension ${\rm
dim}(G)-r$. It parametrizes regular elements of $\frak g_\C$, that
is, elements that commute with only an $r$-dimensional subgroup of
$G_\C$. For example, if $\sigma=0$ and $G=SU(N)$, then the
affiliated regular conjugacy class contains the element $v=V-1$,
where $V$ was defined in \yfo.  For any $\sigma$, the generic
parabolic Higgs bundle has a pole whose residue $\tau$  is in the
regular affiliated conjugacy class ${\frak c}^*$.

We can summarize much of this by saying that conjugacy classes in the Lie algebra
$\frak g_\C$ behave
in many relevant respects like conjugacy classes in the group $G_\C$.

\bigskip\noindent{\it Reformulation}

Going back to the $SU(2)$ example, our result about the polar behavior of $\varphi$
for the case $\beta=\gamma=0$
can be described as follows.  For $G=SU(2)$, the only possible Levi subgroup is $\Bbb{L}=\TT$,
which is the case considered
in the above discussion.  The choice of $\alpha$ determines a parabolic
subgroup $\EUP$ and  a unipotent radical $\EUN$.  In the above example, $\EUP$ is the group
of upper triangular matrices and $\EUN$ is the group whose Lie algebra
$\frak n$ consists of strictly upper triangular matrices.  The result \illo\ says that
the polar residue of $\varphi$ takes values in $\frak n$.  The mechanism by which this came
about is simply that in order for $\phi$ (in the differential geometric description)
to be less singular than $1/r$, its polar residue $\tau$ (in the holomorphic description) must
obey $-i[\alpha,\tau]=\lambda\tau$, $\lambda>0$.  The analog for $G$ of higher rank, still
assuming $\beta=\gamma=0$, is that $\tau$ must be a linear combination of elements of $\frak g$
that obey this condition.  So in other words, $\tau$ takes values in $\frak n$.  This holds
for any choice of Levi subgroup $\Bbb{L}$ and any $\Bbb{L}$-regular $\alpha$.

Going back to $SU(2)$, and without changing $\alpha$, let us perturb $\beta$ and
$\gamma$ to be nonzero, say $\half(\beta+i\gamma)={\rm diag}(q,-q)$
for some $q\in \C$.  What happens in this case?  We must set $\phi=\beta (dr/r)-\gamma\,d\theta$
plus terms that are less singular at $r=0$.  One possibility is to
have $\phi=\beta(dr/r)-\gamma\,d\theta$ exactly.  This corresponds
to $\varphi=(dz/z){\rm diag}(q,-q)$.  But just as in \truro, we can
make an upper triangular deformation\foot{For $q\not=0$, the details
of the solution are more complicated, and it is necessary to also
modify $A$.  The appropriate solution is described by Nahm's
equations; see section \onahm.  It remains true, as at $q=0$, that a
lower-triangular modification of $\varphi$ is more singular than
$1/r$.} of $\varphi$, so the general possibility for the polar part
of $\varphi$ is \eqn\eggo{\varphi\sim{dz\over z}\left(\matrix{q&
*\cr 0 & -q\cr}\right).}
The upper right element denoted $*$ does not affect the conjugacy
class of the residue of the pole if $q\not=0$.  Note that if
$\varphi$ and $\tilde\varphi$ are two Higgs fields with a pole of
this kind, then $\varphi-\tilde\varphi$ has a pole with $\frak n$-valued residue:
\eqn\gruffy{\varphi-\tilde\varphi\sim {dz\over
z}\left(\matrix{0&*\cr 0&0\cr}\right).}

By similar reasoning, this is so not just for $SL(2,\C)$, but
for any gauge group $G$, with any choice of Levi subgroup $\Bbb{L}$ and any $\Bbb{L}$-regular $\alpha$.
The general statement is that the polar residue of $\varphi$ takes values in $\frak p$, the
Lie algebra of $\EUP$, and is equal to $\sigma=\half(\beta+i\gamma)$
modulo an element of $\frak n$.  The last statement can be informally summarized by saying that
the ``eigenvalues'' of the polar residue coincide with those of $\sigma$.

\bigskip\noindent{\it More On Complex Structure $J$}

There is an asymmetry in our discussion of the non-regular behavior in complex structures
$J$ and $I$.  In complex structure $J$, we reasoned somewhat abstractly about the closures
of conjugacy classes, but in complex
structure $I$, we analyzed the behavior of perturbations of Hitchin's equations.
Of course, by analyzing Hitchin's equations, we can be more explicit about what happens in
complex structure $J$.  This will also enable us to get more information.

We carry out the discussion for any gauge group $G$ and Levi subgroup $\Bbb{L}$. To begin with,
we take $\alpha=\gamma=0$, but we take $\beta$ to be generic or in other words $\Bbb{L}$-regular.
This corresponds to a solution of Hitchin's equations with \eqn\remvo{A=0, ~\phi=\beta
{dr\over r}.}
We perturb this to
\eqn\polz{\eqalign{A&=a(r)\,d\theta,\cr \phi&=\beta{dr\over r}+c(r)\,d\theta,\cr}}
where $a$ and $c$ must vanish at $r=0$ (since the deviation from the limiting solution
\remvo\ must be less singular than $1/r$) and
we will work to first order in $a$ and $c$.
The resulting equations can be written
\eqn\zongo{r{d\over dr}(a+ic)=[-i\beta,a+ic].}
For $a+ic$ to vanish at $r=0$, it must take values in $\frak n$, the subspace of $\frak g$
spanned by vectors $\psi$ with $-i[\beta,\psi]=\lambda\psi$, $\lambda>0$.  Hence,
the monodromy of the flat connection ${\cal A}=A+i\phi$,
which in this approximation is $U=\exp(-2\pi(a+ic))$, takes values in $\EUN$,
the unipotent radical of the parabolic subgroup determined by $\beta$.

Every element of $\EUN$ is a unipotent element of $G_\C$. In the theory of semi-simple Lie
groups, it is shown that the generic element of $\EUN$ lies in a unipotent
conjugacy class in $G_\C$ called the Richardson class ${\frak C}_\Bbb{L}$.  {}From the definition,
it seems that the Richardson class depends on $\EUP$, and thus $\beta$, but it can be shown
that the Richardson class is actually determined by $\Bbb{L}$.
(In general, distinct $\Bbb{L}$'s can lead to the same
Richardson class.)  The monodromy $U$ found in the last paragraph is unconstrained except
for taking values in $\EUN$, so generically
it takes values in the conjugacy class ${\frak C}_\Bbb{L}$.

For example, if $\Bbb{L}=\TT$, then ${\frak C}_\Bbb{L}$ is the regular unipotent conjugacy class
described for $SL(N,\C)$ in \yfo.  An $N\times N$
 matrix $U$ such that $U-1$ is strictly upper triangular
is generically in this conjugacy class.  To give another example, for $SL(3,\C)$, if
$\Bbb{L}$ consists of matrices of the form
\eqn\xunpyx{\left(\matrix{* & * & 0 \cr
                        * & * & 0 \cr
                        0 & 0 & * \cr}\right)}
                        and $\alpha$ is such that
                        the unipotent radical $\EUN$ consists of matrices
                        \eqn\xunyx{\left(\matrix{1 & 0 &* \cr
                        0 & 1 & * \cr
                        0 & 0 & 1 \cr}\right),}
then a generic element of $\EUN$ is conjugate under $\Bbb{L}$ to
\eqn\xunlyx{\left(\matrix{1 & 0 & 0 \cr
                        0 & 1 & 1 \cr
                        0 & 0 & 1 \cr}\right).}
                        This is a representative of the Richardson conjugacy class ${\frak C}_\Bbb{L}$.
In general, for $SL(N,\C)$, every unipotent conjugacy class is a Richardson class, but this
is not true for other groups.

This discussion can be generalized to the case that $\alpha,\gamma\not=0$.
If $U=\exp(-2\pi(\alpha-i\gamma))$ is $\Bbb{L}$-regular, then the monodromy $V$ is conjugate to $U$.
In general, it takes the form $V=UN$, where $N$ takes values in the unipotent radical $\EUN$.
Equivalently, the monodromy
lies in a conjugacy class in $\EUP$ whose closure includes $U$.
Moreover, up to conjugacy of $V$,
one can assume that $U$ and $N$ commute.  Generically, $N$ is then
simply a general element of $\EUN$ that commutes with $U$.

If $N$ is of this form, then the orbit in $G_\C$ of the element $V=UN$ of $\EUP$
has the same dimension
as $G_\C/\Bbb{L}_\C$, which is the orbit in $G_\C$ of a generic $\Bbb{L}$-regular element of
$\TT_\C$.  We call elements of $\EUP$ that have this property $\Bbb{L}$-regular, so in particular,
for any $U$, the generic monodromy $V=UN$ of the local system in complex structure $J$ is
$\Bbb{L}$-regular.

For instance, Richardson orbits are $\Bbb{L}$-regular.  In the above $SL(3,\C)$ example,
the dimension of $G_\C/\Bbb{L}_\C$ is 4, which is also the dimension of the Richardson orbit
described in eqn. \xunlyx.  If a given unipotent orbit is the Richardson orbit of several
different Levi subgroups $\Bbb{L}_i$, then it is $\Bbb{L}_i$-regular for each $i$.

\subsec{Parabolic Bundles}
\subseclab\parabolic

At this point, we should perhaps explain a notion that is usually
taken as the starting point in the mathematical theory, but that we
have hidden so far. This is the notion of a parabolic bundle (as
opposed to a parabolic Higgs bundle). All statements have obvious
analogs with parabolic structure at several points, but for
simplicity we consider mainly the case of one point.  Until further notice,
we consider only gauge theory, without the Higgs field.

For motivation, we return to the $SU(2)$ example of section \postpone.  A holomorphic
section $s$  of the bundle ${\rm ad}(E)$ is an ${\rm ad}(E)$-valued function
annihilated by the $\bar\partial_A$ operator.  Given the explicit form \nuro\ of this
operator, this means that near $z=0$,
\eqn\bigeqop{s=\left(\matrix{u & v(\bar z z)^{y}\cr w(\bar z z)^{-y} & -u\cr}\right),}
where $u,v,$ and $w$ are ordinary holomorphic functions. If we want $|s|$ to be bounded
for $z\to 0$, we require that $w(0)=0$.  Hence at $z=0$, $\tilde s=f^{-1}sf$ takes the form
\eqn\onci{\tilde s=\left(\matrix{* & * \cr 0 & * \cr}\right).}

{}From a holomorphic point of view, what is happening is that the
choice of $\alpha$ determines in the fiber of ${\ad}(E)$ at $p$ a
Borel subalgebra $\frak b$,
spanned by vectors $\psi\in \frak g$ with $-i[\alpha,\psi]=\lambda\psi$, $\lambda
\geq 0$.   Eqn. \onci\ says
that $\tilde s(p)$ takes values in $\frak b$.
In this form, the result holds for any $G$: if $\alpha$ is regular, then $\tilde s(p)$
takes values in the Borel subgroup determined by $\alpha$.  More generally, if we pick
a Levi subgroup $\Bbb{L}$ and $\alpha$ is $\Bbb{L}$-regular, then by the same sort of reasoning,
$\tilde s(p)$ takes values in the Lie algebra $\frak p$ of the
parabolic subgroup $\EUP$ determined by the pair $(\Bbb{L},\alpha)$.

A choice of parabolic structure for a $G$-bundle $E\to C$ at a point
$p$ is simply a reduction of the structure group of $E$ at $p$ to a
parabolic subgroup $\EUP$. What we have just seen is that a
bundle with a singularity that in differential geometry is described by
$A=\alpha\,d\theta+\dots$ near a point $p\in C$, for $\Bbb{L}$-regular
$\alpha$, corresponds in complex geometry to a bundle with a choice of parabolic
structure at $p$.

A  theorem of Mehta and Seshadri \ref\mehta{V. B. Mehta and C. S. Seshadri, ``Moduli Of
Vector Bundles On Curves With Parabolic Structures,'' Math. Ann. {\bf 248} (1980) 205-39.}
(which generalizes a theorem of Narashimhan and Seshadri
\ref\nar{M. S. Narasimhan and C. S. Seshadri, ``Stable And Unitary Vector Bundles
On A Compact Riemann Surface,'' Ann. Math. (2) {\bf 82} (1965) 540-67.}
in the absence of parabolic
structure) puts this in a systematic framework. This theorem
establishes a one-to-one correspondence between stable parabolic
$G_\C$-bundles and flat $G$-bundles with the familiar singularity
$A=\alpha\,d\theta+\dots$. On the left hand side of this
correspondence, one considers a holomorphic $G_\C$-bundle with a
reduction of its structure group at a point $p\in C$ to a parabolic
subgroup $\EUP$.  We let $\Bbb{L}$ be the Levi subgroup
of $\EUP$, and we pick an $\alpha$ such that $\EUP$ is determined in the usual way by
the pair $(\Bbb{L},\alpha)$.  We assume that $\alpha$ is generic enough so that the subgroup  of
$G$ that commutes  with $U=
\exp(-2\pi\alpha)$ is precisely $\Bbb{L}$.   We
say that such an  $\alpha$ is strictly $\Bbb{L}$-regular.
For each such $\alpha$,
there is a natural notion of stability for bundles with parabolic structure of type
$\EUP$.\foot{
Moreover, up to
equivalence, this notion is invariant  under shifts of $\alpha$ by a lattice vector.
The equivalence in question involves a Hecke modification of the bundle.}
We will not describe the stability
condition here, though it is fundamental in the mathematical
theory. On the right hand side of the correspondence, one considers flat $G$-bundles on
$C\backslash p$  with monodromy around $p$
conjugate to $U=\exp(-2\pi\alpha)$. Equivalently, one considers
solutions of the familiar equation \eqn\urtu{F=2\pi\alpha\delta_p}
modulo gauge transformations that take values in $\Bbb{L}$ at the point
$p$.  We denote as $\M(\alpha;p)$ the moduli space of such flat
bundles with monodromy.  The theorem of Mehta and Seshadri is
that the moduli space of stable parabolic $G_\C$-bundles is the same
as $\M(\alpha;p)$.  The analogous theorem \nar\ in the absence of
parabolic structure says that the moduli space $\M$ of stable
$G_\C$-bundles over $C$ is the same as the moduli space of flat
$G$-bundles.

Both results are natural from the point of view of the symplectic
quotient of the space of gauge fields by the group of gauge
transformations \ref\abott{M. F. Atiyah and R. Bott, ``The Yang-Mills Equations Over
Riemann Surfaces,'' Phil. Trans. Roy. Soc. London {\bf A308}
(1982) 523-615.}.  The space of connections
(or connections with a singularity $A=\alpha\,d\theta+\dots$) is a
symplectic manifold with symplectic form \eqn\grefo{\omega=-{1\over
4\pi}\int_C\,\Tr\,\delta A\wedge \delta A} and moment map
\eqn\refo{\mu=-{1\over 2\pi}\int_C\,\Tr\,\epsilon F.} The symplectic
quotient of the space of connections (or connections with
singularity) by the appropriate group of gauge transformations is
$\M$ (or $\M(\alpha;p)$).  By reinterpreting the symplectic quotient
as a quotient by the group $\EUG_\C$ of complex-valued gauge
transformations, these spaces can be alternatively interpreted as
moduli spaces of stable bundles, or stable bundles with parabolic
structure.  The reasoning is similar to what we described in detail
for Higgs bundles in sections \hyperquo, \cpxview.

The notation $\M(\alpha;p)$ for the moduli space of flat bundles on $C\backslash p$
with monodromy $U=\exp(-2\pi\alpha)$ around $p$ is slightly misleading, because $\M(\alpha;p)$
does not vary smoothly with $\alpha$.
Its dimension depends on the subgroup of $G$ that commutes with $U$.
$\M(\alpha;p)$ varies smoothly with $\alpha$ only if $\alpha$ is constrained to be
strictly
$\Bbb{L}$-regular for some fixed $\Bbb{L}$.
To emphasize this, we consider $\Bbb{L}$ as part of the definition
and denote this space as $\M_\Bbb{L}(\alpha;p)$.  For given $\Bbb{L}$, the space of strictly
$\Bbb{L}$-regular $\alpha$'s has distinct connected components (which in general are associated with
non-isomorphic  parabolic subgroups, as we learned in section \levi).  When we do not write
$\Bbb{L}$ explicitly, it will mean that we are taking $\Bbb{L}=\TT$.

\bigskip\noindent{\it Parabolic Higgs Bundles}

Now we include the Higgs field and consider the analogous concept for Higgs bundles.

In the mathematical theory, the concept of a parabolic Higgs bundle $(E,\varphi)$ is
usually defined as follows.  $E$ is a parabolic bundle in the above sense, with a reduction
of the structure group to $\EUP$ at the point $p$.  And the differential
$\varphi$ has a pole at
$p$ with a residue that is required to take values in the corresponding Lie algebra $\frak p$;
moreover, this polar residue has the same ``eigenvalues'' as  $\sigma=\half(\beta+i\gamma)$,
and in fact, it equals $\sigma$ modulo $\frak n$.
(See \eggo\ for a concrete
illustration of this.)

If $\alpha$ and $\beta+i\gamma$
are regular, it is not necessary to make explicit the concept of parabolic structure.
For generic $(\alpha,\beta,\gamma)$, it is enough to give
the bundle $E$ together with
the differential $\varphi$, which is required to have a pole at $p$ with residue
conjugate to $\half(\beta+i\gamma)$. Then $\alpha$ is determined in terms of $\varphi$,
since it must commute with $\beta $ and $\gamma$ and its conjugacy class is known.\foot{The
action of the Weyl group introduces no ambiguity, since it acts diagonally on the triple
$(\alpha,\beta,\gamma)$.}  So the parabolic structure is determined, and for generic
parameters, we can define a parabolic
Higgs bundle without ever explaining what it means for a bundle to have parabolic
structure.  That is essentially what we have done in our initial approach to the subject.

\subsec{Topology Of $\M_\Bbb{L}(\alpha;p)$}
\subseclab\topres

We next use some of these ideas to get a rough understanding of the
topology of $\M_\Bbb{L}(\alpha;p)$.   We aim to give
a first orientation to the topology of these spaces for readers who have never encountered
them before.  And we aim to develop the necessary background for certain results
about $\MH$ described in sections \topology\ and \actweyl.
Hopefully, the very incomplete explanations we give will suffice for these
particular goals.

Suppose we are given a particular $G_\C$-bundle $E\to C$, and a parabolic subgroup $\EUP$
of $G_\C$, and we
wish to pick parabolic structure of type $\EUP$ at a point $p\in C$.  We have to
pick at $p$ a subgroup of $G_\C$ that is conjugate to $\EUP$. The
space of all such subgroups is isomorphic to $G_\C/\EUP$. This
suggests that $\M_\Bbb{L}(\alpha;p)$ should be a fiber bundle with fiber
$G_\C/\EUP$ over $\M$, the moduli space of stable $G$-bundles
(without parabolic structure): \eqn\yefto{\matrix{G_\C/\EUP&\to&
\M_\Bbb{L}(\alpha;p)\cr
                                &   & \downarrow \cr
                                & & ~\M. \cr}}
To the extent that it is valid (which we discuss shortly),
this fibration elucidates the complex structure of $\M_\Bbb{L}(\alpha;p)$; the fiber and base
are both complex manifolds, and the fibration is holomorphic.

In addition, $\M_\Bbb{L}(\alpha;p)$ has a natural symplectic structure, which is conveniently
understood from its interpretation as the moduli space of flat $G$-bundles with monodromy.
The symplectic
form is  $\omega=-{1\over 4\pi}\int_C\Tr\,\delta A\wedge\delta A$.  The complex
and symplectic structures of $\M(\alpha;p)$ combine to a Kahler structure.  To
see the symplectic structure of $\M(\alpha;p)$, a variant of \yefto\ is more helpful.
A basic fact in the theory of complex Lie groups is that the
quotient $G_\C/\EUP$ is the same as $G/\Bbb{L}$, where $\Bbb{L}=G\cap \EUP$
is a Levi subgroup of $G$ that is a maximal
compact subgroup of $\EUP$.
So instead of \yefto\ we can exhibit
$\M_\Bbb{L}(\alpha;p)$ as a fibration of symplectic manifolds:
\eqn\zefto{\matrix{G/\Bbb{L}&\to& \M_\Bbb{L}(\alpha;p)\cr
                                &   & \downarrow \cr
                                & & ~\M. \cr}}
Here the base and fiber are symplectic and the fibration will be used below
to describe the symplectic structure of $\M_\Bbb{L}(\alpha;p)$.

In fact, the fibrations \yefto\ and \zefto\ are valid precisely to
the extent that we can assume that every point in $\M$ is
represented by a stable (and not just semi-stable) bundle.
Otherwise, it is possible to have a stable parabolic bundle
$(E,\EUP)$, where the underlying bundle $E$ is not stable.  In
that case, the fibration breaks down, since it is not possible to
construct the moduli space $\M_\Bbb{L}(\alpha;p)$ of stable parabolic
bundles by first picking the bundle $E$ and then endowing it with
all possible parabolic structures.  The fibrations \yefto\ and
\zefto\ do hold away from singularities of $\M$ and give a good
first approximation to the topology of $\M_\Bbb{L}(\alpha;p)$.

There is one important and widely studied
case in which all semi-stable bundles are stable, and therefore
the fibrations \tomzo\ and \omzo\ are precisely valid.
This occurs if $G=PSU(N)$ and $E$ is a bundle whose
characteristic class in $H^2(C;\Z_N)$ is of order $N$. In
general, the codimension at which the fibrations breaks down
increases when the genus $g_C$ of $C$ or the rank $r$ of $G$ is
increased.  For example, the real codimension exceeds 2 if $g_C>2$ or
$g_C=2$, $r>1$.

If the singularities do not play an important role, then
we can use \yefto\ or \zefto\ to describe the second cohomology
group of $\M_\Bbb{L}(\alpha;p)$.  (We do this because it will eventually help us understand
the symplectic structure of $\MH$ and give a concrete illustration of the action of the
affine Weyl group on its cohomology.)  We will do this mainly assuming that $G$ is
simply-connected.  In this case, the Leray spectral sequence for the
cohomology of $\M_\Bbb{L}(\alpha;p)$ begins with
\eqn\uryt{\matrix{
\noalign{\vskip-12pt}
2\cr
1\cr
0\cr
\noalign{\vskip 4pt}
\noalign{\hrule width 14pt}\cr}\hskip-9pt
\left|\matrix{
H^2(G/\Bbb{L};\Z) & 0 & * & *~\cr
                      0 & 0 & 0 & 0~ \cr
                      \Z & 0 & H^2(\M;\Z) & * \cr
\noalign{\vskip 1.4pt}
\noalign{\hrule}\cr
0&1&2&3\cr}\right.}
(We have plotted the $q^{th}$ cohomology of $\M$ with values in the
$p^{th}$ cohomology of $G/\TT$, with $q=0,1,2,3$ running horizontally
and $p=0,1,2$ running vertically.  The precise form of the groups
labeled $*$ will not be important.) In dimension two, as we will
explain later, the differentials in the spectral sequence vanish if
$G$ is simply-connected, so the spectral sequence for the fibration
reduces to an exact sequence \eqn\tomzo{0\to H^2(\M;\Z)\to
H^2(\M_\Bbb{L}(\alpha;p);\Z)\to H^2(G/\Bbb{L};\Z)\to 0.}  Moreover
$H^2(\M;\Z)\cong \Z$ for simply-connected $G$, generated
\nref\kumar{S. Kumar and
M. S. Narasimhan, ``Picard Group Of The Moduli Space Of $G$-Bundles,'' Math. Ann.
{\bf 308} (1997) 155-173.}%
\nref\lasz{Y. Laszlo and C. Sorger, ``The Line Bundles On The Moduli Of
Parabolic $G$-Bundles Over Curves And Their Sections,'' alg-geom/9507002.}%
\refs{\kumar,\lasz}
by the first Chern class of a line bundle $\frak L$ that we
will
loosely call the determinant line bundle (for $G=SU(N)$, $\frak L$ can be defined
as the determinant line bundle of a Dirac operator).   So the exact sequence becomes
\eqn\omzo{0\to \Z\to H^2(\M_\Bbb{L}(\alpha;p);\Z)\to H^2(G/\Bbb{L};\Z)\to 0.}

 For simply-connected  $G$, the spectral
sequence for the fibration
\eqn\befto{\matrix{\Bbb{L}&~\to& G\cr
                                &   & \,\downarrow \cr
                                & & G/\Bbb{L}\cr}}
                                gives
$H^2(G/\Bbb{L};\Z)=H^1(\Bbb{L};\Z)$.  For $\Bbb{L}=\TT$, this gives
\eqn\wefot{H^2(G/\TT;\Z)=H^1(\TT;\Z)=\Lambda_{\rm wt},}
with
$\Lambda_{\rm wt}$ the weight lattice of $G$. This particular result
(with the same lattice $\Lambda_{\rm wt}$) holds whether $G$ is simply-connected
or not, since if we replace $G$ by a finite cover of itself, then
$G/\TT$ is unchanged (the cover extends $\TT$ in the same way, and
cancels out of the quotient $G/\TT$).

By virtue of \wefot,  \omzo\ is equivalent, for $\Bbb{L}=\TT$, to \eqn\bomzo{0\to\Z\to
H^2(\M(\alpha;p);\Z)\to \Lambda_{\rm wt}\to 0.}
For more general $\Bbb{L}$, a similar reasoning gives instead
\eqn\bomzox{0\to\Z\to
H^2(\M_\Bbb{L}(\alpha;p);\Z)\to \Lambda_{{\rm wt},\Bbb{L}}\to 0,}
where $\Lambda_{{\rm wt},\Bbb{L}}$ is the sublattice of $\Lambda_\wt$ that is invariant under
the Weyl group of $\Bbb{L}$.

Like any exact sequence
of lattices, \bomzo\ can be split  to give
\eqn\combo{H^2(\M(\alpha;p);\Z)=\Z\oplus \Lambda_{\rm wt}.}
This result is the theorem stated (in terms of the Picard group)
in section (1.1) of \lasz.  The theorem is stated
there for the case of parabolic structure at several points $p_1,\dots,p_s$.
In this case, one has a fibration like that considered above, with a copy of $G/\TT$
at each of the points $p_i$, so a similar analysis gives
\eqn\zomboz{H^2(\M(\alpha_1,p_1;\dots;\alpha_s,p_s);\Z)=\Z\oplus\left(\oplus_{i=1}^s
\Lambda_{{\rm wt},i}\right),}
with $s$ copies of the weight lattices.

A splitting of lattices such as \combo\ is in general
non-canonical, but in this case there is a canonical splitting.  One way to see this is
to start with the universal bundle $E_{\rm ad}\to \M\times C$ in the adjoint representation.
Upon restriction to $\M\times p$, this gives a $G_{\rm ad}$-bundle $E_{{\rm ad},p}\to \M$.
When pulled back to $\M(\alpha;p)\to \M$, the structure group of this bundle reduces to $\TT$,
so it splits as a sum of line bundles.  The first Chern classes of these line bundles generate
rationally the summand $\Lambda_{\rm wt}$ in \combo.  This gives the splitting, which
is also evident in the interpretation we give below via the affine weight lattice.

In our derivation of \combo, we have made use of the approximate fibration \yefto.
Actually, the singularities that we have neglected are not important for the second cohomology
except at special values of $\alpha$ at which a vanishing cycle collapses and the second
Betti number of $\M(\alpha;p)$ is smaller.  For a description of the values of $\alpha$
at which this happens for  $G=SU(2)$,
see \ref\cpauly{C. Pauly, ``Fibr\'es Parabolique de Rang 2 et Fonctions Th\^eta
G\'eneralis\'ees,'' Math. Z. {\bf 128} (1998) 31-50.}.  Alternatively, if one considers
the cohomology of the ``stack'' of parabolic bundles (rather than the moduli space $\M(\alpha;p)$
of stable parabolic bundles), no jumping occurs.  This is actually the right thing for
gauge theory, since the starting point is the space of all gauge fields,
which is a differential-geometric analog of what in algebraic geometry is the stack of
all bundles or parabolic bundles.  (In sigma models, branes supported on a vanishing cycle
do not disappear when the vanishing cycle collapses; they simply become branes supported on
the resulting singularity.)
In down to earth terms, the drop that occurs in the second cohomology of
$\M_\Bbb{L}(\alpha;p)$ at certain values of $\alpha$ is inessential for our applications, for the
following
reasons.   When we describe the cohomology classes of the symplectic
forms $\omega_I$, $\omega_J$, and $\omega_K$ of $\MH$,
the jumping just means that certain periods
must vanish at certain values of $\alpha$, as will be manifest in the formula we give.
Alternatively,
when we describe the action of the affine Weyl group on the cohomology of $\MH$, we will avoid
the bad values of the parameters.

\def\Weyl{{\cal W}}
\def\AffWeyl{{\cal W}_{\rm aff}}
The above description of the second cohomology of $\M(\alpha;p)$ will
make it possible to usefully describe the cohomology class of the symplectic form of this space.
The symplectic form \eqn\milom{\omega=-{1\over 4\pi}\int_C\,\Tr\,\delta A\wedge
\delta A}
  of $\M(\alpha;p)$
takes values in $H^2(\M(\alpha;p);\R)=H^2(\M(\alpha;p);\Z)\otimes\R=
\R\oplus {\frak t}^\vee$.  Its cohomology
class is described in Theorem 3.2 and Proposition 3.7 of
\ref\jeffrey{L. Jeffrey, ``The Verlinde Formula For Parabolic Bundles,'' math.AG/0003150.};
for a proof for $SU(2)$ using  gluing arguments, see \ref\donaldson{
S. Donaldson, ``Gluing Techniques In The Cohomology Of Moduli Spaces,'' in {\it Topological
Methods In Modern Mathematics} (Publish or Perish Press, Houston, 1993), pp. 137-170.},
 and for a more general argument based on realizing $\M(\alpha;p)$ as a symplectic quotient
 with $\alpha$ as a parameter in the moment map, see \ref\ojeffrey{L.
 Jeffrey, ``Extended Moduli Spaces Of Flat Connections On Riemann Surfaces,'' Math. Annalen
 {\bf 298} (1994) 667-692.}.  We will explain the formula after a few preliminaries.

\bigskip\noindent{\it The Affine Weyl Group}

 We
recall that, up to equivalence, $\alpha$ takes values in
$\TT/{\cal W}$, where ${\cal W}$ is the Weyl group. Alternatively,
we can lift $\alpha$ to $\frak t$, but then there is an
equivalence in transforming $\alpha$ by elements of the
cocharacter lattice $\Lambda_{\rm cochar}$. Since we have assumed
that $G$ is simply-connected, the cocharacter lattice is the same
as the coroot lattice $\Lambda_{\rm cort}$. The combined group
\eqn\notused{\AffWeyl=\Lambda_{\rm cort}\rtimes \Weyl} of lattice
shifts and Weyl transformations is known as the affine Weyl group.
It is the Weyl group of the Kac-Moody or affine Lie algebra of
$G$. So we can think of $\alpha$ as taking values in $\frak
t/\AffWeyl$.

$\frak t$ can be usefully divided as follows into fundamental
domains for the action of $\AffWeyl$. On certain hyperplanes in
$\frak t$, $U=\exp(-2 \pi\alpha)$ is non-regular (it commutes with a
nonabelian subgroup of $G$). Each such hyperplane is a locus of
fixed points of some element of $\AffWeyl$. These hyperplanes divide
$\frak t$ into fundamental domains for the action of the affine Weyl
group.

For example, in the case of $G=SU(N)$, generalizing \neuro, we can
think of $\alpha$ as a diagonal matrix $i\,{\rm diag}(y_1,\dots,y_N)$, with $\sum_a y_a=0$.
${\cal W}$ acts by permutations, and $\Lambda_{\rm cort}$  acts by integer shifts preserving
the vanishing of the sum of the $y_a$.
The condition for $U=\exp(-2\pi\alpha)$ to be non-regular
is that $n=y_a-y_b$ is an integer for some $a$ and $b$, in which case $\alpha$ is invariant
under the affine Weyl transformation $y_a\to y_b+n$, $y_b\to
y_a-n$.  The hyperplanes $n=y_a-y_b$ divide $\frak t$ into
fundamental domains of $\AffWeyl$.

 The ordinary Weyl group has, once we pick a set of positive roots, a distinguished
 fundamental domain in $\frak t$ called the positive Weyl chamber.  It is the
 region in which $\langle \alpha,w\rangle >0$
 for every fundamental weight $w$.  For $G=SU(N)$, this chamber is described by
 $y_1\geq y_2\geq\dots \geq y_N$.  Upon intersecting it with the above-mentioned hyperplanes,
 the positive Weyl chamber decomposes as a union of  infinitely  many
 fundamental domains for the affine Weyl group.  There is a distinguished one, which
 we will call $\eurm D$, whose
 closure contains $\alpha=0$.  For $G=SU(N)$, $\eurm D$
 is characterized by $y_1-y_N\leq 1$ (or $y_1-y_N<1$ if one wishes $U$ to be regular),
 generalizing the condition $y<1/2$ in \neuro.

Now we can explain the above-mentioned formula for the cohomology
class of the symplectic form $\omega$ of $\M(\alpha;p)$.  We write
the formula for $\alpha$ in the distinguished affine Weyl chamber of
$\frak t$, as just described.  Then the cohomology class
$[\,\omega/2\pi]$ is  \eqn\itzo{\left[\,\omega\over 2\pi\right]=e\oplus
(- \alpha^*).} Here we use the fact that $[\,\omega/2\pi]$ takes values in
$\R\oplus \frak t^\vee$.
In \itzo, $e\in H^2(\M;\R)\cong\R$ is the
pullback to $\M(\alpha;p)$ of the first Chern class of the determinant
line bundle $\frak L\to\M$ (equivalently, the pullback of the
cohomology class of the symplectic form $\omega/2\pi$ of $\M$).  And
in the second summand on the right,  $\alpha^*\in \frak t^\vee$ is (as in section \duality) the image
of $\alpha\in \frak t$
under the map from $\frak t$ to $\frak t^\vee$ that comes from the quadratic form
$-\Tr$.

Rather than summarize here the arguments of
\refs{\donaldson,\ojeffrey} leading to this formula, we give a brief explanation using
four-dimensional gauge theory.  To determine the cohomology class of $\omega$, we need
to compute periods $\int_\Sigma\omega$, for $\Sigma\subset \M(\alpha;p)$
a closed two-manifold.  Given a choice of $\Sigma$, we set $M$ to be the four-manifold
$M=\Sigma\times C$, and make our usual construction on $M$.  Let $\Sigma_p=\Sigma \times p$.
Each point
in $\Sigma$ determines a flat $G$-bundle over $C\backslash p$ with monodromy around $p\in C$;
these fit together to a $G$-bundle $E\to
M\backslash \Sigma_p$ with a monodromy around $\Sigma_p$.
($E$ may only exist as a $G_{\rm ad}$-bundle, but this does not affect
the following derivation.) The relation between the symplectic form of $\M(\alpha;p)$
and gauge theory in four
dimensions is
\eqn\realgo{\int_\Sigma{\omega\over 2\pi}=\int_{M}{\Tr\,F\wedge F\over 8\pi^2}}
This is shown exactly like the corresponding statement without parabolic structure;
see eqn. (4.18) of \kapwit.  Now using \bigox\ (and observing that $\Sigma_p\cap \Sigma_p=0$)
we have
\eqn\tealgo{\int_{\Sigma}{\omega\over 2\pi}=\int_{\Sigma_p}\Tr\,{\alpha}{F\over 2\pi}
     =-\langle \alpha^*,\eurm m\rangle~{\rm mod}~\Z,}
where $\eurm m\in H^2(\Sigma_p,\pi_1(\TT))\cong \Lambda_{\cort}$ is the cohomology class of
$F/2\pi$,
and $\langle~,~\rangle$ is the natural pairing between $\frak t$ and $\frak t^\vee$.
This is equivalent to \itzo.  We also see that the homology cycles with which the symplectic
form can naturally be paired are labeled by $\Lambda_{\cort}$, the dual of $\Lambda_{\wt}$.

This description of the symplectic form of $\M(\alpha;p)$
has an obvious similarity to our  claims in section \cpxview\ about Higgs bundles.
For Higgs bundles, we expect the cohomology classes such as $[\,\omega_I]$ to vary
linearly, as claimed in \xoco, while for ordinary bundles we have the linearity seen in \itzo.
We explain the relation between these results in section \topology.

However, in the case of Higgs bundles, the argument leading to
\xoco\ is valid for all $\alpha$, while the analogous statement
\itzo\ for bundles holds only for $\alpha$ in a fundamental affine
Weyl chamber.  For bundles, there is no obvious way to continue the
formula \itzo\ beyond the fundamental affine Weyl chamber, since on
the boundary of $\eurm D$, the manifold $\M(\alpha;p)$ collapses to
a manifold of lower dimension, as we will see in section \topres.  For Higgs
bundles, as we describe in sections \topology\ and \actweyl, we can take
$\beta,\gamma\not=0$ and smoothly continue beyond the boundaries of
the affine Weyl chamber.

In this discussion, we have implicitly taken the Levi subgroup
to be $\Bbb{L}=\TT$.  However, the same result holds for any
$\Bbb{L}$, with the same derivation; one merely has to restrict $\alpha$ to be $\Bbb{L}$-invariant.

\bigskip\noindent{\it The Affine Weight Lattice}

Let us reconsider the description \combo\ of the second cohomology of
$\M(\alpha;p)$ for the case $\Bbb{L}=\TT$: \eqn\crombo{H^2(\M(\alpha;p);\Z)=\Z\oplus
\Lambda_{wt}.} We can describe this by saying that
$H^2(\M(\alpha;p);\Z)$ is the affine weight lattice of $G$, that is,
the weight lattice of the affine Lie algebra or Kac-Moody algebra or centrally extended
loop group associated to $G$.  This description is natural in the existing mathematical
theory \lasz,  where \crombo\ is obtained via the theory of loop groups.
We will sketch the idea in a physical language in
the context of Chern-Simons gauge theory in $2+1$ dimensions \ref\wittenjones{E. Witten,
``Quantum Field Theory And The Jones Polynomial,'' Commun. Math. Phys.
{\bf 121} (1989) 351.}. Let
$k$ be a positive integer, and let $w\in \Lambda_{wt}$ be the
highest weight of an integrable representation of the Kac-Moody
algebra of $G$ at level $k$.  Using the quadratic form $-\Tr$ to
identify $\frak t^\vee={}^L\frak t$ with $\frak t$, $w\in {}^L\frak t$ maps to
 an element $w^*\in \frak t$. We consider
Chern-Simons gauge theory on $C$ with a single marked point $p$
labeled by a representation of $G$ with highest weight $w$.  The
classical phase space is $\M(\alpha;p)$ with $\alpha=w^*/k$.
Quantization is carried out by taking the global sections of a line
bundle $\frak L^k$ whose first Chern class is represented in de Rham
cohomology by $k\omega/2\pi$ where $\omega=-{1\over
4\pi}\int_C\Tr\,\delta A\wedge \delta A$.  ($\frak L^k$ is not
really the $k^{th}$ power of a line bundle unless $w=0$, but we write it as $\frak L^k$ because
of the factor of $k$ in the rational first Chern class.)
Consequently, the class $k\omega/2\pi$ in de Rham cohomology lifts to
an element in the lattice $H^2(\M(\alpha;p);\Z)$. Accordingly, this
lattice must contain a point corresponding to the pair $(k,w)$, that
is, it must contain the affine weight lattice of $G$. These points
have distinct images in de Rham cohomology, in view of the formula
\itzo\ for the cohomology class of $[\,\omega/2\pi]$.   At this point,
it is also clear that the differentials in the spectral sequence
\uryt\ for $H^2(\M(\alpha;p);\Z)$ do vanish (as we assumed in our
above discussion), or that cohomology group could not contain the
affine weight lattice.

Since $H^2(\M(\alpha;p);\Z)$ is the affine weight lattice of $G$, it
admits a natural action of the affine Weyl group of $G$, although
this action has no evident meaning in terms of the geometry of
$\M(\alpha;p)$.  In section \actweyl, we will
explain more conceptually why $H^2(\M(\alpha,p);\Z)$ admits this
action of the affine Weyl group.  We will also describe the action more precisely.

What happens if we relax the assumption that $G$ is simply-connected?
The argument via Kac-Moody algebras
 shows  for any $G$, not necessarily simply-connected, that the lattice
$H^2(\M(\alpha,p;G); \Z)$ modulo torsion contains the affine  character
lattice of $G$ as a sublattice. By the affine character lattice of a
simple but perhaps not simply-connected group $G$, we mean
\eqn\nobo{\Lambda_{\rm aff\,char}=\Z\oplus
\Lambda_{\rm char}(G),} where $\Z$ classifies central extensions\foot{Some care is needed
here.  Let $\bar G$ be the universal cover of $G$.
Related to the fact that instanton number is $\Z$-valued for $\bar G$ but not for $G$,
not every central extension of the loop group of $\bar G$ corresponds to one for $G$.
The summand $\Z$ in \nobo\ is naturally understood as a proper subgroup of the summand
$\Z$ in \crombo.}
of the loop
group of $G$.  We suspect that for generic $\alpha$,
$H^2(\M(\alpha,p;G);\Z)$ mod torsion is always precisely this lattice:
\eqn\zelko{H^2(\M(\alpha;p))=\Z\oplus \Lambda_{\rm char}(G).}
At any rate, we will show in section \actweyl\ that a natural symmetry group
of this lattice acts on $H^2(\M(\alpha,p;G);\Z)$.
It will turn out that, if we restrict ourselves to a connected component of
$\M(\alpha,p;G)$, then the group that acts on the cohomology is the same affine Weyl group
whether $G$ is simply-connected or not.

\bigskip\noindent{\it Alternative Point Of View}

The approximate fibration \zefto\ can also be seen from a purely
topological or symplectic point of view, without mentioning
parabolic structure (for example, see the brief summary in \jeffrey). Pick
on the Riemann surface $C$ a standard set of $A$- and $B$-cycles.
Let $A_i$ and $B_i$ be the monodromies of a flat $G$-bundle around
these cycles.  They obey \eqn\todo{\prod_{i=1}^g[A_i,B_i]=1,} where
$[A,B]=ABA^{-1}B^{-1}$. The moduli space $\M$ of flat bundles is the
space of solutions of this equation, modulo conjugation. Pick an
element $U$ of $G$ that is close to the identity and replace \todo\
by \eqn\zodo{\prod_{i=1}^g[A_i,B_i]=U.} A solution of this equation
describes a flat bundle over $C\backslash p$ with monodromy $U$
around $p$.  As long as we do not encounter singularities of $\M$,
the solution spaces of \todo\ and \zodo\ are topologically the same,
for $U$ sufficiently close to 1.  Now suppose that we want to
specify only the conjugacy class of $U$, which we assume to be
semi-simple and, again, sufficiently close to 1.
Then the conjugacy class of $U$  contains an element $\exp(-2\pi\alpha)$,
where $\alpha$ is an element of $\frak t$ close
to the origin.  Let $\Bbb{L}$ be the subgroup of $G$ that commutes with $\alpha$ or equivalently
with $U$; then $\Bbb{L}$ is
a Levi subgroup.
The possible
choices of $U$ in its conjugacy class
are parametrized by a copy of $G/\Bbb{L}$,
 and so the
solution space of \zodo\ is a $G/\Bbb{L}$-bundle over the solution space
of \todo.  This remains so after dividing both spaces by conjugation
by $G$, so we arrive again at the fibration \zefto:
\eqn\zeftox{\matrix{G/\Bbb{L}&\to& \M_\Bbb{L}(\alpha;p)\cr
                                &   & \downarrow \cr
                                & & ~\M. \cr}}

Suppose that we want to delete points $p_1,\dots,p_s$ from $C$, and denote
as $U_1,\dots,U_s$ the monodromies about these points.  The analog of \todo\ is
\eqn\frodo{\prod_{i=1}^g[A_i,B_i]=\prod_{a=1}^s U_a.}  If the $U_a$ are close enough to
the identity and commute with Levi subgroups $\Bbb{L}_a$,
a similar reasoning to the above leads to a fibration with a factor of $G/\Bbb{L}_a$ for each
puncture  (and this for instance leads to
the description of the second cohomology of $\M(\alpha_1,p_1;\dots;
\alpha_s,p_s)$ in eqn. \zomboz).
However, we want to consider another issue aimed at later applications.
Let $y_a,$ $a=1,\dots,s$ take values in the center of $G$ and suppose that
\eqn\yelfo{\prod_{a=1}^sy_a=1.}
Then eqn. \frodo\ is completely invariant under
\eqn\belfo{U_a\to y_aU_a,~~a=1,\dots,s.}
If $U_a$ is conjugate to $\exp(-2\pi \alpha_a)$ for some $\alpha_a\in \frak t$, and
 $y_a=\exp(-2\pi u_a)$ for some $u_a\in \Lambda_{\rm cowt}$, then the transformation
\belfo\ amounts to
\eqn\minzo{\alpha_a\to \alpha_a+u_a.}
And \yelfo\ is equivalent to
\eqn\winzo{\sum_a u_a\in \Lambda_{\rm cochar}\subset \Lambda_{\rm cowt}.}
This is an illustration of the situation that was described in eqns. \nomo\ and \otimbo.
The shifts $\alpha_a\to \alpha_a+u_a$ individually would
shift the characteristic class $\xi $ of the $G_{\rm ad}$
bundle derived from $E$, and the condition \yelfo\ or \winzo\ ensures that globally
$\xi$ is actually unchanged.

Let us return now for simplicity to the case of one puncture.
If
$U=\exp(-2\pi\alpha)$ is regular, which we can achieve by placing $\alpha$ in the interior
of the distinguished affine Weyl chamber $\eurm D$, then the fibration \zeftox\ takes the
form
\eqn\zeftoxp{\matrix{G/\TT&\to& \M(\alpha;p)\cr
                                &   & \downarrow \cr
                                & & ~\M. \cr}}
Now suppose that $\alpha$ approaches a
boundary point $\bar\alpha$ of $\eurm D$.  Then $U$ ceases to be regular, and its orbit
under conjugation is a copy of $G/\Bbb{L}$ for some $\Bbb{L}$, rather than  $G/\TT$.
So now the fibration takes the form in \zeftox:
\eqn\weftox{\matrix{G/\Bbb{L}&\to& \M_\Bbb{L}(\bar\alpha;p)\cr
                                &   & \downarrow \cr
                                & & ~\M. \cr}}
What happens as $\alpha\to\bar\alpha$?  $G/\TT$ is fibered over $G/\Bbb{L}$ with fiber $\Bbb{L}/\TT$.
As $\alpha\to\bar\alpha$, the orbit of $U$ collapses from a copy of $G/\TT$
to a copy of $G/\Bbb{L}$ and we get a fibration
\eqn\weftox{\matrix{\Bbb{L}/\TT&\to& \M(\alpha;p)\cr
                                &   & \downarrow \cr
                                & & ~\M_\Bbb{L}(\bar\alpha;p). \cr}}
The fibers $\Bbb{L}/\TT$ are symplectic manifolds, and they collapse to
points as $\alpha\to\bar\alpha$.  So this is what happens to $\M(\alpha;p)$ as $\alpha$
approaches a non-regular value $\bar\alpha$: it collapses to a variety of lower dimension,
with vanishing cycles $\Bbb{L}/\TT$.

Generically, a boundary point $\bar\alpha$ is contained in only one of
the hyperplanes that mark the boundary of $\eurm D$, and
$\Bbb{L}/\TT=SU(2)/U(1)=\Bbb{CP}^1=S^2$. For example, for $G=SU(N)$, we
have $\alpha=i\,{\rm diag}(y_1,y_2,\dots,y_N)$, with $y_1\geq
y_2\geq y_3\dots,\geq y_N$, $y_1-y_N\leq 1$, and $\sum y_a=1$. At a
generic boundary point, precisely one of the inequalities is an
equality, and then  $\Bbb{L}=U(2)\times U(1)^{N-3}$, and
$\Bbb{L}/\TT=U(2)/U(1)^2=\Bbb{CP}^1=S^2$. So as $\alpha$ approaches a
generic boundary point of $\eurm D$, $\M(\alpha;p)$ is fibered by
vanishing two-spheres, that is, two-spheres that shrink to points. In general,
the rank of the semi-simple part of $\Bbb{L}$ is the number of boundary hyperplanes that contain
$\bar\alpha$.

More generally, we can consider a pair of Levi subgroups $\Bbb{L}_1$ and $\Bbb{L}_2$, with $\Bbb{L}_1$
a proper subgroup of $\Bbb{L}_2$ but not necessarily equal to $\TT$.  The same sort of reasoning
as above applies.  If $U=\exp(-2\pi\alpha_1)$, where $\alpha_1$
is $\Bbb{L}_1$-regular, we get a fibration
\eqn\zeftoxt{\matrix{G/\Bbb{L}_1&\to& \M_{\Bbb{L}_1}(\alpha_1;p)\cr
                                &   & \downarrow \cr
                                & & ~\M. \cr}}
If $\alpha_2$ is $\Bbb{L}_2$-regular, then for $U=\exp(-2\pi\alpha_2)$, the fibration looks like
\eqn\zeftoxt{\matrix{G/\Bbb{L}_2&\to& \M_{\Bbb{L}_2}(\alpha_2;p)\cr
                                &   & \downarrow \cr
                                & & ~\M. \cr}}
As $\alpha_1$ approaches $\alpha_2$ through $\Bbb{L}_1$-regular values, the orbit $G/\Bbb{L}_1$ of $U$
degenerates to $G/\Bbb{L}_2$.   $G/\Bbb{L}_1$ maps to $G/\Bbb{L}_2$ with fiber $\Bbb{L}_2/\Bbb{L}_1$,
and for
$\alpha_1$ approaching $\alpha_2$ we get, away from singularities, a fibration
\eqn\weftoxt{\matrix{\Bbb{L}_2/\Bbb{L}_1&\to& \M_{\Bbb{L}_1}(\alpha_1;p)\cr
                                &   & \downarrow \cr
                                & & ~\M_{\Bbb{L}_2}(\alpha_2;p). \cr}}
The fibers $\Bbb{L}_2/\Bbb{L}_1$ are vanishing cycles for $\alpha_1\to\alpha_2$.

As one might expect, there is also a natural description of the
limiting behavior in terms of complex geometry.  The pair $(\Bbb{L}_1,\alpha_1)$ determines
a parabolic subgroup $\EUP_1$, and the pair $(\Bbb{L}_2,\alpha_2)$ determines a parabolic
subgroup $\EUP_2$ that contains $\EUP_1$.  {}From a holomorphic point of view, according to the
theorem of Mehta and Seshadri,  $\M_{\Bbb{L}_1}(\alpha_1;p)$
or $\M_{\Bbb{L}_2}(\alpha_2;p)$
parametrizes bundles with parabolic structure at $p$ of type $\EUP_1$ or $\EUP_2$,
respectively.
 After reducing the structure
group of a bundle to $\EUP_2$, the ways of further reducing it to the subgroup $\EUP_1$
are parameterized by $\EUP_2/\EUP_1$.  So from a holomorphic point of view, we get a fibration
\eqn\peftox{\matrix{\EUP_2/\EUP_1&\to& \M_{\Bbb{L}_1}(\alpha_1;p)\cr
                                &   & \downarrow \cr
                                & & \M_{\Bbb{L}_2}(\alpha_2;p). \cr}}
This is in accord with \weftoxt, since $\EUP_2/\EUP_1=\Bbb{L}_2/\Bbb{L}_1$.
This fibration  is Proposition 3.4 in
\ref\kohno{H. U. Boden and Y. Hu, ``Variations Of Moduli Of
Parabolic Bundles,'' Math. Annalen {\bf 301} (1995) 539-559.}.

\subsec{Topology Of $\M_H$}
\subseclab\topology

In section \topres, we explored the topology of the moduli space $\M(\alpha;p)$ of
bundles with parabolic structure.  Here we will consider the analogous questions for
Higgs bundles.

If we simply set $\phi=0$, then Hitchin's equations reduce to the
equations $F=0$ for a flat unitary $G$-bundle.  The moduli space of
such flat bundles is isomorphic, by a theorem of Narasimhan and
Seshadri mentioned above \nar, to the moduli space of stable
$G$-bundles over $C$, which we call $\M(G,C)$, or $\M$ when the
context is clear.

The moduli space $\MH$, viewed in complex structure $I$,
parametrizes stable pairs $(E,\varphi)$.  We can define a
``foliation'' of $\MH$ by forgetting $\varphi$ and remembering only
the holomorphic type of $E$.  For a generic stable (or semistable)
pair $(E,\varphi)$, $E$ is a stable (or semistable) bundle, and the
foliation gives a meromorphic map $\psi:\M_H\to \M$ (in \kapwit, this
was called Hitchin's second fibration).   The fiber of this map is a
linear space parametrized by  $\varphi\in H^0(C,K\otimes{\rm ad}(E))$,
which (if $E$ is stable) is the cotangent space to $\M$ at the point
defined by $E$. Moreover, the  map $\psi$  has a natural  section
(holomorphic in complex structure $I$), because we can embed $\M$ in
$\M_H$ as the space of solutions of Hitchin's equations with
$\varphi=0$. So birationally in complex structure $I$, $\MH$ is the
cotangent bundle $T^*\M$: \eqn\zinko{\MH\cong T^*\M.}
More specifically, $\MH$ contains $T^*\M$ as a dense open set.
 Somewhat like the fibrations
considered in section \topres, this gives a useful first
approximation to the topology of $\MH$ if the genus $g_C$ of $C$ and
the rank $r$ of $G$ are large enough.  For example, it is good
enough for discussing the second cohomology of $\MH$ if $g_C\geq 2$.

This has an analog for ramified Higgs bundles if the parameter
$\alpha$ is $\Bbb{L}$-regular.  Given a point in $\M_{H,\Bbb{L}}(\alpha,\beta,\gamma;p)$
associated with a stable or semistable ramified Higgs bundle
$(E,\varphi)$, we can forget $\varphi$ and simply think of $E$ as a
$G_\C$-bundle with the parabolic structure determined by $\alpha$.
Generically, this parabolic bundle is stable, and thus defines a
point in $\M_\Bbb{L}(\alpha;p)$.   So we get a meromorphic map
$\psi:\M_{H,\Bbb{L}}(\alpha,\beta,\gamma;p)\to \M_\Bbb{L}(\alpha;p)$. If
$\beta=\gamma=0$, the story is exactly as it was above.  The fiber
of the map $\psi$ is a space of Higgs fields with a nilpotent pole
that takes values in $\frak n$, the Lie algebra of the unipotent radical of
the parabolic subgroup $\EUP$ determined by $(\Bbb{L},\alpha)$.
This is precisely the cotangent space to $\M_\Bbb{L}(\alpha;p)$.
Moreover, $\M_\Bbb{L}(\alpha;p)$ can be embedded holomorphically in
$\M_{H,\Bbb{L}}(\alpha,0,0;p)$ as the space of solutions of Hitchin's equations
with $\varphi=0$.  We can think of this as the zero section of the cotangent bundle.
So $\M_{H,\Bbb{L}}(\alpha,0,0;p)$ contains  the cotangent bundle to $\M_\Bbb{L}(\alpha;p)$
as a dense open subspace
\eqn\jury{\M_{H,\Bbb{L}}(\alpha,0,0;p)\cong T^*\M_\Bbb{L}(\alpha;p),}
and the two are birational in complex structure $I$.

For $\beta,\gamma\not=0$, this requires some modification. We can
still forget $\varphi$ and thus define the meromorphic map
\eqn\yellow{\psi:\M_{H,\Bbb{L}}(\alpha,\beta,\gamma;p)\to \M_\Bbb{L}(\alpha;p).}
However, we cannot set $\varphi$ to zero, since its polar part has
eigenvalues determined by $\beta$ and $\gamma$. So the map $\psi$  has no
holomorphic section. Hence, $\M_{H,\Bbb{L}}(\alpha,\beta,\gamma;p)$ is not birational to a
vector bundle over $\M_\Bbb{L}(\alpha;p)$, but to an ``affine bundle.'' This means
that the fibers of $\psi$ are copies of $\C^N$ for some $N$, but the
structure group of the fibration is a group of affine
transformations $x\to ax+b$ (not just linear transformations $x\to
ax$). Related to this, the polar part of $\varphi$ is not nilpotent,
and so $\varphi$ does not represent a cotangent vector to
$\M_{H,\Bbb{L}}(\alpha;p)$. But if $\varphi$ and $\tilde\varphi$ are two Higgs
fields (with the same bundle $E$) then, as we saw in eqn. \gruffy, their difference
$\varphi-\tilde\varphi$ has a  polar part valued in $\frak n$ and hence
represents a cotangent vector to $\MH(\alpha;p)$.  The upshot is
that $\MH(\alpha,\beta,\gamma;p)$, for general $\beta$ and
$\gamma$, and regular $\alpha$,  contains a dense open set that is an ``affine
deformation'' of the cotangent bundle of $\M(\alpha;p)$.

In making this statement, we require $\alpha$ to be $\Bbb{L}$-regular, since
otherwise $\M(\alpha;p)$ collapses to a manifold of lower dimension.
However, as long as $\beta$ and $\gamma$ are generic, the topology
of $\MH(\alpha,\beta,\gamma;p)$ is independent of $\alpha$ whether
$\alpha$ is regular or not \nakajima, so we get a rough description of the topology
for any  $\alpha$ and generic $\beta,\gamma$.
The topology does change in real codimension three when
the triple $(\alpha,\beta,\gamma)$ is non-regular, as we explain
presently.

An immediate application of the relation between  $\MH$ and $\M$ is
that we can describe the second cohomology of $\MH$ and determine
the unknown constants in the formula \xoco\ for the cohomology class
of the symplectic form $\omega_I$. First of all, the second
cohomology of $\MH$ is isomorphic to that of $\M$, because $\MH$,
being an affine bundle, is contractible to $\M$ (away from a
codimension that is too high to affect the second cohomology,
barring special cases of small genus and rank that we will not
consider). So \eqn\gurg{H^2(\MH;\Z)=\Z\oplus\Lambda_{wt}}
 is the affine
weight lattice of $G$, just like $H^2(\M;\Z)$.  Also, according to
eqn. \zixo, $\omega_I=\psi^*(\omega)$, where $\omega$ is the usual
symplectic form of $\M$, whose cohomology class was described in
eqn. \itzo. So we can simply borrow the result of eqn. \itzo\ and
write \eqn\yfosto{\left[{\omega_I\over 2\pi}\right]=e\oplus (-\alpha^*).}
(This formula actually holds for every $\Bbb{L}$; it makes sense because $\alpha^*$
always takes values in the $\Bbb{L}$-invariant part of $\Lambda_\wt\otimes_\Z\R$.)
{}From \oco\ and \noco, we have
therefore
\eqn\dpofo{\left[{\omega_J\over 2\pi}\right]=0\oplus(-\beta^*),~~\left[{\omega_K\over 2\pi}\right]
=0\oplus(-\gamma^*).}   After some discussion of the
singularities of $\MH(\alpha,\beta,\gamma;p)$, we will be able to
draw some interesting conclusions from these formulas.

\bigskip\noindent{\it Singularity At Special Values Of $(\alpha,\beta,\gamma)$}

Even without ramification, the moduli space $\M$ of $G_\C$ bundles and
the moduli space $\M_H$ of Higgs bundles can have singularities at points that correspond
to reducible bundles or Higgs bundles.  Such singularities involve the global behavior on
$C$, and occur in high codimension if the genus of $C$ is large.

Parabolic structure at a point $p\in C$ introduces a new kind of singularity, which
depends only on the parameters characterizing the ramification.  These parameters
are $\alpha\in \TT$ in the case of bundles, or the trio $(\alpha,\beta,\gamma)\in \TT\times
{\frak t}\times {\frak t}$ in the case
of Higgs bundles.  These singularities depend only on the local behavior at $p$ and
their codimension is independent of the genus of $C$.  They will be much more prominent
in applications to the geometric Langlands  program than the singularities that are global
in nature.

For bundles, we described this kind of singularity in section \topres.  The result
is described in eqns. \weftox\ and \peftox.  Suppose that $\alpha$ approaches a  value
$\bar\alpha$ that is not $\Bbb{L}$-regular, but is $\Bbb{L}'$-regular, where $\Bbb{L}'$ is a Levi group that contains
$\Bbb{L}$ as a proper subgroup.  And let the corresponding parabolic subgroups be $\EUP$ and $\EUP'$.
Then as $\alpha\to\bar\alpha$,  $\M_\Bbb{L}(\alpha;p)$
is fibered by vanishing cycles of the form $\Bbb{L}'/\Bbb{L}$ (or
$\EUP'/\EUP$).  The fibers collapse everywhere, so the
codimension of the singularity is zero.
The picture looks like
\eqn\ury{\matrix{\EUP'/\EUP&\to & \M_\Bbb{L}(\alpha;p)\cr
                           & &        \downarrow\cr
                           & &      \M_{\Bbb{L}'}(\bar\alpha;p),\cr}}
                           with the fibers collapsing as $\alpha\to\bar\alpha$.
This is a local singularity, in the sense that it has only to do
with the value of $\alpha$ and the behavior near the point $p$.  It
has nothing to do with a global problem such as finding a semistable
bundle or a reducible flat connection.

Now what does this imply for ramified Higgs bundles?
First let us see what happens if $\beta$ and
$\gamma$ have the most special possible values, namely 0. But we begin with
an $\Bbb{L}$-regular value of $\alpha$.  (See \ref\whox{H. U. Boden and K. Yokogawa,
``Moduli Spaces Of Parabolic Higgs Bundles And Parabolic $K(D)$ Pairs,'' Internat. J.
Math. {\bf 7} (1996) 573-598.} for a mathematical discussion of these issues, justifying
many statements below from a different point of view.) As we have
discussed, $\M_{H,\Bbb{L}}(\alpha,0,0;p)$ is generically the cotangent bundle
of $\M_\Bbb{L}(\alpha;p)$.  Likewise, if $\Bbb{L}'$ is a Levi subgroup that properly
contains $\Bbb{L}$ and
$\bar\alpha$ is $\Bbb{L}'$-regular,
then $\M_{H,\Bbb{L}'}(\bar\alpha,0,0;p)$ is generically the cotangent bundle of
$\M_{\Bbb{L}'}(\bar\alpha;p)$.
To go from parabolic bundles to ramified Higgs bundles, we just replace everything in \ury\ with
its cotangent bundle.  So for $\alpha$ near $\bar\alpha$,
$\M_{H,\Bbb{L}}(\alpha,0,0;p)$ is generically fibered over $\M_{H,\Bbb{L}'}(\bar\alpha,0,0;p)$:
\eqn\peftoxical{\matrix{T^*(\EUP'/\EUP)=T^*(\Bbb{L}'/\Bbb{L})&\to& \M_{H,\Bbb{L}}(\alpha,0,0;p)\cr
                                &   & \downarrow \cr
                                & & \M_{H,\Bbb{L}'}(\bar\alpha,0,0;p). \cr}}
We have used the fact that $\EUP'/\EUP=\Bbb{L}'/\Bbb{L}$.  When $\alpha$ approaches
$\bar\alpha$, $\Bbb{L}'/\Bbb{L}$ becomes a ``vanishing cycle'' and collapses to a
point.  The codimension of $\Bbb{L}'/\Bbb{L}$ inside its cotangent bundle is
equal to the dimension of $\Bbb{L}'/\Bbb{L}$. So this is the codimension of the
vanishing cycle that $\M_{H,\Bbb{L}}(\alpha,0,0;p)$ acquires as $\alpha$ approaches a
nonregular value $\bar\alpha$. At $\alpha=\bar\alpha$, the vanishing cycle collapses to a point,
and $\MH$ acquires a singularity of whose codimension is twice as great.
We call this a local singularity, since it only
depends on the local behavior near $p$ (which is determined by the parameters
$(\alpha,\beta,\gamma)$) and not on solving any global problem.
 For example, if $\Bbb{L}=\TT$ and if $\bar\alpha$ is a
generic nonregular value, then $\Bbb{L}'/\Bbb{L}=\Bbb{CP}^1=S^2$, as we have
seen in section \topres. So the vanishing cycle in this case has real codimension
two.

Because $\M_{H,\Bbb{L}}(\alpha,0,0;p)$ is hyper-Kahler, the geometry near the
vanishing cycle is highly constrained.  The highly curved geometry
near an almost vanishing cycle (for $\alpha$ near
$\bar\alpha$) must itself  be hyper-Kahler, in order for it to be
possible for $\M_{H,\Bbb{L}}(\alpha,0,0;p)$ to be hyper-Kahler.  It is possible
to see explicitly how this happens. In fact, a family of
hyper-Kahler metrics on $T^*(\Bbb{L}'/\Bbb{L})$ can be constructed \kron\
using Nahm's or Hitchin's equations.  This is part of the construction of hyper-Kahler
metrics on coadjoint orbits of the complex Lie group $\Bbb{L}'_\C$; we review this construction
in section \onahm.

Rather than use this full machinery, we will consider in some detail the case that $\Bbb{L}=\TT$,
and $\bar\alpha$ is a generic non-regular value, so that $\Bbb{L}'/\Bbb{L}=\Bbb{CP}^1$.  In this case, the
relevant family of hyper-Kahler metrics on
$T^*(\Bbb{L}'/\Bbb{L})=T^*(\Bbb{CP}^1)$ is a more familiar family of metrics
first constructed by Calabi and by Eguchi and Hansen.

As the area of $\Bbb{CP}^1$ converges to zero, $T^*\Bbb{CP}^1$
converges, in the Calabi-Eguchi-Hansen metric, to $\R^4/\Z_2=\C^2/\Z_2$, that
is, to an $A_1$ singularity. Hence, in the limit $\alpha=\bar\alpha$,
$\M_{H,\Bbb{L}}(\alpha,0,0;p)$ (which is the same as $\MH(\alpha,0,0;p)$, since we have set
$\Bbb{L}=\TT$) has a family of $A_1$ singularities.  The singular locus is
precisely the hyper-Kahler manifold $\M_{H,\Bbb{L}'}(\bar\alpha,0,0;p)$,
embedded inside $\M_H(\bar\alpha,0,0;p)$.
This follows from  the fibration \peftoxical, since when the fiber $T^*(\Bbb{L}'/\Bbb{L})$ degenerates
to $\C^2/\Z_2$, which has an isolated singularity, the singular locus of $\M_H(\alpha,0,0;p)$
becomes a section of the fibration or in other words a copy of $\M_{H,\Bbb{L}'}(\bar\alpha,0,0;p)$.
We will also explain below
a slightly different approach to this result (see the discussion of eqn. \dufmo).
If $\alpha$ is close to but not equal to $\bar\alpha$,
or $\alpha=\bar\alpha$ but $\beta$ and $\gamma$ are not quite zero, then the local singularity
is deformed or resolved, and the behavior near the singularity is described by the
Calabi-Eguchi-Hansen metric.
These assertions will hopefully become clear below and in section \onahm.

It will help to recall a few facts about the $\R^4/\Z_2$
singularity. If we single out one complex structure (which in our
application corresponds to the complex structure $I$ of $\MH$), then
$\R^4/\Z_2$ can be described as the complex singularity
$a_1^2+a_2^2+a_3^2=0$. It can be deformed to the smooth complex manifold
\eqn\yerto{a_1^2+a_2^2+a_3^2=\epsilon.} It also can be resolved to make
the cotangent bundle $T^*\Bbb{CP}^1$, with the exceptional cycle having an area $r$.
Moreover, the deformation and
resolution can be made simultaneously. The Calabi-Eguchi-Hansen
metric depends on the real parameter $r$ as well as the complex
parameter $\epsilon$.  The $A_1$
singularity appears precisely if $r=\epsilon=0$; otherwise the
manifold is smooth. The picture can be described very naturally by
constructing $\R^4/\Z_2$  and its smooth deformations  via
hyper-Kahler quotients of a Euclidean space \ref\krono{P.
Kronheimer, ``The Construction Of ALE Spaces As Hyper-Kahler Quotients,''
J. Diff. Geom. {\bf 29} (1989) 665-83.}.  Alternatively, it is a special case
of the construction based on Nahm's equations that we review in section \onahm.

Now let us apply this to $\MH$.
In complex structure $I$, $\alpha$ is a Kahler parameter, as
asserted in the table in section \geomint. So when $\alpha$ is
varied, $\MH$ can change only by a birational transformation, and
this only when a vanishing cycle appears (as occurs
when the triple $(\alpha,\beta,\gamma)$ becomes nonregular).
Hence varying $\alpha$ will give
us the blowup parameter $r$ of the $A_1$ singularity.  To see the
complex parameter $\epsilon$, we must vary $\beta$ and $\gamma.$

\bigskip\noindent{\it Analysis For $G=SU(2)$}

To see what happens in varying $\beta$ and $\gamma$,
we will, to keep things simple, take $G=SU(2)$. A
point in $\MH$ corresponds in complex structure $I$ to a Higgs
bundle $(E,\varphi)$.  $\varphi$ takes values in $V=H^0(C,K_C\otimes
{\rm ad}(E)\otimes \CO(p))$; thus, it is a section of $K_C\otimes
{\rm ad}(E)$ with a possible pole at $p$.  The space $V$ has
dimension $3g_C$.  $\varphi$ is constrained to obey
\eqn\helmo{\Tr\,\varphi^2=\Tr\,\sigma^2 (dz/z)^2+\dots,}
where
$\sigma=\half(\beta+i\gamma)$.

Pick a basis $b_i$, $i=1,2,3$, of the fiber of ${\rm ad}(E)$ at $p$, normalized
 so that $\Tr\,b_ib_j=\delta_{ij}$.  For
$i=1,2,3$, pick an element $\varphi_i\in V$ whose polar part is
$\varphi_i\sim b_i(dz/z)$.  And complete the $\varphi_i$ to a basis
$\varphi_1,\dots,\varphi_{3g_C}$ of $V$, such that the $\varphi_i$,
$i>3$, have no pole at $p$.  Now introduce complex parameters $a_i$,
$i=1,\dots,3g$, and expand $\varphi$ as
$\varphi=\sum_{i=1}^{3g_C}a_i\varphi_i$.
$\MH(\alpha,\beta,\gamma;p)$ is parametrized by the choice of a
bundle $E$ ($3g_C-3$ parameters) and the coefficients $a_i$ ($3g_C$
parameters), subject to the equation \helmo, so its dimension is $6g_C-4$.
The equation \helmo\  tells us that
\eqn\dufmo{\sum_{i=1}^3a_i^2=\Tr\,\sigma^2.}
This equation
describes the deformation of the $A_1$ singularity, with
$\Tr\,\sigma^2$ playing the role of the parameter $\epsilon$ in \yerto.

At $\sigma=0$, we potentially recover the $A_1$ singularity, but now
we must remember the parameter $\alpha$ that controls the Kahler
structure.  For generic $\alpha$, we get the resolution of the $A_1$
singularity.  The reason that this occurs is simply that when the polar part of
$\varphi$ is nonzero,
it (plus the specification of the conjugacy class of $\alpha$) fixes what $\alpha$ must
be.  But when $\varphi$ has no pole, there is a family of possible choices of $\alpha$
that (since the conjugacy class of $\alpha$ is fixed) is parametrized by $\Bbb{CP}^1$.
Since the locus at which $\varphi$ has no pole is precisely where the $A_1$ singularity
would be, taking into account the choice of $\alpha$ replaces the $A_1$ singularity
in each transverse slice with a copy of $\Bbb{CP}^1$.  This operation is the
resolution of the singularity.  If, however, $\alpha=0$, then $\MH$ does
develop a family of $A_1$ singularities.

If we set $\alpha=\sigma=0$, then the locus of $A_1$ singularities
is given by $a_1=a_2=a_3=0$.  This condition ensures that $\varphi$
has no pole, so that $(E,\varphi)$ is an ordinary Higgs bundle,
without ramified structure.  Thus, the locus of $A_1$ singularities
in this example is simply $\M_H\subset \M_H(0,0,0;p)$.

This is a special case of the general description of local singularities of
$\M_{H,\Bbb{L}}(\alpha,\beta,\gamma;p)$.  If we set $(\alpha,\beta,\gamma)$ to a triple
$(\bar\alpha,\bar\beta,\bar\gamma)$
that is not $\Bbb{L}$-regular but is $\Bbb{L}'$-regular, where $\Bbb{L}'$
properly contains $\Bbb{L}$,
  then $\M_{H,\Bbb{L}}(\bar\alpha,\bar\beta,\bar\gamma;p)$ contains
$\M_{H,\Bbb{L}'}(\bar\alpha,\bar\beta,\bar\gamma;p)$ as a locus of local singularities.  For given
$\Bbb{L}$, all possible $\Bbb{L}'$ that properly include $\Bbb{L}$ can occur, including $\Bbb{L}'=G$.

\subsec{Action Of The Affine Weyl Group}
\subseclab\actweyl

To keep things simple, we begin this section with the assumption that $G$ is simply-connected
and the Levi group is $\Bbb{L}=\TT$.

The parameters $(\alpha,\beta,\gamma)$ in general
take values in $\TT\times \frak t\times \frak t$,
modulo the action of the Weyl group ${\cal W}$.  Equivalently, we can take
a slightly different point of view and think of $(\alpha,\beta,\gamma)$ as taking
values in $\frak t\times \frak t\times \frak t$ modulo the action of the affine Weyl
group $\AffWeyl$.  We recall that $\AffWeyl$ is an extension of ${\cal W}$ by the coroot
 lattice of $G$:
\eqn\durgo{0\to \Lambda_{\rm cort}\to\AffWeyl\to {\cal W}\to 1.}
We let $\AffWeyl$ act on the trio $(\alpha,\beta,\gamma)$ by acting on $\alpha$ in the
natural fashion, while acting on $\beta$ and $\gamma$ via the quotient ${\cal W}$.
With this action of $\AffWeyl$, we have
\eqn\noxo{(\frak t\times \frak t\times \frak t)/\AffWeyl=(\TT\times\frak t\times \frak t)/
{\cal W}, }
simply because $\TT={\frak t}/\Lambda_{\rm cort}$ for simply-connected $G$.
We call an element of a space acted on by $\cal W$ regular if it is  not left invariant
by any element of $\cal W$ except the identity, and likewise for an element of a space
acted on by $\AffWeyl$.
$(\alpha,\beta,\gamma)\in {\frak t}^3$ is regular for the action of $\AffWeyl$
if and only if its projection to $\TT\times \frak t\times \frak t$ is regular for the
action of $\Weyl$.

\def\Weyl{{\cal W}}
Now before dividing by $\AffWeyl$, we would like to omit from $\frak t\times \frak t\times
\frak t$ the points on which $\MH(\alpha,\beta,\gamma;p)$ develops a singularity.
We analyzed in section \topology\ the local singularities, which depend only on the behavior
near $p$.
These occur precisely when the triple $(\alpha,\beta,\gamma)\in \frak t\times \frak t\times
\frak t$
is non-regular for the action of $\AffWeyl$ (in the discussion in section \topology,
the more natural  criterion is the equivalent one that the projection
to $\TT\times\frak t\times \frak t$ is nonregular in the usual sense).
This happens in real codimension three, because being invariant under some element
$x\in \AffWeyl$ that has a nonzero image in $\Weyl$ places a non-trivial condition
on each of $\alpha$, $\beta,$ and $\gamma$ separately.
(And an element of $ \Lambda_{\rm cort}$ acts freely on $\alpha$ and hence
has no fixed points at all.)

There also are global singularities, which arise at values of $(\alpha,\beta,\gamma)$
at which there are  reducible solutions of Hitchin's equations. Even after
we have suitably adjusted the triple $(\alpha,\beta,\gamma)$ so that such solutions
exist, they occur on
$\MH$ only in high codimension if the genus of $C$ is large.  (This contrasts with local
singularities, which generically are $A_1$ singularities, of real codimension four, as we
saw in section \topology.)
The values of $(\alpha,\beta,\gamma)$ at which these singularities occur
can be described precisely \nakajima.
They arise on a discrete set of affine
linear spaces in $\frak t^3$ of real codimension three. (Actually, the analysis shows that
for the case of precisely one parabolic point, there are no global singularities for a
regular triple $(\alpha,\beta,\gamma)$, but for two or more parabolic points, there can be
global singularities with each triple being regular.)  The basic reason that the global
singularities are in real codimension three is the same as for the local singularities: the
hyper-Kahler nature of $\M_H$.  Viewing $\MH$ as a complex manifold
in one of its complex structures, to obtain a singularity one must always adjust at least one
complex parameter that controls the complex structure and one real parameter that controls
the Kahler metric, making three real parameters in all.

\def\eurx{\eurm X}
Let us omit from $\frak t^3$ all of the codimension three affine linear spaces on
which a local or global singularity occurs.  Call what remains $\eurx$.   $\eurx$
is connected and  simply-connected, since $\frak t^3$ is a linear space, and what we
have omitted is of codimension three. The topology of $\MH(\alpha,\beta,\gamma;p)$
does not change when we vary the parameters without meeting a singularity.
So it does not change if we vary the parameters in $\eurx$.  Since $\eurx$ is connected,
 the varieties $\MH(\alpha,\beta,\gamma;p)$
are independent of $\alpha,\beta$, and $\gamma$ topologically, as long as we restrict
ourselves to $(\alpha,\beta,\gamma)\in \eurx$.  This statement is one of
the main results of \nakajima.

We can learn more by observing that the group $\AffWeyl$ acts freely on $\eurx$, since
 all triples $(\alpha,\beta,\gamma)\in \eurx$ are regular.  So $\eurx/\AffWeyl$
is a smooth manifold, with fundamental group $\AffWeyl$.  The cohomology of
$\MH(\alpha,\beta, \gamma;p)$
varies as the fiber of a flat bundle over $\eurx/\AffWeyl$.  Taking the monodromy of the flat
bundle, we get an action of $\AffWeyl$ on the cohomology of $\MH(\alpha,\beta,\gamma;p)$.
Similarly, $\AffWeyl$ acts on, for example, the $K$-theory of this space.

This result is somewhat analogous to the Springer representations of the Weyl
group \springer, which are also naturally understood, as suggested in \atbiel,
by varying the parameters of hyper-Kahler metrics on coadjoint orbits and
their Slodowy slices \okrontwo.  (We review the framework for this in section
\onahm.) The Springer representations have been generalized to an action of
the affine Hecke algebra on equivariant $K$-theory \refs{\klo,\kl} and more
recently to an action of the affine braid group on certain derived categories
\bez\ of sheaves on the Springer resolution.
See the book by Chriss and Ginzburg
\chrissginz\ for an exposition of some of these results.  The affine Weyl group
action on cohomology or $K$-theory of $\MH$ is enriched to an affine braid group action on
the categories of $A$-branes or $B$-branes, as we will see in section \opmon.  As for
whether one can see the affine Hecke algebra in the context of $\MH$, to attempt to do so,
we would set $\beta=\gamma=0$, whereupon $\MH$ admits an action of $\C^*$, and one can define
its equivariant cohomology or $K$-theory.  These may well admit an action
of the affine Hecke algebra, which would improve the analogy between the ``global'' problem
involving $\MH$ and the ``local'' problem involving complex coadjoint orbits.

\bigskip\noindent{\it Example}

The affine Weyl group acts on the cohomology of $\MH$ in all dimensions.
However, we can describe this action explicitly if we restrict to the two-dimensional
cohomology,
which we described in eqn. \gurg:
\eqn\zombo{H^2(\MH(\alpha,\beta,\gamma;p);\Z)=\Z\oplus \Lambda_{\rm wt}.}
The right hand side is the affine weight lattice of $G$, and so
admits a natural action of $\AffWeyl$.
To justify the obvious guess that $\AffWeyl$ actually does act on $H^2(\MH)$ in this
natural way, we use
the result \yfosto\ for the cohomology class of the symplectic form
$\omega_I$:
\eqn\trombo{\left[{\omega_I\over 2\pi}\right]=e\oplus (-\alpha^*).}
The cohomology class of $\omega_I$ must be invariant under the combined action of
$\AffWeyl$ on $\alpha$ and on $H^2(\MH(\alpha,\beta,\gamma;p);\Z)$.
This uniquely determines the action of $\AffWeyl$ on
$H^2(\MH(\alpha,\beta,\gamma;p);\Z)$ to be its natural action on the affine weight lattice.
This means that the subgroup $\Weyl\subset \AffWeyl$ acts trivially on $\Z$ and in the
usual fashion on $\Lambda_{\rm wt}$.  And  $\eurm m\in \Lambda_{\rm cort}$ acts by
\eqn\wombo{e\to e\oplus {\eurm m}^*,}
while acting trivially on $\Lambda_{\rm wt}$.  Clearly, the right hand side of \trombo\
is invariant under this transformation together with $\alpha\to\alpha+\eurm m$.

\bigskip\noindent{\it Several Ramification Points}

We can readily generalize this to the case of several ramification points $p_1,\dots,p_s$.
Associated with each such point is a triple $(\alpha_i,\beta_i,\gamma_i)\in {\frak t}^3$,
with its own action of $\AffWeyl$.  Thus a group $(\AffWeyl)^s$ acts on the collection
of $s$ triples, taking values in $(\frak t^3)^s={\frak t}^{3s}$.

The corresponding
 moduli space $\MH(\alpha_1,\beta_1,\gamma_1,p_1;
\dots;\alpha_s,\beta_s,\gamma_s,p_s)$ of ramified Higgs bundles is a
smooth manifold
if the triples $(\alpha_i,\beta_i,\gamma_i)$ take values in a suitable
parameter space $\eurx_s$.
This space is obtained from $\frak t^{3s}$ by omitting certain affine linear spaces of
codimension three.  To avoid local singularities, one must require that each
triple $(\alpha_i,\beta_i,\gamma_i)$ is separately regular.  To avoid global singularities,
one must omit certain additional affine linear spaces described in \nakajima.

The group $(\AffWeyl)^s$ acts freely on $\eurx_s$, since in defining $\eurx_s$, we require
each triple to be separately regular.  Hence, by the same logic as before, we get
an action of $(\AffWeyl)^s$ on the cohomology, $K$-theory, etc., of $\MH$.

As before, we can describe this action explicitly
if we specialize to the two-dimensional cohomology, which was described in \zomboz:
\eqn\womboz{H^2(\M(\alpha_1,p_1;\dots;\alpha_s,p_s);\Z)=\Z\oplus\left(\oplus_{i=1}^s
\Lambda_{wt,i}\right).}
$s$ copies of the ordinary Weyl group act on the $s$ copies of $\Lambda_{wt}$.
And the lattices act by a generalization of \wombo,
\eqn\gomboz{e\to e\oplus\left(\oplus_{i=1}^s {\eurm m}_i^*\right).}
That this is the right action of $(\AffWeyl)^s$ actually follows from the analog of
\trombo, which is $[\,\omega_I/2\pi]=e\oplus\left(\oplus_i(-\alpha^*_i)\right)$.

One might be puzzled by these results, since a slightly larger
group could act on the lattice \womboz.  Instead of shifting $e$
by $\oplus_{i=1}^s {\eurm m}^*_i$ for a collection of coroots
$\eurm m_i$, why not simply shift it by $\oplus_{i=1}^s w_i$ for
an arbitrary set of weights $w_1,\dots,w_s\in \Lambda_{\rm wt}$,
without worrying about whether $w_i$ is of the form ${\eurm
m}^*_i$ for a coroot $\eurm m_i$?  In fact, in the case of several
ramification points, a group smaller than this but larger than we
have so far described does act naturally on the cohomology of
$\MH( \alpha_1,\dots,p_s)$.  According to \minzo\ and \winzo,
there is a symmetry $\alpha_i\to \alpha_i+u_i$ for any family of
coweights $u_i$, $i=1,\dots,n$, such that $\sum_{i=1}^su_i$ is a
coroot.  (We demonstrated the symmetry for parabolic bundles, but
the same reasoning applies for ramified Higgs bundles.) So the
cohomology of $\MH(\alpha_1,\dots,p_s)$ must admit the action of
the group  $\Lambda^*\rtimes (\Weyl)^s$, where $\Lambda^*$ is the
sublattice of $\oplus_{i=1}^n \Lambda_{{\rm cowt},i}$ consisting
of elements $u_1\oplus\dots \oplus u_s$, with all $u_i\in
\Lambda_{\rm cowt}$ and $\sum_iu_i\in \Lambda_{\rm cort}$.  Since
$\Lambda^*$ has $\oplus_{i=1}^s\Lambda_{{\rm cort},i}$, whose
action we have already described, as a sublattice of finite index,
its action must be given by the same formula: \eqn\gombozo{e\to
e\oplus\left(\oplus_{i=1}^s u^*_i\right).} This makes sense
because the map $u\to  u^*$ does map $\Lambda_{\rm cowt}$ to
$\Lambda_{\rm wt}$, as explained in Appendix A.

\bigskip\noindent{\it Non-Simply-Connected $G$}

What happens if $G$ is not simply-connected?  $\alpha$ now takes
values in $\TT=\frak t/ \Lambda_{\rm cochar}$, and it might seem
that the group that would act on cohomology or $K$-theory of $\MH$
would be now $\Lambda_{\rm cochar}\rtimes \Weyl$.

Whether this is correct depends on precisely what one means.  When
$G$ is not simply-connected, $\MH$ has $\# \pi_1(G)$ components,
labeled by the value of the characteristic class $\xi(E)$ that
measures the obstruction to lifting $E$ to a bundle with
simply-connected structure group $\bar G$.  The action of
$\Lambda_{\rm cochar}$ permutes the components, as discussed in
eqn. \nomo.  The subgroup that acts on the cohomology of one given
component is $\Lambda_{\cort}$. Hence if we restrict our attention
to one fixed component, the  group that acts on the cohomology,
for the case of one parabolic point,  is
$\AffWeyl=\Lambda_{\cort}\rtimes \Weyl$, just as if $G$ is
simply-connected. Similarly, with $s$ parabolic points, the group
that acts on the cohomology of a single component of $\MH$ is
$\Lambda^*\rtimes (\Weyl)^s$, whether $G$ is simply-connected or
not. But the larger group acts if one wants to include
transformations that permute the components.

\bigskip\noindent{\it Ramification Of Type $\Bbb{L}$}

We can consider in a similar fashion a point $p$ endowed with a singularity
labeled by an arbitrary Levi subgroup $\Bbb{L}$.
The parameters $(\alpha,\beta,\gamma)$ labeling such a point are invariant under $\Bbb{L}$, and
a local singularity is avoided precisely if this triple is $\Bbb{L}$-regular.
The group that naturally acts is the subgroup of $\AffWeyl$
that commutes with $\Bbb{L}$.  Let us call this group ${\cal W}_{{\rm aff},\Bbb{L}}$.

The same reasoning as above shows that $\Weyl_{{\rm aff},\Bbb{L}}$ acts on the cohomology
(or $K$-theory, etc.) of $\M_{H,\Bbb{L}}$.  In the extreme case that $\Bbb{L}=G$, this statement
becomes trivial, as $\Weyl_{{\rm aff},\Bbb{L}}$ is then the trivial group.

\bigskip\noindent{\it Generalization To Four-Dimensional Gauge Theory And Sigma Models}

In applications to the geometric Langlands program, we really care not about the
variety $\MH$, but about a two-dimensional sigma model in which the target space is $\MH$.  This
sigma model arises as a low energy approximation to a four-dimensional gauge theory.
More relevant than whether the classical variety is smooth is whether the sigma model
is smooth.  We call the sigma model smooth when its spectrum, correlation functions, etc., vary
smoothly with the parameters.

When $\MH$ is smooth, the sigma model is certainly smooth. However, the sigma model
may remain smooth even when classically $\MH$ has a singularity.
Typically, in two-dimensional sigma models with $(4,4)$ supersymmetry, to get a
singularity of the quantum theory, one must adjust four parameters, not just three.
Three parameters control the classical geometry, and the fourth controls a theta-angle
of the sigma model.

As we learned in section \thetang, in addition to the parameters
$(\alpha,\beta,\gamma) \in \TT\times \frak t\times \frak t$, the
gauge theory depends on another parameter, a theta-angle $\eta\in
\,^L\neg\TT$.    So the parameters labeling a point with ramified
structure are really a quartet\foot{If we wish, we can lift $\alpha$
and $\eta$ to be $\frak t$-valued.  Then the group by which we must
divide is not the affine Weyl group but
 an extension of $\Weyl$ by the product of a pair
of lattices, which act by shifting $\alpha$ and $\eta$, as we
discuss in section \opmon.} $(\alpha,\beta,\gamma,\eta)\in \TT\times
\frak t\times \frak t\times ^L\neg\TT$, with an equivalence  under
the action of the Weyl group on all four variables. This makes sense
because $G$ and $^LG$ have the same Weyl group!

By a local singularity of the sigma model with target
$\MH$, we mean a singularity whose position only depends on the
parameters $(\alpha,\beta,\gamma,\eta)$ labeling a single ramification point.  Global
singularities are those whose positions depend on the parameters of two or more points.

We recall that a point in a
 space acted on by $\Weyl$ is regular if no non-trivial element of $\Weyl$ leaves
the point fixed.  To keep this discussion simple, we begin with the case of one
ramification point, and use the fact that in this case $\MH$ is singular only for
non-regular triples $(\alpha,\beta,\gamma)\in \TT\times\frak t\times \frak t$.
It follows that the sigma model is smooth if $(\alpha,\beta,\gamma)$ is regular.
However, the gauge theory also has the $S$-duality transformation $\tau\to -1/\tau$
which exchanges $\alpha$ and $\eta$, as we discussed in section \duality.  The smoothness
of the sigma model must be invariant under this transformation, so we learn that the
sigma model is smooth if the triple $(\eta,\beta,\gamma)$ is regular.

More generally, we have an infinite discrete
duality group $\Gamma$ acting on $\alpha,\eta$ as in \zolpo.  If $G$ is simply-laced or
we restrict to an index two subgroup of $\Gamma$, then  $\beta,\gamma$ transform by multiplication
by a positive real number
(note the discussion following \oobus).   Such a rescaling does not affect the question of whether
a triple $(\alpha,\beta,\gamma)$ is regular.
 For example, for $G=E_8$, the duality group is $SL(2,\Z)$, and contains a
transformation that maps $\alpha$ to $m\alpha+n\eta$, for any relatively prime integers
$m,n$.  So the sigma model is smooth if $(m\alpha+n\eta,\beta,\gamma)$ is regular.
The fact that this is so for all pairs $m,n$ implies that actually, the sigma model
is smooth if the quartet $(\alpha,\beta,\gamma,\eta)$ is regular.  We would still
reach the same conclusion if we replace $SL(2,\Z)$ by a congruence subgroup.  For
any $G$, the duality group contains a congruence subgroup of $SL(2,\Z)$ that acts as just
described, so it is always
the case that the sigma model is smooth if the quartet $(\alpha,\beta,\gamma,\eta)$ is
regular.

For the case of ramification at several points $p_i$, we can similarly consider
global singularities, which we can define to be simply singularities
whose positions depend upon the parameters
$(\alpha_i,\beta_i,\gamma_i,\eta_i)$ associated with more than one point.  The
duality symmetry can now be used to show that the conditions found in \nakajima\ can
be extended by an additional condition involving the $\eta$'s, so the global singularities
now occur in real codimension four.

\subsec{Nahm's Equations And Local Singularity Of $\MH$}
\subseclab\onahm

Gauge theory and Nahm's equations
can be used to obtain hyper-Kahler metrics on coadjoint orbits
of complex Lie groups.  See
\refs{\kron,\okrontwo}
for the original constructions, \refs{\biquard,\kovalev} for generalizations to arbitrary
orbits, and
\nref\bielawski{R.
Bielawski, ``Lie Groups, Nahm's Equations, And Hyper-Kahler Manifolds,'' math.DG/0509515.}%
\refs{\bielawski,\atbiel}
for reviews and further references.  Our interest in these metrics is that they
give the behavior of $\MH(\alpha,\beta,\gamma;p)$ near a local singularity.

There are several routes to the construction of these hyper-Kahler metrics.
For us, it is most convenient to
consider Hitchin's equations on a punctured disc $C$, defined as the region of the complex
$z$-plane with $|z|\leq 1$, and $z\not= 0$.  We write as usual $z=re^{i\theta}$.
On $C$, we consider solutions of Hitchin's
equations that are invariant under rotations of the disc and have the familiar singularity
near $r=0$:
\eqn\porgox{\eqalign{A & = \alpha \,d\theta +\dots\cr
           \phi & = \beta\,{dr\over r}-\gamma\,d\theta+\dots,\cr}}
We suppose that the triple $(\alpha,\beta,\gamma)$ is regular, that is, that the subgroup
of $G$ that commutes with this triple is precisely the torus $\TT$.
We let $\EUG_{C,p}$ be the group of rotation-invariant gauge transformations
$g:C\to G$ that equal 1 for $|z|=1$ and take  values in $\TT$ at $z=0$.
The space of rotation-invariant solutions of Hitchin's equations, with the boundary
condition \porgox, and modulo the action
of $\EUG_{C,p}$, is a hyper-Kahler manifold that we will call ${\cal Q}(\alpha,\beta,\gamma)$.

In fact, it can be constructed as the hyper-Kahler
quotient by $\EUG_{C,p}$ of the space of rotation-invariant pairs $(A,\phi)$.
Such a pair is given in general by
\eqn\yofgo{\eqalign{A & = a(r)\,d\theta+h(r){dr\over r}
\cr\phi &= b(r){dr\over r}-c(r)\,d\theta\cr}}
with $\frak g$-valued functions $a,b,c$, and $h$.  The functions $a,b,$ and $c$ are
related in a fairly obvious way to the usual parameters $\alpha,\beta,$ and $\gamma$, but
rotational
symmetry allows a fourth function $h$.  Though $h$ can be gauged away, it is more convenient
not to do so for the moment.
The space of solutions  of Hitchin's
equations that are of this form has a hyper-Kahler structure that
is the usual one appropriate to Hitchin's equations, specialized
to this case.  One way to describe it is to think of the functions $(h,a,b,c)$
as giving a map from the open unit interval to the quaternions $\Bbb{H}\cong\R^4$, tensored
with $\frak g$.
The hyper-Kahler structure comes from the hyper-Kahler structure on $\Bbb{H}$.  Concretely,
in one complex structure, which we will call $I$, the complex variables are $h-ia$
and $b+ic$.  The others can be obtained by applying an $SO(3)$ rotation to the triple
$(a,b,c)$.

If we set $s=-\ln r$ and $D/Ds=d/ds+[h,\,\cdot\,]$,
then Hitchin's equations become
\eqn\bogfo{\eqalign{{Da\over Ds}& = [b,c] \cr
                    {Db\over Ds}&=  [c,a] \cr
                     {Dc\over Ds}&= [a,b]. \cr}}
These become Nahm's equations  \ref\nahm{W. Nahm, ``The Construction Of All
Self-Dual Multimonopoles By The ADHM Method,'' in {\it Monopoles And Quantum Field Theory},
ed. N. S. Craigie et. al. (World Scientific, 1982) 87-95.} if we set $h$ to zero, which
we can do locally by a gauge transformation.

To elucidate the nature of the moduli
space ${\cal Q}(\alpha,\beta,\gamma)$ of solutions of these equations, first note
that a linear combination of two of the equations gives
\eqn\ogfo{{d\over ds}(b+ic)=-[h-ia,b+ic].}  This implies that the conjugacy class
of $b+ic$ in $\frak g_\C$ is independent of $s$.
We also have the boundary condition
$\lim_{s\to\infty}(b+ic)=\beta+i\gamma$.  If $\sigma=\half(\beta+i\gamma)$ is regular, this
implies that $\tau=\half(b+ic)$ is everywhere in the conjugacy class $\frak c$ that contains
$\sigma$.  In particular, \eqn\zumo{\tau=\half(b(0)+ic(0))}
is contained in this conjugacy class.  $\tau$  is also gauge-invariant (since we
only allow gauge transformations that equal 1 at $s=0$).  So by mapping a solution of
Nahm's equations to the corresponding value of $\tau$, we get a map
$\Phi:{\cal Q}\to \frak c$
that is holomorphic in complex structure $I$. By interpreting the remaining part of
Nahm's equations as a moment map condition, it is shown in \kron\ that this map is
an isomorphism.

By definition,
$\frak c$ is the orbit under conjugation of the vector $\sigma\in \frak g_\C$
(or equally well in the dual space
$\frak g_\C^\vee$, since the quadratic form $-\Tr$ gives a $G_\C$-invariant
identification between $\frak g_\C$ and $\frak g_\C^\vee$).
It is known as an adjoint (or co-adjoint)
orbit.  For regular $\sigma$, the
subgroup in $G_\C$ that leaves $\sigma$ invariant is precisely
$\TT_\C$, the complexification of the torus $\TT$.  Hence $\frak c$ is
isomorphic to $G_\C/\TT_\C$.

If $\sigma$ is not regular, then $\tau$
is in one of the affiliated orbits $\frak c_\lambda$, described in section
\postpone.  All possibilities occur, but for a generic point in ${\cal Q}$,
$\tau$ takes values in the affiliated orbit
of maximal dimension, which is the regular orbit $\frak c^*$.
To get some insight about what happens when $\sigma$ is non-regular, let us go
to the extreme case $\sigma=0$.  Then the equations and boundary conditions enable
us to find solutions in which $b$ and $c$ identically vanish.  The equations collapse
to $Da/Ds=0$, so the conjugacy class of $a$ is independent of $s$ and equal to that of
$\alpha$.  This conjugacy class
 must be regular, since we have assumed that $(\alpha,\beta,\gamma)$ is regular
and we have set $\beta=\gamma=0$.
We can now reason somewhat as before. $a(0)$ is gauge-invariant and is conjugate
to $\alpha$.  Moreover, modulo the gauge group, $a(0)$ is the only invariant
of a solution with $b=c=0$.  Finally, $a(0)$ can be any element of the orbit of
$\alpha\in \frak g$.  That orbit is (for regular $\alpha$) a copy of $G/\TT$, which
is a Kahler manifold, known as the flag manifold.
To include $b$ and $c$, we note that from a holomorphic point of view in complex structure
$I$, $b+ic$ is characterized by the linear equation \ogfo.  Thus ${\cal Q}$ is a holomorphic
vector bundle over $G/\TT$.  For ${\cal Q}$ to be a complex symplectic manifold (and
actually hyper-Kahler), this bundle must be the cotangent bundle.

So when $\beta=\gamma=0$
but $\alpha$ is generic, ${\cal Q}$ in complex structure $I$ is the cotangent bundle
of $G/T$:
\eqn\milro{{\cal Q}\cong T^*(G/\TT).}
If we deform to $\beta,\gamma\not=0$, the cotangent bundle is deformed to an affine
bundle over $G/\TT$.  When $\sigma=\half(\beta+i\gamma)$ is regular, the affine bundle
is isomorphic to the $G_\C$-orbit of $\sigma\in \frak g_\C$.

\bigskip\noindent{\it Analog Of Type $\Bbb{L}$}

This construction can be repeated with $\TT$ replaced by any Levi subgroup $\Bbb{L}$ of
$G$.

We require the triple $(\alpha,\beta,\gamma)$ to be $\Bbb{L}$-regular (that is, it commutes
precisely with $\Bbb{L}$),
and we modify the definition of $\EUG_{C,p}$ so that it comprises gauge transformations that
at $p$ take values in $\Bbb{L}$.  We write ${\cal Q}_\Bbb{L}(\alpha,\beta,\gamma)$ for the space
of solutions of the equations \bogfo, with boundary conditions set by $\alpha$, $\beta$,
and $\gamma$, modulo the action of $\EUG_{C,p}$.

The same reasoning as above shows that if $\sigma=\half(\beta+i\gamma)$ is $\Bbb{L}$-regular, then
\eqn\yerot{{\cal Q}_\Bbb{L}\cong G_\C/\Bbb{L}_\C}
is the orbit of $\sigma\in \frak g_\C$ under the adjoint action of $G_\C$.
At the other extreme, if $\sigma=0$ but $\alpha$ is generic,
\eqn\zerot{{\cal Q}_\Bbb{L}\cong T^*(G/\Bbb{L}).}
In general, ${\cal Q}_\Bbb{L}$ is an affine deformation of $T^*(G/\Bbb{L})$.
If we replace $G$ by a general semi-simple Lie group $\Bbb{L}'$ containing $\Bbb{L}$ as a Levi subgroup,
the same construction based on gauge theory with gauge group $\Bbb{L}'$
 gives a family of hyper-Kahler metrics on
\eqn\berot{{\cal Q}_{\Bbb{L}',\Bbb{L}}\cong T^*(\Bbb{L}'/\Bbb{L})}
and its affine deformations.

\bigskip\noindent{\it Local Model Of $\M_H$}

Many properties that we have described for Hitchin's moduli space $\M_H$ have
local analogs involving the hyper-Kahler metrics ${\cal Q}$.

For example,
the cohomology classes of the symplectic forms of $\M_H$ are linear functions of
the parameters $\alpha,\beta,$ and $\gamma$.  A similar linearity holds for ${\cal Q}$,
as stated in Theorem 2.6 of \kron.

To give another example, just as the affine Weyl group acts on the cohomology of $\MH$,
the ordinary Weyl group similarly acts on the cohomology
of   ${\cal Q}$.  The framework for proving this is described in section 5 of \atbiel, using
the action of the Weyl group on the parameters $(\alpha,\beta,\gamma)$ and the fact
that the singularities are in codimension three -- in other words, the same facts
that we used in section \actweyl\ to construct an affine Weyl group action on cohomology
of $\MH$. Actually, what is considered in \atbiel\ is
a somewhat larger class of hyper-Kahler varieties constructed in \okrontwo\ and involving
Slodowy slices.  (These more general varieties are constructed by solving the same
equations as above but with different asymptotic behavior at $s=0$.)
The Weyl group representations that arise for these varieties are
known as the Springer representations. They can be understood geometrically in a relatively
elementary construction \ref\kl{D. Kazhdan and G. Lusztig, ``A Topological Approach
To Springer's Representations,'' Adv. Math. {\bf 38} (1980) 222-228.} without hyper-Kahler
metrics;  perhaps this has also  an analog for $\MH$.

The basic reason that the hyper-Kahler manifolds ${\cal Q}$ give local models
for many properties of $\MH$ is that they do in fact describe the behavior of
the moduli space of ramified Higgs bundles near a local singularity.  For example,
as in \peftoxical, the local behavior of $\M_{H,\Bbb{L}}(\alpha,0,0;p)$ when $\alpha$ approaches
a  value $\bar\alpha$ that is not $\Bbb{L}$-regular is modeled by $T^*(\Bbb{L}'/\Bbb{L})$ for some Levi
group $\Bbb{L}'$ that properly contains $\Bbb{L}$.  To get a local
model of the situation, we need a suitable family of hyper-Kahler metrics on $T^*(\Bbb{L}'/\Bbb{L})$
(and its affine deformations with $\beta,\gamma\not=0$).  This is what we get from
the construction summarized above, as noted in \berot, if we take the gauge group
to be $\Bbb{L}'$.  There is
a simple rationale
for using $\Bbb{L}'$ gauge theory rather than $G$ gauge theory to construct the local model:
the boundary conditions at the singularity, defined in this case by the trio
$(\bar\alpha,0,0)$, are invariant only under $\Bbb{L}'$, not $G$, so a theory with $\Bbb{L}'$ as the
gauge symmetry suffices for describing the singularity.

\bigskip\noindent{\it Detailed Analysis For $SU(2)$}

To  understand more fully why the varieties ${\cal Q}$ give a good description of local
singularities of $\MH$, we will examine more closely the behavior of ${\cal Q}$ near a
non-regular point.  To keep things simple, we concentrate on $G=SU(2)$ for illustration.

The only nonregular value of the triple $(\alpha,\beta,\gamma)$ for $G=SU(2)$ is
$\alpha=\beta=\gamma=0$.  At $\alpha=\beta=\gamma=0$, ${\cal Q}$ describes solutions
of Hitchin's equations on the disc $|z|\leq 1$ that are rotation-invariant and less singular
than $1/r$ at $r=0$.

Equivalently, taking the gauge $h=0$, we need solutions of Nahm's equations
\eqn\bigskin{\eqalign{{da\over ds}&=[b,c]\cr
                      {db\over ds}&=[c,a]\cr
                      {dc\over ds}&=[a,b],\cr}}
on the half-line $[0,\infty)$,  with $a,b,c\to 0$ for $s\to \infty$.

Obviously, one such solution is $a=b=c=0$.  Another
simple solution, which is the starting point in \okrontwo, is
\eqn\yeflo{\eqalign{a&=-{1\over s}t_1\cr
                    b&=-{1\over s}t_2\cr
                    c&=-{1\over s}t_3,\cr}}
where $t_1,t_2,$ and $t_3$ are fixed elements of ${\frak sl}(2,\C)$ obeying
$[t_1,t_2]=t_3$, and cyclic permutations thereof.

In complex structure $I$, this solution describes a Higgs bundle $(E,\varphi)$ in which
$\varphi$ has a pole with nilpotent residue.
(Notice that $b+ic=-s^{-1}(t_2+it_3)$
is in fact nilpotent for all $s$.)
In complex structure $J$, it describes a flat bundle with monodromy around the point $r=0$
that is unipotent but not equal to 1.  These results are what one might expect from
section \postpone.

A slight generalization is to introduce a positive constant $f$ and take
\eqn\zeflo{\eqalign{a&= -{1\over s+f^{-1}}t_1\cr
                    b&=-{1\over s+f^{-1}}t_2\cr
                    c&=-{1\over s+f^{-1}}t_3,\cr}}
We have parametrized the solutions in this particular way, because \zeflo\ actually
has a limit for $f\to 0$, namely the trivial solution $a=b=c=0$.  So $f$ takes
values in $\R_{\geq 0}=[0,\infty)$.
In our application, we will be concerned with small $f$.
We can also
pick an element $R\in SO(3)$ and generalize \zeflo\ to
\eqn\heflo{\eqalign{a&= -{1\over s+f^{-1}}Rt_1R^{-1}\cr
                    b&=-{1\over s+f^{-1}}Rt_2R^{-1}\cr
                    c&=-{1\over s+f^{-1}}Rt_3R^{-1}.\cr}}
The parameter space of this family is thus $\R_{\geq 0}\times SO(3)=\R^4/\Z_2$,
where $\Z_2$ acts by a reflection on all four coordinates of $\R^4$.
This is the $A_1$ singularity.
In particular, the natural
metric on this family ${\cal Q}$ of solutions of Nahm's or Hitchin's equations, obtained by
integrating the ${\eurm{L}}^2$ norm of the variation of the fields $(A,\phi)$ over the disc
$r\leq 1$, is the flat metric on $\R^4/\Z_2$; all modes are square-integrable, thanks
to the factor of $1/s=1/(-\ln r)$, and $f=0$
is at finite distance.  For nonzero $\alpha,\beta,\gamma$, we get instead the Eguchi-Hansen
metric, describing the deformation and resolution of the $A_1$ singularity.

Now we can explain the basic reason that Nahm's equations and Kronheimer's construction
give a good model for local singularities of $\MH$.  The singularity of ${\cal Q}$ occurs
at $f=0$ and corresponds to the trivial solution $a=b=c=0$.  A point in ${\cal Q}$
near the singularity
corresponds to a solution with very small $f$.  When Nahm's equations
are embedded in Hitchin's equations, there is an extra factor of $1/r$,
visible in eqn. \yofgo.
This means that for small $f$, all fields are very small except for $r\lesssim f$.
This continues to be the case if we perturb
$\alpha,\beta,\gamma$ to be nonzero but of order $f$.  Because of this, we can start
with an arbitrary Higgs bundle $(E,\varphi)$ without ramification, and ``glue in''
the above family of solutions, to get a family of ramified Higgs bundles.
We use the old solution for $r>>f$, and the exact solution \heflo\ for $r\lesssim f$.
The family of solutions obtained this way acquires an $A_1$ singularity if one sets $f=0$,
whereupon the ``new'' solutions reduce to the old ones.

Of course, the gluing operation does not give an exact solution.  The set of fields
obtained by gluing must be modified to get an exact solution, but the requisite
modification is small if $f$ is small.
This is somewhat analogous to the construction
\ref\taubes{C. H. Taubes, ``Self-Dual Connections
On 4-Manifolds With Indefinite Intersection Matrix,'' J. Diff. Geom. {\bf 19} (1984)
517-560.} of Yang-Mills instanton solutions on a four-manifold
by gluing in an exact solution from $\R^4$ that has its support mainly on a very small
region in $\R^4$.  However, in contrast to instanton moduli space, the deformation theory
of Higgs bundles is unobstructed (as long as we keep away from reducible Higgs bundles)
so there is no analog of the topological conditions described in \taubes\ that can potentially
obstruct the process of deforming the glued fields to an exact solution.

Thus we get a more precise way to see what was argued in section \topology: for $\alpha,
\beta,\gamma\to 0$, $\MH(\alpha,\beta,\gamma;p)$ develops an $A_1$ singularity, the
singular locus being precisely $\MH$, the moduli space of unramified Higgs bundles.
This specific result is of course special to $G=SU(2)$.
For $G$ of higher rank, as we have argued,
a similar construction leads in general
to more complicated singularities whose resolution is $T^*(\Bbb{L}'/\Bbb{L})$ for various $\Bbb{L}'$ and $\Bbb{L}$.

\subsec{The Hitchin Fibration}\subseclab\hitchfib

\nref\beau{A. Beauville, ``Jacobiennes des Courbes
Spectral et Syst\`emes Hamiltoniens Compl\`etement Int\'grables,'' Acta Math. {\bf 164}
(1990) 211-35.}%
\nref\markdon{E. Markman and R. Donagi, ``Spectral Curves, Algebraically Completely
Integrable Hamiltonian Systems, And Moduli Of Bundles,'' in Lecture Notes in Math. vol. 1620
(Springer, 1996) pp.  1-119,  alg-geom/9507017.}%
Now we describe the Hitchin fibration and complete integrability in the context of
Higgs bundles with ramification.  All of these matters have been understood
in the literature in much more detail; for example, see \refs{\beau,\markdon}.

We begin with the example of $SL(2,\C)$, from section \postpone.
Up to conjugacy, the local behavior of $\varphi$ near a parabolic point with $\sigma\not=0$
is
\eqn\gilo{\varphi={dz\over z}\sigma(1+{\cal O}(z)).}  This implies that
\eqn\bilo{\Tr\,\varphi^2=\Tr\,\sigma^2 \left({dz\over z}\right)^2(1+\CO(z)).}
This statement is actually true uniformly for all $\sigma$, zero or not.
Indeed,
for $\sigma=0$,
the polar part of $\varphi$ is nilpotent, but in a general solution
$\varphi$ also has regular terms.  After trivializing the holomorphic structure near
$z=0$, we have
up to conjugacy
\eqn\filo{\varphi={dz\over z}\left(\left(\matrix{0&1\cr 0&0\cr}\right)+az+bz^2+\dots\right),}
with $a,b\in {\frak sl}(2,\C)$.
Hence the quadratic differential $\Tr\,\varphi^2$
may have a simple pole at $z=0$, but no double
pole, showing that \bilo\ also holds for $\sigma=0$.

The reason that this is important is that the quadratic differential $\Tr\,\varphi^2$
is the key to the complete integrability of $\MH$ as a complex manifold in complex
structure $I$.  In the absence of ramification, $\MH$ has (complex)
dimension equal, for $G_\C=SL(2,\C)$,
to $6g_C-6$, where $g_C$ is the genus of $C$.  To establish complete integrability,
one requires $3g_C-3$ commuting Hamiltonians.  These are precisely the components of
$\Tr\,\varphi^2$, which takes values in the space of quadratic differentials on $C$. The
 dimension of that space is $3g_C-3$.
 (For a very brief explanation of complete integrability of Hitchin systems,
 see section 4.3 of \kapwit.)

Let us carry out the analogous computation in the presence of ramification.
For simplicity in the exposition, we suppose that there is just one ramified
point $p$.  A Higgs bundle is a pair $(E,\varphi)$, where $E$ is an
$SL(2,\C)$-bundle over $C$ and $\varphi\in H^0(C,K_C\otimes {\rm ad}(E)\otimes \CO(p))$.
Here we include the factor of $\CO(p)$ (the bundle whose sections are functions that
may have a simple pole at $p$), since $\varphi$ is allowed to have a pole at $p$.
The number of parameters required to specify the bundle $E$ is $3g_C-3$.  By Riemann-Roch,
the  dimension of
$H^0(C,K_C\otimes {\rm ad}(E)\otimes \CO(p))$ is $3g_C$.  (Indeed, $H^0(C,K_C\otimes{\rm ad}(E))$
is the cotangent space to the moduli space of stable bundles and has dimension $3g_C-3$;
tensoring with $\CO(p)$ adds 3 to the dimension, since $K_C\otimes {\rm ad}(E)$ has rank
3.)  However, $\varphi$ obeys the one constraint \filo.  So the choice of $\varphi$
depends on $3g_C-1$ parameters, and the dimension of $\MH(\alpha,\beta,\gamma;p)$ is
$(3g_C-3)+(3g_C-1)=6g_C-4$.  So we need $3g_C-2$ commuting Hamiltonians to establish
complete integrability.

These are precisely the components of $\Tr\,\varphi^2$.  In general, $\Tr\,\varphi^2$
is a quadratic differential with a double pole.  The space of such quadratic differentials
has dimension $3g_C-1$, but one parameter is determined by \bilo, so $\Tr\,\varphi^2$
lives in a $3g_C-2$ dimensional space.  This gives the $3g_C-2$ parameters needed for
complete integrability of $\MH(\alpha,\beta,\gamma;p)$.

All generalities about the Hitchin fibration of $\MH$, complete integrability, etc.,
have natural analogs for Higgs bundles with ramification.  We write $\EUBB$ for
the space of quadratic differentials with double pole at $p$  obeying \bilo.
The Hitchin fibration is a map $\pi:\MH(\alpha,\beta,\gamma;p)\to\EUBB$ which is holomorphic
in complex structure $I$. It is defined by mapping a Higgs bundle $(E,\varphi)$
to the point in $\EUBB$ defined by $\Tr\,\varphi^2$.  The functions on $\EUBB$ are the commuting
Hamiltonians, and the fibers are complex Lagrangian submanifolds
(that is, they are Lagrangian from the point of view of the holomorphic two-form $\Omega_I$).
The generic fiber is an abelian variety.  Moreover, all this remains true if one has
ramification at several points, not just one.

For any $G$, the analog of this is as follows.   Let $r$ be the rank of $G$.
The ring of invariant polynomials on the Lie algebra $\frak g$ is freely generated by $r$
polynomials ${\cal P}_i$, which are homogeneous of degree $d_i$ for some integers $d_i$.
These integers obey
\eqn\yeo{\sum_{i=1}^r(2d_i-1)={\rm dim}(G).}
Instead of the quadratic differential $\Tr\,\varphi^2$, we consider its analogs
${\cal P}_i(\varphi)$, which are holomorphic sections of $K_C^{d_i}$.  They obey
\eqn\nilo{{\cal P}_i(\varphi)=
{\cal P}_i(\sigma)\left({dz\over z}\right)^{d_i}(1+{\cal O}(z)).}
These conditions hold for all $\sigma$, and, for $\Bbb{L}=\TT$, the differentials ${\cal P}_i(\varphi)$
otherwise have arbitrary poles of degree $d_i$ at $z=0$.
For other $\Bbb{L}$, there are additional conditions on the poles.  We will concentrate here
on the case $\Bbb{L}=\TT$.
$\EUBB$ is defined by saying that a point in $\EUBB$ is a collection of
$d_i$-differentials, $i=1,\dots, r$, that are holomorphic away from ramification points and
behave like \nilo\ near such a point.  The dimension of
$\EUBB$ (for the case of one ramified point) is
$(g_C-1){\rm dim}(G)+\sum_{i=1}^r(d_i-1)$, as one can prove with the help
of \yeo.

The dimension of $\MH$ is computed as we did for $SU(2)$.  The choice of  a $G$-bundle $E$
depends on $(g_C-1){\rm dim}(G)$ parameters.  The dimension of
$H^0(C,K_C\otimes {\rm ad}(E)\otimes\CO(p))$
is $g_C{\rm dim}(G)$, but $\varphi$ is subject to $r$ constraints \nilo, so it
depends on $g_C{\rm dim}(G)-r$ parameters.  The dimension of $\MH(\alpha,\beta,\gamma;p)$
is thus
\eqn\defrop{\eqalign{{\rm dim}(\M_H(\alpha,\beta,\gamma;p))
=&\,(2g_C-1)\,{\rm dim}(G)-r =(2g_C-2)\,{\rm dim}(G)+\sum_{i=1}^r(2d_i-2)\cr=&
\,{\rm dim}(\MH)+{\rm dim}(G)-r.\cr}}
The number of commuting Hamiltonians needed to establish complete integrability is therefore
$(g_C-1){\rm dim}(G)+\sum_{i=1}^r(d_i-1)$.  This is precisely the dimension of $\EUBB$.
The Hitchin fibration $\pi:\MH(\alpha,\beta,\gamma;p)\to \EUBB$, defined by mapping
a pair $(E,\varphi)$ to the point in $\EUBB$ defined by
$({\cal P}_1(\varphi),\dots,{\cal P}_r(\varphi))$, has all
the key properties that we described for $G=SL(2,\C)$.

The formula \defrop\ for the (complex) dimension of $\MH(\alpha,\beta,\gamma;p)$ is,
of course, compatible with the existence, in a suitable limit, of the generic fibration
\peftoxical.    For a more general Levi subgroup $\Bbb{L}$, there is a natural
modification of this, with a suitable extension of the constraints \nilo.

\newsec{Geometric Langlands With Tame Ramification}
\seclab\tame

\subsec{Review Of Unramified Case}
\subseclab\unramreview

The basic steps to get from ${\cal N}=4$ Yang-Mills theory to geometric
Langlands are described in \kapwit, beginning in section 3.  We briefly review
them, since the same procedure applies in the presence of ramification.

One first makes
a certain topological twist of the ${\cal N}=4$ theory, the GL twist.  The twisting
depends on a complex parameter $t$, and leads to a family of four-dimensional topological
field theories parametrized by $\Bbb{CP}^1$.  It is convenient to think of this
$\Bbb{CP}^1$ as the complex $\Psi$ plane plus a point at infinity.  $\Psi$ is a certain
combination of $t$ and the gauge coupling parameter $\tau=\theta/2\pi+4\pi i/e^2$.
A duality transformation of the ${\cal N}=4$ theory that acts on $\tau$ by
$\tau\to (a\tau+b)/(c\tau+d)$ acts likewise on $\Psi$ by $\Psi\to (a\Psi+b)/(c\Psi+d)$.
In particular, the duality operation $S:\tau\to -1/n_{\frak g}\tau$ maps $\Psi=\infty$
to $\Psi=0$.  The usual form of the geometric Langlands
program involves these two values of $\Psi$ and the duality between them.

The next step \refs{\vafa,\hm}
is to consider the theory on a four-manifold $M=\Sigma\times C$, where $C$
is the Riemann surface on which one wishes to study the geometric Langlands program.
In an appropriate limit of
$\Sigma$ much larger than $C$, or in the topological field
theory (in which distances are irrelevant), the compactified theory can be described in
terms of a sigma model on $\Sigma$ in which the target is the moduli space $\MH(G,C)$
of Higgs bundles on $C$.

At $\Psi=\infty$, as explained in section 5 of \kapwit,
the resulting two-dimensional theory is the $B$-model of $\MH$
in complex structure $J$.  It varies holomorphically in $\alpha+i\gamma$, and is locally
independent of $\beta$ and $\eta$.
At $\Psi=0$, it is the $A$-model with symplectic structure \eqn\olpo{\omega=({\rm Im}\,\tau)\omega_K,}
where the factor of ${\rm Im}\,\tau={4\pi/e^2}$ comes from the normalization of the kinetic energy.
We call these models the $B$-model of type $J$ and the $A$-model of type $K$.
Each of these models is independent of the four-dimensional gauge coupling $e^2$,
and hence can be studied at weak coupling.

The next step is to let $\Sigma$ have a boundary, labeled by a
brane $\cal B$. $S$-duality automatically exchanges $B$-branes of
type $J$ with $A$-branes of type $K$. This is the basic geometric
Langlands duality.  To get the usual formulation of geometric
Langlands duality, one must incorporate Wilson and 't Hooft
operators and the duality between them (sections 6,8,9, and 10 of
\kapwit), and one must reinterpret the $A$-model of type $K$ in
terms of ${\cal D}$-modules on $\M(G,C)$ (section 11).

\subsec{Sigma Model With Ramification}
\subseclab\ramcase

We now begin to describe an analogous program in the presence of
ramified  structure at a point $p\in C$.  (To keep the notation
simple, we consider the case of one ramification point, except
when it is essential to allow several.) We consider gauge theory
on $M=\Sigma\times C$, with gauge group $^LG$, in the presence of
a ``surface operator'' supported on $\Sigma_p=\Sigma\times p$. For
tame ramification, which is the subject of this paper, we take the
surface operator to be of the familiar type, labeled by the
parameters
$(\,^L\neg\alpha,^L\neg\beta,^L\neg\gamma,^L\neg\eta)\in\,
^L\TT\times \frak t^\vee\times \frak t^\vee\times \,\TT$. $^L\TT$ and
$\TT$ are exchanged, of course, relative to most of our previous
discussion, since we take the gauge group to be $^LG$.  Until further notice,
we take the Levi subgroup to be $\Bbb{L}=\TT$.

The duality transformation $S:\tau\to -1/n_{\frak g}\tau$ maps the
$^LG$ theory with coupling $^L\neg \tau$ to a theory with gauge group $G$ and coupling
$\tau=-1/n_\frak g{}^L\neg\tau$.  To get the basic geometric Langlands duality, it is convenient
to take $^L\tau$ to be imaginary and to take the twisting parameter of the $^LG$ theory to be $t=i$.
This gives the $B$-model at $\Psi=\infty$, and is mapped by $S$ to a model with $t=1$ and imaginary
$\tau$.  The latter gives a convenient description of the $A$-model at $\Psi=0$.

A surface operator with parameters $(\,^L\neg\alpha,^L\neg\beta,
^L\neg\gamma,^L\neg\eta)$ maps to a surface operator
of the same type but with different parameters
$(\alpha,\beta,\gamma,\eta)$. The relation among the parameters was
analyzed in section \duality:
\eqn\relad{\eqalign{(\alpha,\eta)&=(^L\neg\eta,-^L\neg\alpha)\cr
                     (\beta,\gamma)&=({\rm Im}\,^L\neg\tau)\,\,(^L\beta^*,^L\neg\gamma^*).\cr}}
                     The second formula is eqn. \oobus, and can be inverted as in \zoobus:
                     \eqn\elad{({}^L\neg\beta,{}^L\neg\gamma)=({\rm Im}\,\tau)(\beta^*,\gamma^*).}

Our goal is to adapt the steps that were summarized in section
\unramreview. The four-dimensional theory on $M=\Sigma\times C$
reduces in this situation to a two-dimensional sigma model with
target the moduli space of ramified Higgs bundles. Just as in the
absence of ramification, this sigma model has $(4,4)$
supersymmetry, since the target space is hyper-Kahler.

The surface parameters $(\alpha,\beta,\gamma)$ enter the classical
geometry of the moduli space. The remaining surface parameter $\eta$
was introduced in section \thetang\ by analogy with  theta-angles of
two-dimensional gauge theory. Indeed, for $G$ of rank $r$ (and
barring some exceptional cases of small $r$ and small $g_C$),
$\MH(\alpha, \beta,\gamma;p)$ has second Betti number $r+1$, so the
sigma model has room for $r+1$ theta-angles.  One such angle
descends from the theta-angle $\theta$ of the underlying
four-dimensional gauge theory, by precisely the reasoning described
in eqn. (4.18) of \kapwit.  The other $r$ theta-angles of the
effective two-dimensional theory are derived from $\eta$.  As we see
in \relad, $S$-duality exchanges theta-angles and geometrical
parameters.

\bigskip\noindent{\it Discrete Electric and Magnetic Fluxes}

The  sigma model additionally depends on  discrete electric and
magnetic fluxes (see section 7 of \kapwit, where a somewhat
different point of view is taken). We describe them in some detail,
since otherwise it is impossible to give a completely precise
description of the action of electric-magnetic duality.

The bundle $E\to M=\Sigma\times C$ has a characteristic class
$\xi(E)\in H^2(M,\pi_1(G))$.
We write ${\bf m}_0$ for the restriction of this class to $q\times C$, for $q$ a generic
point in $\Sigma$.  ${\bf m}_0$ takes values in $H^2(C,\pi_1(G))\cong \pi_1(G)$,
and $\MH$ has components labeled by ${\bf m}_0$.

Discrete electric flux needs more explanation. We introduce a slight twist into
 $G$ gauge theory to allow bundles  $E\to M=\Sigma\times C$ with structure group
$G_{\rm ad}$ but with the property that, when restricted to $q\times C$ (for $q$ a point
in $\Sigma$), $E$ can
be lifted to a $G$-bundle.  Thus, locally along $\Sigma$ but globally along $C$,
the theory has gauge group $G$, but globally along $\Sigma$ there may be a ``twist'' that
prevents lifting $E$ to a $G$-bundle.
A bundle of this type, restricted to $\Sigma\times r$ for generic $r\in C$, has a
characteristic class ${\bf a}_0\in H^2(\Sigma,\pi_1(G_{\rm ad}))$.

Now in performing the path integral, we introduce a discrete
theta-angle ${\bf e}_0$ and include in the path integral a phase
factor \eqn\enons{\exp(2\pi i\langle {\bf e}_0,{\bf a}_0\rangle).}
Thus, ${\bf e}_0$ (though written additively) is a character of
$\pi_1(G_{\rm ad})$.  However, in $G$ gauge theory, it is natural
to consider only ${\bf e}_0$ that annihilate $\pi_1(G)\subset
\pi_1(G_{\rm ad})$ (so that ${\bf e}_0$ is only sensitive to the
``twist''). Thus ${\bf e}_0$ is a character of $\pi_1(G_{\rm
ad})/\pi_1(G)$ (this is the same as the center of $G$, ${\cal
Z}(G)$). Equivalently, \eqn\helj{{\bf e}_0\in \Lambda_{\rm
char}/\Lambda_{\rm rt}.}
  If we restrict
${\bf e}_0$ in this way, we can project ${\bf a}_0$ to \eqn\elj{
\pi_1(G_{\rm ad})/\pi_1(G)=\Lambda_\cowt/\Lambda_\cochar.} The
duality transformation $S:\tau\to -1/n_{\frak g}\tau$ exchanges
${\bf e}_0$ and ${\bf m}_0$.  This is possible because
$\pi_1(\,^L\neg G)=\pi_1(G_\ad)/\pi_1(G)$, and similarly with
$^L\neg G$ and $G$ exchanged.

Ramification does not substantially change the definition of ${\bf
m}_0$ and ${\bf e}_0$. However, their role is qualitatively
different in the presence of ramification. As we observed in
discussing eqn. \nomo, ${\bf m}_0$ can be changed by shifting
$\alpha$. Dually, as we will now explain, ${\bf e}_0$ can be
changed by shifting $\eta$. The effect of $\eta$ in the path
integral is a factor \eqn\bnons{\exp(2\pi i\langle\eta,\eurm
m\rangle),} where $\eurm m$ is the characteristic class of the
$\TT$-bundle obtained by restricting $E$ to $\Sigma\times p$, for
$p\in C$. In standard $G$ gauge theory, $\eurm m$ takes values in
$\Lambda_{\rm cochar}$, but the ``twist'' means that in the
present context $\eurm m$ takes values in $\Lambda_{\rm cowt}$.
Indeed, we have \eqn\ytropo{\eurm m \cong {\bf a}_0~~{\rm
mod}~\Lambda_{\rm cochar}.} This is a general fact about the
two-dimensional characteristic class of  a $G_{\rm ad}$-bundle
whose structure group reduces to the torus $\TT_{\rm ad}$.

Because $\eurm m$ does not necessarily take values in $\Lambda_{\rm cochar}$, the
usual symmetry $\eta\to\eta+v$ for $v\in \Lambda_{\rm char}$ does not necessarily hold.
Rather, comparing \enons\ to \bnons\ and using \ytropo, we find that
$\eta\to \eta+v$ is equivalent to ${\bf e}_0\to {\bf e}_0+\bar v$, where
$\bar v$ is the image of $v$ in $\Lambda_{\rm char}/\Lambda_{\rm rt}$.
The shifts of $\eta$ that do not change ${\bf e}_0$ are by $\Lambda_{\rm rt}$.

This mirrors the corresponding statement for $\alpha$, which is
that the shifts that do not change ${\bf m}_0$ are precisely those
by $\Lambda_{\rm cort}$.  The parallelism remains if there are
several ramification points $p_1,\dots,p_s$.  As in the discussion
of eqn. \gombozo, the shifts of $(\alpha_1,\dots,\alpha_s)$ that
do not change ${\bf m}_0$ are $\alpha_i\to\alpha_i+u_i$, with
$u_i\in \Lambda_{\cowt}$ and $\sum_i u_i\in \Lambda_\cort.$
Likewise, the shifts in $(\eta_1,\dots,\eta_s)$ that do not change
${\bf e}_0$ are $\eta_i\to \eta_i+v_i$, where $v_i\in \Lambda_\wt$
and $\sum_iv_i\in \Lambda_\rt$.

\subsec{Branes} \subseclab\brano

Now we specialize to the topological field theories that arise at
$\Psi=\infty$ with gauge group $^L\neg G$  or at $\Psi=0$ with gauge
group $G$.

At $\Psi=\infty$, we get the $B$-model in complex structure $J$. In
this complex structure,
$\MH(^L\neg\alpha,^L\neg\beta,^L\neg\gamma;p)$ parametrizes flat
$^LG_\C$-bundles  $E\to C\backslash p$ whose monodromy $V$ around
the point $p$ obeys a condition that depends on $^L\neg\alpha$ and
$^L\neg\gamma$.

As explained most fully in section \postpone, the condition is that
the orbit of $V$ under conjugation contains in its closure the
element $U=\exp(-2\pi(^L\neg\alpha-i^L\neg\gamma))$.  If $U$ is
regular, this means that $V$ is conjugate to $U$.  If, at the other
extreme, $U=1$, this means that $V$ is unipotent. At any rate, for
any $V\in\, ^L\neg G_\C$, there is a choice of $^L\neg\alpha$ and $^L\neg\gamma$,
unique up to a Weyl transformation, such that a flat connection with
monodromy $V$ represents a point in
$\MH(^L\neg\alpha,^L\neg\beta,^L\neg\gamma;p)$. We simply choose
$^L\neg\alpha$ and $^L\neg\gamma$ so that $U$ is contained in the
closure of the orbit of $V$.

So if we want to use the $B$-model in complex structure $J$ to say
something about a flat bundle with a given monodromy, then
$^L\neg\alpha$ and $^L\neg\gamma$ are uniquely determined.  But
$^L\neg\beta$ and $^L\neg\eta$ are arbitrary, since from the point
of view of complex structure $J$, they are Kahler parameters. The
$B$-model in complex structure $J$, of course, varies
holomorphically with $^L\neg\gamma+i^L\neg\alpha$, but it is locally
independent of $^L\neg\beta$ and $^L\neg\eta$.  Globally, when we
vary $^L\neg\beta$ and $^L\neg\eta$, the $B$-model has monodromies,
which we will study in section \opmon.

In the geometric Langlands program with tame ramification, we
begin with a flat $^L\neg G_\C$-bundle $E\to C\backslash p$ with
monodromy $V$.  Roughly speaking, at the right value of
$^L\neg\alpha$ and $^L\neg\gamma$, $E$ determines a zerobrane
${\cal B}_E$ on
$\MH(^L\neg\alpha,\,^L\neg\beta,\,^L\neg\gamma;p)$.  (When $V$ is
non-regular, this statement requires some elaboration, which we
provide in section \repbrane.)

Now we apply the $S$-duality transformation $S:\Psi\to -1/n_{\frak
g}\Psi$.  The gauge group is transformed from $^LG$ to $G$. $\Psi$
is mapped from $\infty$ to 0; the resulting model at $\Psi=0$ is
the $A$-model with symplectic structure $\omega=({\rm Im\,\tau})\omega_K$. The parameters
$(\alpha,\beta,\gamma,\eta)$ of the model with gauge group $G$ are
expressed in \relad\ in terms of the $^L\neg G$ parameters
$(^L\neg\alpha,\,^L\neg\beta,\,^L\neg\gamma,\,^L\neg\eta)$. In
particular, $^L\neg\gamma+i^L\neg\alpha$, on which the  model at
$\Psi=\infty$ depends holomorphically, is equal according to \elad\ to $({\rm Im}\,\tau) \gamma^*-i\eta$.
So we expect the $A$-model at $\Psi=0$ to vary holomorphically
in that variable.  Indeed, $({\rm Im}\,\tau)\gamma^*-i\eta$ is the complexified
Kahler class from the standpoint of complex structure $K$.
(The symplectic form $\omega=({\rm Im}\,\tau)
\omega_K$ has cohomology class proportional to $({\rm Im}\,\tau)\gamma^*$, in view of
\dpofo, while $\eta$ supplies the imaginary part of the complexified Kahler class.)
Likewise, since the model at $\Psi=\infty$ with gauge group
$^L\neg G$ is locally independent of $^L\neg\beta$ and $^L\neg\eta$, we
expect the dual model at $\Psi=0$ with gauge group $G$ to be
locally independent of $\alpha$ and $\beta$.  This is in accord
with the fact that, as a symplectic variety with symplectic
structure $\omega_K$, $\MH$ is independent of $\alpha$ and
$\beta$.  The $A$-model depends only on this symplectic variety
with its complexified Kahler form $({\rm Im}\,\tau)\gamma^*-i\eta$. On the other
hand, the complex structure $K$, which is irrelevant in the
$A$-model, varies holomorphically in $\alpha+i\beta$.

\def\CMF{{\cmmib F}}
The duality transformation $S$ maps branes in the $B$-model of
type $J$ to branes in the $A$-model of type $K$.  So in
particular, the zerobrane ${\cal B}_E$ that is determined by a
ramified flat bundle $E$ is mapped to an $A$-brane $\hat{\cal
B}_E$. We can see quite concretely what sort of $A$-brane this
will be. With or without ramification, $\MH$ admits the Hitchin
fibration $\pi:\MH\to \EUBB$, as we described in section
\hitchfib. The generic fibers of the Hitchin fibration are complex
tori, holomorphic in complex structure $I$. $S$-duality acts via
$T$-duality on the fibers of the Hitchin fibration, according to
the same reasoning as in sections 5.4 and 5.5 of \kapwit.  (This
duality has been studied mathematically from several points of
view, including applications to the geometric Langlands program
\nref\thadhau{
M. Thaddeus and T. Hausel, ``Mirror Symmetry, Langlands Duality, And
The Hitchin System,'' math.AG/0205236.}%
\nref\bravbez{A. Braverman and R. Bezrukavnikov, ``Geometric Langlands Correspondence
For ${\cal D}$-Modules In Prime Characteristic: The $GL(N)$ Case,'' math.AG/0602255.}%
\nref\donpan{R. Donagi and T. Pantev, ``Langlands Duality For
Hitchin Systems,'' math.AG/0604617.}%
\refs{\thadhau-\donpan}.) $T$-duality on the fibers of a torus
fibration maps a zero-brane to a brane wrapped on a fiber, and
endowed with a flat Chan-Paton line bundle ${\cal L}$. So just as
in the absence of ramification, the $S$-dual of a zerobrane is a
brane of this type. Such a brane  is called in \kapwit\ a brane of
type $\CMF$.

Now we can make a preliminary statement of the geometric Langlands duality.  In stating it,
we think of $\MH$ not just as a classical space but as defining a quantum sigma model,
so we will include $\eta$ when we list its parameters.
Also, in making the statement, we make
the gauge groups explicit, recalling that electric-magnetic duality exchanges $G$ and
$^L\neg G$.

Our first statement of geometric Langlands duality is that for every
zerobrane in the sigma model with target
$\MH(^L\neg\alpha,^L\neg\beta,^L\neg\gamma,^L\neg\eta,p;\,^L\neg
G)$, there is a corresponding brane of type $\CMF$ in the sigma
model with target $\MH(\alpha,\beta,\gamma,\eta,p;G)$; as usual, the
parameters are related by
\relad.
More generally, for every $B$-brane of type $J$ (that is, every brane of the $B$-model
in complex structure $J$) in the sigma model with target
$\MH(^L\neg\alpha,^L\neg\beta,^L\neg\gamma,^L\neg\eta,p;\,^L\neg
G)$, there is a naturally corresponding $A$-brane of type $K$ in the
sigma model with target $\MH(\alpha,\beta,\gamma,\eta,p; G)$.  This
correspondence extends to a natural correspondence between all of
the structures of the $B$-model on one side and all of the
structures of the $A$-model on the other side.

The analogs of Wilson and 't Hooft operators are described in section \opmon,
and some details will be clarified in section \repbrane. But the most
pressing problem is that the geometric Langlands correspondence is
usually stated with  $B$-branes of type $J$ on the left hand
side (just as in the last paragraph), but with ${\cal D}$-modules
of an appropriate sort, rather than $A$-branes of type $K$, on the
right hand side. To reconcile the two points of view, we will
follow the approach of section 11 of \kapwit, which the reader may
want to consult.

\subsec{Twisted ${\cal D}$-Modules}

In our formulation, the right hand side of the geometric Langlands correspondence involves
a brane in the topological field theory at $\Psi=0$.
As in \kapwit, a convenient way to get to $\Psi=0$ is
to take the twisting parameter $t$ to equal 1, and to take the
four-dimensional theta-angle $\theta=2\pi\,{\rm Re}\,\tau$ to vanish.
The branes at $\Psi=0$ are branes of the $A$-model of type $K$.

The usual statement of the geometric Langlands duality involves ${\cal D}$-modules
on the moduli space $\M(\alpha,p;G)$ of parabolic bundles.
To reconcile the two formulations, we must associate to every $A$-brane
of type $K$ on $\MH(\alpha,\beta,\gamma,\eta,p;G)$ a ${\cal
D}$-module of a suitable type  on $\M(\alpha,p;G)$. When confusion
is unlikely, to make the notation less clumsy, we will generally
omit to specify the gauge group $G$.

As explained in section \topology, precisely if $\beta=\gamma=0$ and
$\alpha$ is regular, $\MH(\alpha,\beta,\gamma,\eta;p)$ is birational
in complex structure $I$ to  the cotangent bundle of $\M(\alpha;p)$.
More precisely, for sufficiently large $g_C$, away from high codimension, $\MH$ contains the
cotangent bundle as a dense open set:
\eqn\bogup{T^*\M(\alpha,p;G)\subset\MH(\alpha,0,0,\eta;p).} Setting
$\beta$ to zero and assuming that $\alpha$ is regular are not severe
restrictions, since the $A$-model of type $K$ is independent of
these parameters.  But the $A$-model does depend on $\gamma$, so we
will want to restore the $\gamma$ dependence later.

Also, taking $\alpha$ to be regular
means choosing an affine Weyl chamber that contains $\alpha$.  This
in a sense reduces the symmetry of the model.
So one could argue that the description of the duality in terms of ${\cal D}$-modules, which
depends on this choice, is less natural than the description in terms of mirror symmetry
between an $A$-model and a $B$-model.\foot{One could argue the same based on the fact that the
mirror symmetry preserves the full supersymmetry of branes on both sides of the duality,
while the description in terms of ${\cal D}$-modules does not.  For an explanation of this
point, see the introduction to section \thooft.}  However, the description by ${\cal D}$-modules
is motivated by an analogy with number theory, and we want to see how it comes about.

Let $X$ be any hyper-Kahler manifold with complex structures $I,J,K$ and symplectic
forms $\omega_I,\omega_J,\omega_K$ such that $[\,\omega_J]=0$
(we recall that $[\,\omega]$ denotes the cohomology class of a
closed form $\omega$).  Pick a positive real number ${\rm Im}\,\tau$,
and consider the $A$-model on $X$ with symplectic structure $\omega=({\rm Im}\,\tau)\,\omega_K$. There
is as explained in section 11.1 of \kapwit\ a distinguished
space-filling $A$-brane.  Its Chan-Paton line bundle is a trivial
line bundle endowed with a connection whose curvature form is
$F=({\rm Im}\,\tau)\,\omega_J$.
This formula ensures that
$(\omega)^{-1}F$ is a complex structure, indeed the one that we have called $I$.  This
is the condition for
 a coisotropic brane in the sense of \ref\kapor{A.
Kapustin and D. Orlov, ``Remarks On $A$-Branes, Mirror Symmetry,
and the Fukaya Category,'' J. Geom. Phys. {\bf 48} (2003) 84-99,
arXiv:hep-th/019098.}.  We call this brane the c.c. or  canonical
coisotropic brane, and denote it as ${\cal B}_{\rm c.c.}$. For
${\cal B}_{\rm c.c.}$ to exist, the cohomology class
$[\,\omega_J]$ must vanish, so that it is possible for a trivial
line bundle to have a connection form $F=({\rm Im}\,\tau)\,\omega_J$. Indeed,
$[\,\omega_J]=0$ precisely if $\beta=0$, as was explained most
directly at the end of section \hyperquo.  At any rate, we have
already set $\beta=0$ to ensure the relationship of $\MH$ to
$T^*\M$.

\def\CB{{\cal B}}
\def\cc{{\rm c.c.}}
For any brane $\cal B$, the $(\CB,\CB)$ strings form a ring. To all orders in
sigma model perturbation theory, one can ``localize'' the
$(\CB,\CB)$ strings, by considering wavefunctions that are regular
in an open set ${\cal U}\subset X$.
The open strings that are regular in  ${\cal U}$ form a ring, and by
letting ${\cal U}$ vary, one gets, to all orders in perturbation theory, a sheaf of rings over $X$.
In general, this construction of a sheaf of rings is only valid to all orders in perturbation
theory, not as an exact statement.
However, as explained in \kapwit\  (and as is certainly known in the mathematical literature; for
a much deeper analysis, see \ref\konts{M. Kontsevich, ``Deformation Quantization Of
Algebraic Varieties,'' math.AG/0106006.}),
one can under certain conditions go beyond perturbation theory in a
very special way in the case of the canonical coisotropic    brane
$\CB_\cc$ of a hyper-Kahler manifold $X$. The requisite conditions
are that $X$
should contain as a dense open set a cotangent bundle $T^*{\cal M}$ (for some Kahler
manifold ${\cal M}$),
where this identification is holomorphic in complex structure $I$
and $\omega_K$ is the imaginary part of the natural holomorphic
two-form of the cotangent bundle. In this case, one can
associate a ring of $(\CB_\cc,\CB_\cc)$ strings to each open set
in $X$ that is of the form $T^*{\cal U}$ for ${\cal U}$ an open set in
$\M$.  This association gives a sheaf of rings over $\M$.  This sheaf of rings is
precisely
the sheaf of differential operators on $\M$ acting on sections of
some line bundle\foot{More generally, as reviewed in section 11.1
of \kapwit, ${\cal L}$ may be a complex power of a line bundle, or
a tensor product of such complex powers. The sheaf of rings ${\cal
D}_{\cal L}$ is invariant under twisting ${\cal L}$ by a flat line
bundle, and as a result makes sense in this greater generality
\ref\beibern{
A. Beilinson and J. Bernstein, "A Proof of Jantzen Conjectures,"
  Advances in Soviet Mathematics 16, Part 1, pp. 150, AMS, 1993.}.
For the same reason, in what follows it does not matter if
$K_\M^{1/2}$ exists or is unique as a line bundle.  It actually
does exist but is not always unique.} $\cal L$. We write ${\cal
D}_{\cal L}$ for this sheaf of rings.

${\cal L}$ can be identified
in a particularly simple way if the sigma model of $X$ has
time-reversal symmetry, which is the case if its theta-angles
vanish.  If $K_\M$ denotes the canonical line bundle of $\M$, then
time-reversal symmetry implies that ${\cal L}=K_\M^{1/2}$ and the
sheaf of rings is ${\cal D}_{K_\M^{1/2}}$.

To apply this to our problem, we want the theta-angles of the sigma
model with target $\MH(\alpha,\beta,\gamma,\eta;p)$ to vanish. We
have already set the four-dimensional parameter $\theta$ to zero;
the remaining theta-angles of the sigma model are precisely $\eta$.
So we achieve the time-reversal symmetry by setting $\eta=0$.  We
have already simplified the problem by setting $\gamma=0$, so our
combined condition is equivalent to
the vanishing of the complexified Kahler class
$({\rm Im}\,\tau)\gamma^*-i\eta$.  The condition $\eta=0$ is equivalent, of course, to $^L\neg\alpha=0$.

The fact that the sheaf of $(\CB_\cc,\CB_\cc)$ strings coincides with the sheaf of rings
${\cal D}_{K_\M^{1/2}}$ makes possible a very general construction.
Let $\CB'$ by  any $A$-brane on $X$ of type $K$.
To $\CB'$, we can associate the sheaf of $(\CB_\cc,\CB')$ strings.
It is automatically a sheaf of modules for the $(\CB_\cc,\CB_\cc)$ strings, that is,
for ${\cal D}_{K_\M^{1/2}}$.  So we get a natural way to associate a
${\cal D}_{K_\M^{1/2}}$-module to every $A$-brane of type $K$.

Combining this with what we learned from $S$-duality, we see that
a $B$-brane of type $J$ on $\MH(0,{}^L\neg\beta,0,{}^L\neg\eta,
p;{}^L\neg G)$ is naturally associated to a sheaf of modules for
the sheaf of rings ${\cal D}_{K_\M^{1/2}}$ over $\M(\alpha,p;G)$
(with as usual $\alpha=\,^L\neg\eta$). This is
essentially\foot{The usual statement involves the sheaf of rings
${\cal D}$, that is the sheaf of differential operators acting on
functions. This sheaf of rings is Morita-equivalent to ${\cal
D}_{K_\M^{1/2}}$.  There are some subtleties that we will not consider here involving
the dependence of this Morita-equivalence on a choice of spin structure on $C$.}
the usual statement of the geometric Langlands
program with tame ramification, for the case
$^L\neg\alpha=\,^L\neg\gamma=0$, or in other words for flat
$^L\neg G$ bundles with unipotent monodromy.

As in \kapwit, the fact that the $(\CB_\cc,\CB_\cc)$ strings are the
differential operators on $K_\M^{1/2}$ (rather than on some more
general line bundle) can be seen more explicitly, without relying on
time-reversal symmetry. For this, we make use of another important
brane on $\MH$: the brane ${\cal B}'$ supported at $\varphi=0$, that
is, on the zero section of the cotangent bundle $T^*\M$, and endowed
with a trivial Chan-Paton line bundle. Quantization of the
$(\CB_\cc,\CB')$ strings shows that they are sections of
$K_\M^{1/2}$.  Since they also furnish a sheaf of modules for the
$(\CB_\cc,\CB_\cc)$ strings, the latter are the differential
operators acting on sections of $K_\M^{1/2}$.

\bigskip\noindent{\it Restoring The Parameters}

The case $\gamma=\eta=0$ is the case that $U=\exp(-2\pi(^L\neg
\alpha-i^L\neg \gamma))$ is equal to 1. Being limited to this case
would mean describing geometric Langlands only for the case of
unipotent monodromy.  To get beyond this case, we must restore the
dependence on $\gamma$ and $\eta$.

As we will see, incorporating either $\gamma$ or $\eta$ has the
effect of replacing differential operators that act on
$K_\M^{1/2}$ by differential operators that act on ${\cal S}$,
where loosely speaking ${\cal S}$ is a more general line bundle --
more precisely, it is a tensor product of complex powers of line
bundles. The sheaf of differential operators acting on such an
${\cal S}$ makes sense, though ${\cal S}$ itself cannot be defined
as a line bundle.
  We aim to
{\it (i)} justify the claim that for
any $\gamma$ and  $\eta$, the $A$-branes on $\MH$ of type $K$ are
naturally associated with modules for some sheaf of rings of this
type, and {\it (ii)} identify ${\cal S}$.

Let us first justify the claim {\it (i)}.  Let us  consider the
effect of having $\gamma\not=0$, with $\eta$ still vanishing.  The
result of this, as explained in section \topology, is that $\MH$ is
no longer birational to $T^*\M$. Rather, it contains as a dense open
set an affine deformation of $T^*\M$ (giving us a situation similar to the case
of $\theta\not=0$, as analyzed in section 11.3 of \kapwit).   By itself, replacing the cotangent
bundle with an affine deformation of it does not
modify the definition of the c.c.  brane
$\CB_\cc$ (as long as we keep $\beta=0$; see below).  Nor does it modify the argument
that the sheaf of $(\CB_\cc,\CB_\cc)$ strings is the sheaf of
differential operators acting on some ``line bundle'' ${\cal S}$.
However, it does spoil the
use of time-reversal symmetry to show that ${\cal S}=K_\M^{1/2}$.
Indeed, the definition of the relevant time-reversal operation
requires that $\MH$ should have a symmetry $\varphi\to-\varphi$
(acting holomorphically in complex structure $I$); this symmetry is
absent when $T^*\M$ is deformed to an affine bundle. The alternative
argument for the identification ${\cal S}\cong K_\M^{1/2}$ uses the brane ${\cal B}'$
supported on $\M\subset T^*\M$, which is precisely the zero-section of the
cotangent bundle.  This argument also fails for $\gamma\not=0$,
since the affine deformation of $T^*\M$ admits no holomorphic
section.

Now let us
consider the effect of having $\eta\not=0$, with $\gamma$ zero or
nonzero. Because of holomorphy in $({\rm Im}\,\tau)\gamma^*-i\eta$, the effect of $\eta$ must
be similar to that of $\gamma$, but it is interesting to see how this comes
about.  Having $\eta\not=0$ causes the $B$-field of the sigma
model, which we simply call $B$, to be nonzero.  As a result, some care is
needed in defining the canonical coisotropic brane. Let $F$ denote
the curvature of the Chan-Paton bundle ${\cal L}$ of this brane. In
the presence of a $B$-field, the condition for a coisotropic brane is that
$\omega^{-1}(F+B)$ should be an integrable complex structure, which we will take to be
$I$.  We achieve this by taking $F+B=({\rm Im}\,\tau)\omega_J$.
We will still assume that ${\cal L}$ is topologically trivial.  This
requires at the level of cohomology classes that $[F]=0$, so $[B]=({\rm Im}\,\tau)[\,\omega_J]$.
Since $[\,\omega_J/2\pi]=-\beta^*$, as we learned in \dpofo, and
$[B/2\pi]=\eta$, the c.c brane can be defined only for
$\beta^*=-({\rm Im}\,\tau)^{-1}\eta$.  So for nonzero $\eta$, we must have $\beta\not=0$, and hence
again $\MH$ is birational to an affine deformation of $T^*\M$, not to $T^*\M$ itself.

We assumed here that ${\cal L}$ is topologically trivial,
but we can also construct branes with non-trivial ${\cal L}$.  As we explain below,
the results that can be obtained this way are Morita-equivalent to what we can learn from the
case that ${\cal L}$ is topologically trivial.

As before, once the brane $\CB_\cc$ is defined, the standard
arguments show that the sheaf of  $(\CB_\cc,\CB_\cc)$ strings is the
sheaf of differential operators acting on some ``line bundle''
${\cal S}\to \M$. However, as in the case of deforming $\gamma$, we cannot
argue that ${\cal S}=K_\M^{1/2}$. We cannot use time-reversal invariance to
prove that ${\cal S}=K_\M^{1/2}$, because having $\eta\not= 0$ violates time-reversal
symmetry. What happens if we try to argue using the brane ${\cal B}'$ is described later.

To determine ${\cal S}$, we proceed as follows. We already know that
at $\gamma=\eta=0$, ${\cal S}=K_\M^{1/2}$.  So we write ${\cal S}=
K_\M^{1/2}\otimes {\cal S}'$, where ${\cal S}'$ is trivial for
$\gamma=\eta=0$. As a tensor product of complex powers of line bundles over
$\M$, ${\cal S}'$ has a first Chern class $w\in H^2(\M,\C)$.
(Concretely, if ${\cal S}'=\otimes_i{\cal L}_i^{g_i}$ with ordinary
line bundles ${\cal L}_i$ and $g_i\in \C$, then
$w=\sum_ig_ic_1({\cal L}_i)$.) We formally write ${\cal S}'={\cal
L}^w$ to express the statement that the first Chern class of ${\cal
S}'$ is $w$.  We wish to determine $w$.

Actually, $w$ is linear in the Kahler parameter $z=({\rm Im}\,\tau)\gamma^*-i\eta$.  This follows by
precisely the reasoning that was used in section 11.3 of \kapwit\
to show that an analogous exponent $f(\theta)$ depends linearly on
the four-dimensional parameter $\theta$.  The idea is to calculate
for weak coupling, taking $e^2=4\pi/{\rm Im}\,\tau$ to be small with $z$ fixed.
Then $w$ can be computed in perturbation theory in
$z/{\rm Im}\,\tau$.  Concretely, perturbation theory
is used to compute the cocycle that enters in an explicit
description of the ring structure of the sheaf of
$(\CB_\cc,\CB_\cc)$ strings. The structure of perturbation theory is such that only a
linear term in $z$ can appear. For more detail, see the discussion of
eqn. (11.39) in \kapwit.

It remains to determine exactly which linear function of $({\rm Im}\,\tau)\gamma^*-i\eta$ is equal to $w$.
$w$ takes values in the second cohomology of $\M(\alpha,p)$ with complex coefficients.
According to \combo\ and \zelko,  this is $(\Z\oplus \Lambda_\char)\otimes\C$.
As $\eta+i({\rm Im}\,\tau)\gamma^*$
takes values in $\Lambda_\char\otimes \C$, there is a naturally defined
linear function
\eqn\nolgo{w(\gamma,\eta)=0\oplus (-\eta-i({\rm Im}\,\tau)\gamma^*),}
and we claim that this is the right result.

By analogy with the treatment of dependence on $\Psi$ or $\theta$ in section 11.3 of \kapwit,
we will justify this result by showing that it holds when $\gamma=0$ and $\eta$ is equal
to a lattice vector $v\in \Lambda_\char$.  The key step is to define the brane ${\cal B}'$
that is, roughly speaking, supported on the zero section of the cotangent bundle
$\M\subset T^*\M\subset \MH$.  Once we find this brane, the $({\cal B}_{c.c.},{\cal B}')$ strings
give a natural sheaf of modules for the sheaf of rings provided by the
$({\cal B}_{c.c.},{\cal B}_{c.c.})$ strings.
By identifying this module, we can identify the sheaf of rings.

In carrying out this program, we run into an important detail.  To define the brane
${\cal B}_{c.c}$ at $\eta\not=0$, assuming that its Chan-Paton bundle ${\cal L}$ is trivial,
we need to set $\beta^*=-({\rm Im}\,\tau)^{-1}\eta$,
as we explained above.  This means that from a holomorphic
point of view in complex structure $I$, $\MH$ is not birational to $T^*\M$, but to an
affine deformation thereof, and hence we do not have a holomorphic embedding $\M\subset\MH$.
This may appear to obstruct the definition of the brane ${\cal B}'$,  which is supported on
$\M$.

However, for our present purposes, we are not interested in what happens holomorphically
in complex structure $I$.  Rather, our concern is the $A$-model of type $K$, in which the
target space is $\MH$ understood as a symplectic variety with symplectic structure a multiple
of $\omega_K$.
As we explained in section \cpxview, as a symplectic variety with
symplectic structure $\omega_K$, $\MH$ is naturally independent of $\beta$.
The $\beta$-independence is established by thinking of $\MH$ as a moduli space of flat
bundles in complex structure $J$.

For $\beta\not=0$, we do not have an embedding $\M\subset
\MH$ that is holomorphic in complex structure $I$, but we do have such an embedding
that is Lagrangian with respect to the symplectic structure $\omega_K$.
To get such an embedding,
we start at $\beta=0$ with the usual embedding $\M\subset T^*\M\subset \MH$, holomorphic
in complex structure $I$.  Defined this way, $\M$ is a complex Lagrangian submanifold in complex
structure $I$ -- that is, it is holomorphic in complex structure $I$ and Lagrangian for
the holomorphic symplectic form $\Omega_I=\omega_J+i\omega_K$.  In particular, $\M$ is
Lagrangian with respect to $\omega_K$.

Now we change our point of view and think of $\MH$ as a complex symplectic manifold with
complex structure $J$.  {}From this point of view, $\MH$ is canonically independent of $\beta$.
(We encounter no singularities in varying $\beta$, since we are taking $\alpha$ regular in
order to aim for an answer involving ${\cal D}$-modules on $\M(\alpha,p)$.)
When $\beta $ is varied, we do not change the holomorphic symplectic structure
$\Omega_J=\omega_K+i\omega_I$.  So in particular, we vary $\beta$ keeping fixed the symplectic
structure $\omega_K$.  So the submanifold $\M\subset \MH$ found in the last paragraph,
if understood merely as a Lagrangian submanifold of type $K$, is naturally defined for any
$\beta$.  We take it to be literally independent of $\beta$, if $\beta$ is varied keeping
fixed the ramified flat bundle determined (in complex structure $J$) by a point in $\MH$.

To get an $A$-brane ${\cal B}'$ supported on $\M$, we need a unitary
line bundle ${\cal L}'\to \M$
whose curvature $F$ obeys $F+B|_{\cal M}=0$. ${\cal L}'$ will be the Chan-Paton
line bundle of the brane ${\cal B}'$. Here $B$ is the $B$-field which (since we
assume that $\theta=0$) is determined by $\eta$.  In fact, $B$ is a closed two-form on
$\MH$ whose cohomology class is $0\oplus \eta\subset
(\Z\oplus\Lambda_\char)\otimes_\Z\R=H^2(\M(\alpha,
p);\R)$.  For generic $\eta$, this cohomology
class is not a lattice vector, and hence no suitable line bundle
${\cal L}'$ or brane ${\cal B}'$ exists.
However, if $\eta=v$ for some $v\in \Lambda_\char$, then we can take
\foot{The minus sign in the exponent of the following formula arises
as follows.  There is no minus sign in the relation between $B$ and $\eta$: $[B/2\pi]=\eta$.
So as $[F+B]=0$, we have $[F/2\pi]=-\eta$.  Hence if $\eta=v$, the first Chern class of the Chan-Paton
line bundle is $-v$.}
${\cal L}'={\cal L}^{-v}$,
that is, ${\cal L}'$ is a line bundle with first Chern class
$0\oplus (-v)$.

Having thus defined an $A$-brane ${\cal B}'$ of type $K$, we now consider the $({\cal B}_{c.c.},
{\cal B}')$ strings, which will furnish a sheaf of modules for the sheaf of rings derived
from the $({\cal B}_{c.c.},{\cal B}_{c.c.})$ strings.  In quantizing the
$({\cal B}_{c.c.},{\cal B}')$ strings, the only thing that is new, compared to the case $\eta=0$,
is that the Chan-Paton line bundle of the brane ${\cal B}'$, instead of being trivial,
is now
${\cal L}'={\cal L}^{-v}$.  This means that the sheaf of $({\cal B}_{c.c.},{\cal B}')$
strings, instead of being the sheaf of sections of $K_\M^{1/2}$, is now the sheaf of sections of
$K_\M^{1/2}\otimes {\cal L}^{-v}$.  This shows that \nolgo\ is valid for $\gamma=0,$ $\eta
\in\Lambda_\char$, and hence for all $\gamma,\eta$.

We can now restate the geometric Langlands
duality for tamely ramified flat bundles
in terms of ${\cal D}$-modules rather than mirror symmetry.
On the left hand side of the geometric Langlands correspondence, we consider
 a ramified flat bundle $E\to C$ with structure
group $^L\neg G$ and monodromy whose orbit contains in its closure the semisimple
element $U=\exp(-2\pi({}^L\neg\alpha-i{}^L\neg\neg\gamma))$.  On the right hand side,
the dual of such a ramified flat bundle is a sheaf of modules for
the sheaf of differential operators on $\M(\alpha,p;G)$ twisted by $K_\M^{1/2}\otimes
{\cal L}^{-(\eta+i({\rm Im}\,\tau)\gamma^*)}$, which is the same as
$K_\M^{1/2}\otimes
{\cal L}^{^L\neg\alpha-i{}^L\neg\gamma}$.  Thus, the ``exponent'' of the line bundle is the same
as the ``eigenvalue'' of the monodromy.
$S$-duality establishes a natural correspondence between objects of these two kinds.
(For a flat bundle with non-regular monodromy, this statement needs some clarification,
which we defer to section \repbrane.)
This version of the geometric Langlands duality has been conjectured
mathematically.  See the survey in section 9.4 of \ofrenkel, and additional references in the
introduction.

\bigskip\noindent{\it Informal Explanation}

The explanation that we have just given is the most precise one that we know.
However, some readers may prefer an alternative explanation that we will just present informally.

Consider any complex variety ${\cal M}$ and let $X=T^*{\cal M}$, endowed with its natural holomorphic
symplectic form $\Omega_X$.  Quantization of $X$ leads to differential operators on ${\cal M}$ twisted by
$K_\M^{1/2}$.  Now let us replace $X$ by a variety $Y$ that is an affine deformation of the
cotangent bundle, meaning that there is a holomorphic fibration $\pi:Y\to {\cal M}$, whose fibers
are those of the cotangent bundle, but $\pi$ admits no holomorphic section.  Locally, one can pick
a holomorphic section and identify $Y$ with $X$; globally, the obstruction to this is determined
by an element $w\in H^1({\cal M},T^*{\cal M})$.  Let us furthermore require that $Y$ admits
a holomorphic symplectic form $\Omega_Y$ that locally (once we pick a local holomorphic section of $\pi$,
giving a local identification of $Y$ with $X$) coincides with $\Omega_X$.  This is equivalent to
saying that $w$ can be represented by a complex-valued $(1,1)$-form on ${\cal M}$ that is annihilated
by both the $\bar\partial$ and $\partial$ operators and hence in particular by $d=\partial+\bar\partial$.
This is automatically so if ${\cal M}$ is a compact Kahler manifold.
In this situation, it is natural to claim that quantization of $Y$ with symplectic form $\Omega_Y$
leads to differential operators on ${\cal M}$ twisted by $K_{\cal M}^{1/2}\otimes {\cal L}^w$,
where ${\cal L}^w$ is symbolically a ``line bundle'' with first Chern class $w$.

A convenient way to characterize the twisting parameter $w$ is as follows.  The
holomorphic symplectic form $\Omega_X$ has vanishing cohomology class.  Indeed, if $q^\alpha$ are local
coordinates on ${\cal M}$ and $p_\alpha$ are the canonical momenta, then $\Omega_X=\sum_\alpha
\,dp_\alpha\wedge dq^\alpha=d(\sum_\alpha\,p_\alpha\,dq^\alpha)$.  Here the one-form $\lambda=
\sum_\alpha \,p_\alpha\,dq^\alpha$ is globally defined, so $[\Omega_X]=0$.  By contrast, it is a
standard result that the cohomology class of $\Omega_Y$ is equal to the class in
$H^2({\cal M},\Bbb{C})$  of the closed $(1,1)$-form $w$ (or more precisely, the pullback of this
class to $Y$).  We will simply write $w$ for this cohomology class.

Now in our problem, $\MH$ in complex structure $I$ is (away from a set of high codimension)
an affine deformation of the cotangent bundle of $\M$,
with holomorphic symplectic form
$\Omega=({\rm Im}\,\tau)\Omega_I=({\rm Im}\,\tau)(\omega_J+i\omega_K)$.
According to \dpofo, the cohomology class of $\Omega$ is
$w=-({\rm Im}\,\tau)(\beta^*+i\gamma^*)$.  To construct the c.c. brane that is needed to relate the
category of $A$-branes to ${\cal D}$-modules, we had to take $\beta^*=-({\rm Im}\,\tau)^{-1}\eta$, so the
twisting parameter is actually $w=\eta-i({\rm Im}\,\tau)\,\gamma^*$, as found above.

The reason that we consider this explanation heuristic, even if perhaps more understandable for some
readers, is that we are not entitled to arbitrarily postulate what properties the c.c. brane should
have. This should be deduced as part of the standard framework of quantum field theory.  The arguments
given above, together with those in section 11 of \kapwit, are an attempt to do this.

\bigskip\noindent{\it Symmetry Of Lattice Shift}

One might now ask  why have we not, in the above derivation, seen the symmetry
of shifting $\eta$ by a lattice vector, say $\eta\to\eta+v_0$ for $v_0\in\Lambda_\char$.
In fact, we lost this symmetry by assuming that the Chan-Paton bundle ${\cal L}$ of the
c.c. brane was trivial.  The lattice shift $\eta\to\eta+v_0$ acts on all branes
by tensoring their Chan-Paton bundle with ${\cal L}^{-v_0}$.  To restore the symmetry,
we simply add additional coisotropic branes ${\cal B}_{c.c.}^{v_0}$ constructed from
${\cal B}_{c.c.}$
by shifting $\eta$ by $v_0$ and tensoring the Chan-Paton bundle with ${\cal L}^{-v_0}$.
We do not need to introduce any more ${\cal B}'$ branes.
In fact, the full set of ${\cal B}'$ branes constructed above already  has the shift symmetry,
as these branes were defined for all possible lattice vectors $v$ and all $\beta.$

{}From a mathematical point of view, what we learn by contemplating the shift
symmetry is something called Morita-equivalence.
If ${\cal L}^{v_0}$ is an honest line bundle (rather than a complex power of line bundles),
then the sheaf of differential operators
acting on sections of
$K_\M^{1/2}\otimes {\cal L}^w\otimes {\cal L}^{v_0}$ is Morita-equivalent to the
sheaf of differential operators acting on sections of
$K_\M^{1/2}\otimes {\cal L}^w$.  The Morita-equivalence
is established by considering a ``bimodule'' which in the present context is the
sheaf of $({\cal B}_{c.c.},{\cal B}_{c.c.}^{v_0})$ strings.

\bigskip\noindent{\it Analog of Type $\Bbb{L}$}

Everything that we have said  has a direct analog for an arbitrary Levi
subgroup $\Bbb{L}$, as opposed to the case $\Bbb{L}=\TT$ that we have considered so far.
We write ${\cal W}_\Bbb{L}$ for the Weyl group of $\Bbb{L}$.

For general $\Bbb{L}$, the parameters $(\alpha,\beta,\gamma,\eta)$ are restricted to
be ${\cal W}_\Bbb{L}$-invariant.  With this understood, the statement of the duality as a relation
between a $B$-model at $\Psi=\infty$ and an $A$-model at $\Psi=0$ requires no essential
modification.  The duality establishes a natural correspondence between $B$-branes
of $\M_{H,{}^L\neg\Bbb{L}}({}^L\neg\alpha,{}^L\neg\beta,{}^L\neg\gamma,{}^L\neg\eta;p,{}^L\neg G)$
and $A$-branes of $\M_{H,\Bbb{L}}(\alpha,\beta,\gamma,\eta;p,G)$, with the usual relation
among the parameters.

To express the duality in terms of ${\cal D}$-modules, we must assume that $\alpha$ is
$\Bbb{L}$-regular.  Then we denote as $\EUP$ the parabolic subgroup determined by the pair
$(\Bbb{L},\alpha)$, and we introduce $\M_\Bbb{L}(\alpha;p,G)$, the moduli space of $G$-bundles over
$C$ with parabolic structure of type $\EUP$ at the point $p$.  For $\beta=\gamma=0$,
the Higgs bundle moduli space\foot{We omit $\eta$ here as it does not enter the classical
geometry.} $\M_{H,\Bbb{L}}(\alpha,\beta,\gamma;p,G)$ contains
$T^*{\M}_\Bbb{L}(\alpha;p,G)$
as a dense open subspace, as we observed in \jury.  For $\beta,\gamma\not=0$,
$\M_{H,\Bbb{L}}(\alpha,\beta,\gamma;p,G)$ contains as a dense open subspace an affine deformation
of the cotangent bundle.

These geometrical facts and the existence of a canonical coisotropic brane can be used
exactly as above to get a
natural map from an $A$-brane of $\M_{H,\Bbb{L}}(\alpha,\beta,\gamma,\eta;p,G)$
to a ${\cal D}$-module.  All the key statements and arguments are the same.  The only
changes are that the statements now refer to $\M_{H,\Bbb{L}}$ and $\M_\Bbb{L}$ (rather than
$\MH$ and $\M$), and the parameters
are ${\cal W}_\Bbb{L}$-invariant.

In one sense, the statement for general $\Bbb{L}$ is slightly
less natural than the statement for $\Bbb{L}=\TT$.
Different choices of $\Bbb{L}$-regular $\alpha$ (which
correspond of course to different choices of $^L\neg\eta$) lead to different choices of
$\EUP$. They  therefore lead to ${\cal D}$-modules
over moduli spaces of bundles with parabolic structures of different types.
The parabolic structure is of type $\EUP$,
where $\EUP$ can be any parabolic subgroup of $G_\C$ that contains $\Bbb{L}$ as a Levi subgroup.
These different spaces, however, are birationally equivalent.

\subsec{Line Operators And Monodromies}\subseclab\opmon

In the absence of ramification, line operators, supported at a point
$q\in C$ (times a line in $\Sigma$ that runs along its boundary),
act in a natural way on branes.  This is explained in section 6.4 of
\kapwit.

At $\Psi=\infty$, the natural line operators are the Wilson
operators, which are classified by a choice of representation of the
gauge group.  We take the gauge group at $\Psi=\infty$ to be $^L\neg
G$, so a Wilson operator is labeled by a representation of that
group. In general, Wilson operators can change the discrete electric
flux ${\bf e}_0$. The Wilson operators that leave fixed ${\bf e}_0$
are those that are derived from representations of $^L\neg G_\ad$,
the adjoint form of $^L\neg G$.

At $\Psi=0$, the natural line operators  are 't Hooft operators,
and the gauge group is $G$.  The 't Hooft operators of $G$ gauge
theory are constructed using singular BPS monopoles and again are
classified by representations of $^LG$. In general, 't Hooft
operators change the discrete magnetic flux ${\bf m}_0$ (into
which ${\bf e}_0$ transforms under the duality). The 't Hooft
operators that leave ${\bf m}_0$ fixed are those associated with
representations of $^LG_{\rm ad}$.

So at either $\Psi=\infty$ or $\Psi=0$, the line operators that act
at a given point $q\in C$, in the absence of ramification,
correspond to representations of $^L\neg G$, or  of $^L\neg G_\ad$
if we want only operators that leave the discrete fluxes fixed. The
composition of these Wilson or 't Hooft operators corresponds to the
tensor product of representations.  A central statement of the geometric Langlands
program is that the duality between $^L\neg G$ and $G$ maps a Wilson operator
labeled by a representation of $^L\neg G$ to an 't Hooft operator labeled by the
same representation.

Now suppose that $p\in C$ is a ramification point. We want to
determine what structure at $p$ replaces the action of the Wilson
or 't Hooft line operators at a generic point. We will show that
the answer to this question can be described in terms of the
monodromies in the space of ramification parameters.

\bigskip\noindent{\it Duality-Symmetric Monodromies}

We begin by repeating the reasoning of section \actweyl\ in a
duality symmetric way. For simplicity, we start with the case that
the Levi subgroup is $\Bbb{L}=\TT$.

For gauge group $G$, the definition of ramified structure at a
point $p$ depends on a choice of parameters
$(\alpha,\beta,\gamma,\eta)\in \frak t^3\times \frak t^\vee$,
modulo the action of a certain group of equivalences. This group
is generated by {\it (i)} the translations of $\alpha$ by the
lattice $\Lambda_\cochar$; {\it (ii)} the translations of $\eta$
by the dual lattice $\Lambda_\char$; and {\it (iii)} the Weyl
group $\Weyl$.  So the group that acts on the parameters is an
extension \eqn\nory{\hat {\cal
V}=\left(\Lambda_\cochar\oplus\Lambda_\char\right)\rtimes\Weyl.}
The quotient $(\frak t^3\times \frak t^\vee)/\hat{\cal V}$ is the
same as $(\TT\times \frak t\times\frak t\times ^L\neg\TT)/\Weyl$.

If we want to restrict to translations of $\alpha$ and $\eta$ that
do not shift the discrete fluxes ${\bf m}_0$ and ${\bf e}_0$, we
should replace $\Lambda_\cochar$ by $\Lambda_\cort$ and
$\Lambda_\char$ by $\Lambda_\rt$.  Then we get a smaller group
\eqn\nory{{\cal
V}=\left(\Lambda_\cort\oplus\Lambda_\rt\right)\rtimes\Weyl.}
${\cal V}$ is a sort of duality-symmetric version of the affine
Weyl group $\AffWeyl$, since the affine Weyl group of $G$ is
$\Lambda_\cort\rtimes \Weyl$, and the affine Weyl group of $^LG$
is $\Lambda_\rt\rtimes \Weyl$.  The relationship between ${\cal
V}$ and $\hat {\cal V}$ is extremely simple. The quotient
$\Lambda_\cochar/\Lambda_\cort$ is  $\pi_1(G)$, and likewise
$\Lambda_\char/\Lambda_\rt=\pi({}^L\neg G)$. The Weyl group acts
trivially on each of these, so we have a group extension
\eqn\zory{1\to {\cal V}\to \hat{\cal V}\to \pi_1( G)\times
\pi_1({}^L\neg G)\to 1.}

\def\eury{{\eurm Y}}
We say that  a point $(\alpha,\beta,\gamma,\eta)\in \frak
t^3\times \frak t^\vee$ is ${\cal V}$-regular if it is not invariant
under any non-trivial element of $\cal V$. According to section
\topology, the  points that are not regular in this sense are the
points at which the sigma model with target
$\MH(\alpha,\beta,\gamma,\eta;p)$ has a local singularity. We let
$\eury$ denote the complement of the non-${\cal V}$-regular points
in $\frak t^3\times\frak t^\vee$.  (With two or more ramification
points, we would also omit the locus of global singularities in
defining $\eury$; this has real codimension four and will not affect
the argument.) The group ${\cal V}$ acts freely on $\eury$ so the
quotient $\eury/{\cal V}$ is a smooth manifold.

We have therefore a family of smooth $(4,4)$ sigma models
parametrized by the manifold $\eury/{\cal V}$ (as well as other
data, such as the gauge coupling parameter $\tau$). We can now make
an argument that is a sort of quantum version of the reasoning in
section \actweyl.  A basic fact about branes is that every brane has
a  charge or $K$-theory class, taking values roughly speaking in the
complex $K$-theory of the target space $\MH$.  $K(\MH)$ varies with
$(\alpha,\beta,\gamma,\eta)$ as the fiber of a flat bundle over
$\eury/{\cal V}$.  Taking monodromies, we get an action of $\cal V$
on $K(\MH)$.

To be more precise about this, we must recall that $\MH$ has
distinct topological components labeled by ${\bf m}_0$. Also, for
each ${\bf e}_0$, one defines a space of twisted branes, whose
charge takes values in a twisted version\foot{This is the
$K$-theory of twisted vector bundles, which are described in the
present context in section 7.1 of \kapwit. The idea is that ${\bf e}_0$ defines a
flat gerbe over $\MH$, and $K_{{\bf e}_0}(\MH)$ is the twisted
$K$-theory defined relative to this gerbe.}
 of $K$-theory that we might
call $K_{{\bf e}_0}(\MH)$.  Let $K_{{\bf e}_0}(\MH;{\bf m}_0)$ be
the ${\bf e}_0$-twisted $K$-theory of the component of $\MH$
defined by ${\bf m}_0$.  The above argument shows that ${\cal V}$
acts on $K_{{\bf e}_0}(\MH;{\bf m}_0)$ for each choice of ${\bf
e}_0$ and ${\bf m}_0$.  The larger group $\hat {\cal V}$ also
acts, by a slight extension of this reasoning; its action, of
course, permutes the possible values of ${\bf e}_0$ and ${\bf
m}_0$.  In particular, as an abelian group, $K_{{\bf
e}_0}(\MH;{\bf m}_0)$ is independent of ${\bf e}_0$ and ${\bf
m}_0$ up to isomorphism.

We can understand in a relatively concrete way how ${\cal V}$ (or
its extension $\hat{\cal V}$) acts on $K(\MH)$.  The action of the
subgroup ${\cal W}_{\rm aff}=\Lambda_{\cort}\rtimes\Weyl$ was
already described in section \actweyl.

On the other hand, the shift $\eta\to \eta+v$, for $v\in
\Lambda_\rt$, is a shift of the world-sheet $B$-field that changes
its periods by integer multiples of
 $2\pi$.  Such a shift acts on a brane $\cal
B$ by tensoring the Chan-Paton vector bundle of $\cal B$ by the
line bundle ${\cal L}^v$ whose first Chern class is $v$. The
associated action on $K$-theory is thus simply the tensor product
with ${\cal L}^v$, for $v\in \Lambda_\rt$. Since this action is
clear classically,  the fact that ${\cal V}$ acts on $K(\MH)$ does
not by itself tell us much beyond the action of $\AffWeyl$, which
we already know from section \actweyl.

What does appear to give more information is, however, the action
of $S:\tau\to -1/n_{\frak g}\tau$. This exchanges $K(\MH(^L\neg G,C))$ with
$K(\MH(G,C))$ while exchanging $\Lambda_{\rm rt}$ with
$\Lambda_\cort$ (and ${\bf e}_0$ with ${\bf m}_0$).  So it
exchanges the action of $\Lambda_{\rm rt}$ via twisting by a line
bundle with the action of $\Lambda_\cort$ via monodromy.  This
symmetry of the $K$-theory of $\MH$ is not obvious classically.

We will briefly indicate a simple example of how this works.  The
identity element of the multiplicative structure of $K(\MH)$ is
the $K$-theory class of a brane ${\cal B}$ whose support is all of
$\MH$ and whose Chan-Paton line bundle is trivial.  This brane
varies continuously when we vary $\alpha$, $\beta$, and $\gamma$,
so its $K$-theory class is invariant under the action of the
affine Weyl group $\AffWeyl=\Lambda_\cort\rtimes\Weyl$.  So the
duality operation $S$ must transform it into a brane $\tilde{\cal
B}$ that is invariant under the action of $\Lambda_\rt$.  In other
words, the $K$-theory class of $\tilde {\cal B}$ must be invariant
under the operation of tensoring the Chan-Paton bundle of $\tilde
{\cal B}$ by a line bundle of the form ${\cal L}^v$. This
statement means that ${\cal L}^v$  must be (topologically) trivial
when restricted to the support of $\tilde{\cal B}$. Actually,
$\tilde{\cal B}$ is a brane supported on a section of the Hitchin
fibration, so its support
 is isomorphic to the base $\EUBB$ of this fibration. As
this is a contractible space (isomorphic to $\C^N$ for some integer $N$),
any line bundle ${\cal L}^v$ is indeed trivial when restricted to
the support of $\tilde{\cal B}$.

\bigskip\noindent{\it Monodromies Of The $A$-Model}

What we have said so far is hopefully interesting, but may not seem to
get us very close to finding the analog of Wilson and 't Hooft
operators.

To get farther, we must discuss, not branes in general, but specific
kinds of branes.  For example, let us discuss the branes of the
$A$-model of type $K$, with gauge group $G$.  This is the relevant
model at $\Psi=0$. The definition of this model depends on the
parameters $\gamma$ and $\eta$.  So  to study branes of the
$A$-model, we should keep $\gamma$ and $\eta$ fixed. On the other
hand, the $A$-model is locally independent of $\alpha$ and $\beta$.
So the group of monodromies that we obtain by varying $\alpha$ and
$\beta$ will be a symmetry group of the $A$-model.  This group will
depend on $\gamma$ and $\eta$.

The process of varying $\alpha$ and $\beta$ while keeping $\MH$
fixed as a symplectic variety with symplectic structure $\omega_K$
was discussed in section \cpxview.  {}From this discussion, we know
that when one goes around a loop in $\alpha-\beta$ space, avoiding
singularities and keeping $\gamma$ fixed, $\MH$ changes by a
symplectomorphism.  Thus, the group of symmetries of the $A$-model
that we are about to analyze is simply a group of classical
symplectomorphisms (or rather a group of components of the group of
symplectomorphisms). This contrasts with the dual monodromies of the
$B$-model, which involve varying the quantum parameter $\eta$, and
so only make sense quantum mechanically.

Now let us discuss the action of ${\cal V}$. Since $\Lambda_\rt$
acts only on $\eta$, which is held fixed in studying the $A$-model
and its monodromies, it will play no interesting role. We may as
well divide by $\Lambda_\rt$ and replace the pair $\gamma,\eta$ by
$Y=\exp(-2\pi(\eta-i\gamma))$.  $Y$ takes values in, roughly
speaking, the complex maximal torus $^L\TT_\C$ of $ ^LG_\C$. To be
more precise, since we have only divided by $\Lambda_\rt$, not
$\Lambda_\char$, $Y$ takes values in the maximal torus of $^L\bar
G_\C$, the simply-connected cover of $^LG_\C$. Since the $A$-model
depends on $Y$, we are not interested in monodromies that involve
varying $Y$. So the relevant part of the Weyl group is the
subgroup that leaves $Y$ fixed.  We call this ${\cal W}_Y$.

\def\eurz{\eurm Z}
The subgroup of ${\cal V}$ that acts only on $\alpha$ and $\beta$
is an extension of $\Weyl_Y$ by $\Lambda_\cort$:
 \eqn\nombo{{\cal V}_Y=\Lambda_\cort\rtimes \Weyl_Y.}
 So if $Y=1$, ${\cal V}_Y$ is just the affine Weyl group.
 We say
that a pair $(\alpha,\beta)\in \frak t\times \frak t$ is
${\cal V}_Y$-regular if it is not left fixed by any element of ${\cal V}_Y$
other than the identity.  The pairs that are not ${\cal V}_Y$-regular
correspond to non-regular quadruples $(\alpha,\beta,\gamma,\eta)$,
and therefore to local singularities of $\MH$.  We define
$\eurz_Y$ to be the space of ${\cal V}_Y$-regular pairs.  The group ${\cal
V}_Y$ acts freely on $\eurz_Y$, so we get a family of smooth sigma
models parametrized by $\eurz_Y/{\cal V}_Y$.  If we pass from the
sigma model to the associated $A$-model of type $K$, the family
becomes locally constant, since this $A$-model is locally
independent of the parameters $\alpha$ and $\beta$.  So the
fundamental group $B_Y=\pi_1(\eurz_Y/{\cal V}_Y)$ acts as a
group of automorphisms of the $A$-model.  ($B$ loosely stands for ``braid,'' as will
be clearer in a moment.)

This reasoning is similar to what we presented earlier in discussing the
actions of both $\AffWeyl$ and ${\cal V}$.  However, there are two
notable differences:

{\it (A)} The pairs $(\alpha,\beta)$ that are not ${\cal V}_Y$-regular are
in general of real codimension two, simply because we are
considering only two variables $\alpha$ and $\beta$.  (By
contrast, nonregular triples $(\alpha,\beta,\gamma)$ or quadruples
$(\alpha,\beta,\gamma,\eta)$ are of real codimension three or
four, respectively.)  As a result, the space $\eurz_Y$ of
${\cal V}_Y$-regular pairs is not necessarily simply-connected, and the
fundamental group $B_Y$ of the quotient $\eurz_Y/{\cal V}_Y$ is in
general not equal to ${\cal V}_Y$.  We do, however, get an exact
sequence \eqn\zilot{1\to \pi_1(\eurz_Y)\to B_Y\to {\cal V}_Y\to
1.}

{\it (B)} Related to this, in the case of two or more ramification
points, the global singularities are of real codimension two and
might play a role in a complete treatment.  They can be avoided if
one considers only branes supported for Higgs bundles
$(E,\varphi)$ such that the underlying bundle $E$, forgetting its
parabolic structure, is stable.  In this paper, we will not be
concerned with the global singularities.

\def\eurq{{\eurm Q}}
A useful
alternative description is to divide first by the action of $\Lambda_\cort$ on $\alpha$.
Modulo this action,
 the pair $(\alpha,\beta)$ combine to the
element $T=\exp(-2\pi(\alpha+i\beta))\in \bar\TT_\C$, the maximal
torus of the simply-connected cover $\bar G_\C$ of $G_\C$.  We say
that $T$ is ${\cal W}_Y$-regular if it is not invariant under any
non-trivial element of ${\cal W}_Y$.  We write $\eurq_Y$ for the
space of ${\cal W}_Y$-regular elements of $\bar\TT_\C$.  Then ${\cal W}_Y$
acts freely on $\eurq_Y$, and $B_Y=\pi_1(\eurq_Y/\Weyl_Y)$.

As usual, to get some understanding of the group $B_Y$, it is
helpful to consider the cases that $Y$ is regular or $Y=1$.  Other
cases are intermediate between these two.  In the extreme cases,
we have:

(1) If $Y$ is regular, then $\Weyl_Y=1$, and $\eurq_Y=\bar\TT_\C$.
Then $B_Y=\pi_1(\eurq_Y/\Weyl_Y)=\pi_1(\eurq_Y)=\Lambda_\cort.$

(2) Alternatively, if $Y=1$, then $\Weyl_Y=\Weyl$.  So $\eurq_Y$
is the space of regular points in $\bar\TT_\C$, and
$\eurq_Y/\Weyl_Y=\eurq_Y/\Weyl$ is the space of regular conjugacy
classes in $\bar G_\C$.  The fundamental group of this space is
called the affine braid group\foot{The motivation for the name
``affine braid group'' is as follows.  The ordinary braid group on
$n$ letters is the fundamental group of the space of unordered
$n$-plets of distinct points in $\C$.  If we replace $\C$ by
$\C^*$, we get the affine braid group of $GL(n,\C)$, since an
unordered $n$-plet in $\C^*$ is equivalent to a regular semisimple
conjugacy class of $GL(n,\C)$.} of $^L\neg G$.  We denote it as
$B_{\rm aff}(^L\neg G)$. So the monodromy group for $Y=1$ is $B_1=
B_{\rm aff}(^L\neg G) $. Now, for $Y=1$, ${\cal
V}_Y=\Lambda_\cort\rtimes \Weyl$ is the affine Weyl group of
$^L\neg G$.   \zilot\ in this case gives an exact sequence
\eqn\pilot{1\to \pi_1(\eurz)\to B_{\rm aff}(^L\neg G)\to
\AffWeyl(^L\neg G)\to 1,} where $\eurz=\eurz_1$ is the space of
regular pairs  $(\alpha,\beta)\in\bar\TT\times \frak t$, or
equivalently $\AffWeyl$-regular pairs   $(\alpha,\beta)\in\frak
t\times \frak t$.

The affine braid group was introduced in \bez\ in the context of,
roughly speaking, a local version of the present problem.  We have
discussed the relationship in section \onahm\ and return to it in
section \local.

The group $B_Y$ that we have defined is the group of monodromy
transformations of the ramification parameters of the $A$-model
that leave fixed ${\bf m}_0$.  We can relax this condition and
consider an extended group $\hat B_Y=B_Y\rtimes {\cal Z}(G)$ that
includes transformations that shift ${\bf m}_0$. For regular $Y$,
$\hat{B}_Y=\Lambda_\cochar$, and for $Y=1$, it is an extension
\eqn\zombo{\hat{B}_{\rm aff}({}^L\neg G)= B_{\rm aff}(^L\neg
G)\rtimes {\cal Z}(G).}

We have carried out this analysis at $\Psi=0$.  Obviously we could
make a similar analysis for the $B$-model with complex structure
$J$, which corresponds to $\Psi=\infty$.  This exchanges the roles
of $\alpha$ and $\eta$ along with $G$ and $^LG$. The net effect is
that the group of monodromies acting on the branes of the
$A$-model of type $K$ with gauge group $G$ is the same as the
group of monodromies of the $B$-model of type $J$ with gauge group
$^L\neg G$. (However, as we noted above,  the monodromy group
of the $A$-model acts by classical symplectomorphisms, while that of the
$B$-model is highly non-classical.)  As we will now discuss, this
generalizes the correspondence between 't Hooft operators in one
case and Wilson operators in the other case.

\bigskip\noindent{\it Relation To Line Operators}

We have expressed all this in terms of a group ${\cal V}_Y$ that
acts on the $B$-branes at $\Psi=\infty$ or $A$-branes at $\Psi=0$.
More generally, since we can take the direct sum of two branes or
multiply a brane by a positive integer, positive integer linear
combinations of elements of ${\cal V}_Y$ can act on branes. Such
linear combinations form what we will call the
group semiring of ${\cal V}_Y$.

We will here mainly consider the case of regular $Y$.  Then ${\cal
V}_Y=\Lambda_\cort$, and $\hat {\cal V}_Y=\Lambda_\cochar$.  Let
${\cal B}$ be a brane with Chan-Paton bundle ${\cal U}$.  We found
above that  $\Lambda_\cort$ and $\Lambda_\cochar$ act on ${\cal
B}$ by ${\cal U}\to {\cal L}^v\otimes {\cal U}$, for
$v\in\Lambda_\cort$ or $v\in\Lambda_\cochar$. So the group
semiring acts by tensor product with a direct sum of such line
bundles \eqn\yero{{\cal U}\to \left(\oplus_{i=1}^w{\cal
L}^{v_i}\right)\otimes {\cal U}.}

Now we are going to show, for the case of regular $Y$, exactly how
this structure is related to the line operators that exist in the
absence of ramification.  We express this in the language of the
$B$-model. (For background, see sections 7 and 8 of \kapwit.) Let
$^L\neg R$ be a representation of $^L\neg G$. Let $({\cal
E},\hat\varphi)$ be the universal Higgs bundle over $\MH(^L\neg
G,C)\times C$, and let ${\cal E}_{^L\neg R}$ be the associated
bundle\foot{If the center of $^L\neg G$ acts nontrivially in the
representation $^L\neg R$, then ${\cal E}_{^L\neg R}$ is a twisted
vector bundle rather than an ordinary one, and the corresponding
Wilson operator shifts ${\bf e}_0$.} in the representation $^L\neg
R$.  Let ${\cal E}_{^L\neg R}|_q$ denote the restriction of ${\cal
E}_{^L\neg R}$ to $\MH\times q$ for $q$ a point in $C$. Now
consider a Wilson operator $W_q(^L\neg R)$ at the point $q$ in the
representation $^L\neg R$.  Its action on a brane ${\cal B}$ is as
follows. If ${\cal U}$ is the Chan-Paton vector bundle of a brane
${\cal B}$, then
 the brane obtained by acting on ${\cal B}$ with
$W_q(^L\neg R)$ has Chan-Paton bundle ${\cal E}_{^L\neg R}|_q\otimes {\cal
U}$.

Now let $q$ approach a ramification point $p$ for which $Y$ is
regular. At $p$, the structure group of the $^L\neg G$-bundle $E$
is reduced to the maximal torus $^L\TT$, and accordingly the fiber
at $p$ of the $^L\neg G$ bundle $E\to C$ splits as a direct sum of
subspaces corresponding to representations of $^L\TT$.
Accordingly, the vector bundle ${\cal E}_{^L\neg R}|_q$ breaks up,
for $q=p$, as a direct sum of line bundles ${\cal L}^v$ for $v\in
\Lambda_\char$. The tensor product with such a sum of line bundles
is an example of the action of an element of the
group semiring, as described in \yero.

So we have shown that, at least for regular $Y$, what we get at a
generic point $q$ from Wilson or 't Hooft operators is, in the
limit that $q$ approaches a ramification point $p$, a special case
of what we get from monodromies of the $B$- or $A$-model. The
exact sequence \zilot\ shows that it is tricky to generalize this
to non-regular $Y$.   The Wilson line operator $W(^L\neg R)$
gives, similarly to what we have seen above, an element of the
group semiring of ${\cal V}_Y =\Lambda_\char({}^L\neg G)\rtimes
{\cal W}_Y$. But there is no unique way to lift this to the  group
semiring of $B_Y$.  In fact, the ``direction'' in which $q$
approaches $p$ (on the Riemann surface $C$) affects the limit of
the Wilson operator, as  has been analyzed in another language
\ref\gaitnear{D. Gaitsgory, ``Construction Of Central Elements In
The Affine Hecke Algebra Via Nearby Cycles,'' math.AG/9912074.}.

The explanation we have given may seem rather abstract.  In section \thooft, we
re-examine these questions more directly in gauge theory.  We will
get a fairly satisfactory description for the $A$-model, and some insight for the $B$-model.

\bigskip\noindent{\it Generalization For Any $\Bbb{L}$}

All this has a natural generalization to any Levi subgroup $\Bbb{L}$ of $G$.  All we have to
do is to include $\Bbb{L}$ in all statements.

To begin with, of course, we require the parameters
$(\alpha,\beta,\gamma,\eta)$ to be invariant under the Weyl group
${\cal W}_\Bbb{L}$ of $\Bbb{L}$.  Focusing for simplicity on the
$A$-model, to describe its monodromies we must keep fixed
$Y=\exp(-2\pi(\eta-i\gamma))$.  We denote as ${\cal
W}_{Y,\Bbb{L}}$ the subgroup of the Weyl group of $G$ that
normalizes ${\cal W}_\Bbb{L}$ and commutes with $Y$. This is the
subgroup that acts on $(\alpha,\beta)$, preserving their
$\Bbb{L}$-invariance and keeping $Y$ fixed.   In shifting
$\alpha$, we should restrict ourselves to the ${\cal
W}_\Bbb{L}$-invariant sublattice $\Lambda_\cort^\Bbb{L}\subset
\Lambda_\cort$.  So the analog of \nombo\ is that the group that
acts only on $\alpha$ and $\beta$ is now \eqn\hujer{{\cal
V}_{Y,\Bbb{L}}=\Lambda_\cort^\Bbb{L}\rtimes{\cal W}_{Y,\Bbb{L}}.}

We denote as $\eurz_{Y,\Bbb{L}}$ the space of $\Bbb{L}$-invariant pairs
$(\alpha,\beta)\in \frak t\times \frak
t$ that are not left fixed by any element of ${\cal V}_{Y,\Bbb{L}}$.  The group ${\cal V}_{Y,\Bbb{L}}$
acts freely on $\eurz_{Y,\Bbb{L}}$, so we get a family of smooth $A$-models parametrized
by the quotient $\eurz_{Y,\Bbb{L}}/{\cal V}_{Y,\Bbb{L}}$.  This family is locally constant,
so the fundamental group $B_{Y,\Bbb{L}}=\pi_1(\eurz_{Y,\Bbb{L}}/{\cal V}_{Y,\Bbb{L}})$
acts as a group of
automorphisms of the $A$-model.  This statement has an immediate analog for the $B$-model.

\subsec{Representations And Branes}
\subseclab\repbrane

In the absence of ramification, an irreducible flat $^LG$-bundle $E\to C$ corresponds
to a smooth point $x_E\in\MH$.  The $B$-model of $\MH$ is independent of the Kahler metric,
which is controlled by the gauge coupling parameter $\tau=\theta/2\pi+4\pi i/e^2$.
By taking ${\rm Im}\,\tau$ large, we can go to a region in which the $B$-model can
be treated semiclassically.  Then there is a zerobrane ${\cal B}_E$ supported at the
smooth point $x_E\in\MH$.  It is an eigenbrane for the action of the Wilson line operators,
and is a primary object of study in the geometric Langlands
program.

The role here of irreduciblity of $E$ is that it keeps us away
from the singularities of $\MH$.  If $E$ is reducible (but
semistable), it still corresponds to a point  $x_E\in\MH$, but  a
singular point.  We can still consider branes that are supported
at $x_E$, but the theory of such branes is more complicated. There
is not, in general, a unique, canonically determined, brane
associated with $E$.  Instead, we can define a space  or
``category'' of  branes supported at $x_E$; it can be argued that
this space is closed under the action of the Wilson operators.  This complication
is a geometrical analog of a phenomenon that is known in number theory.

We want to discuss the analogous question for the ramified case, that is,
for a flat bundle over $C\backslash p$
(or more generally over $C\backslash\{p_1,\dots,p_s\}$ for some finite set $\{p_1,\dots,p_s\}$),
 with monodromy around
$p$.  Let $E$ be such a bundle, and let $V$ denote its monodromy around $p$.  We assume
that $E$ is irreducible, as otherwise we would meet a global singularity just as
in the absence of ramification.  Our goal here is to consider
the role of local singularities that depend only on the conjugacy class of $V$.

For any $V$, we pick $^L\neg\alpha$ and $^L\neg\gamma$ such that $U=\exp(-2\pi({}^L\neg\alpha-i
{}^L\neg\gamma))$
is in the closure of the orbit of $V$.  After making some choice of $^L\neg\beta$,
we want to associate a brane on $\MH({}^L\neg\alpha,{}^L\neg\beta,{}^L\neg\gamma;p)$
 with the given flat bundle $E\to
C\backslash p$.
If $V$ (and therefore $U$) is regular and semisimple, then there is no problem.
$\MH({}^L\neg\alpha,{}^L\neg\beta,{}^L\neg\gamma;p)$
is smooth (for any ${}^L\neg\beta$),
and we can treat it semi-classically by taking ${\rm Im}\,\tau$
large.  The flat bundle $E$ determines a zerobrane ${\cal B}_E$, for any choices of
${}^L\neg\beta$ and ${}^L\neg\eta$.

Since $U$ is semi-simple, the monodromies studied in section \opmon\
are abelian and act by tensoring the Chan-Paton bundle of a brane with a line bundle
${\cal L}^v$.
But ${\cal L}^v$ is trivial when restricted to a point.  So ${\cal B}_E$
is an eigenbrane for the abelian group of monodromies.  It likewise is an eigenbrane
for Wilson line operators acting at a generic point $q\in C$.  This follows by the
same reasoning as in the absence of ramification; see section 8.2 of \kapwit.  As
 an eigenbrane for all relevant operations, ${\cal B}_E$
is a good analog of a zerobrane supported at a smooth
point in the absence of ramification.

So in short, for regular semi-simple $V$, we are in essentially
the same situation as in the absence of ramification.
Now let us suppose that $V$ ceases to be semi-simple, but is still regular.
An example to keep in mind is that $V$ might be a regular unipotent element of $^LG_\C$,
for example, the element
\eqn\yfo{V=\left(\matrix{1& 1& 0&\dots&0\cr
                         0&1&1&\dots&0\cr
                           & & &\vdots&\cr
                           0&0&0&\dots&1\cr}\right)}
of $SL(N,\C)$.  If $V$ is not semi-simple, then the pair $({}^L\neg\alpha,{}^L\neg\gamma)$
is non-regular;
for example, ${}^L\neg\alpha={}^L\neg\gamma=0$ if
$V$ is unipotent.  Then $\MH({}^L\neg\alpha,{}^L\neg\beta,\gamma;p)$ may,
depending on $^L\neg\beta$, have a local singularity as described in sections \topology\ and
\onahm.  This will happen precisely if the triple
$({}^L\neg\alpha,{}^L\neg\beta,{}^L\neg\gamma)$ is  nonregular.

Even if $\MH({}^L\neg\alpha,{}^L\neg\beta,{}^L\neg\gamma;p)$
has a local singularity, the point $x_E$ is away
from this singularity if $V$ is regular.  This means that there is no problem in defining
the brane ${\cal B}_E$.  Since it is defined in a natural and unique way, it is also
an eigenbrane for all of the monodromy operations.  So again, for regular $V$,
we are in essentially the same situation as in the absence of ramification.

A problem does occur if $V$ is non-regular.  Then, if we take
${}^L\neg\beta=0$, the point $x_E$ is contained in the locus of
local singularities.  This gives us a problem in defining the brane
${\cal B}_E$.  We can resolve the singularity by taking
${}^L\neg\beta\not=0$ and generic enough so that the triple
$({}^L\neg\alpha,{}^L\neg\beta, {}^L\neg\gamma)$ is regular.
${}^L\neg\beta$ is a Kahler parameter in complex structure $J$, and
taking sufficiently generic ${}^L\neg\beta$ has the effect of
blowing up the locus of local singularities. (This gives us a global
version of what locally is the Springer resolution of the nilpotent
cone.) This makes $\MH({}^L\neg\alpha,{}^L\neg\beta,{}^L\neg\gamma)$
non-singular, but the blow-up replaces the point $x_E$ by a variety
$\Upsilon_E$ of positive dimension. The best we can do is to
associate to $E$ the whole family or ``category'' of branes
supported on $\Upsilon_E$. The monodromy group acts on this
category, possibly in a way that is in some sense irreducible.

Alternatively, we can keep ${}^L\neg\beta=0$ (or
sufficiently special to avoid blowing up the point $x_E$), and
make the sigma model smooth by taking ${}^L\neg\eta$ to be sufficiently
generic.   Such a choice of ${}^L\neg\eta$ may enable us to define a
zerobrane ${\cal B}_{E,{}^L\neg\eta}$.  But the possible choices of sufficiently generic
${}^L\neg\eta$ are divided into ``chambers.'' We suspect that if it is possible to use a choice
of ${}^L\neg\eta$ to define a zerobrane ${\cal B}_{E,{}^L\neg\eta}$, then this
zerobrane depends on the
``chamber'' containing ${}^L\neg\eta$.

For example, in the case ${}^L\neg\alpha={}^L\neg\gamma=0$ of unipotent monodromy,
if we also keep ${}^L\neg\beta=0$ to avoid resolving singularities, then
the condition on ${}^L\neg\eta$ to avoid a singularity of the sigma model
is that ${}^L\neg\eta$ must be regular.  But the space of regular ${}^L\neg\eta$'s
is not connected, as the non-regular ${}^L\neg\eta$'s divide ${}^L\neg\TT$
into affine Weyl chambers.
 If we try to pass from one chamber to another
by varying ${}^L\neg\beta$ so as to go around the singularities that
separate between the different chambers, we will run into the
monodromies that were used to define the group ${\cal V}_U$.

The result is that for non-regular $V$, we can define a family or ``category''
of branes on Higgs bundle moduli space,
associated with the flat bundle $E\to C\backslash p$, and supported at $x_E$ or its blowup.
This category is acted on by the nonabelian
group ${\cal V}_U$.   All branes in this category are eigenbranes for Wilson operators
acting at points away from $p$.  But there is no apparent
way to pick a particular brane
in this family.  It might be that in some sense the group ${\cal V}_U$ acts
irreducibly on this family.

Applying the duality transformation $S:\tau\to -1/n_{\frak g}\tau$,
we get the same sort of picture in the $A$-model of gauge group $G$: a family of $A$-branes
that are acted on by ${\cal V}_U$, and are eigenbranes for 't Hooft/Hecke operators acting
away from $p$.   Finally, using the c.c. brane, we can map this family to a family
of twisted ${\cal D}$-modules over $\M(\alpha,p;G)$, related in the same way to the
Hecke operators and the group ${\cal V}_U$.

Accordingly,
for non-regular $V$, the geometric Langlands program can be expressed as a duality that
maps a family of $B$-branes, acted on by ${\cal V}_U$, to a family of
$A$-branes or ${\cal D}$-modules, acted on by the same group.

\bigskip\noindent{\it Searching For The Canonical Zerobrane}

We have just seen that the duality statement that one can deduce using the
standard surface operator associated with the maximal torus $\TT\subset G$ becomes
more involved when the ramified flat bundle $E$ has a non-regular monodromy $V$.  Although this
complication will probably surprise most physicists who have gotten this far, if any,
 it will
come as no surprise to geometers, since it is expected based on an analogy with number theory.

We can, however, ask the following question.  Given a flat bundle $E\to C\backslash p$
with non-regular monodromy $V$, can we find another duality statement involving this
flat bundle that involves a canonical zerobrane rather than a whole category?  It is not
clear that there is anything wrong if the answer is ``no,'' but if the answer is ``yes,''
we would like to find out.

For some conjugacy classes of $V$, we can answer this question simply
by using the surface operator based on a general
Levi subgroup $\Bbb{L}$, rather than the generic surface operator that we have used so far.
In the presence of  a surface operator of type $\Bbb{L}$, a solution of Hitchin's equations
describes in complex structure $J$ a flat bundle $E\to C\backslash p$ whose possible monodromy
$V$ was analyzed at the end of section \postpone.  This monodromy is conjugate to an element
of  $\EUP$, the parabolic subgroup that is determined by the pair $(\Bbb{L},{}^L\neg\alpha)$.
 Generically the
monodromy is $\Bbb{L}$-regular (in a sense described in section \postpone).  Precisely
if $V$ is $\Bbb{L}$-regular, the ramified flat bundle
$E$ corresponds to a smooth point on $\M_{H,\Bbb{L}}$, and
hence to a canonical zerobrane ${\cal B}_{E}$
in the sigma model of target $\M_{H,\Bbb{L}}$.
This canonical zerobrane is an eigenbrane for all of the relevant Wilson operators and
monodromies by the same arguments as above.  Applying to it $S$-duality, we get a brane of
the $A$-model with symplectic structure $\omega_K$ (and gauge group $G$ rather than $^L\neg G$),
and then using the relation between $A$-branes and ${\cal D}$-modules, we associate to $E$
a ${\cal D}$-module on the moduli space ${\cal M}_\Bbb{L}(\alpha;p,G)$ of parabolic bundles.

So in short, if the monodromy is $\Bbb{L}$-regular for some $\Bbb{L}$,
we can reduce to a situation as simple as
the unramified case by considering the theory with a surface operator of type $\Bbb{L}$.
For instance, referring back to the Levi subgroup described in eqn. \xunpyx\ with $G=SL(3,\C)$,
typical examples of $\Bbb{L}$-regular conjugacy classes are the semi-simple conjugacy class
containing the element ${\rm diag}(\lambda,\lambda,\lambda^{-2})$ with $\lambda^3\not=1$,
and the Richardson conjugacy class with a representative given in eqn. \xunlyx.

For $G=SL(N,\C)$, every conjugacy class is $\Bbb{L}$-regular for some $\Bbb{L}$.  Given $V\in SL(N,\C)$,
we simply let $U\in \TT$ be contained in the closure of the orbit of $V$, and we let $\Bbb{L}$
be the subgroup of $SL(N,\C)$ that commutes with $U$.  However, for other groups,
this is not the case. For example, semi-simple Lie groups other than $SL(N,\C)$ contain
rigid  noncentral orbits. (An orbit is called rigid if it cannot be deformed to any nearby
orbit, usually because all nearby orbits have greater dimension.)
Such an orbit is not $\Bbb{L}$-regular for any $\Bbb{L}$.

We suspect that it may be possible to define additional supersymmetric surface operators
in ${\cal N}=4$ super Yang-Mills theory with the following properties.
For every ramified flat bundle $E\to C\backslash p$,
we hope to be able to pick a surface operator
such that, in the presence of this surface operator, $E$ corresponds to a canonical
zerobrane ${\cal B}_E$ in the $B$-model at $\Psi=\infty$.  Moreover, this surface operator should
have an $S$-dual (which would be a surface operator of a roughly similar type).
$S$-duality applied to the zerobrane ${\cal B}_E$ would
give an $ A$-brane in the dual model at $\Psi=0$.
However, the new surface operators will not be related to parabolic
subgroups, so in the presence of such an operator, we do not
expect $\MH$ to be related in the usual fashion to a cotangent bundle.  Consequently, it would
not be possible to relate the dual $A$-brane to a ${\cal D}$-module, and the duality will
have to be expressed as a mirror symmetry or $S$-duality
between the $B$-model of $\MH({}^L\neg G)$
and an $A$-model of $\MH(G)$, rather than
a relation between coherent sheaves (or $B$-branes)
on $\MH$ and ${\cal D}$-modules on some other space.
The lack of an interpretation via ${\cal D}$-modules might mean that this more general
duality, if it exists, will not be relevant to number theory, but it might still give
an elegant application of $S$-duality and mirror symmetry.

\newsec{Line Operators And Ramification}
\seclab\thooft

In section \opmon, we described the analog of Wilson and 't Hooft operators at
ramification points.  The description may have seemed rather abstract, and the
answer -- especially in the case of unipotent monodromy --
is surprisingly complicated.
By contrast, the Wilson and 't Hooft operators that act at a generic point on $C$ are usually
defined using quite different methods of gauge theory.

It is unsatisfying to describe the two cases with completely different methods, so in the
present section we will attempt to give a gauge theory definition in the ramified case.
This has another advantage: it will help us understand the full supersymmetry
of these operators.

The topological field theory
that arises at $\Psi=\infty$, after compactifying to two dimensions, is the $B$-model
with target $\MH$ in complex structure $J$.
This supersymmetry is preserved by the monodromies considered
in section \opmon.  However, the relevant Wilson operators at an unramified point preserve
a greater supersymmetry.  They have supersymmetry of type $(B,B,B)$;
that is, they preserve the supersymmetry of the
$B$-model in any of the complex structures that make up the hyper-Kahler structure of $\MH$.

Similarly, at $\Psi=0$, the relevant topological field theory is the $A$-model of $\MH$ with
symplectic structure $\omega_K$.  The appropriate supersymmetry is preserved by the
monodromies studied in section \opmon.  However, the relevant 't Hooft operators actually
preserve supersymmetry of type $(B,A,A)$; that is, they preserve the supersymmetry of the
$B$-model of type $I$ and of the $A$-model of type $J$ or $K$.

We will aim to use gauge theory to show that the appropriate
operators acting at a ramification point really have the same supersymmetry as the Wilson
and 't Hooft operators at a generic point. We will be able to achieve this for the $A$-model,
and partially for the $B$-model.

\subsec{General Framework}\subseclab\genframe

For simplicity, we will work on a four-manifold $M=\Sigma\times C$.  As usual, $C$ is
the Riemann surface on which we study the geometric Langlands program.  We take $\Sigma
=\R\times I$, where $\R$ parametrizes the ``time,'' and $I$ is a closed interval.  We thus
can also write $M=\R\times W$, where $W=I\times C$ is a three-manifold.  Branes are chosen
to define boundary conditions on the boundary of $W$.

\ifig\trunko{\bigskip A time zero slice of a time-independent configuration. The support
$D$ of a surface operator intersects the time zero slice on the line $S$.  The supports $L_i$
of several line operators intersect the time zero slice at points $y_i$, $i=1,2,3$. }
{\epsfxsize=4in\epsfbox{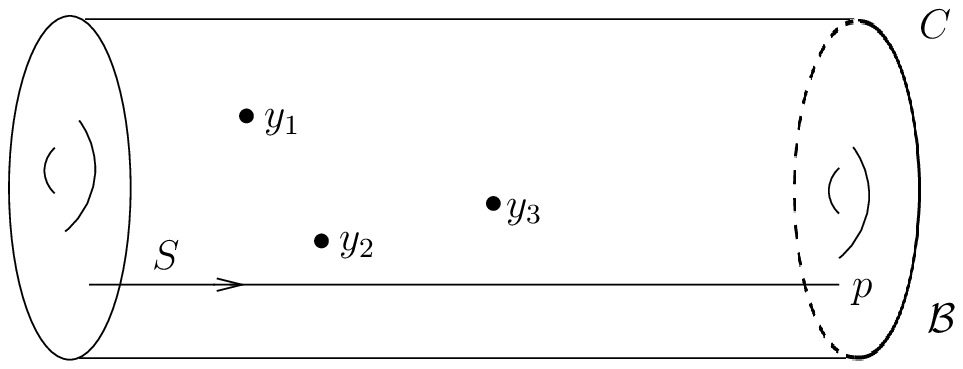}}

For simplicity, we take our line and surface operators to have time-independent support.
Therefore, we can describe any configuration of line and surface operators by indicating
what is happening in $W$.   A time-independent surface operator is supported on a two-manifold
$D=\R\times S$, for $S$ a curve in $W$.  For applications to the ramified case of geometric
Langlands, we take $S=I\times p$ for $p$ a point in $C$.  We also include line
operators with time-independent support $L_i=\R\times y_i$ for some points $y_i\in W$.
We will denote these line operators simply as $L_i$.  This configuration is sketched in \trunko.

The line operators can be understood as operators acting on branes, as explained in detail in
section 6.4 of \kapwit.  For example, using the topological invariance, we can move the points
$y_i$ to approach the right boundary in \trunko, which is labeled by the brane ${\cal B}$.
As $y_i$ approaches the boundary, the corresponding line operator $L_i$ acts on ${\cal B}$ to
give a new brane ${\cal B}_i$ which we describe symbolically as $L_i\cdot {\cal B}$.
By contrast, because $S=I\times p$ ends on
the boundary of $W$, the surface operator with support on $D=\R\times S$ must be included
as part of the definition of what we mean by a brane; it does not really give in this situation
an operator acting on branes.\foot{A line or surface operator whose support is of finite
extent in the time direction will give an operator acting on branes.}

Topological invariance means that there is a natural flat connection such that
a line operator supported at a point $y_i\in W$ is naturally equivalent to
one supported at a nearby point $y_i'$.
This implies that line operators commute with each other, since
the space of configurations of distinct
points in a small open set in a three-manifold is connected and simply-connected.  The
connectedness means that we can move two line operators past each other without any singularity,
and the simple connectivity means that there is (up to homotopy) no ambiguity about how to do
this.

\bigskip\noindent{\it Line Operators Supported On A Surface}

\ifig\flunko{\bigskip  Line operators $L_i$ supported on a surface $D$. Sketched is the time
zero slice of a time-independent situation.  At time zero, $D$ is represented by the indicated
line $S$, and the line operators are represented by points $y_i\in S$. }
{\epsfxsize=4in\epsfbox{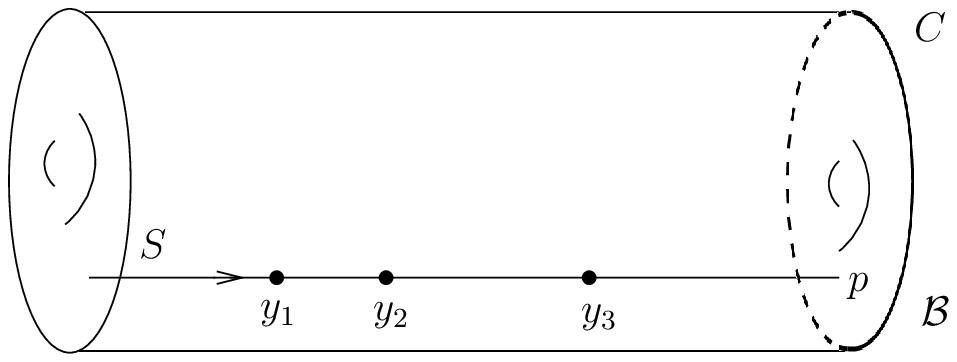}}

Now, however, consider the case that some of the points $y_i$ are actually contained in the
curve $S$.  In the situation just considered, this will occur if we take $y_i=u_i\times p$
where $u_i\in I$ and $p$ is the ramification point in $C$
(\flunko).  In this case, of course,  the line $L_i=\R\times y_i$ is contained
in the surface $D=\R\times S$, so what we have is a line operator supported on a
surface.  Of course, what kinds of line operator are possible  on the support of
 a given surface operator
is one question that we will have to address.

Line operators supported on a surface are quite different from generic line operators in
four-space, because, granted suitable orientations,
they have a natural ordering.
This is shown in \flunko, in which the ordering is $y_1<y_2<y_3$.
A line operator with support  $\R\times y_i$
gives us an operator that acts on branes, since we can move the points $y_i$ to the right,
whereupon the corresponding line operators act on the brane ${\cal B}$.  However,
in general, we should not expect line operators supported on branes to commute with each other,
since they have a definite ordering and there is no way to move them past one another without
meeting a singularity.

Thus, at a generic point $q\in C$, we get a commutative algebra ${\cal S}_{q}$
of Wilson or 't Hooft operators
that act on branes.  But at a ramification point $p$, we get instead in general
a noncommutative algebra
$\hat {\cal S}_p$.

Although $\hat {\cal S}_p$ is in general noncommutative, it commutes with the algebra
${\cal S}_{q}$ of Wilson or 't Hooft operators
supported at a generic point $q\in C$.  This is so simply because, as long as $q$ is
not a ramification point, a line operator that acts at $q$ can be freely moved to the left
or right, without encountering a singularity.

Of course, a noncommutative algebra may contain a large center.  We can map
${\cal S}_{q}$ to the center of $\hat {\cal S}_p$ by taking the limit as $q$ approaches
$p$.  In a topological field theory, this limit will exist since the distance between $q$
and $p$ is anyway irrelevant.  However, generically there will be a monodromy
as $q$ is circled around $p$, and if so, the limit of an element of ${\cal S}_{q}$ as
$q\to p$ depends on the direction from which the limit is taken.
This is obvious for the $B$-model, as we note in section \bmodel, and
has been analyzed in \gaitnear\ for the $A$-model (in a language very different from that
of this paper).

So the usual algebra of Wilson or 't Hooft operators can be mapped to the center of the
noncommutative algebra that acts at a ramification point, but not quite in a canonical way.

\bigskip\noindent{\it Relation To Monodromies}

Now let us discuss how this formulation is related to the
description by monodromies in section \opmon.  In that approach, one
considers operations on branes that preserve only one topological
supersymmetry -- the symmetry of the $B$-model in complex structure
$J$ or of the $A$-model with symplectic structure $\omega_K$.
For simplicity, we will just express the following argument in terms of the $B$-model.
For the $A$-model, the story is the same with $\alpha$ and $\eta$ exchanged.

In the $B$-model in complex structure $J$, the
dependence of the Lagrangian on the Kahler parameters $(\beta,\eta)$  is
entirely of the form $\int_M d^4x\sqrt g \{Q,V\}$, where
$Q$ is the appropriate topological supercharge and $V$ is suitably
chosen.   The dependence on the  parameters $(\alpha,\gamma)$ which control the complex
structure  cannot be put in this form.

With an appropriate choice of $V$, we can let the pair
$(\beta,\eta)$  be non-constant functions along the support
$D$ of a surface operator.  For our purposes, we take these functions to be time-independent,
but to vary as one moves from left to right along the curve $S$ of
\flunko.  We must keep away from values of the pair $(\beta,\eta)$
at which the quantum theory would become singular.  Letting
$(\beta,\eta)$ vary along $S$, keeping away from the locus of
singularities, gives us back exactly the monodromies that were
described in section \opmon.

Now we can see the relation between monodromies and line
operators.  In describing a monodromy, we can let the pair $(\beta,\eta)$ be constant
outside of a small interval $I_0\in S$.  In a topological field theory, the size of the interval
$I_0$ does not matter.  If we think of this interval as being essentially pointlike,
the monodromy is just a particular way to describe a line operator.

However, the description of line operators by
monodromies has a drawback if we wish to see the full
topological supersymmetry of the $B$-model at $\Psi=\infty$.  We recall that this symmetry is
of type $(B,B,B)$.
To preserve $B$-type supersymmetry in complex structure $J$, we must keep $\alpha$ and $\gamma$
fixed (as we did above), since this complex structure depends on those parameters.
Likewise, to preserve $B$-type supersymmetry in complex structure $I$, we must keep $\beta$
fixed, as complex structure $I$ depends on $\beta$.
So altogether, to preserve the full topological supersymmetry of type $(B,B,B)$,
we can only vary $\eta$, and at best we will
only be able to see abelian monodromies that correspond to shifting
$\eta$ by a lattice vector.

The most interesting nonabelian singularities are expected if
$\alpha=\beta=\gamma=0$.  In this case, to avoid a singularity,
$\eta$ must be regular.  Thus, $\eta$ is confined to the interior of
an affine Weyl chamber.  With this restriction, we cannot observe
any monodromies at all.

There is a similar problem, of course, in the $A$-model at $\Psi=0$ if we wish to see
the full topological supersymmetry of type $(B,A,A)$.
If we want to preserve the full symmetry of the problem,  monodromies are not an adequate
framework.  Our goal in the rest of this section
is to see how much better we can do by thinking in terms of line operators rather than
monodromies -- that is, by defining discrete operations
analogous to Wilson and 't Hooft operators in the unramified case.

\subsec{The $B$-Model}
\subseclab\bmodel

The easy case to discuss is the $B$-model, but this is also the case
in which we will not be able to get a fully satisfactory answer.

In the unramified case, with gauge group $^L\neg G$, the important
operators that act on branes are Wilson operators in a
representation $^L\neg R$ of $^L\neg G$.  They are defined by
parallel transport using the complex-valued connection
$\CA=A+i\phi$.  If $L$ is a closed loop in the four-manifold $M$, we
define the Wilson operator \eqn\tofog{W({}^L\neg
R;L)=\Tr\,P\exp\left(-\int_L\CA\right),} where $P\exp\left(-\int_L\CA\right)$ is the
holonomy of the connection $\CA$ along the line $L$, and the trace
is taken in the representation $^L\neg R$.  If instead $L$ is an
open line (connecting boundaries or ends of the four-manifold $M$),
we define $W({}^L\neg R;L)$ as this holonomy, regarded as an operator
acting between initial and final states in the representation $^L\neg R$.

For the geometric Langlands program in the unramified case, we take
$M=\R\times I\times C$, where $\R$ parametrizes the ``time,'' $I$ is
an interval, and $C$ is a Riemann surface.  To get a Wilson operator
that can act on branes, we take $L=\R\times u\times q$, with chosen
points $u\in I$ and $q\in C$.

If $L$ is contained in the support $D$ of a surface operator, as in \flunko, we can carry out
much the same construction.  There is one important difference. On the support of a surface
operator, the
group $^L\neg G$ is reduced to its maximal torus $^L\neg \TT$.
Hence, if $L$ is contained in $D$, we can define Wilson
operators associated with a choice of representation of $^L\neg
\TT$.  This is more general than a choice of representation of $^L\neg G$, since a representation
of $^L\neg G$ can be decomposed as a direct sum of representations of $^L\neg \TT$, but
in general not the other way around.

So in general, at a ramification point the representation ring of $^L\neg \TT$ acts on
branes of the model at $\Psi=\infty$, extending the action of representations of $^LG$ that occurs
at a generic point.  What happens is simply that a Wilson
operator at a generic point $q$, associated with a representation of
$^LG$, splits up, in the limit that $q$ approaches $p$, as a sum of
Wilson operators associated with representations of  $^L\neg \TT$.

Because the ramified flat bundle of the $B$-model has a monodromy around the support of a surface
operator,
there is a subtlety in  the limit $q\to p$ in this situation.  The limit depends on
the path via which one takes $q$ to approach $p$.

\bigskip\noindent{\it Interpretation}

The representation ring of $^L\neg \TT$ is the coweight lattice
$\Lambda_\cowt$.  This contains a sublattice $\Lambda_\cort$
consisting of Wilson operators that do not change the discrete
electric flux ${\bf e}_0$.

We similarly encountered in section \opmon\ an action on branes of the
lattice $\Lambda_\cort$ or $\Lambda_\cowt$.  However, the
explanation given there was somewhat different: in that approach,
the lattice acts by shifts of $\eta$, that is, by $2\pi$ shifts of
the theta-angles that are defined on the impurity surface.

It is, however, a classic result \ref\coleman{S. Coleman, ``More
About The Massive Schwinger Model,'' Annals Phys. {\bf 101} (1976)
239. } that a Wilson operator in abelian gauge theory in two
dimensions causes a shift in the theta-angle by an integer multiple
of $2\pi$.  Let us recall how this comes about. We consider a gauge
theory with gauge group $H=U(1)$ on a two-manifold $D$.  We let $L$
be a one-manifold on $D$ of a suitable type.  The simplest case to
consider is that $L$ is a closed one-manifold that is the boundary
of a region $R\subset D$. Let $A$ be an abelian gauge field\foot{We consider $A$ as
a connection on a principal $U(1)$ bundle, represented locally by a real one-form.} on $D$
with curvature $F$. Then for the holonomy of $A$ around $L$, we have
the identity
\eqn\nilfox{\exp\left(i\oint_LA\right)=\exp\left(i\int_RF\right).}
This says that  the theta-angle jumps by $2\pi$ in crossing the line on which a charge
1 Wilson operator is supported.  With more
care in the analysis, one can reach the same conclusion even if the one-manifold $L$ is not
closed.

\bigskip\noindent{\it Discussion}

So for the case of regular semi-simple monodromy, the lattice action
on branes that we have found from gauge theory is all that we expect
to see. However, according to the analysis of section \opmon, a larger
noncommutative group should act on branes in case the pair
$(\alpha,\gamma)$ is non-regular. Unfortunately, as we will now
explain, it seems difficult at $\Psi=\infty$ to see this larger symmetry by
semi-classical gauge theory methods.  Thus we will not really be able with these
methods to do better than we did with monodromies.

The monodromies that we want to see arise by varying $\beta$ and
$\eta$, and are absent if $\beta$ is constrained so that the triple
$(\alpha,\beta,\gamma)$ is always regular.  So the existence of
these monodromies depends on what happens when that triple becomes
non-regular. At this point, the classical moduli space $\MH$ gets a
local singularity, in the language of section \topology.

The monodromies still make sense, because the quantum theory remains
smooth if $\eta$ is generic. However, it is hard to use this fact about the quantum theory in
a semi-classical gauge theory construction of line operators.
Any such construction begins with a classical
construction which is then
implemented quantum mechanically.  Operators whose definition
depends on quantum properties of the theory are difficult to see
using gauge theory methods.

The situation is different in the $A$-model, because the roles of
$\alpha$ and $\eta$ are exchanged.  The relevant monodromies are
found by varying $\alpha$ and $\beta$, with $\gamma$ and $\eta$
fixed. Even if $\gamma=\eta=0$, one can see all the monodromies in
the region in which  the pair $(\alpha,\beta)$ is regular, which
means that the classical geometry of $\MH$ is smooth.

This suggests that in the $A$-model, we might be more successful in
using gauge theory methods to describe the full set of operators
that act on branes and all the  supersymmetry that they preserve.
That will be our next goal.

\subsec{The $A$-Model}
\subseclab\amodel

In the absence of ramification, the natural operators acting on branes in the $A$-model are
't Hooft operators, defined by prescribing a singularity that the
fields are required to have.
So in the presence of ramification, we will look for 't Hooft-like operators.  To keep
things simple, we will take the Levi subgroup defining our surface operator to be $\Bbb{L}=\TT$.

It is convenient to get to $\Psi=0$, that is, the $A$-model with symplectic structure $K$,
by taking the twisting parameter of the underlying four-dimensional theory
to be $t=1$ and the four-dimensional theta-angle to vanish.
On a general four-manifold $M$, the conditions to preserve the topological symmetry of the
$A$-model were described in
eqns. (9.1), (9.2) of \kapwit.  If we set $\CA=A+i\phi$, and write $\CF=d\CA+\CA\wedge \CA$
for the curvature of $\CA$, and set $d_A=d+A$, then the conditions are
\eqn\ijon{\eqalign{\CF+i\star \overline\CF & = 0\cr
                    d_A\star \phi & = 0.\cr}}

We specialize to $M=\R\times W$, with $\R$ parametrized by a ``time'' coordinate $s$.
We restrict the equations \ijon\ to the time-independent case.  This is motivated
by the fact that the usual 't Hooft operators are defined by a time-independent singularity
(and by arguments given in the introductory part of section 9 of \kapwit, which are also
relevant in the presence of ramification).
In a time-independent situation, the four-dimensional connection reduces to $A=A'+A_0\,ds$
with $A'$ a connection on a $G$-bundle $E\to W$ and $A_0$ an ${\rm ad}(E)$-valued zero-form.
Similarly we write $\phi=\phi'+\phi_0\,ds$, where $\phi'$ and $\phi_0$ are ${\rm ad}(E)$-valued
forms on $W$.  We set $\CA'=A'+i\phi',$ $\CF'=d\CA'+\CA'\wedge \CA'$, and $\Phi_0=\phi_0-iA_0$.
Since the whole subsequent discussion will occur in three dimensions, we omit the primes.
The first equation in \ijon\ becomes
\eqn\bijon{\CF=\star\bar {\cal D}\Phi_0,}
where  $\bar{\cal  D}=d+\bar\CA$ and of course $\star $ is now the three-dimensional Hodge $\star$
operator.  The second equation becomes
\eqn\ijon{D\star \phi={i\over 2}[\bar\Phi_0,\Phi_0].}

\bigskip\noindent{\it Scaling Symmetry}

An 't Hooft-like operator is defined using a singular solution of these equations.
For example, ordinary 't Hooft operators are defined by using a singular solution of
these equations that has a singularity in codimension three, that is, at an isolated point
$r\in W$.  The construction  was described in section 6.2 of \kapwit, and will be generalized
below.  In the present paper, our main topic has been surface operators, which arise by specifying
a codimension two singularity -- the familiar singularity with parameters
$(\alpha,\beta,\gamma)$.

\ifig\dunko{\bigskip  A line operator supported on a surface is represented in this
time zero slice by a point $y$ on a line $S$.  In the $A$-model, the fields have a prescribed
singularity along $S$.  Near $y$, the singular
behavior is different from what it is near a generic point on $S$.  The generic singular
behavior along $S$ may in turn be different on the two sides of  $y$. }
{\epsfxsize=4in\epsfbox{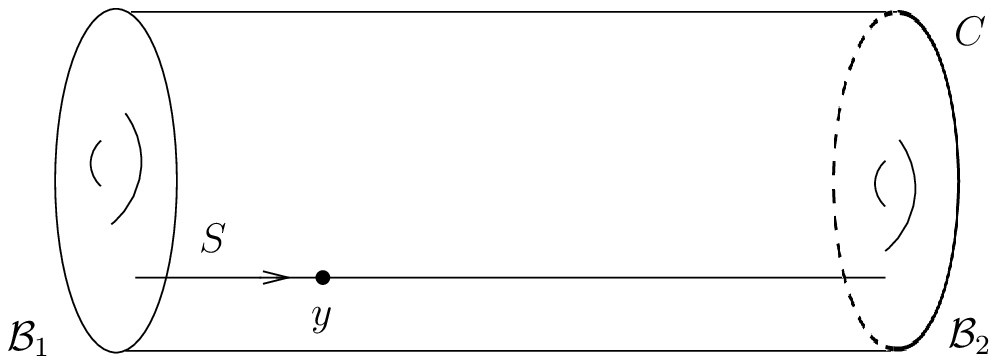}}
We are interested in a more complicated situation in which a curve $S$ of codimension two
singularities contains an isolated point $y$ at which the singular behavior is ``worse.''
Thus, near a generic point of $S$, we see the usual codimension two singularity, but
a new singularity occurs at the point $y\in S$.  {\it A priori}, to the left of
$y$, the line $S$ is labeled by parameters $(\alpha,\beta,\gamma)$, and to the right it is
labeled by parameters $(\alpha',\beta',\gamma')$.  If we are to get a surface operator
that preserves the full supersymmetry of type $(B,A,A)$, we must have $(\beta,\gamma)=(\beta',
\gamma')$, because $\beta$ and $\gamma$ are physical parameters in the $B$-model of type $I$,
and cannot jump if we are to preserve the supersymmetry of this model. However, there
is no reason to consider only the case that $\alpha=\alpha'$.  The topological field
theories of type $(B,A,A)$ are locally independent of $\alpha$, so we should expect to
get results that are essentially independent of $\alpha$ and $\alpha'$.

\ifig\dulco{\bigskip  To describe the singularity locally, we replace the three-manifold $W$
by $\R^3$, and the line $S$ by a straight line $\R\subset \R^3$.  The point $y$ becomes the
origin in $\R^3$, marked here by the black dot.  ${\eurm V}$ is a unit sphere
surrounding the origin.  It intersects the line
$S$ at two points, $p$ and $p'$.}
{\epsfxsize=3in\epsfbox{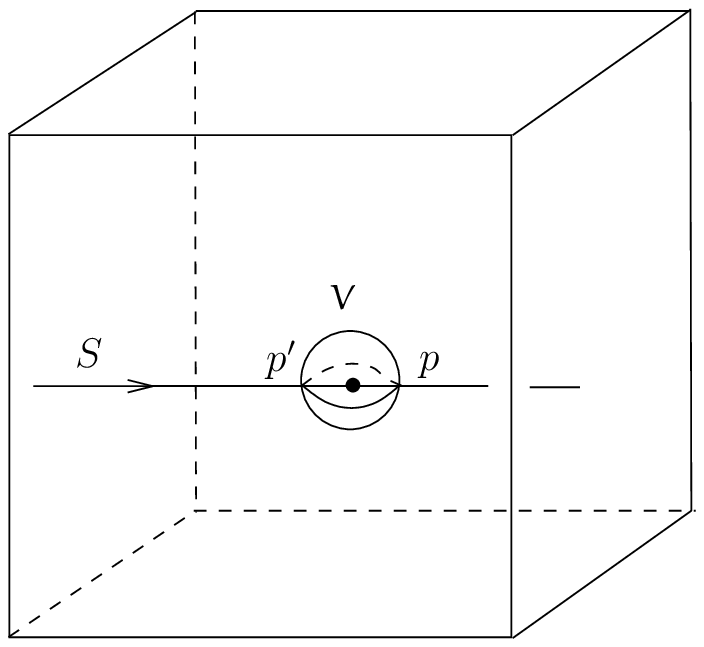}}
To simplify things, we will
make an assumption about the nature of the singularities that we are looking
for.  We will assume that they are scale-invariant,
like the singularities used to define ordinary 't Hooft operators.
Indeed, since we are only interested in the local singular
behavior, we can replace $W$ by $\R^3$, with $y$ as the origin in $\R^3$.
We choose Euclidean
coordinates $x^1,x^2,x^3$ on $\R^3$, and we take the line $S$ of \dulco\
to be the $x^1$ axis, defined by $x^2=x^3=0$.
  $\R^3$ admits a scaling symmetry
$\vec x\to \lambda \vec x$, for real positive $\lambda$.  The equations of interest are
invariant under this scaling symmetry (if we take $\Phi_0$ to scale with dimension 1), and
the required singularities along $S$ away from the point $y$ are compatible with scale
invariance.  So it makes sense to look for a scale-invariant singular solution, and this
will suffice for our purposes here.

\def\EURV{\eurm V}
If we remove the point $y$ from $\R^3$ and divide by scaling, we get
$(\R^3\backslash y)/\R^*_+=S^2\cong \Bbb{CP}^1$. We will call this
quotient $\EURV$. The interesting solutions can be largely described
in terms of data on $\EURV$.  The fields on $\EURV$ will be singular
at two points sketched in \dulco, namely the points $p$ and $p'$ at
which $\EURV$ meets the line $S$.

\bigskip\noindent{\it The Case $\phi=0$}

As we learned in section \opmon, the operators acting on branes are most interesting if
$\beta=\gamma=0$.  In this case, it is possible to have $\phi=0$ (since $\phi$ has no
singularity along a generic point of $S$).  It turns out that the 't Hooft-like operators
that we are seeking can be defined using a singular solution with $\phi=0$, and also $A_0=0$.

This leads to a drastic simplification in the equations \bijon\ and
\ijon.  They reduce to the Bogomolny equations
\eqn\plijon{F=\star d_A\Phi_0,} where now $F=dA+A\wedge A$ and
$\Phi_0=\phi_0$ take values in the real Lie algebra $\frak g$ of
$G$. We will make an assumption that is stronger than
scale-invariance: we assume that $A$ and $\Phi_0$ are  pullbacks
from $\EURV=S^2$. We will  think of $\EURV$ as the unit sphere in
$\R^3$, defined by $r=1$ where $r=|\vec x|$.  The equations on $S^2$
come out to be \eqn\lijon{\eqalign{F&=-\star\Phi_0\cr
                  d_A\Phi_0&=0,\cr}}
                  where now these are equations on the two-manifold $\EURV$ and
                  $\star$ is meant in the two-dimensional sense.
For example, to get the first equation, we use the fact that the
dependence of $\Phi_0$ on $r$ is precisely a factor of $1/r$ (since $\Phi_0$ scales
with weight 1), so
that at $r=1$ we have $\partial\Phi_0/\partial r = -\Phi_0$.  The
second equation is obtained using the fact that $F$ vanishes when
contracted with $\partial/\partial r$ (because we assume the
solution to be a pullback from $\EURV$).

Obviously, the two equations combine to \eqn\blijon{d_A\star F=0.}
These are the two-dimensional Yang-Mills equations, that is, they
are the Euler-Lagrange equations that can be derived from the
Yang-Mills action \eqn\zureto{{\cmmib I}=-{1\over
2}\int_\EURV\,\,\Tr \,F\wedge \star F.}

Another interpretation of these equations is as follows. Given a
Riemann surface $\EURV$, every connection $A$ on a $G$-bundle $E\to
\EURV$ endows $E$ with a holomorphic structure, defined by the
$\bar\partial $ operator $\bar \partial_A=d\bar z(\partial_{\bar z}+A_{\bar
z})$. Turning this around, we can begin with a holomorphic
$G_\C$-bundle $E\to \EURV$ and ask what sort of unitary connections
$E$ admits that are compatible with its holomorphic structure (in
the sense that the holomorphic structure induced by the connection
agrees with the given holomorphic structure of $E$). Stable and
semistable bundles are precisely the ones that admit a flat unitary
connection  compatible with their holomorphic structure \nar.  More
generally, if we endow $\EURV$ with a Kahler metric and thus a
$\star$ operator (so that the Yang-Mills equations and action are
defined), then \abott\ every $G_\C$-bundle $E\to \EURV$, stable or
not, admits a unitary connection, unique up to gauge transformation,
that obeys the Yang-Mills equations and is compatible with the
holomorphic structure of $E$. Thus, in particular, solutions of the
Yang-Mills equations up to gauge transformation are in natural
correspondence with equivalence classes of holomorphic
$G_\C$-bundles.

This result has a natural interpretation in terms of the gradient flow of the Yang-Mills
action \abott, but we will not need this  here.   More important for us is the
generalization of the above result to allow parabolic structure \ref\nits{N. Nitsure,
``Cohomology Of The Moduli Of Parabolic Vector Bundles,'' Proc. Indian Acad. Sci. Math. Sci.
{\bf 95} (1986) 61-77.}.  We will state this theorem for any number of ramification points
with arbitrary choices of parabolic subgroups.
 We remove from the Riemann surface $\EURV$ a finite
set of points $p_1,\dots,p_s$.  We label each of the $p_i$ by a
parabolic weight $\alpha_i \in \frak t$, and an associated conjugacy
class $U_i=\exp(-2\pi\alpha_i)$.  We assume $\alpha_i$ to be generic
enough that for each $i$, the Levi subgroup $\Bbb{L}_i\subset G$
that commutes with $\alpha_i$ is the same as the subgroup that
commutes with $U_i$; we write ${\EUP}_i$ for the parabolic subgroup
determined by the pair $(\Bbb{L},\alpha_i)$. We consider  a
connection $A$ on the $G$-bundle $E\to \EURV\backslash
\{p_1,\dots,p_s\}$ that obeys the Yang-Mills equations and whose
holonomy around the point $p_i$ is conjugate to $U_i$.  The claim is
that such connections, up to gauge transformation, are in one-to-one
correspondence with holomorphic $G_\C$-bundles $E\to \EURV$ with
parabolic structure (of type $\EUP_i$) at the points $p_i$.  A
parabolic bundle is stable or semi-stable if and only if the
corresponding solution of the Yang-Mills equations is flat;
otherwise, it is unstable.

In applying this theorem, since we are taking our surface operators to be of type $\TT$,
we are for the moment interested in the case that the $\Bbb{L}_i$ are
all conjugate to the maximal torus $\TT$.  Hence, until the end of this section, where we
briefly discuss surface operators of general type $\Bbb{L}$, the phrase ``parabolic structure''
means a reduction of the structure group to a Borel subgroup $\EUB$.

\bigskip\noindent{\it Parabolic Bundles For $SL(2,\C)$}

For constructing 't Hooft-like operators, we are interested in
$G_\C$-bundles over $\EURV=\Bbb{CP}^1$ with parabolic structure at
two points,
 $p$ and $p'$.  The parabolic weights at the two points are $\alpha$ and $-\alpha'$.
The reason for the minus sign is that in \dulco, the two-sphere
$\EURV$ intersects the line $S$ with opposite orientation at the
points $p$ and $p'$.

Parabolic bundles
in this situation can be described very explicitly.   To explain the key ideas, we take
 $G_\C=SL(2,\C)$,
and carry out the analysis both holomorphically
and using the relation to the Yang-Mills equations.

Looking at things from a holomorphic point of view, the $SL(2,\C)$
bundle $E\to \EURV$ (regarded as a rank two vector bundle of trivial
determinant) must be $\CO(m)\oplus \CO(-m)$ for some integer $m$,
which we may as well assume to be non-negative.  In addition, we
must endow $E$ with parabolic structure at the two points $p,p'\in
\EURV$.  Parabolic structure for a rank two bundle $E$ at a given
point on a Riemann surface is simply a choice of a one-dimensional
subspace of the fiber of $E$ at that point. We denote these fibers
as $E|_p$ and $E|_{p'}$. So the parabolic structures at the points
$p$ and $p'$ are given by choices of
 subspaces $E_p\subset E|_p$ and $E_{p'}\subset E|_{p'}$.
The isomorphism classes of such parabolic bundles can be classified as follows:

(1) Suppose first that $m=0$.  Then the bundle $E$ is trivial, and the group $G_\C\cong
SL(2,\C)$ acts on it.  We say that the parabolic
structures at $p$ and $p'$ agree if there is a non-zero global section of $E$
whose restriction
to $p$ takes values in $E_p$ and whose restriction to $p'$ takes values in $E_{p'}$.  Otherwise
we say they disagree.
Up to isomorphism, that is, modulo the action of $SL(2,\C)$,
there are precisely two choices: the parabolic structures at $p$ and $p'$
may agree or disagree.

(2) Now suppose that $m>0$.  The bundle $E=\CO(m)\oplus \CO(-m)$ has a unique sub-bundle
isomorphic to $\CO(m)$ (characterized by the fact that any global holomorphic section of $E$
is actually a section of this sub-bundle).  By constrast, the embedding of $\CO(-m)$ in $E$
and hence the splitting $E=\CO(m)\oplus \CO(-m)$ is not canonical.  Up to an automorphism of $E$,
the only invariant information in the parabolic structure is whether the subspaces $E_p$ and
$E_{p'}$ coincide or do not coincide with the fibers of $\CO(m)$ at $p$ and $p'$.  So
for given $m$, there are $2\times 2=4$ choices.  We organize the four possibilities as follows.
We label the parabolic bundle $E$ by the integer $m$ if $E_p=\CO(m)|_p$,
and otherwise we label it by
$-m$.  And we say that the parabolic structure at $p'$ agrees with that at $p$ if $E_p$ and
$E_{p'}$ both coincide with, or both differ from, the relevant fiber of $\CO(m)$.  Otherwise,
we say that the parabolic structures disagree.

Now let us look at things from the point of view of the Yang-Mills equations.  In this
case, we need to pick parabolic weights at the points $p$ and $p'$.  (The weights did not
enter the holomorphic description, since we were not concerned with the question of which
parabolic bundles were stable.)  We take
\eqn\pkop{\eqalign{\alpha&=iy\,\left(\matrix{1 & 0\cr 0&-1\cr}\right)\cr
\alpha'&=iy'\,\left(\matrix{1 & 0\cr 0&-1\cr}\right),\cr}}
with $0<y,y'<1/2$.  It is also convenient to take initially $y\not=y'$.

The Yang-Mills equations \blijon\ tell us that the curvature is
covariantly constant, so we can pick a gauge in which the curvature
is actually constant \eqn\nikop{F=i\left(\matrix{f & 0\cr 0 &
-f\cr}\right),} with a real constant $f$ that we may as well take to
be nonnegative.  We will see momentarily that $f\not=0$ if $y\not=
y'$.  Since $F$ is covariantly constant, it commutes with the
holonomies $U$ and $U'$ around the points $p$ and $p'$, and
therefore (assuming $f\not=0$) these holonomies are diagonal
matrices. Since the holonomies must be conjugate to
$\exp(-2\pi\alpha)$ and $\exp(-2\pi (-\alpha'))$, respectively, it
follows that $U=\exp(- 2\pi\epsilon \alpha)$,
$U'=\exp(2\pi\epsilon'\alpha')$, with $\epsilon,\epsilon'=\pm 1$.
Finally, a bundle $E$ with the connection and curvature that we have
just described exists if and only if
\eqn\tinzo{\exp\left(\int_{\EURV}F\right)=UU'.} This condition
is equivalent to \eqn\binzo{\int_\EURV{f\over 2\pi}+\epsilon
y-\epsilon' y'= m',} for some integer $m'$.

The Weyl group acts by changing the sign of $F$, $y$, $y'$, and $m'$. So by a Weyl
transformation, we can fix
$\epsilon=1$, after which the solutions of the Yang-Mills equations are labeled by
an integer $m'$ and the variable $\epsilon'\in\{\pm 1\}$.   The classification we have
given of the solutions of the Yang-Mills equations is also valid if $y=y'$, but in that
case one needs a little more care in treating the case $\epsilon'=1$, $m'=0$.

Thus, the classification of Yang-Mills solutions agrees with the classification of parabolic
bundles.
In this comparison, $m'$ corresponds to $m$,
while $\epsilon'=1$ (or $-1$) corresponds to the case that
the two parabolic structures agree (or disagree).

Each such solution or bundle enables us to define an 't Hooft-like operator.
Thus, for every pair $(m,\epsilon')\in \Z\times \Z_2$, we get an 't
Hooft-like operator $T_{m,\epsilon'}$.

These operators do preserve the full topological supersymmetry of type $(B,A,A)$.
To see this, we first note that, since they are derived from solutions of the requisite
equations \bijon\ and \ijon, they preserve the topological symmetry of the $A$-model of type
$K$ (the one with symplectic structure $\omega_K$).  As these solutions are invariant
under $\varphi\to i\varphi$, which rotates the symplectic structure $\omega_K$ into $\omega_J$,
they also preserve the topological symmetry of the $A$-model of type $J$.  Linear combinations
of these two supercharges generate the full topological supersymmetry of type $(B,A,A)$.

\bigskip\noindent{\it Action On Branes}

\ifig\kulco{\bigskip  Action of a line operator, here inserted at the point $y$, on
branes.  The fields $(A,\phi)$
 determine a ramified Higgs bundle $E_u\to C_u=\{u\}\times C$ in complex structure $I$ for
every $u\in S$.  In a solution of the supersymmetric equations, the holomorphic type of $E_u$
is constant except when $C_u$ crosses the point $y$, at which point it jumps.}
{\epsfxsize=3in\epsfbox{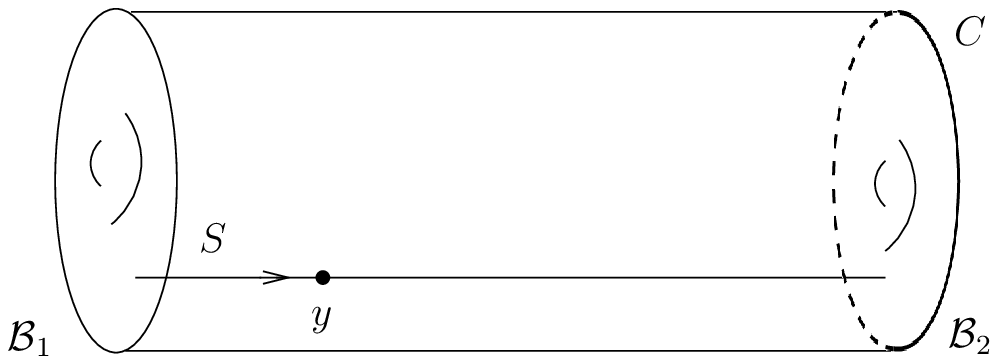}}
In broad outline, the action of one of these 't Hooft-like operators on branes
is analogous to the action of ordinary 't Hooft operators in the unramified case,
as described in  \kapwit.  We will hence be rather brief.

In \kulco, we consider fields $(A,\phi)$ on a $G$-bundle $E$ over the
three-manifold $W=I\times C$ with the usual singularity along the line $S$.
For any $u\in I$, the restriction of the bundle $E$ and fields $A,\phi$ to $C_u=\{u\}\times C$
give a ramified Higgs bundle  $E_u\to C_u$.  If the Bogomolny equations \plijon\ are obeyed,
the holomorphic type of $E_u$ is independent of $u$, except in crossing the position
of a line operator.  If one crosses a line operator supported at a generic point in $C$
(times a point in $I$), the bundle $E_u$ undergoes an ordinary Hecke modification, and if
as in \kulco\ one crosses a line operator supported on $S$, one gets a more general
Hecke-like modification that can involve the parabolic structure.  The justification
of these statements is as in \kapwit.

Now suppose that we want to see how the line operator $L_y$ at the point $y$ in \kulco\
acts on, say, the brane ${\cal B}_1$ on the left of the figure.
What we have just described means that the operator $L_y$ determines a ``correspondence''
${\EUQ}\subset \MH\times \MH$.  $\EUQ$ parametrizes pairs $(\tilde E_u,E_u)$, where $\tilde E_u$
can be obtained from $E_u$ in ``jumping'' over the line operator $L_y$.  $\EUQ$ can be
regarded as a brane in $\MH\times \MH$ of type $(B,A,A)$, since it is obtained by solving equations
that preserve this symmetry.  We are mainly interested in it as an $A$-brane of type $K$.

In fact, most  parabolic bundles over $\EURV\cong \Bbb{CP}^1$ that
may be used to define the 't Hooft-like operator $L_y$ are unstable.
When this is the case, we really want to define $\EUQ$ using a
compactification of the space of pairs $(\tilde E_u, E_u)$ by
including Hecke-like modifications defined using a less unstable
parabolic bundle (for $G_\C=SL(2,\C)$ this simply means that $|m|$
is smaller). This is just as in the unramified case.  One would
expect that in principle use of such a compactification could be
justified based on the underlying four-dimensional gauge theory.

Now let $\pi_i$, $i=1,2$, be the two projections $\pi_i:\MH\times\MH\to \MH$.
Given a brane ${\cal B}_1$ over $\MH$, we ``pull it back'' to $\MH\times \MH$ via
$\pi_1^*$, tensor it with $\EUQ$, and ``push forward'' to $\MH$ via $p_{2}$, to get the new
brane $L_y\cdot {\cal B}_1$.  Thus $L_y\cdot {\cal B}_1=\pi_{2\,*}(p_1^*({\cal B}_1)\otimes
\EUQ)$.  If ${\cal B}_1$ is a brane of type $(B,A,A)$, these operations can be carried out
using complex geometry in complex structure $I$, but more generally, if ${\cal B}_1$
is simply an $A$-brane of type $K$, one must use the analogs of these operations in the $A$-model,
that is, in Floer cohomology.

Here is a  possibly more down-to-earth explanation. To construct the
physical Hilbert space in the situation of \kulco, with branes
defining the boundary conditions at the ends and the operator $L_y$
in the interior, one proceeds as follows.  One considers the
compactified moduli space $\bar{\cal N}$ of supersymmetric
configurations, with boundary conditions determined at the two ends
by the branes ${\cal B}_1$, ${\cal B}_2$, and $L_y$ in the interior.
The physical Hilbert space is the cohomology of $\bar{\cal N}$. To
define the action of $L_y$ on the branes, the idea is that the
cohomology of $\bar{\cal N}$ is the same as it would be if the
singularity due to $L_y$ were omitted and ${\cal B}_1$ replaced by
some other brane $L_y\cdot{\cal B}_1$, which will not depend on
${\cal B}_2$. The Chan-Paton sheaf of $L_y\cdot{\cal B}_1$ has a
fiber at a point $E_u\in\MH$ that is equal to the cohomology of the
space $\Xi(E_u)$ obtained as follows: $\Xi(E_u)$ parametrizes Higgs
bundles $\tilde E_u$ such that  the pair $(\tilde E_u,E_u)$
represents a point in $\EUQ$.  (In this oversimplified explanation,
we have ignored the singularities of $\EUQ$.)  This recipe can be
justified by thinking through how one would compute the physical
Hilbert spaces with or without insertion of the operator $L_y$.

\bigskip\noindent{\it Affine Weyl And Braid Groups}

Now let us return to the example $G_\C=SL(2,\C)$, in which we classified the 't Hooft-like
operators by pairs $(m,\epsilon')\in \Z\times\Z_2$.
The affine Weyl group $\AffWeyl$ of $SL(2,\C)$, which was introduced in  eqn. \notused, is an
extension \eqn\yero{0\to \Z\to \AffWeyl\to\Z_2\to 0,} and
elements of this group are fairly naturally labeled by pairs
$(m,\epsilon')$, just like the 't Hooft-like operators.  Actually, because $\Z_2$ acts on $\Z$ by
reversing the sign, there is a subtlety about the sign of $m$, just as there was for the 't
Hooft operators.  The correspondence of 't Hooft operators with elements of $\AffWeyl$ is more
natural than the correspondence of either one with pairs $(m,\epsilon')$.

This has an analog for any simply-connected $G$: bundles
on $\Bbb{CP}^1$ with parabolic structure at precisely two points
correspond naturally to elements of $\AffWeyl$.  This is fairly easy
to see in the Yang-Mills approach, by generalizing the arguments we gave for $SL(2,\C)$,
and is left to the reader.  It is
also a standard result in the holomorphic approach.\foot{It is equivalent to the statement
that bundles on $\Bbb{CP}^1$ with
parabolic structure at two points correspond to orbits in the action of the Iwahori
group on the affine flag manifold.}

\ifig\blunko{\bigskip  Two line operators living on the support of a  surface operator
can be composed
by simply moving the lines together.  In the static case depicted here, this
is done by taking the limit as $y_1\to y_2$.  In topological field theory, line operators
generate
an associative algebra  that acts on branes. }
{\epsfxsize=3in\epsfbox{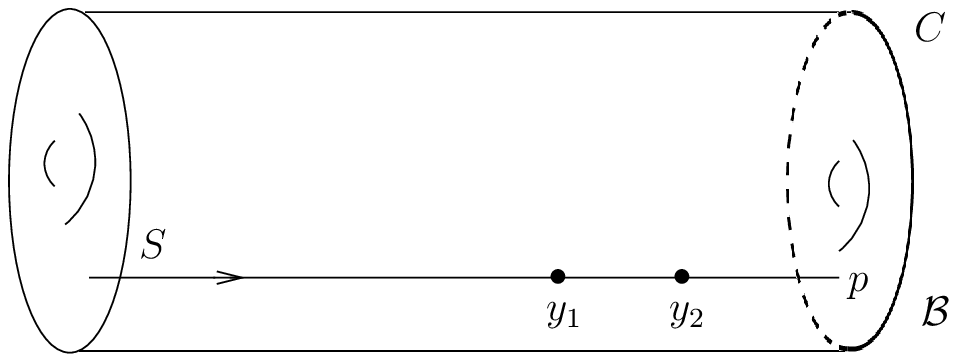}}
The 't Hooft-like operators that we have defined can be composed in
a natural fashion by moving two lines together, as  sketched in \blunko.  One might expect
that these operators would compose according to the multiplication law of the
 affine Weyl group. However,
according to \bez\ as well as our arguments in section \opmon, one
expects that the group that acts on branes of the $A$-model at a ramification point
$p\in C$ should
be the affine braid group, not the affine Weyl group. As in \pilot,
the two groups are related by an exact sequence \eqn\ptilot{1\to
\pi_1(\eurz)\to B_{\rm aff}\to \AffWeyl\to 1,} where $\eurz$ is the
space of  pairs $(\alpha,\beta)\in \frak t\times \frak t$ on which $\AffWeyl$ acts freely.
This is a non-trivial extension, so $\AffWeyl$ can be lifted to
$B_{\rm aff}$, but not in a way that respects the group law.
Therefore, we do not expect the operators $T_{m,\epsilon'}$ acting
on branes to respect the relations of the affine Weyl group.

We can be slightly
more specific.  For $G$ of rank $r$, $\AffWeyl$ is generated by elementary reflections
$T_i$, $i=1,\dots,r+1$, one for each node of the Dynkin diagram.  They obey $T_i^2=1$
and certain additional relations.  For example, for simply-laced $G$ of rank greater than 1,
one has
\eqn\udz{T_iT_jT_i=T_jT_iT_j}
if the vertices $i$ and $j$ are connected by an edge, and
\eqn\nudz{T_iT_j=T_jT_i}
otherwise.  To get the affine braid group, one omits the relation $T_i^2=1$, but keeps the others.
A word of shortest length in the affine Weyl group can be lifted in a fairly natural way to
the ``same'' word in the affine braid group, but this does not give a group homomorphism.

We will not try here to describe via gauge theory methods the operations on branes that
correspond to $\pi_1(\eurz)$.  In examples, it appears that $\pi_1(\eurz)$ acts by tensoring
the Chan-Paton bundle of a brane with certain vector bundles with a non-trivial action of the
ghost number.

\bigskip\noindent{\it Incorporation Of $\beta$ And $\gamma$}

So far we have taken the Higgs field $\phi$
to vanish, which only makes sense if $\beta=\gamma=0$.
If $(\beta,\gamma)\not=0$, then $\phi$ has prescribed singularities, and cannot vanish.

It is surprisingly  straightforward to incorporate $\beta$ and $\gamma$ in the analysis,
for the following
reason.  The solutions of the Yang-Mills equations that we have used to describe the
operators $T_{m,\epsilon'}$ are actually all abelian, that is, the structure group reduces
from $G$ to its maximal torus $\TT$.  This reflects the fact that the fundamental group of
the twice-punctured two-sphere is abelian.

This being so, we can incorporate $\beta$ and $\gamma$
 by simply adding an explicit abelian solution for
$\phi$.  Given $(\beta,\gamma)\in \frak t$, we define
\eqn\nobo{\varphi =\half(\beta+i\gamma){dz\over z},} where
$z=x^2+ix^3$, with $x^2$ and $x^3$ being the normal coordinates to
the line $S\subset W=\R^3$. ($\varphi$ is not a pullback from
$\EURV$.) Then we obey  the equations \bijon\ and \ijon\ by setting
$\phi=\varphi+\bar\varphi$ (still with $A_0=0$). The singularity of
$\varphi$ is independent of $x^1$, so as expected
 this construction is only possible given that
 $\beta$ and $\gamma$, unlike $\alpha$, do not jump in crossing the support of a line operator.

The line operators obtained this way do preserve topological symmetry of type $(B,A,A)$.
In adapting the argument that we gave at $\varphi=0$, care is needed only at one point.
The transformation $\varphi\to i\varphi$ is not a symmetry of the solution just described,
but rather maps this solution into a solution of
the same form with different $(\beta,\gamma)$.  This is enough to justify the argument.

Including $\phi$ in this way, even though it may seem like a trivial operation,
actually places an interesting restriction on the part of the solution involving the gauge field.
(For an analogous statement without ramification, see section 10.5 of \kapwit.)
With $\beta=\gamma=0$, the parabolic structures at the points $p$ and $p'$ are related by
a relative Weyl transformation.  For example, in the Yang-Mills construction, this happened
because a knowledge of the conjugacy classes of the monodromies $U_1$ and $U_2$, plus the
fact that they commute with the curvature $F$,
 fixes them only up to  Weyl transformations.  However, the Weyl group acts diagonally
on the triples $(\alpha,\beta,\gamma)$ and $(\alpha',\beta,\gamma)$.  So since $\beta$ and
$\gamma$ are constant, we cannot act on $\alpha$ and $\alpha'$ with arbitrary Weyl transformations.
We are restricted to Weyl transformations that act trivially on $\beta$ and $\gamma$.
Quantum mechanically, if we include $\eta$, which likewise cannot jump because it is a physical
parameter of the $A$-model of type $K$, the Weyl transformation must commute with $(\beta,\gamma,
\eta)$. Hence, a surface operator of this type is determined by the characteristic class
$\eurm m$ and a Weyl transformation that commutes with $(\beta,\gamma,\eta)$.
This is the result that one would expect
from  section \opmon.\foot{The analysis there
gave a slightly different answer (the
Weyl transformation had to commute with $(\gamma,\eta)$) because, as only the
$A$-model of type $K$ was considered, $\beta$ was free to vary. Here we are constructing
't Hooft-like operators that preserve the full $(B,A,A)$ symmetry.}

In particular, if the triple
$(\beta,\gamma,\eta)$ is regular, there is no freedom at all to make a Weyl
transformation.  The 't Hooft-like operators that act at a ramification point are hence
classified entirely by the characteristic class $\eurm m\in \Lambda_{\rm cort}$
(or $\Lambda_{\rm cowt}$ if one wishes to include operators that change the characteristic
class of the bundle).  This agrees with the analysis in section \opmon, where we found
that in the  case of a ramification point with regular semi-simple monodromy,
what acts on branes is precisely this lattice.

\bigskip\noindent{\it Analog Of Type $\Bbb{L}$}

As usual, we can generalize all this to the case of a surface
operator based on an arbitrary Levi subgroup $\Bbb{L}$.  The details
are fairly obvious and are left to the reader. We will just note
that 't Hooft-like operators supported on a surface of type
$\Bbb{L}$ with $\beta=\gamma=0$ are still classified by
$G_\C$-bundles over $\EURV\cong \Bbb{CP}^1$ with parabolic structure
at the two points $p$ and $p'$.  Now, however, the parabolic
structures at $p$ and $p'$ are respectively of type $\EUP$ and
$\EUP'$, where $\EUP$ and $\EUP'$ are the parabolic subgroups of
$G_\C$ determined by the pairs $(\Bbb{L},\alpha)$ and
$(\Bbb{L},-\alpha')$.  Bundles with parabolic structure of this type
can be easily classified using the Yang-Mills equations and
correspond to elements of the group called ${\cal W}_{{\rm
aff},\Bbb{L}}$ in section \actweyl.  This is the group that acts on
the cohomology of $\M_{H,\Bbb{L}}$ via monodromies in the space of
ramification parameters. The composition of such 't Hooft-like
operators is expected to give the braid-like group described in
section \opmon.  Including $\beta$ and $\gamma$ has again the effect
that the Weyl group element that enters in comparing the two
parabolic structures must commute with $\beta$ and $\gamma$.

\newsec{Local Models And Realizations By String Theory}
\seclab\local
\subsec{Overview}\subseclab\overview

The concluding section of this paper is devoted to alternative approaches to a few of the
topics in this paper.  We first provide an overview.

Instead of describing a surface operator by simply postulating that the fields
have the familiar singularity
\eqn\yongo{\eqalign{A & =  \alpha \,d\theta \cr
\phi & = \beta\,{dr\over r}-\gamma\,d\theta\cr}}
along a codimension two surface $D$, it is tempting to describe such an operator
by introducing new degrees of freedom supported on $D$ and coupling them to the gauge
fields of the ${\cal N}=4$ super Yang-Mills theory in bulk.  For example, hypermultiplets
or vector multiplets with $(4,4)$ supersymmetry in two dimensions
can be coupled to ${\cal N}=4$
super Yang-Mills theory in bulk.
This will give a theory somewhat like the one that
we have described ``by hand'' in the present paper.

The physics literature of course contains many analyses of conformal field theory coupled to
defects  of various kinds, that is, fields supported on a submanifold of positive
codimension.  The Kondo model in condensed matter physics is a classic example.  Closer to
our present concerns, hypermultiplets localized in codimension one have been coupled to
${\cal N}=4$ super Yang-Mills theory via  brane constructions, for example  in
\nref\kapseth{A. Kapustin and S. Sethi,  ``The Higgs Branch Of Impurity Theories,''
Adv. Theor. Math. Phys. {\bf 2} (1998) 571-591,
hep-th/9804027.}%
\nref\dew{O. DeWolfe, D.
Z. Freedman, and H. Ooguri, ``Holography And Defect Conformal Field Theories,'' Phys. Rev. {\bf
D66} (2002) 025009, hep-th/0111135.}%
\refs{\kapseth,\dew}, and hypermultiplets of codimension two, the case most
relevant to us, have been described in \coner.

{}From section \onahm, we actually can see what sort of impurity model would reproduce some
properties of the surface operators that we have considered.  One of the important themes
in the present paper is the ``local singularity'' of the moduli space $\MH$ of Higgs bundles
when the triple $(\alpha,\beta,\gamma)$ becomes non-regular.  The hyper-Kahler resolution of
such a local singularity is conveniently described by hyper-Kahler metrics \kron\
on certain complex
coadjoint orbits, or equivalently (in a different complex structure) hyper-Kahler metrics
on $X=T^*(G/\Bbb{L})$, for various Levi subgroups $\Bbb{L}\subset G$.
Therefore, in fact, some topics explored in this
paper can be expressed in terms of a combined system consisting of ${\cal N}=4$ super Yang-Mills
theory on a four-dimensional spacetime  together with a  supersymmetric sigma model supported on
a two-dimensional submanifold of spacetime.
We call this sigma model the local model.

Despite the usefulness of this approach, it has (as far as we know) a fundamental limitation.
The theory developed in the present paper depends on the parameters\foot{For brevity, we
describe here the parameters for the case that the Levi subgroup is $\Bbb{L}=
\TT$.  In general, all parameters are required to be $\Bbb{L}$-invariant.}
 $(\alpha,\beta,\gamma,\eta)
\in \TT\times \frak t\times\frak t\times {}^L\neg\TT$.  The sigma model depends on somewhat
similar parameters $(\alpha,\beta,\gamma,\eta)$, where $(\alpha,\beta,\gamma)$
are geometrical parameters that enter in the hyper-Kahler metric on $X$, and $\eta$ incorporates
the theta-angles of the sigma model.  The parameter $\eta$ takes values in $^L\neg\TT$, just
as in the gauge theory (see \wefot), and likewise $\beta$ and $\gamma$ are $\frak t$-valued in
each case.  But $\alpha$, which is $\TT$-valued in the full geometry of $\MH$,
 is $\frak t$-valued (and related to $\beta$ and $\gamma$ by an $SO(3)$ symmetry) in the
 hyper-Kahler geometry of $T^*(G/\Bbb{L})$,
as we have reviewed in section \onahm.  Consequently, the parameters of the full model and
the local model do not quite match.

This leads to no contradiction with the claim that the local model captures the behavior near
a local singularity, since for that purpose, the behavior when $\alpha$ becomes large is not
important.   However, it does apparently mean that the local model cannot be $S$-dual.
Since $S$-duality exchanges $\alpha$ and $\eta$, it hardly can hold in the sigma model of
target $T^*(G/\Bbb{L})$,
given that in this model $\eta$ is an angular variable but $\alpha$ takes values in a linear
space.  Hence, we cannot expect to maintain $S$-duality if we couple the ${\cal N}=4$
theory to this local model.

Although this may seem surprising, another point of view perhaps makes the conclusion more
natural.  The spaces $T^*(G/\Bbb{L})$ are not torus fibrations and do not admit mirror symmetry,
at least not in a $G$-invariant way.  (A mirror symmetry of the sigma model that is not
$G$-invariant would not help much when the sigma model is coupled to four-dimensional gauge
theory with gauge group $G$.)
So expecting to maintain $S$-duality when coupling  ${\cal N}=4$ super Yang-Mills theory
to the local model may be unrealistic.

\def\BAff{{B_{\rm aff}}}
Nonetheless, as we have stressed, we do believe that the local model is useful
for studying the behavior
near a local singularity of $\MH$, that is, near a non-regular triple $(\alpha,\beta,\gamma)$.
For example, consider the action of the affine braid group $\BAff$ on the branes of the $B$-model
at $\Psi=\infty$.  In this action, $\alpha$ and $\gamma$ are held fixed (and the most
interesting case is that they are near zero), and one studies monodromies in $\beta$ and $\eta$.
Whether $\alpha$ is $\frak t$-valued or $\TT$-valued is unimportant, and $\BAff$ acts on
the $B$-branes of the local model for the same reason that it acts on the branes on $\MH$.
Indeed, the analysis of the $\BAff$ action in \bez\ is equivalent, from a physical point of
view, to studying the monodromy action on the $B$-branes of the local model.  This seems
to be a good local model for the $\BAff$ action on $B$-branes of $\MH$.

By contrast, the local model is not a good model for studying the action of the $\BAff$ on
the $A$-branes at $\Psi=0$.  This involves monodromy action on the pair $(\alpha,\beta)$,
and this monodromy action is lost if $\alpha$ takes values in $\frak t$ rather than $\TT$.
To study the monodromy action on  $A$-branes of $\MH$,
it is essential, as far as we know, to use the full sigma model of target $\MH$
(or the full four-dimensional gauge theory).   There is a slightly ironical role reversal here.
The monodromy action of the $A$-model involves classical symplectomorphisms, while that
of the $B$-model is highly nonclassical; but the monodromy action of the $B$-model can be studied
in the local model, while that of the $A$-model cannot.

\bigskip\noindent{\it Contents Of This Section}

Now we can give a brief overview of the contents of this section.

As we have discussed, despite its limitations, the local model does have its uses.
The question arises of whether there is some physical approach to the local model by which
we can learn more than we have gleaned already.
Though we do not have a good  answer to this question for general gauge groups,
there is a useful description of the local model in the case of $G_\C=SL(2,\C)$.
This will be the topic of section \linsig.

The analysis in section \linsig\ will reveal the importance of the fact that the instanton
action (for certain supersymmetric instantons of the local model) vanishes when the triple
$(\alpha,\beta,\gamma)$ becomes nonregular.  This fact is actually inherited from a similar
fact in the four-dimensional gauge theory, as we explain in section \instac.

Finally, it is not very satisfactory to merely say that the local model lacks duality symmetry.
What can we do instead that is similar but does
preserve the duality symmetry?  In fact, an answer to
this question has already been given in string theory \coner\ in an interesting special
case ($G_\C=SL(N,\C)$ with a minimal coadjoint orbit).  We explain this in section \stringco.

\subsec{Linear Sigma Model For $G_\C=SL(2,\C)$} \subseclab\linsig

To use the local model -- the supersymmetric sigma model with target $T^*(G/\Bbb{L})$ --
to learn something that is not obvious from four-dimensional
gauge theory, we need a different way of studying it.  We do not have a general approach
that works uniformly for all groups, but we will explain an interesting approach for
$G=SU(2)$.  Here we rely upon the fact that the only relevant choice of $\Bbb{L}$ is $\Bbb{L}=U(1)$,
and that $T^*(SU(2)/U(1))=T^*\Bbb{CP}^1$ is the resolution of an $A_1$ singularity.
As a result, it can be constructed as a hyper-Kahler quotient of a finite-dimensional
linear space
\krono.  This has many applications in brane physics
\ref\dougmoore{M. Douglas and G. Moore, ``$D$-Branes, Quivers, And ALE Instantons,''
hep-th/9603167.}
 and is relevant for our purposes because it can be
used to construct a linear sigma model
\ref\wittencom{E. Witten, ``Some Comments On String Dynamics,'' in I. Bars et. al., eds. (1996),
{\it Future Perspectives In String Theory},
hep-th/9507121.} that gives a simple way to understand
the vacuum structure of the supersymmetric sigma model with target $T^*\Bbb{CP}^1$.

Complex coadjoint orbits of all the classical Lie groups can similarly be constructed
as hyper-Kahler quotients of finite-dimensional linear spaces \ref\kobsw{P. Z. Kobzk and A. F.
Swann, ``Classical Nilpotent Orbits As Hyper-Kahler Quotients,'' Int. J. Math. {\bf 7} (1996)
193-210.}.  These more general constructions could  be taken as the starting point
in constructing a linear sigma model relevant to surface operators for any classical group.
However, the case of $SU(2)$ has an important advantage,
which is that the action of the Weyl group is transparent in the linear sigma model.  This
is not the case for $G$ of higher rank, which is why we will in this analysis
consider only $SU(2)$.

To get the $A_1$ singularity as a hyper-Kahler quotient, we start
with $\Bbb{H}^2\cong \R^8$, regarded as a flat hyper-Kahler
manifold.  It admits the action of $Sp(2)$, which in our notation is
the group of $2\times 2$ unitary matrices of quaternions. Two kinds
of quaternionic matrices will play an important role.  $Sp(2)$
contains the group $SU(2)$ consisting of quaternion multiples of the
identity \eqn\hefrox{\left(\matrix{v&0\cr 0&v\cr}\right),} where $v$
is a quaternion of modulus 1, and also the group $O(2)$ of unitary
quaternion
 matrices whose entries are actually all real:
\eqn\lembox{\left(\matrix{a&b\cr c&d\cr}\right).}
These two groups commute with each other, but the global structure is not a product, because
$SU(2)$ and $O(2)$ contain the common central element $-1$.  So the global structure is
\eqn\dembox{\left(SU(2)\times O(2)\right)/\Z_2\subset Sp(2).}

The connected component of $O(2)$ is $U(1)$.
Let $\vec \mu=(\mu_1,\mu_2,\mu_3)$ be the hyper-Kahler
moment map for the action of this $U(1)$.  We introduce real parameters $(\alpha,\beta,\gamma)$,
and define $X(\alpha,\beta,\gamma)$ to be the hyper-Kahler quotient $\Bbb{H}^2/\neg/\neg/U(1))$
taken at $\vec\mu=(\alpha,\beta,\gamma)$.  In other words, $X(\alpha,\beta,\gamma)$ is defined
by setting $\vec\mu=(\alpha,\beta,\gamma)$ and dividing by $U(1)$. For $\alpha=\beta=\gamma=0$,
this gives the $A_1$ singularity $\R^4/\Z_2$; in general, it gives the hyper-Kahler resolution
of the $A_1$ singularity.

{}From a physical point of view, we construct a  supersymmetric linear sigma model on a two-manifold
$D$ in which the fields are a $U(1)$ vector multiplet
and a pair of hypermultiplets $H_1,\, H_2$ (whose bosonic
components parametrize $\Bbb{H}^2$, acted on by $U(1)$ as above).  The parameters
$(\alpha,\beta,\gamma)$ are the $D$-terms or Fayet-Iliopoulos parameters of the vector multiplet.
The moduli space of classical vacua is precisely the hyper-Kahler quotient
$X(\alpha,\beta,\gamma)$ described in the last paragraph, with one caveat: if $\alpha=\beta
=\gamma=0$, this moduli space has a second branch, as described in \wittencom.

The linear sigma model has an $SO(3)$ symmetry that rotates the parameters $(\alpha,\beta,\gamma)$.
This arises as follows.   If we think of an element of $\Bbb{H}^2$ as a column of quaternions
$\left(\matrix{u_1\cr u_2\cr}\right)$, then the action of $Sp(2)$ on the left commutes with a right
action of $ Sp(1)$.  This $Sp(1)$ acts by rotating the complex structures of the hyper-Kahler
manifold $\Bbb{H}^2$, and likewise rotating the parameters $(\alpha,\beta,\gamma)$.

In addition to $(\alpha,\beta,\gamma)$, the quantum theory also depends on the theta-angle
of the $U(1)$ vector multiplet, which as usual we will call $\eta$.  Clearly, $\eta$ is
angle-valued.  It appears in a factor in the path integral of the form
\eqn\dolfus{\exp\left(i\eta\int_DF\right).}  Here $F$ is the $U(1)$ curvature, which we represent by a
real two-form.
This means, in the usual form of the model, that there is a symmetry
\eqn\medigo{\eta\to \eta+1.}
As we explain later, there is a certain sense in which the basic symmetry is $\eta\to\eta+2$.

 The parameters $(\alpha,\beta,\gamma,\eta)$  take values in
$\eurm K_0=\R^3\times S^1$. But this is not quite the natural
parameter space of the model. We must recall from \lembox\ that the
$U(1)$ symmetry of the model can be extended to $O(2)$. Let $w$ be
an element of the disconnected conjugacy class of $O(2)$.  As we
will see momentarily, $w$ will play the role of  the non-trivial
element of the Weyl group of $SU(2)$.  (This fact does not have a
good analog for similar models for coadjoint orbits of groups of
higher rank.  That is why we limit ourselves here to $SU(2)$.) $w$
acts by an outer automorphism of the gauge group $U(1)$ of the
linear sigma model, and acts on the coupling parameters by changing
all signs: \eqn\iki{w:(\alpha,\beta,\gamma,\eta)\to
(-\alpha,-\beta,-\gamma,-\eta).} This is the action that in the rest
of the present paper has always come from the Weyl group of $SU(2)$.
The parameter space of the  model is thus really $\eurm K=\eurm
K_0/\Z_2$. At its fixed points, which will be described momentarily,
 $w$ becomes a symmetry of the linear sigma model.

The question of where in parameter space the model becomes singular was addressed in
\wittencom.  The answer is that it becomes singular precisely at the point
$P_0:\alpha=\beta=\gamma=\eta=0$, when the vacuum becomes unnormalizable because of the role of
the ``second branch'' of classical vacua.

Obviously, $P_0$ is one of the two fixed points for the action of
$\Z_2$ on $\eurm K_0$.  The second is
$P_1:\alpha=\beta=\gamma=0$, $\eta=1/2$.  What happens there? {}From
the point of view of classical geometry, one does not see the
parameter $\eta$. The hyper-Kahler quotient at
$\alpha=\beta=\gamma=0$ is the orbifold singularity $\R^4/\Z_2$.
There is a soluble conformal field theory associated with this
orbifold \ref\orbifold{L. Dixon, J. Harvey, C. Vafa, and E. Witten,
``Strings On Orbifolds,'' Nucl. Phys. {\bf B261} (1985) 678-686.}.
One may guess that for some value of $\eta$, the linear sigma model
will reduce to this orbifold theory at low energies.  This in fact
occurs \ref\aspinwall{P. Aspinwall, ``Enhanced Gauge Symmetries And
K3 Surfaces,'' hep-th/9507012.} at $\eta=1/2$, or more exactly at
$P_1:(\alpha,\beta,\gamma,\eta)=(0,0,0,1/2)$.  The theory at this
point is a perfectly smooth, well-behaved quantum field theory, and
in fact a simple and exactly soluble one.  Since $P_1$ is  a fixed
point of $w$, the orbifold theory has $w$ as a symmetry. In fact,
$w$ is the ``quantum symmetry'' of the orbifold (which acts as $-1$
for strings in the twisted sector and $+1$ on untwisted strings).

Now let us focus on the $A$-model or $B$-model derived from this theory (upon picking one
of the symplectic or complex structures), and the associated monodromies as the parameters
are varied.  Actually, as was explained in the overview of this section,
it is not very interesting to consider the $A$-model, because in the
present local model $\alpha$ takes values in a linear space, preventing the
existence of interesting monodromies. So we will concentrate on the
$B$-model in one of the complex structures.
Because of the $SO(3)$ symmetry that rotates
the complex structures, it does not matter which one
we pick.  To agree with the terminology in the rest of this paper,
we will consider the $B$-model in complex structure $J$.

The monodromies of this model come as in section \opmon\ by varying $\beta$ and $\eta$ while
keeping $\alpha$ and $\gamma$ fixed.  If the pair $(\alpha,\gamma)$ is nonzero, then the relevant
parameters are $\beta$ and $\eta$ modulo the symmetry $\eta\to \eta+1$.
We do not get anything new from the ``Weyl transformation'' $w$, since it does not leave
$(\alpha,\gamma)$ fixed.  Thus, in this situation, the appropriate parameter space is simply  the
quotient of the $\beta-\eta$ plane by $\eta\to \eta+1$, or
$\R^2/\Z=\R\times S^1$. The monodromy group is simply $\Z$.  This answer is familiar from
section \opmon.

\def\eura{\eurm A}
\def\eurr{\eurm R}

\def\eurb{\eurm B}
\def\eurt{\eurm T}
As usual, the most interesting case is $(\alpha,\gamma)=(0,0)$.  Here after dividing
by $\Z$ to get $\R\times S^1\cong\C^*$,
we must still divide by $w:(\beta,\eta)\to (-\beta,-\eta)$.
$w$ acts on $\lambda=\exp\left(2\pi(\beta+i\eta)\right)$
by $\lambda\to\lambda^{-1}$.  There are two fixed points, at $\lambda=1$ and $\lambda=-1$.
The fixed point at $\lambda=1$ is the point $P_0$ at which the theory becomes singular.  We
omit this point from the parameter space and call what remains $\tilde\C^*$.  We still
have the action of $\Z_2$ on $\tilde \C^*$, with a fixed point at $P_1:\lambda=-1$.  We do not
want to remove this fixed point, since the quantum field theory is well-defined there.
Rather, we think of the parameter space on which the quantum field theory depends as
an orbifold $\tilde \C^*/\Z_2$, and the relevant monodromy group is the fundamental
group in the orbifold sense, $\pi_1^{\rm orb}(\tilde \C^*/\Z_2)$.  This group is generated
by $\eura$, the monodromy around $\lambda=1$, and $\eurt$, the monodromy around $\lambda=-1$, with
the sole relation being $\eurt^2=1$, expressing the fact that $\lambda=-1$ is a $\Z_2$ orbifold
point.

This answer for the monodromies seen in the local model is essentially equivalent to the answer
obtained in section \opmon, but some explanation is required.  First of all, let us consider
the maximal torus $\TT$ of the group $SO(3)$.  It consists of elements of the following form:
\eqn\utut{U=\left(\matrix{\cos(2\pi\eta)&\sin(2\pi\eta)&0\cr -\sin(2\pi\eta)&
\cos(2\pi\eta)&0\cr 0&0&1\cr}\right).}
$\eta$ is real if we want the compact group $SO(3)$, or should be replaced by its
complexification $\eta-i\beta$ in the case of the complex
Lie group $SO(3,\C)=SL(2,\C)/\Z_2$.  In that case, we can write
\eqn\butut{U=\left(\matrix{{1\over 2}(\lambda+\lambda^{-1})&{i\over 2}(\lambda-\lambda^{-1})
&0\cr -{i\over 2}(\lambda-\lambda^{-1})&{1\over 2}(\lambda+\lambda^{-1})&0\cr 0&0&1\cr}\right),}
with $\lambda=\exp(2\pi i(\eta-i\beta))\in \C^*$.
We call an element $U\in\TT$ very regular if the subgroup of $SO(3)$
that commutes with $U$ is precisely $\TT$, and regular if the connected component of this
group is equal to $\TT$.  This criterion does not depend on whether we work in $SO(3)$ or
$SO(3,\C)$.  The element $U=1$ corresponding to $\lambda=1$
is nonregular.   By contrast, the element $U$ that corresponds to
 $\lambda=-1$ is regular but it is not very regular, since it commutes with
the element
\eqn\nubtut{w=\left(\matrix{1&0&0\cr 0&-1&0\cr 0&0&-1\cr}\right)}
which acts by  $\lambda\to \lambda^{-1}$,
 and, in fact, generates the Weyl group of $SO(3)$.

We get a regular conjugacy class for any $\lambda\in \tilde\C^*$.  However,
the conjugacy classes in $SO(3)$
that correspond to $\lambda$ and $\lambda^{-1}$ are conjugate by the action of $w$.
So the moduli space of regular conjugacy classes in $SO(3)$ is the orbifold $\tilde\C^*/\Z_2$,
and, as we have seen, its orbifold fundamental group $\pi_1^{\rm orb}(\tilde \C^*/\Z_2)$
is also the monodromy group of our $B$-model for the case $(\alpha,\gamma)=(0,0)$.

In section \opmon, it was convenient to consider the subgroup of the monodromy group that
consists of transformations that act trivially on the discrete electric flux ${\bf e}_0$.
So let us identify that subgroup in the local model.  As we explained in section \unramreview,
to define ${\bf e}_0$ in $SU(2)$ gauge theory, one allows a ``twist'' so that the gauge
bundle $E\to M=\Sigma\times C$ has structure group $SO(3)=SU(2)_{\rm ad}$, but lifts
to an $SU(2)$ bundle if restricted to $q\times C$ for $q$ a point in $\Sigma$.  In the local
model, we do not necessarily have such a lifting to $SU(2)$, since the
support of the local model is not $q\times C$ but is $D=\Sigma\times p$ for some point
$p\in C$.  In defining ${\bf e}_0$, we include $SO(3)$-bundles $E\to D$
with $\int_Dw_2(E)\not=0$.

What does this mean in the local model?  The group that acts on the hypermultiplets of the
local model is not $SO(3)$ but $SU(2)$, so at first sight one might think it is impossible
to twist the local model by an $SO(3)$ bundle $E$ with non-zero $w_2(E)$.
However, the local model has a gauge group $U(1)$, and by restricting \dembox\ to the connected
component, we see that the global form of the symmetry group is not $SU(2)\times U(1)$
but $(SU(2)\times U(1))/\Z_2$.  An $SO(3)$ bundle $E\to D$ with $w_2(E)\not=0$ cannot
be lifted to an $SU(2)$ bundle, but it can be lifted to a bundle with structure group
$(SU(2)\times U(1))/\Z_2$.  When we do this, the $U(1)$ curvature $F$ obeys
\eqn\hyro{\int_D{F\over 2\pi}={1\over 2}\int_Dw_2(E)~{\rm mod}~\Z.}
Looking back to \dolfus, we see that this means that the effect of $\eta\to \eta+1$ is to
multiply the integrand of the path integral by
\eqn\yro{(-1)^{\int_Dw_2(E)}.}
This operation is equivalent to shifting ${\bf e}_0$.

The result is that the fundamental group $\pi_1^{\rm orb}(\tilde\C^*)$ is the full
monodromy group of the model, including transformations that shift ${\bf e}_0$.
If we want to identify the subgroup of monodromies that keep ${\bf e}_0$ fixed, which will
facilitate the comparison to the result of section \opmon, we should consider the pair
$(\beta,\eta)$ subject to the symmetries $w:(\beta,\eta)\to (-\beta,-\eta)$ and
\eqn\trog{\eta\to\eta+2.}
These are the symmetries that keep ${\bf e}_0$ fixed.
Identifying $\eta$ mod 2 is equivalent to lifting $U$ to an element of the maximal
torus of $SU(2)$, rather than $SO(3)$.  As such it is conjugate to
\eqn\rog{\hat U=\left(\matrix{\exp({i\pi}(\eta-i\beta))& 0\cr 0&\exp(-{i\pi}(\eta-i\beta))
\cr}\right).}
The Weyl group still acts on the pair $(\beta,\eta)$ with two fixed points, but there
is an essential difference.  The fixed points are now $P_0:(\beta,\eta)=(0,0)$ and
$P_1':(\beta,\eta)=(0,1)$.  The essential difference is that the two fixed points of the
Weyl group are now both points at which the sigma model is singular (since its singularities
 are invariant under $\eta\to\eta+1$).  In contrast,
 before lifting to $SU(2)$, one fixed point was a singularity
 of the sigma model and one was an orbifold point.

The non-regular values of $\hat U$ are $1$ and $-1$, which correspond precisely to the fixed
points $P_0$ and $P_1'$.  So when we omit the points at which the local model is singular,
and divide by the Weyl group and by $\eta\to\eta+2$, we get precisely the moduli space of
regular conjugacy classes in $SL(2,\C)$.  Its fundamental group
 is known as the affine braid group of
$SL(2,\C)$, and denoted $B_{\rm aff}(SL(2,\C))$.   This group is
freely generated by elements $\eura$ and $\eurb$ that we
can regard as the monodromies around the points $1$ and $-1$.  They obey no relations at all.

The full monodromy group of the local model, including transformations that change ${\bf e}_0$,
is an extension:
\eqn\uneo{1\to B_{\rm aff}\to \pi_1^{\rm orb}(\tilde\C^*)\to \Z_2\to 1.}
Here $B_{\rm aff}$ is extended by an outer automorphism of order 2 that acts by $\eta\to\eta+1$
and exchanges the points
$\hat U=1$ and $\hat U=-1$, which are equivalent in $SO(3)$.  We call this automorphism
$\eurt$; it obeys $\eurt^2=1$, $\eurt\eura\eurt=\eurb$, $\eurt\eurb\eurt=\eura$.  Since $\eurb$
is the same as $\eurt\eura\eurt$, the extension
$\pi_1^{\rm orb}(\tilde\C^*)$ is generated by $\eura$ and $\eurt$ with the sole relation $\eurt^2=1$.
This is the result we gave earlier for the full monodromy group.

All of this is in accord with the analysis of section \opmon.  The subgroup of the monodromy
group that acts trivially on ${\bf e}_0$ is the affine braid group, and the full monodromy
group is an extension of the affine braid group by the center of the simply-connected form
of the gauge group $G$.  In the present example, $G=SL(2,\C)$, and the center is $\Z_2$.

To tie up a loose end, we should perhaps mention that $SL(2,\C)$
is an exception to the description of the affine braid group given
in eqns. \udz\ and \nudz.  The reason for the exception is that in
$SL(N,\C)$ for $N>2$, the adjacent nodes of the affine Dynkin
diagram correspond to vectors at an angle $2\pi/3$, but this is
not so for $SL(2,\C)$ (where the angle is $\pi$). To describe what
happens for $SL(2,\C)$, we go back to the definition of the affine
Weyl group as an extension $\Lambda_\cort\rtimes \Weyl$, with
$\Weyl$ the Weyl group. For $SU(2)$, $\Lambda_\cort\cong\Z$, so a
vector in $\Lambda_\cort$ is just an integer $n$. The Weyl group
is $\Z_2$, generated by ${\eurm A}:n\to -n$.  The affine Weyl
group is generated by $\eura$ together with $\eurr:n\to n+1$. They
obey ${\eura}^2=1$ and $\eura \eurr\eura=\eurr^{-1}$.
Equivalently, $\AffWeyl$ is generated by $\eura$ and $\eurb=\eura
\eurr$ with $\eura^2=\eurb^2=1$ and no other relations. $\eura$
and $\eurb$ are the reflections corresponding to the two nodes of
the extended Dynkin diagram. To get the affine braid group of
$SL(2,\C)$, we just drop the conditions $\eura^2=\eurb^2=1$, so
$B_{\rm aff}(SL(2,\C))$ is simply a free group with the two
generators $\eura$ and $\eurb$.

\subsec{Instantons And The Local Singularity}
\subseclab\instac

In this discussion, the fact that the local model is singular at $\alpha=\beta=\gamma=\eta=0$
(and non-singular elsewhere) played a crucial role.

One explanation of this is given in \wittencom, using the fact that
a second branch of classical vacua becomes relevant precisely when
these parameters all vanish. There is, however, another standard
explanation of the significance of having
$\alpha=\beta=\gamma=\eta=0$. When $(\alpha,\beta,\gamma)\not=0$,
the $A_1$ singularity undergoes a hyper-Kahler resolution to produce
a smooth manifold with the topology of $T^*\Bbb{CP}^1$.  The zero
section of this cotangent bundle is holomorphic in the complex
structure \eqn\joplo{{\cal I}={(\alpha I+\beta J+\gamma K)\over
\sqrt{\alpha^2+\beta^2+\gamma^2}}.}  It is antiholomorphic in the
opposite complex structure. Its area $A$ is proportional to
$\sqrt{\alpha^2+\beta^2+\gamma^2}$.  A holomorphic or
antiholomorphic map of $D$ to $\Bbb{CP}^1\subset T^*\Bbb{CP}^1$ is
holomorphic in complex structure $\cal I$ or $-\cal I$.  We interpret it as
an instanton of the sigma model.  In
other complex structures, there are no compact holomorphic or
anti-holomorphic curves.  An instanton of the sigma model that is defined by a degree $d$
holomorphic
mapping of $D$ to the zero section of $T^*\Bbb{CP}^1$ makes a
contribution to the path integral that is proportional to $q^d$, where
$q=\exp(- kA+2\pi i \eta)$. (Here $k$ is a constant that depends on
the gauge coupling of the linear sigma model.)  Upon summing over $d$, the instanton series
has a pole at $q=1$, or in other words when $(\alpha,\beta,\gamma,\eta)$ all
vanish. This pole reflects the singularity of the sigma model.

\def\EUV{\eusm V}
What has just been summarized is a standard analysis in two dimensions.
We want to consider here how this story looks  in four dimensions.
The twisted ${\cal N}=4$ super Yang-Mills theory
that  underlies the present discussion depends on a parameter $t$.   The analog of the instanton
equation in four dimensions, according to section 3.2 of \kapwit,
is the condition $\EUV^+(t)=\EUV^-(t)=\EUV^0=0$, where
\eqn\albo{\eqalign{\EUV^+(t)&=(F-\phi\wedge\phi+td_A\phi)^+\cr
                   \EUV^-(t)&=(F-\phi\wedge\phi-t^{-1}d_A\phi)^-\cr
                    \EUV^0&=D\star \phi.\cr}}
Moreover, according to eqn. (5.28) (or eqns. (5.7) and (5.11)) of
\kapwit, if we specialize to $M=\Sigma\times C$, then the
four-dimensional topological field theory reduces on $\Sigma$ to a
two-dimensional $A$-model with target space $\MH(G,C)$, and
symplectic structure a multiple of \eqn\nalbo{\omega_t={1-t^2\over
1+t^2}\omega_I-{2t\over 1+t^2}\omega_K.} The instantons in this
$A$-model are holomorphic curves for the corresponding complex structure
\eqn\zalbo{I_t={1-t^2\over 1+t^2}I-{2t\over 1+t^2}K.}

Now let us incorporate a surface operator supported on
$D=\Sigma_p=\Sigma\times p$, for $p$ a point in $C$.  We endow the
surface operator with parameters $(\alpha,\beta,\gamma)$. {}From the
point of view of the local model, we expect holomorphic instantons
only in the complex structure $\cal I$.  The local model should be
adequate at least for describing those instantons whose action
goes to zero as $(\alpha,\beta,\gamma)\to 0$.  We will call these
the instantons with small action. Four-dimensional instantons of small action
should correspond to instantons of the
sigma model with complex structure $I_t$.  The sigma model only has instantons
in complex structure ${\cal I}$.  Therefore, we expect the
four-dimensional gauge theory to have instantons -- or at least instantons
of small action -- only if $I_t={\cal I}$.  The condition to have $I_t={\cal I}$,
in view of \joplo\ and \zalbo,
is that \eqn\balbo{{\alpha\over\gamma}={t-t^{-1}\over 2},~~\beta=0.}
We want to derive these conditions directly in four dimensions.

First of all, from a four-dimensional point of view, $\EUV^0$ cannot
vanish if $\beta\not=0$.  Indeed, having $\beta\not=0$ causes
$d_A\star\phi$ to have a delta function along $D$, so it cannot
vanish.  This is one of the desired results;
it remains to derive the first equation in \balbo. We will
do this by generalizing the vanishing arguments of section 3.3 of
\kapwit.

What in that reference is called Vanishing Theorem 1  asserts that in any solution of
the equations \albo\ (in the absence of surface operators) the
Pontryagin number vanishes.  This follows from the following
identity: \eqn\instid{\int_M\,\Tr\left(\EUV^+(t)\wedge
\EUV^+(-t^{-1})+\EUV^-(t)\wedge \EUV^-(-t^{-1})\right)=\int_M\Tr
\,F\wedge F.} The derivation of this identity depends on integration
by parts.  The left hand side vanishes in a solution of the
equations, and of course the right hand side is a multiple of the
Pontryagin number. So the identity implies vanishing of the
Pontryagin number.

A few things are different in the presence of a surface operator supported on $D\subset M$. As
in the derivation of \hombo, we interpret the integral in \instid\
as an integral over the complement of $D$ in $M$.   With that
understood, the Pontryagin number is not simply a multiple of
$\int_M\Tr\, F \wedge F$, but has the extra terms indicated in
\hombo.  Second, in the integration by parts that is needed to
derive \instid, one runs into a term $(t-t^{-1})\int_M d\left({\rm Tr}\,
\phi\wedge F\right)$.  For $\gamma\not= 0$, despite being the integral of an exact
form, this does not vanish, even for
compact $M$.  Instead, there is a sort of surface contribution
localized on $D$, of the form $-2\pi(t-t^{-1})\int_D\Tr\,\gamma F$.
With this understood, the generalization of \instid\ is
\eqn\tongox{\eqalign{\int_M &\,\Tr\left(\EUV^+(t)\wedge
\EUV^+(-t^{-1})+\EUV^-(t)\wedge \EUV^-(-t^{-1})\right)\cr &
 =-8\pi^2\eurm
N+4\pi^2(D\cap D)\Tr\,\alpha^2
+4\pi\int_D\Tr\,\left(\alpha-{t-t^{-1}\over 2}\gamma\right)F.\cr}}

The left hand side vanishes in any solution of the supersymmetric
equations.  So obviously, if such a solution is to exist for small
$(\alpha,\beta,\gamma)$, the integer $\eurm N$ must vanish. In addition, in our
application with $M=\Sigma\times C$ and $D=\Sigma\times p$, we have
$D\cap D=0$.  So all terms on the right vanish except the last one, which hence must
also vanish.
The $G$-bundle $E\to M$, when restricted to $D$, has
structure group $\TT$, and is characterized topologically by the
characteristic class $\eurm m$ of its curvature $F$.  For $G=SU(2)$, as
assumed in the derivation of \balbo, $\eurm m$ is equivalent to the
instanton number of the sigma model (the degree of the map $D\to
\Bbb{CP}^1$), and the singularity of the sigma model at
$(\alpha,\beta,\gamma)$ approaching zero is supposed to come from instantons
with $\eurm m\not=0$ and small action. But \tongox\ shows us
that if $\eurm N=D\cap D=0$, then an instanton with $\eurm m\not=0$
must have $\alpha-{1\over 2}(t-t^{-1})\gamma=0$.  This is the desired condition
in eqn. \balbo, so we have explained  from a four-dimensional point of view the fact that
instantons or supersymmetric field configurations appear only when this condition is obeyed.

It is similarly possible to derive the same condition by
generalizing Vanishing Theorem 2 of \kapwit\ to incorporate surface
operators.

\subsec{Some String Theory Constructions}
 \subseclab\stringco

The assertion that the local model does not preserve $S$-duality
is a little perplexing.  In order to understand this better, we
will consider a string theory construction in which $S$-duality is
manifest, and see how this construction fails to give the local
model.

As in \coner, we will consider a much studied situation in Type IIB
superstring theory: a stack of $N$ $D3$-branes, giving a $U(N)$
gauge theory in four dimensions, that intersects in codimension two
another $D3$-brane which we will call $D3'$.  Thus everything
happens in a six-dimensional subspace of the ten-dimensional
spacetime of the Type IIB theory.  We will take this six-dimensional
subspace to be a product $Z=\Sigma\times C\times\tilde C$ of three
Riemann surfaces.  The $N$ $D3$-branes are supported on $M=\Sigma\times C\times r$, with
$r$ a point in $\tilde C$,
and the $D3'$-brane is supported on $\tilde M=\Sigma\times p\times \tilde C$,
with $p$ a point in $C$.
$\tilde C$ may be either the complex plane $\C$
or a compact Riemann surface.  In our usual application, $C$ is the
Riemann surface on which we consider the geometric Langlands
program, and $\Sigma$ is the spacetime of the two-dimensional
effective field theory that results from compactification on $C$.
One can also replace $M=\Sigma\times C$ with a more general
four-manifold, and presently this will be convenient.

The stack of $N$ $D3$-branes wrapped on $M$
produces a $U(N)$ gauge theory on $M$ with ${\cal N}=4$ supersymmetry,
and the $D3'$-brane produces a $U(1)$ gauge theory on $\tilde M$
also with ${\cal N}=4$ supersymmetry.
On the intersection $M\cap \tilde M = \Sigma\times p\times r$, one gets hypermultiplets
in the $N$-dimensional representation of $U(N)$ and with charge 1 under $U(1)$.  To try to make
contact with the situation explored in this paper, one might hope  that the
hypermultiplets, together with the $U(1)$ gauge multiplet supported on $\tilde M$, would be
equivalent at low energies to a sigma model with target $\Bbb{H}^N/\neg/\neg/U(1)$,
where $\Bbb{H}^N\cong \C^{2N}$ is parametrized by the hypermultiplets and $\Bbb{H}^N/\neg/\neg/
U(1)$ is its hyper-Kahler quotient by $U(1)$.

This hyper-Kahler quotient depends on the constant values chosen for
the three components of the hyper-Kahler moment map.  Physically,
the constants will have to arise from the Fayet-Iliopoulos $D$-terms
of the $U(1)$ gauge theory on $\tilde M$, which one would like to
somehow derive from parameters of the string theory construction.
For non-zero $D$-terms, the hyper-Kahler quotient is equivalent in
one complex structure to $T^*\Bbb{CP}^{N-1}$, and in a different
complex structure it is equivalent to the orbit of an element
\eqn\nxonp{\delta= \left(\matrix{N-1&&&\cr
                 & -1&& \cr
                 &&\ddots&\cr
                  &&&-1\cr}\right)}
of the Lie algebra ${\frak{ gl}}(N)$.  What we get here is a special case of the
local model discussed in section \overview, namely the case with $G=U(N)$, $\Bbb{L}=U(N-1)$,
and so $T^*(G/\Bbb{L})=T^*(U(N)/U(N-1))=T^*\Bbb{CP}^{N-1}$.  The parameters $(\alpha,\beta,\gamma)$
must each, in this case, be real multiples of $\delta$, with the proportionality determined
by the Fayet-Iliopoulos parameters.  Thus,  each depends on only one real parameter,
an assertion that will remain true in the Anti de Sitter construction explained below.

This looks like a plausible way to get from branes an example of the situation discussed
in this paper.  However,
since the brane construction (like any configuration of $D3$-branes in Type IIB superstring
theory) has $SL(2,\Z)$ $S$-duality, and the model consisting of the $N$ hypermultiplets
plus $U(1)$ gauge field does not, the brane configuration cannot really be equivalent to
the coupling of ${\cal N}=4$ super Yang-Mills theory to the two-dimensional system of
hypermultiplets and gauge fields.  What fails depends on the choice of $\tilde C$.

If we take $\tilde C=\Bbb{C}$, then we do not quite get the reduction
we want, since the infinite area of $\tilde C$ means
that one cannot treat the $D3'$-brane in two-dimensional terms. More
fundamentally, the dynamics of this problem in general involves
brane recombination, in which one of the $N$ $D3$-branes combines
with the $D3'$-brane to make a smooth (but deformed) brane. Locally,
if $C$ is defined by a complex equation $x=0$ and $\tilde C$ by
$y=0$, the recombination can be described by deforming an
intersection $xy=0$ to a smooth curve\foot{For small $\epsilon$, in the space $\C^2$
parametrized by $x$ and $y$, there is a supersymmetric disc, of area proportional to
$\epsilon$, whose boundary is on the curve $xy=\epsilon$.  One can also consider a multiple
covering of this disc.
 This gives a stringy model of the
instantons of small action that were discussed in section \instac.}
\eqn\deqn{xy=\epsilon.} The low energy effective physics cannot then
be described without taking into account the non-compactness of
$\tilde C$, and does not lead to the sort of surface operator
considered in this paper.  On the plus side of the ledger, the
noncompactness of $\tilde C$ means that whatever we get from this
construction depends on parameters $(\alpha,\beta,\gamma)$ that can
be defined in terms of the behavior at infinity on $\tilde C$.  For
example, $\alpha$ is the holonomy at infinity on $\tilde C$ of the
$U(1)$ gauge field, and $\beta+i\gamma$ is the parameter $\epsilon$
of eqn. \deqn, which can also be measured at infinity
(in terms of the way the brane is ``bent'').

If we take $\tilde C$ to be compact, we have the opposite problem.
Compactness of $\tilde C$ means that all fields on $\tilde C$ other
than zero modes can be eliminated in a low energy description. So we
get some sort of  description in terms of four-dimensional fields defined
on $M=\Sigma\times C$ and impurity fields supported on $\Sigma\times
p$.  In particular, the $U(1)$ gauge field on the $D3'$-brane can be
eliminated at low energies by taking a hyper-Kahler quotient. But
when $\tilde C$ is compact, all variables describing the physics of
the brane intersection and the fields on $\tilde C$ are dynamical
(rather than being ``frozen'' at infinity). So there is no apparent
way to introduce parameters corresponding to
$(\alpha,\beta,\gamma,\eta)$. It would be interesting, however, to know what kind of surface
operator one does get from the case of compact $\tilde C$.

\bigskip\noindent{\it An Alternative}

There is an alternative to this, explained in \coner.  This alternative enables us to
maintain the $S$-duality and to see the parameters $(\alpha,\beta,\gamma,\eta)$.  It is
not equivalent to the local model (it hardly can be, as the local model is not $S$-dual),
but it is an elegant way to use branes and string theory
to describe an example of the construction studied in this paper.

The alternative is merely to use the AdS/CFT duality
\ref\malda{J. Maldacena, ``The Large $N$ Limit Of
Superconformal Field Theories And Supergravity,'' Adv. Theor. Math. Phys. {\bf 2} (1998) 231-252,
hep-th/9711200.} to replace the $D3$-branes with
 a description involving the Type IIB geometry that these $D3$-branes
create.  The $D3'$-brane will remain as part of the Type IIB description, and this will
give the desired surface operator.

To simplify things, we will take $M=S^4$, in which case the simplest
relevant Type IIB spacetime is $AdS_5\times S^5$.  The simplest
choice of the $D3'$-brane is $AdS_3\times S^1$, where $AdS_3\times
S^1$ is embedded in $AdS_5\times S^5$ in the obvious way.  The
embedding of $S^1$ in $S^5$ is unstable topologically, but
energetically it is stable and in fact leads to a supersymmetric
membrane on $AdS_5$.  The conformal boundary of $AdS_3$ is
$D\cong S^2$, embedded in $ M\cong S^4$ in the obvious way (the $S^1$ factor   of $AdS_3\times S^1$
effectively shrinks to a point near the conformal boundary). $D$ is, of course, a two-dimensional
submanifold of $M$.  By quantizing in the presence of a $D3'$-brane  whose asymptotic behavior
is required to coincide with that of
 $AdS_3\times S^1$, one defines in the ${\cal N}=4$ super Yang-Mills theory on $M$
a supersymmetric surface operator
supported on $D$. This is precisely in parallel with the use
\nref\maldat{J. Maldacena, ``Wilson Loops In Large $N$ Field
Theories,'' Phys. Rev. Lett. {\bf 80}
(1998) 4859-4862, hep-th/9803002.}%
\nref\rey{S.-J. Rey and J.-T. Yee, ``Macroscopic Strings As Heavy Quarks
in Large $N$ Gauge Theory And Anti-de Sitter Supergravity,'' Eur. Phys. J. {\bf C22} (2001) 379-394,
hep-th/9803001.}%
\refs{\maldat,\rey} of strings in $AdS_5\times S^5$ to describe supersymmetric
Wilson and 't Hooft operators on $M$.  One can modify this to consider more general
$M$ or $D$, but this is not essential for what we will say.

The surface operator obtained this way
has the familiar parameters $(\alpha,\beta,\gamma,\eta)$.  To
see them, we consider the ${\cal N}=4$ super Yang-Mills theory, with gauge group $U(1)$,
supported on the $D3'$-brane.  The world-volume of that brane, in our simple example,
is $X=Y\times S^1$,
with $Y=AdS_3$.  And in general this gives the right asymptotic
behavior, which is all that one needs in describing the parameters
$(\alpha,\beta,\gamma,\eta)$. They arise as follows.  The parameter
$\alpha$ is the holonomy of the $U(1)$ gauge field around $S^1$, measured near the conformal
boundary of $Y$. The
part of the $U(1)$ gauge field on $X$ that is invariant under
rotations around $S^1$ is a $U(1)$ gauge field on $Y$ that can be
dualized to give an angle-valued scalar field.  The expectation
value of this scalar field near the boundary of $Y$ is $\eta$.    Finally, to find $\beta$
and $\gamma$, we use the ``Higgs field'' $\varphi$ of the gauge
multiplet on the $D3'$-brane.  In Poincar\'e coordinates, the metric
of $AdS_3$ can be written as ${1\over r^2}(dr^2+dx^2+dy^2)$, where
$0\leq r\leq \infty$, and the $(x,y)$-plane is the two-sphere $D$ with
a point at infinity removed. The metric of $X=Y\times S^1$ is thus ${1\over
r^2}(dr^2+dx^2+dy^2)+d\theta^2$ where $\theta$ is an angle. Upon
setting $z=re^{i\theta}$, we introduce $\beta$ and $\gamma$ by
requiring that $\varphi$ should be asymptotic at $r=0$ (that is, on
the conformal boundary of $Y$) to $(\beta+i\gamma)(dz/2z)$.  One can
verify that the underlying $S$-duality of the $U(1)$ gauge theory,
which we reviewed in section \duality, acts in the expected fashion
on these parameters.

\bigskip\noindent{\it Another Variant}

\nref\corr{D. Berenstein, R. Corrado, W. Fischler, and J. M. Maldacena, ``The Operator
Product Expansion For Wilson Loops And Surfaces In The Large $N$ Limit,'' Phys. Rev.
{\bf D59} (1999) 105023, hep-th/9809188.}%
\nref\graw{C. R. Graham and E. Witten, ``Conformal Invariance Of Submanifold Observables In
AdS-CFT Correspondence,'' Nucl. Phys. {\bf B546} (1999) 52-64, hep-th/9901021.}%
For another variant of this construction, we begin with the $(0,2)$ conformal field theory
in six dimensions \wittencom.  This theory on $S^6$
can be described \malda\ in terms of $M$-theory on $AdS_7\times S^4$ (or more precisely on
spacetimes with that asymptotic behavior) with $N$ units of flux on $S^4$.  The $(0,2)$ model
has surface operators \refs{\corr,\graw}, which
can be defined by considering $M2$-branes in $AdS_7\times S^4$
whose world-volume is asymptotic at infinity to a two-dimensional surface $D\subset S^6$.
One can replace $S^6$ by a more general six-manifold $Z$ by replacing $AdS_7\times S^4$
with spacetimes with the appropriate asymptotic behavior.

Now take the six-manifold on which the (0,2) model is formulated
to be $Z=M\times T^2$ with $M$ a four-manifold.  The (0,2) model formulated on $Z$
is equivalent at low energies to ${\cal N}=4$ super Yang-Mills theory on $M$, with gauge
group $U(N)$.  A surface operator on $Z$ whose support is a circle in $M$ times one of the two
factors in $T^2=S^1\times S^1$ gives a Wilson or 't Hooft operator on $M$.  (More generally,
any choice of a geodesic circle in $T^2$ gives a mixed Wilson-'t Hooft operator.) However,
a surface operator on $Z$ whose support is a surface $D\subset M$ times a point $r\in T^2$
gives a surface operator on $M$, with $\alpha$ and $\eta$ determined by the choice of $r$.
$S$-duality comes from the geometrical action of $SL(2,\Z)$ on $T^2$.
This construction is closely related to the previous one, since $M$-theory on $T^2$
is dual to Type IIB on a circle.

\appendix{A}{Review Of Duality}

In this appendix, we will give a brief review of the relation between a Lie group $G$
and its Langlands or GNO dual $^L\neg G$.  For general background, see chapter 3 of \ref\humphreys{J. E.
Humphreys, {\it Introduction To Lie Algebras And Representation Theory} (Springer-Verlag, New York,
1972).} or chapter 8 of \ref\simon{B. Simon, {\it Representations Of Finite And Compact Groups}
(American Mathematical Society, 1996).}.  For brevity, we describe only the case that $G$ is
simple, though the theory extends to any compact Lie group.

\def\eurme{\eurm E}
The root system $\Phi$ of  $G$ is a finite set of nonzero vectors in a fixed
Euclidean space $\eurme\cong \Bbb{R}^r$ (where $r$ will be the rank of $G$).
$\eurme$ will eventually be interpreted as $\frak t^\vee$, the
dual of the Lie algebra $\frak t$ of a maximal torus $\TT$ of $G$.  We consider $\eurme$ to be endowed
with a metric $(~,~)$ which to begin with we consider to be defined only up to multiplication by
a positive real scalar.  Later, we will choose a particular metric.

$\Phi$ is required to obey certain axioms.
The vectors in $\Phi$ generate a rank $r$ lattice $\Lambda_\rt$ (eventually interpreted as the root
lattice of $G$), which moreover has no decomposition as a direct sum of orthogonal sublattices.
Additionally,
for $\mu\in \Phi$,  we require that the
 multiples of $\mu$ that are contained in $\Phi$ are precisely
 the vectors $\pm \mu$.  This will ensure
that there is an $\frak{sl}(2)$ subalgebra of $\frak{g}$ with nonzero roots $\pm \mu$.  We call
this algebra $\frak{sl}(2)_\mu$.
We also ask that for any other root $\nu$, the expression
\eqn\noggo{\langle\nu,\mu\rangle ={2(\nu,\mu)\over (\mu,\mu)} }
is an integer.  Note that the symbol $\langle~,~\rangle$, which
 is independent of the normalization of $(~,~)$, is only linear in the first variable.
The integrality of $\langle\nu,\mu\rangle$ is interpreted as the condition
that each root $\nu$ has integer or half-integer weight with respect to $\frak{sl}(2)_\mu$.
Finally, we ask that  the set of roots should be
closed under the reflection with respect to $\mu$, which
is the operation
\eqn\poggo{\nu\to \nu-\langle\nu,\mu\rangle \mu.}
This operation is eventually interpreted as the Weyl transformation of the algebra $\frak{sl}(2)_\mu$.

For any such root system $\Phi$, one defines a dual root system $\Phi^\vee$ in the following way.
For $\mu\in \Phi$, we define
\eqn\zono{\mu^\vee={2\mu\over (\mu,\mu)}.}
$\Phi^\vee$ is defined to consist precisely of the vectors $\mu^\vee$ for $\mu\in\Phi$.
A short computation shows that $\Phi^\vee$ obeys the same axioms as $\Phi$.  It will be interpreted
as the root system of the dual group.

However, the definition of $\mu^\vee$ depends on the normalization of the metric $(~,~)$, and
therefore the vectors $\mu^\vee$ are not really naturally defined as vectors in $\eurme$.
They actually are more naturally understood as vectors in the dual space $\eurme^\vee$.  The reason
for this claim is that for $\mu,\nu\in \Phi$, the pairing
\eqn\bongo{\mu^\vee(\nu)=(\nu,\mu^\vee)={2(\nu,\mu)\over (\mu,\mu)}=
\langle \nu,\mu\rangle }
is independent of the normalization of the metric $(~,~)$.
Hence, while the original root system  $\Phi$ lies in $ \eurme\cong \frak t^\vee$, the dual root
system  $\Phi^\vee$ lies in $\eurme^\vee\cong \frak t$.

The Weyl   group $\cal W$ of $\Phi$ is the group generated by the reflections with respect to the roots
$\mu$,
namely the operations
$\nu\to\nu-\mu\langle\nu,\mu\rangle$.
Similarly, the Weyl group $\cal W^\vee $ of $\Phi^\vee$ is the group generated
by the reflections with respect to the coroots $\mu^\vee$, namely $\nu\to\nu-\mu^\vee\langle\nu,
\mu^\vee\rangle$.  However, as $\mu^\vee$ is a real multiple of $\mu$, we have
$\mu\langle\nu,\mu\rangle=\mu^\vee\langle\nu,\mu^\vee\rangle$, and the reflections with respect to
$\mu$ or $\mu^\vee$ coincide.  Hence the groups $\cal W$ and $\cal W^\vee$ likewise coincide.
This leads to the fact that $G$ and its dual group $^L\neg G$ have the same Weyl group.

Just as the vectors $\mu\in\Phi$ generate a rank $r$ lattice $\Lambda_\rt$, the vectors
$\mu^\vee\in\Phi^\vee$ generate a rank $r$ lattice $\Lambda_\cort$ (the coroot lattice).
We write $\Lambda_\cowt$ (the coweight lattice)
for the dual to $\Lambda_\rt$ and $\Lambda_\wt$ (the weight lattice)
for the dual of $\Lambda_\cort$.  Thus (momentarily
regarding all these lattices as embedded in $\eurme$, using the metric $(~,~)$), we have
$\nu\in\Lambda_\cowt$ if and only if $(\nu,\mu)\in\Z$ for all $\mu\in\Phi$, and
$\nu\in\Lambda_\wt$ if and only if $(\nu,\mu^\vee)\in\Z$ for all $\mu^\vee\in\Phi^\vee$.
Integrality of the pairing $(\nu,\mu^\vee)$ for $\nu\in \Lambda_\rt$, $\mu^\vee\in\Lambda_\cort$
means that we have inclusions
\eqn\forgo{\eqalign{\Lambda_\rt & \subset \Lambda_\wt \cr
                    \Lambda_\cort & \subset\Lambda_\cowt.\cr}}
These lattices all have the same rank $r$, so the inclusions are of finite index.

Now let $\Lambda_\char$ (the character lattice)
be any lattice that is intermediate between $\Lambda_\rt$ and $\Lambda_\wt$:
\eqn\orgo{\Lambda_\rt\subset \Lambda_\char\subset\Lambda_\wt.}
The dual of $\Lambda_\char$ is a lattice $\Lambda_\cochar$ (the cocharacter lattice) that
lies between the coroot and coweight lattices:
\eqn\corgo{\Lambda_\cort\subset \Lambda_\cochar\subset\Lambda_\cowt.}
Of course, the three lattices in \corgo\ are naturally regarded as embedded in $\eurme^\vee$;
their embedding in $\eurme$ depends on the choice of metric $(~,~)$.

The classification of compact simple Lie groups states that such groups correspond to character lattices
that can arise in a construction of this type.
For every choice of $\Lambda_\char$, there is a compact Lie group $G$ such that the highest
weight of a representation of $G$ is an element of $\Lambda_\char$.
In fact, $\Lambda_\char$ is the group of characters, that is of homomorphisms
of the maximal torus $\Bbb{T}$ of $G$ to $U(1)$.
Thus \eqn\horyt{\Lambda_\char={\rm Hom}(\Bbb{T},U(1)),} and dually,
\eqn\oryt{\Lambda_\cochar={\rm Hom}({}^L\neg\TT,U(1))={\rm Hom}(U(1),\Bbb{T}).}
If $\Lambda_\char=\Lambda_\wt$, then $G$ is simply-connected.  If $\Lambda_\char=\Lambda_\rt$,
then $G$ is of adjoint type.

More generally,
let ${\cal Z}(G)$ denote the center of $G$.  For any locally compact abelian group $A$, we write
$A^\vee$ for ${\rm Hom}(A,U(1))$; according to Pontryagin duality, if $B=A^\vee$, then $A=B^\vee$.
In an irreducible representation $R$ of $G$, ${\cal Z}(G)$ acts centrally, or in other words
via a homomorphism to $U(1)$. The highest weight of $R$ is an element  $w\in\Lambda_\char$,
and ${\cal Z}(G)$ acts trivially if $w\in\Lambda_\rt$.  The quotient $\Lambda_\char/\Lambda_\rt$
is  ${\rm Hom}({\cal Z}(G),U(1))$:
\eqn\forgox{{\cal Z}(G)^\vee = \Lambda_\char/\Lambda_\rt.}
Using the duality between the inclusions in \orgo\ and \corgo, this is equivalent to
\eqn\eforgox{{\cal Z}(G)=\Lambda_\cowt/\Lambda_\cochar.}
Similarly, an element of the fundamental group $\pi_1(G)$, if lifted to the universal cover $\bar G$ of
$G$, determines a path from the identity to a central element $\xi$ of $\bar G$.  In a representation
$R$ of $\bar G$, $\xi$ acts by multiplication by an element of $U(1)$.  By this construction,
 $R$ determines
an element of $\pi_1(G)^\vee$, and this element vanishes if $R$ is actually a representation of $G$,
that is, if its highest weight is in $\Lambda_\char$.  This leads to a relation
\eqn\bforgox{\pi_1(G)^\vee=\Lambda_\wt/\Lambda_\char.}
Again, the duality between \orgo\ and \corgo\ leads to an alternative version
\eqn\cforgox{\pi_1(G)=\Lambda_\cochar/\Lambda_\cort.}

 The Langlands or GNO dual group $^L\neg G$ is related to the
lattices $\Lambda_\cort,$ $\Lambda_\cochar$, and $\Lambda_\cowt$
exactly as $G$ is related to $\Lambda_\rt$, $\Lambda_\char$, and
$\Lambda_\wt$.    All of the above statements have obvious duals.
For example, $^L\neg G$ is
 simply-connected if $\Lambda_\cochar=\Lambda_\cowt$, and is of adjoint type if $\Lambda_\cochar
=\Lambda_\cort$.  And more generally,
\eqn\forgex{\eqalign{
{\cal Z}({}^L\neg G)^\vee&=\Lambda_\cochar/\Lambda_\cort\cr
\pi_1({}^L\neg G)^\vee& = \Lambda_\cowt/\Lambda_\cochar.\cr}}
Comparing \forgex\ to  \eforgox\ and \cforgox, we learn that
\eqn\yorgox{\eqalign{\pi_1({}^LG)&={\cal Z}(G)^\vee\cr
                     {\cal Z}({}^LG)&=\pi_1(G)^\vee.\cr}}

The whole construction is completely symmetric under exchange of $\eurme$
with $\eurme^\vee$, the lattices in \orgo\ with the dual lattices in \corgo, and the group
$G$ with the dual group $^L\neg G$.  In the theory of Lie algebras, the roots of $G$ take values in
$\frak t^\vee$ (the dual of the Lie algebra $\frak t$
of the maximal torus of $G$) and the coroots in $\frak t$,
so we identify $\eurme$ with $\frak t^\vee$ and $\eurme^\vee$ with $\frak t$.
The results \eforgox\ and \cforgox\ have more direct explanations using the fact that
the lattices $\Lambda_\cort,$ $\Lambda_\cochar$, and $\Lambda_\cowt$ are all naturally embedded in
$\frak t$.

 \bigskip\noindent{\it Choice of Metric}

At this point, it is useful to make a convenient choice for the metric $(~,~)$ on $\eurme$.

The group $G$ is said to be simply-laced if all roots $\mu\in\Phi$ have the same length squared.
This is so precisely if $G$ is of type $ {A}$, $ {D}$, or $ {E}$.
In this case, it is convenient to pick a metric such that the roots actually have length squared equal
to 2, that is $(\mu,\mu)=2$ for all $\mu$.  Looking back to the definition
of $\mu^\vee$ in eqn. \zono, we see that this condition ensures that $\mu^\vee=\mu$ for
all $\mu$.  This causes the construction summarized above to simplify.  The distinction between
$\eurme$ and $\eurme^\vee$ can be omitted without any loss of symmetry,
 and we then have $\Lambda_\rt=\Lambda_\cort$, $\Lambda_\wt=\Lambda_\cowt$.  In this situation,
the groups $G$ and $^L\neg G$ have the same Lie algebra, since their root lattices
are the same, but they are not necessarily isomorphic, since
the lattices $\Lambda_\char $ and $\Lambda_\cochar$ may differ.

Now let us consider the case that $G$ is not simply-laced.  In this case, it is shown via
the classification
of simple Lie algebras that there are precisely two values for the length squared of a root.
We denote the
ratio of these two values
as $n_{\frak g}$; it equals 2 for groups of type $ {B},$ $ {C}$, or $ {F}_4$,
and 3 for $ {G}_2$.  From $\mu^\vee=2\mu/(\mu,\mu)$, we find
\eqn\pinfo{(\mu^\vee,\mu^\vee)={4\over (\mu,\mu)}.}
So the function $(\mu^\vee,\mu^\vee)$ again takes two values for $\mu^\vee\in\Phi^\vee$,
and these values have
the same ratio $n_{\frak g}$.  Moreover, from \pinfo, we see that if $\mu$ is a short root,
then $\mu^\vee$ is a long coroot, and if $\mu$ is a long root, then $\mu^\vee$ is a short
coroot.

\ifig\vilunko{\bigskip
Shown here are the roots of the Lie group $ {G}_2$ (left) and the coroots (right).
  The map $\mu\to\mu^*$ (defined using the metric in which short roots have length squared 2) maps
a vector on the left to the ``same'' vector on the right.
A short root maps to a long coroot, and a long root maps to 3 times a short coroot.
To get an isomorphism between the root diagram and the coroot diagram,
we can compose the map $\mu\to\mu^*$ with rotation  by an angle $\pm \pi/6$
and multiplication by $1/\sqrt 3$.}
{\epsfxsize=3.5in\epsfbox{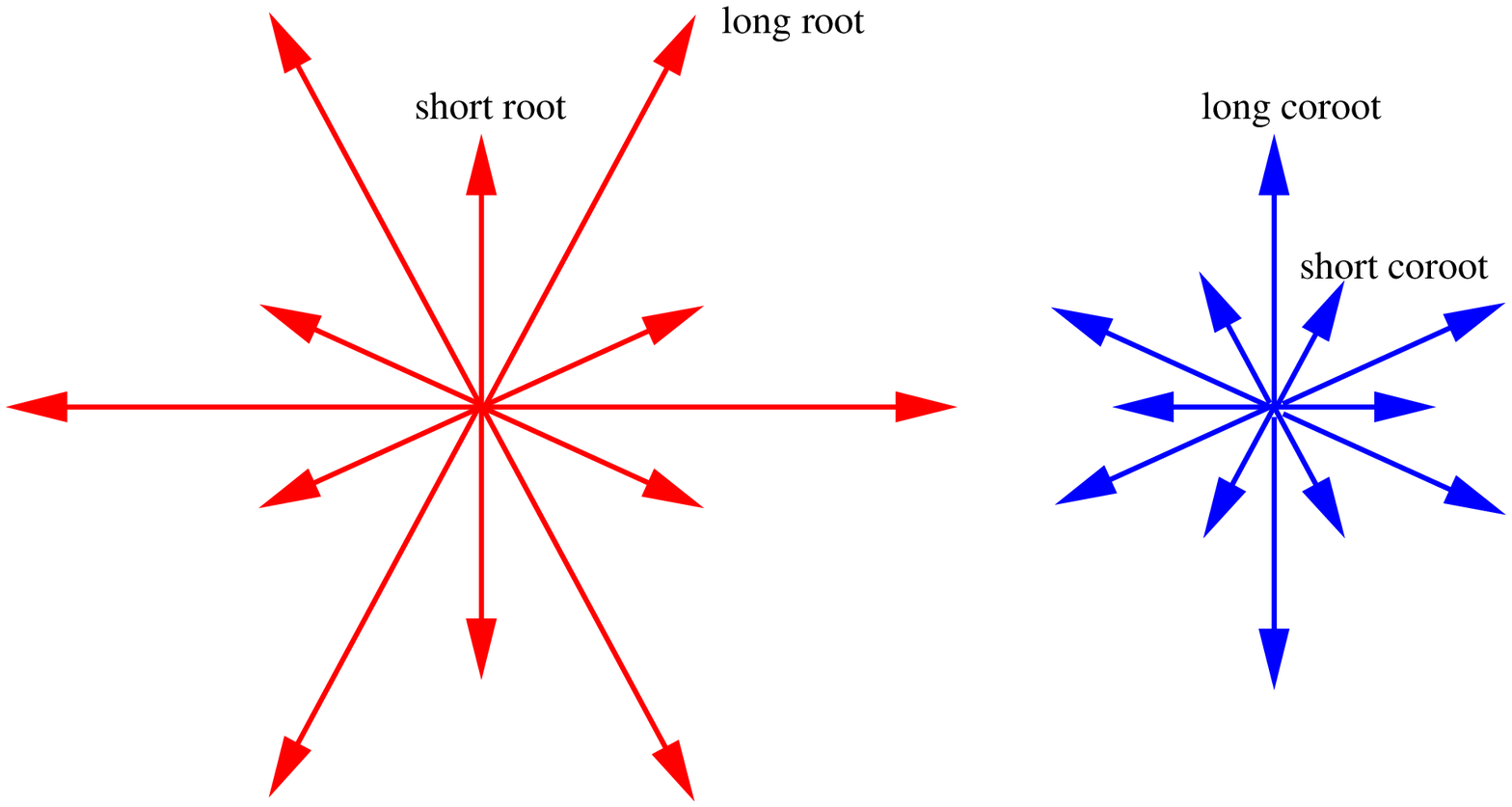}}

We obviously cannot now normalize the metric $(~,~)$ on $\eurme$ so that all roots have length
squared 2.  However, it is convenient to normalize the metric so that short roots have length
squared 2, and therefore long roots have length squared $2n_{\frak g}$.
When this choice is made, we have
\eqn\winfo{\mu^\vee  =\cases{\mu & if $\mu$ is a short root;\cr
                                 {\mu/ n_{\frak g}}& if $\mu$ is a long root.\cr}}
Reciprocally,
\eqn\winfo{\mu  =\cases{\mu^\vee & if $\mu$ is a short root;\cr
                                 n_{\frak g}{\mu^\vee}& if $\mu$ is a long root.\cr}}

Once we pick the metric, we get a natural identification of $\eurme$ with $\eurme^\vee$.  This
gives a map from $\eurme$ to $\eurme^\vee$, mapping $\nu\in\eurme$ to the linear form on
$\eurme$ defined by $\nu(\mu)=(\nu,\mu)$.  We denote this map as $\nu\to\nu^*$.

From \winfo, we can be more precise.
If $\mu\in\Phi$ is a short root, then $\mu^*$ (which is the same as $\mu$, but interpreted
as an element of $\eurme^\vee$) is equal to a long coroot (namely $\mu^\vee$).  But if $\mu$
is a long root, then $\mu^*$ is $n_{\frak g}$ times the short coroot $\mu^\vee$.  This is
illustrated in \vilunko\ for the case of $ {G}_2$.
In particular, $\mu^*$
is not always a coroot but is always an element of the coroot lattice, so the linear transformation
$\mu\to\mu^*$ maps $\Lambda_\rt$ into $\Lambda_\cort$.

Of course, there is also a dual of this construction.  We can pick on $\eurme^\vee$ a metric
in which a short coroot  has length squared 2.  This leads to a natural map $\eurme^\vee\to \eurme$,
which we denote $\mu^\vee\to (\mu^\vee)^*$.  By applying the same reasoning, we see
that this operation maps a short coroot to a long root, or a long coroot to $n_{\frak g}$ times a short
root.  In particular, it maps $\Lambda_\cort$ into $\Lambda_\rt$.

The composition of the two maps is multiplication by $n_{\frak g}, $ that is
$(\mu^*)^*=n_{\frak g}\mu$ for all $\mu$.  Indeed, a short root maps to a long coroot
and thence to $n_{\frak g}$ times the original
 short root, while a long root maps to $n_{\frak g}$ times
a short root, and thence to $n_{\frak g}$ times the original long root.

\bigskip\noindent{\it Dynkin Diagrams}

A collection of $r$ roots $\mu_i\in\Phi$ is called a set of simple positive roots if every root
$\mu$ has an expansion $\mu=\sum_i a_i\mu_i$, with coefficients $a_i$ that are all non-negative or
all non-positive.   Such a set always exists and is unique up to a Weyl transformation.

If $\mu_i$ are a set of simple positive roots of $\Phi$, then the dual roots $\mu_i^\vee\in\Phi^\vee$
are a set
of simple positive roots of $\Phi^\vee$.  Indeed, the $\mu_i^\vee$ are positive multiples of the $\mu_i$,
and any dual root
$\mu^\vee$ is similarly related to $\mu$.  So in an expansion $\mu^\vee=\sum_ib_i \mu_i^\vee$,
the coefficients $b_i$  have the same signs as the coefficients $a_i$ in $\mu=\sum_ia_i\mu_i$.

The Dynkin diagram of $G$ has a node for every simple positive root.  The nodes are connected in
a way that encodes the angles among the simple positive roots.  When $G$ is not simply-laced,
the diagram is also commonly
labeled with an arrow that points from long to short roots.

\ifig\zulco{\bigskip
Dynkin diagrams of the non-simply-laced Lie groups.  Duality has the effect of reversing
the arrow that points from long to short roots.  Thus, the groups $B_n$ and $C_n$ are exchanged
by duality.  However, $F_4$ and $G_2$ are self-dual, since for those groups,
a reversal of the arrow is equivalent to exchanging the two ends of the  diagram.
}
{\epsfxsize=2.3in\epsfbox{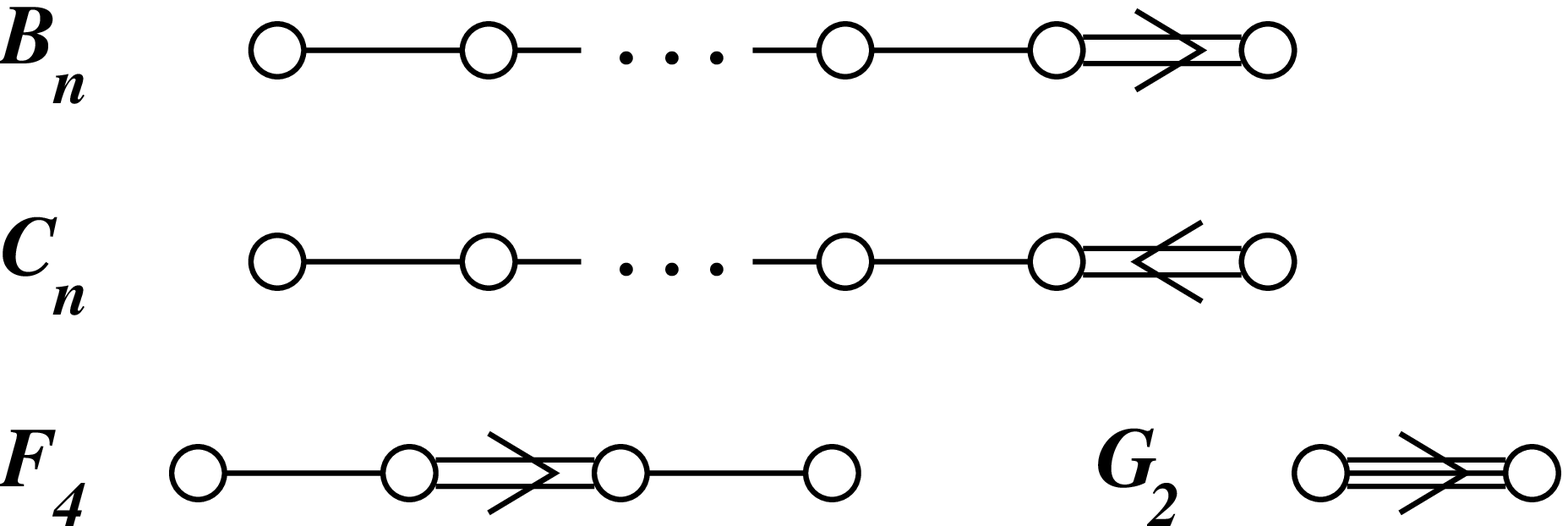}}

The angles between the simple positive roots $\mu_i$ of $\Phi$ are the same as the angles between the
simple positive roots $\mu_i^\vee$ of $\Phi^\vee$, since the $\mu_i^\vee$ are positive multiples of the
$\mu_i$.  However, the transformation $\mu\to\mu^\vee$ maps short roots to long
ones, and vice-versa, as we have seen above.  So the root systems
$\Phi$ and $\Phi^\vee$ have Dynkin diagrams that look just the same, except that, in the
non-simply-laced case, the arrows point
in opposite directions.  This is illustrated in \zulco.

A Levi subgroup $\Bbb{L}$ of $G$ has a Dynkin diagram obtained by simply removing
some nodes from the Dynkin diagram of $G$.  $^L\neg G$ has a  corresponding Levi subgroup $^L\Bbb{L}$
whose Dynkin diagram consists of the ``same'' nodes.  $\Bbb{L}$ and $^L\Bbb{L}$ have
the same Weyl group for the same reason that $G$ and $^L\neg G$ do.

\bigskip\noindent{\it Comparison To Gauge Theory}

For simply-laced $G$, the metric in which
the roots have length squared 2 is invariant under duality,
since in this metric the roots and coroots coincide.

When $G$ is not simply-laced, if we identify $\eurme$ with its dual using a metric in which the roots
have length squared $2$ or $2n_{\frak g}$, then the coroots have length squared $2$ or $2/n_{\frak g}$,
so this recipe is not invariant under duality.  The factor of $n_{\frak g}$ in the formula
$(\mu^*)^*=n_{\frak g}\mu$ reflects this discrepancy, as does the factor of $n_{\frak g}$ in
the $S$-duality transformation $\tau\to -1/n_{\frak g}\tau$.

Of course, we could restore duality by simply dividing the metric by $\sqrt{n_{\frak g}}$.  However,
this would lead to awkwardness in other formulas.  Indeed,
the metric on $\frak t\cong \eurme^\vee$ such that a short coroot has length squared 2
is widely used in gauge theory, for example in the present paper beginning
in sections \duality\ and \thetangle, because
it leads to  definitions of the instanton number \migox\ and theta-angle \migo\ that are uniformly
valid for all $G$.

In addition, normalizing the metrics as we have done has the virtue that it leads to  maps
$\mu\to\mu^*$ and $\mu^\vee\to (\mu^\vee)^*$ that, although not  isomorphisms of the
relevant lattices, do
 map $\Lambda_\rt$ into $\Lambda_\cort$,
and vice-versa.  This is lost if we divide by $\sqrt{n_{\frak g}}$.

\bigskip\noindent{\it Relation Of $G$ and $^L\neg G$}

If $G$ is simply-laced, then $G$ and $^L\neg G$ have the same Lie algebra, and this
fact is manifest in the above construction, in the sense that the map $\mu\to\mu^*$ is an
isomorphism from $\Phi$ to $\Phi^\vee$.

If $G$ is not simply-laced, the story is more complicated.  For $G$ of type $ {B}$ or $ {C}$,
the Lie algebras of $G$ and $^L\neg G$ are distinct, as is clear from the diagrams in \zulco.
For $G$ of type $ {G}_2$ or $ {F}_4$, the groups
$G$ and $^L\neg G$ are actually isomorphic.
This is obvious from the Dynkin diagrams; replacing $\Phi$ with $\Phi^\vee$ has the effect
of reversing the arrow that points from long to short roots, but for $G_2$ and $F_4$, this is
equivalent to exchanging the two ends of the diagram, as one can see in \zulco.

For $G_2$ and $F_4$, the isomorphism between the group and the dual group
 does not simply come from  the maps $\mu\to\mu^*$ and $\mu^\vee\to
(\mu^\vee)^*$, which are not isomorphisms between $\Phi$ and $\Phi^\vee$.
To get such an isomorphism, one can compose the map $\mu\to\mu^*$ with a linear
transformation ${\frak R}:\eurme^\vee\to\eurme$.
Modulo a Weyl transformation, ${\frak R}$ is determined by the way it must exchange the two
ends of the Dynkin diagram.
For $ {G}_2$, ${\frak R}$ can be chosen to be the composition of a rotation by an angle
$\pm \pi/6$ and multiplication by $1/\sqrt 3$, as one can see in \vilunko.  For $ {F}_4$, it
is the composition of a rotation and multiplication by $1/\sqrt 2$.

In  the physics literature \refs{\dorey,\kapsei},
$S$-duality is often defined so that, in a vacuum in which $G$ is
spontaneously broken to an abelian subgroup, the action of ${\frak R}$ is incorporated as part of
the definition of $S$.      At the expense of treating $G_2$ and $F_4$ as exceptional cases,
this makes the self-duality of those groups manifest.
We instead prefer, as for example in \donpan, to treat all groups uniformly, omitting the factor of
${\frak R}$. In this way, formulas such as \noobus\ can be written uniformly for all groups.

\appendix{B}{Index Of Notation}

\halign to 3.5in{\tabskip=3em plus2em
minus.5em#\hfill&#\hfill\tabskip=0pt\cr
\multispan2{\hskip 1cm {\it I. Gauge Theory}\hfill}\cr
\noalign{\medskip}
$M$&four-manifold\cr
$W$&three-manifold\cr
$C$&Riemann surface on which we do geometric Langlands\cr
$\Sigma$&Riemann surface to which we compactify\cr
$G$& compact gauge group, usually simple\cr
$\bar G$& simply-connected cover of $G$\cr
$E$&a $G$-bundle\cr
$A$& the gauge field; a connection on $E$\cr
$d_A$& the gauge-covariant exterior derivative $d+[A,\,\cdot\,]$\cr
$\phi$&${\rm ad}(E)$-valued one-form; the Higgs field\cr
$\varphi$&the $(1,0)$ part of $\phi$\cr
${\cal A}$&the complex-valued connection $A+i\phi$\cr
$V$ & monodromy of ${\cal A}$\cr
$U$ & semi-simple element in closure of orbit of $V$\cr
$\tau$& the gauge coupling parameter $\theta/2\pi+4\pi i/e^2$\cr
$\eurm N$&instanton number\cr
$\eurm m$&magnetic flux; characteristic class of a $\Bbb{T}$-bundle\cr
$\xi$&characteristic class that obstructs lifting a $G$-bundle $E$ to a $\bar G$-bundle\cr
$D$&support of a surface operator \cr
$L$& support of a line operator\cr
$(\alpha,\beta,\gamma,\eta)$ &parameters of a surface operator\cr
${\bf e}_0$ & discrete electric flux\cr
${\bf m}_0$ & discrete magnetic flux\cr
$\MH$ & moduli space of Higgs bundles\cr
$\M$ & moduli space of  $G$-bundles\cr
$\MH(\alpha,\beta,\gamma)$ & moduli space of ramified Higgs bundles\cr
$\M(\alpha)$ & moduli space of  parabolic bundles \cr
$\EUBB$ & base of the Hitchin fibration\cr
$\CMF$ & fiber of the Hitchin fibration\cr
${\cal B}$ & a brane\cr
${\cal L}$ & Chan-Paton line bundle of a brane\cr
$\EUG$ & the group of gauge transformations\cr
$\star$ & Hodge star operator \cr
$\Gamma$ & duality group \cr
\noalign{\medskip}
\multispan2{\hskip 1cm{\it II. Hyper-Kahler Structure}\hfill}\cr
\noalign{\medskip}
$X$&generic hyper-Kahler manifold\cr
$I,J,K$&complex structures on $X$\cr
$\omega_I,\omega_J,\omega_K$&Kahler structures on $X$\cr
$\Omega_I,\Omega_J,\Omega_K$&holomorphic two-forms on $X$\cr
$\vec\mu=(\mu_I,\mu_J,\mu_K)$ & hyper-Kahler moment map\cr
\cr
\noalign{\medskip}
\multispan2{\hskip 1cm{\it III. Group Theory}\hfill}\cr
\noalign{\medskip}
$G$& compact Lie group, usually simple\cr
$\bar G$ & universal cover of $G$\cr
$G_{\rm ad}$ & adjoint form of $G$\cr
$^L\neg G$ & dual group\cr
$\Bbb{T}$ & maximal torus of $G$\cr
$\Bbb{L}$ & Levi subgroup of $G$\cr
$G_\C,\,\Bbb{T}_\C,$ etc. & complexifications\cr
$\EUB$ & Borel subgroup of $G_\C$\cr
$\EUP$ & parabolic subgroup of $G_\C$\cr
$\EUN$ & unipotent radical of $\EUP$\cr
$\frak g$ & Lie algebra of $G$\cr
$\frak t$ & Lie algebra of $ \Bbb{T}$\cr
$\frak t^\vee$ & dual of $\frak t$\cr
$\frak b$ & Lie algebra of $\EUB$\cr
$\frak p$ & Lie algebra of $\EUP$\cr
$\frak n$ & Lie algebra of $\EUN$\cr
${\cal W}$ & Weyl group \cr
${\cal W}_{\rm aff}$ & affine Weyl group\cr
$\eurm D$ & affine Weyl chamber\cr
$B_{\rm aff}$ & affine braid group\cr
${\cal Z}(G)$ & center of $G$\cr
$\pi_1(G)$ & fundamental group of $G$\cr
$\frak C$ & conjugacy class in $G_\C$\cr
$\frak c$ & conjugacy class in $\frak g_\C$\cr
$\frak C_{\Bbb{L}}$ & Richardson conjugacy class associated to Levi subgroup $\Bbb{L}$\cr
 \noalign{\medskip}
\multispan2{\hskip 1cm{\it IV. Lattices}\hfill}\cr
 \noalign{\medskip}
 $\Lambda_\rt$ & root lattice\cr
 $\Lambda_\cort$ & coroot lattice\cr
 $\Lambda_\wt$ & weight lattice\cr
 $\Lambda_\cowt$ & coweight lattice\cr
 $\Lambda_\char$ & character lattice\cr
 $\Lambda_\cochar$ & cocharacter lattice\cr
 }
 \listrefs\end